%% file: thesis.tex
\documentclass{mydiss}
\pdfoutput=1 
\titlepagefile{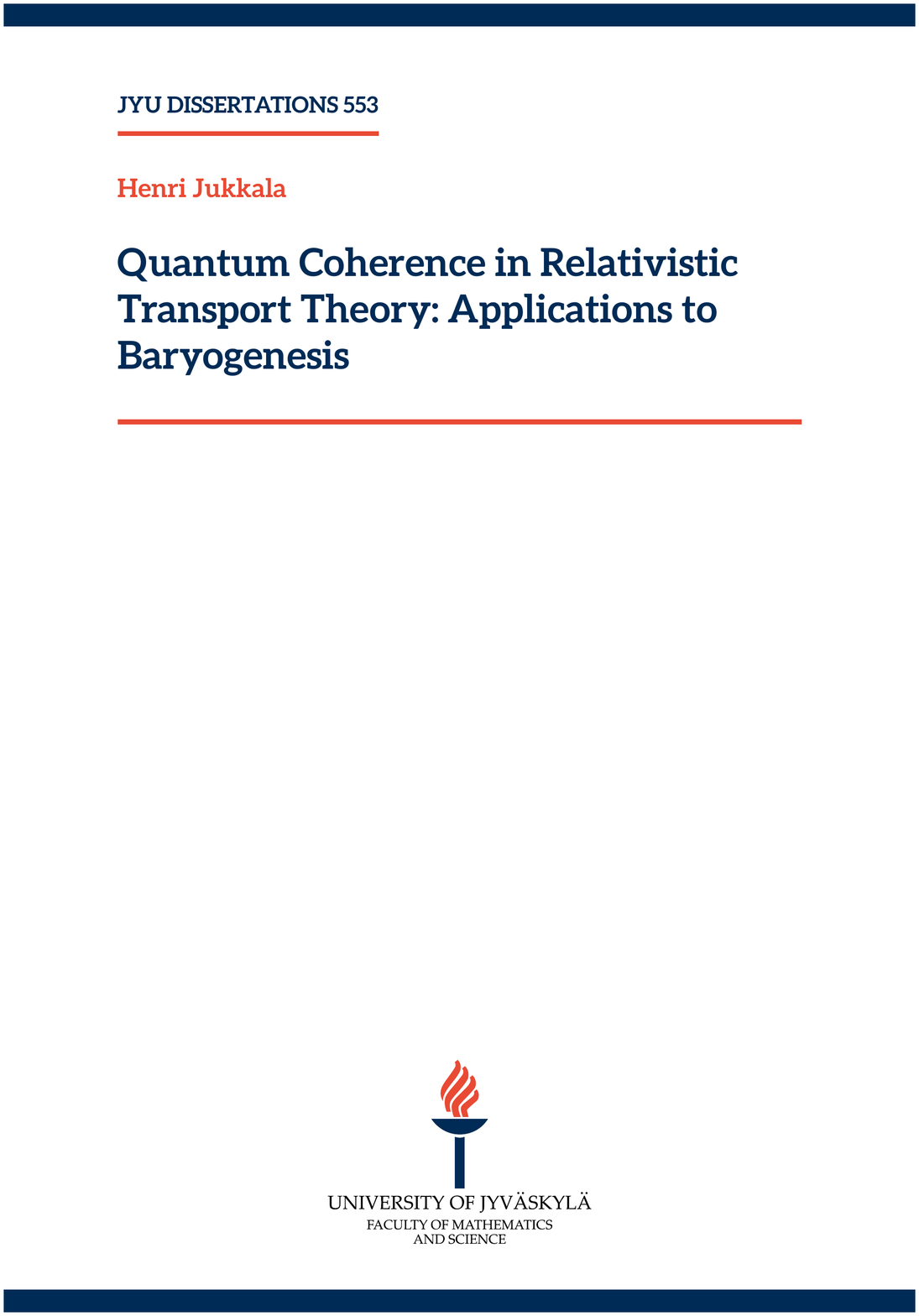}
\title{Quantum Coherence in Relativistic\\ Transport Theory: Applications to\\ Baryogenesis}%
      {Quantum coherence in relativistic transport theory: applications to baryogenesis}
\author{Henri}{Jukkala}
\defensedate{September 23, 2022}
\year{2022}
\series{JYU Dissertations}
\serialnumber{553}
\totpages{103}
\issn{2489-9003}
\isbn{978-951-39-9190-6 (PDF)}
\subject{Theoretical particle physics and cosmology}
\hyperaddtitleinfo 

\input{macros.tex}

\thesisarticles{Jukkala:2019slc,Jukkala:2021sku} 
\extracontribution{henri_jukkala_2021_5025929} 

\begin{document}

\frontmatter

\maketitle
\abstract{\input{frontmatter/abstract.tex}}
\finnishabstract{\input{frontmatter/abstractfin.tex}}
\listofpeople{\input{frontmatter/people.tex}}
\preface{\input{frontmatter/preface.tex}}
\listofarticles{\input{frontmatter/contribution.tex}}
\tableofcontents

\mainmatter

\input{chapters/ch1_introduction.tex}
\input{chapters/ch2_cosmology.tex}
\input{chapters/ch3_nonequilibrium.tex}
\input{chapters/ch4_coherent_QKEs.tex}
\input{chapters/ch5_applications.tex}
\input{chapters/ch6_conclusions.tex}


\backmatter

\references

\end{document}

%% file: macros.tex
\newcommand{\widebar}[1]{\mkern 1.3mu\overline{\mkern-1.3mu#1\mkern-1.3mu}\mkern 1.3mu}
\renewcommand{\bar}{\widebar}

\newcommand{\eqdef}{\equiv}

\newcommand{\im}{{\mathrm{i}}} 
\newcommand{\e}{\mathop{}\!\mathrm{e}} 

\newcommand{\evec}[1]{{\bm{#1}}} 
\newcommand{\Uevec}[1]{{\hat{\bm{#1}}}} 

\newcommand{\dd}{\mathop{}\!\mathrm{d}} 

\newcommand{\idmat}{\bm{1}} 
\newcommand{\transp}{{\mathsf{T}}} 

\newcommand{\sfrac}[2]{{\textstyle{\frac{#1}{#2}}}} 

\newcommand{\convol}{\mathbin{\ast}} 
\newcommand{\nconvol}{\mathbin{\!\convol\!}} 

\newcommand{\lt}{{\scriptscriptstyle <}} 
\newcommand{\gt}{{\scriptscriptstyle >}} 

\newcommand{\tin}{{t_{\mathrm{in}}}} 
\newcommand{\tf}{{t_{\mathrm{f}}}} 
\newcommand{\cw}{c_{\mathrm{w}}} 

\newcommand{\eg}{e.g.}
\newcommand{\ie}{i.e.}
\newcommand{\cf}{cf.}
\newcommand{\etc}{etc.}

\DeclareMathOperator{\sgn}{sgn} 
\DeclareMathOperator{\tr}{tr} 
\DeclareMathOperator{\Tr}{Tr} 
\DeclareMathOperator{\diag}{diag} 
\DeclareMathOperator{\Det}{Det} 

\renewcommand{\Re}{\operatorname{Re}} 
\renewcommand{\Im}{\operatorname{Im}} 
\DeclareMathOperator{\Arg}{Arg} 

\newcommand{\reals}{\mathbb{R}} 
\newcommand{\naturals}{\mathbb{N}} 

\newcommand{\SU}{\operatorname{SU}} 
\newcommand{\U}{\operatorname{U}} 

\DeclarePairedDelimiter{\abs}{\lvert}{\rvert} 
\DeclarePairedDelimiter{\comm}{[}{]} 
\DeclarePairedDelimiter{\anticomm}{\{}{\}} 
\DeclarePairedDelimiter{\expectval}{\langle}{\rangle} 

\DeclareOldFontCommand{\rm}{\normalfont\rmfamily}{\mathrm}
\DeclareOldFontCommand{\sf}{\normalfont\sffamily}{\mathsf}
\DeclareOldFontCommand{\tt}{\normalfont\ttfamily}{\mathtt}

\let\dummyepsilon\epsilon 
\let\epsilon\varepsilon
\let\varepsilon\dummyepsilon

\makeatletter
\DeclareRobustCommand{\cev}[1]{%
  {\mathpalette\do@cev{#1}}%
}
\newcommand{\do@cev}[2]{%
  \vbox{\offinterlineskip
    \sbox\z@{$\m@th#1 x$}%
    \ialign{##\cr
      \hidewidth\reflectbox{$\m@th#1\vec{}\mkern4mu$}\hidewidth\cr
      \noalign{\kern-\ht\z@}
      $\m@th#1#2$\cr
    }%
  }%
}
\makeatother


%% file: frontmatter/abstract.tex
We derive field-theoretic local quantum transport equations which can describe quantum coherence.
Our methods are based on Kadanoff--Baym equations derived in the Schwinger--Keldysh closed time path formalism of non-equilibrium quantum field theory.
We focus on spatially homogeneous and isotropic systems and mixing fermions with a time-dependent mass and a weak coupling to a thermal plasma.

We introduce a new local approximation (LA) method and use it to derive quantum kinetic equations which can describe coherence and also include effects of the spectral width. The method is based on a local ansatz of the collision term.
We also improve the earlier coherent quasiparticle approximation (cQPA) by giving a straightforward derivation of the spectral ansatz, a new way of organizing the gradient expansion, and a transparent way to derive the coherence-gradient resummed collision term.
In both methods the transport equations can describe flavor coherence and particle--antiparticle coherence, and the related oscillations, of the mixing fermions.

In addition to formulating the local equations we apply them to baryogenesis in the early universe. More specifically, we study the details of CP-asymmetry generation in resonant leptogenesis and the evolution of the axial vector current in electroweak baryogenesis (in a time-dependent analogue).
We solve the equations numerically, and perform extensive analysis and compare the results to semiclassical (Boltzmann) equations.
The results cover known semiclassical effects.
We find that dynamical treatment of local quantum coherence is necessary for an accurate description of CP-asymmetry generation. When these details are known they can then be partially incorporated into simpler (\eg~semiclassical) approaches. However, coherent quantum kinetic equations are needed for accurate results across different scenarios or wide parameter ranges.

%% file: frontmatter/abstractfin.tex
Tässä väitöskirjassa johdetaan kenttäteoreettiset lokaalit kvanttikuljetusyhtälöt, joilla voidaan kuvailla kvanttikoherenssia.
Kehittämämme menetelmät pohjautuvat Kadanoffin--Baymin yhtälöihin, jotka on johdettu käyttäen suljettua Schwinger--Keldysh-aikapolkua epätasapainon kvanttikenttäteoriassa.
Keskitymme spatiaalisesti homogeeniseen ja isotrooppiseen tapaukseen ja sekoittuviin fermioneihin, joilla on ajasta riippuva massa ja heikko kytkentä termiseen hiukkasplasmaan.

Muotoilemme uuden lokaalin approksimaatiomenetelmän ja johdamme sen avulla kvantti-kineettiset yhtälöt, joilla voidaan kuvailla koherenssia sekä ottaa huomioon spektraalisen leveyden vaikutus. Menetelmä perustuu vuorovaikutustermien lokaaliin yritteeseen.
Parannamme myös aiemmin kehitettyä koherenttia kvasihiukkasapproksimaatiota antamalla suoraviivaisen johdon sen spektraaliyritteelle, uuden muotoilun gradienttikehitelmälle ja selkeän tavan johtaa gradienttiresummatut vuorovaikutustermit.
Molempien menetelmien kuljetusyhtälöillä voidaan kuvailla sekoittuvien fermionien makutilojen välistä koherenssia ja hiukkas--antihiukkas-koherenssia sekä näihin liittyviä oskillaatioita.

Lokaalien yhtälöiden johtamisen lisäksi sovellamme niitä varhaisen maailmankaikkeuden baryogeneesiin. Tutkimme erityisesti CP-epäsymmetrian synnyn yksityiskohtia resonantissa leptogeneesissä sekä aksiaalivektorivirran kehitystä sähköheikossa baryogeneesissä (analogisessa ajasta riippuvassa tapauksessa).
Ratkaisemme yhtälöt numeerisesti, analysoimme tuloksia kattavasti ja vertaamme niitä semiklassisiin (Boltzmannin) yhtälöihin.
Tulokset kattavat tunnetut semiklassiset ilmiöt. Lokaalin kvanttikoherenssin dynaaminen tarkastelu on tarpeellista, kun tavoitteena on CP-epäsymmetrian kehittymisen tarkka kuvailu. Kun tämän yksityiskohdat tunnetaan, ne voidaan osittain sisällyttää yksinkertaisempiin (esim. semiklassisiin) menettelytapoihin. Koherentteja kvantti-kineettisiä yhtälöitä kuitenkin tarvitaan tarkkoihin tuloksiin eri skenaarioissa ja laajasti vaihtelevilla parametreilla.

%% file: frontmatter/people.tex
\item[Author]
Henri Jukkala \\
Department of Physics\\
University of Jyväskylä\\
Jyväskylä, Finland

\item[Supervisor]
Prof. Kimmo Kainulainen\\
Department of Physics\\
University of Jyväskylä\\
Jyväskylä, Finland

\item[Reviewers]
Prof. Mikko Laine\\
Institute for Theoretical Physics\\
University of Bern\\
Bern, Switzerland

Prof. Aleksi Vuorinen\\
Department of Physics\\
University of Helsinki\\
Helsinki, Finland

\item[Opponent]
Priv.-Doz. Dr. rer. nat. habil. Mathias Garny\\
Department of Physics\\
Technical University of Munich\\
Munich, Germany

%% file: frontmatter/preface.tex
The research leading to this thesis was carried out at the Department of Physics in the University of Jyväskylä between the years 2012 and 2022.
I gratefully acknowledge the financial support from the Väisälä Fund of the Finnish Academy of Science and Letters, from the Academy of Finland (projects 278722 and 318319), from the University of Jyväskylä, and from the Helsinki Institute of Physics.

There are several people who were integral to this research.
First of all, I would like to thank my supervisor Prof. Kimmo Kainulainen for excellent and enduring guidance during all these years. You were always very encouraging. For a long time, during my studies earlier, I could not decide between majoring in physics or mathematics. Under your guidance I finally found my own path between them. It was a priviledge to share a part of your enthusiasm and deep understanding of physics.
I would like to warmly thank also Mr. Pyry Rahkila and Mr. Olli Koskivaara for collaboration. I am especially grateful to Pyry for the countless discussions which were indispensable to the research process and also to my understanding of the subject.
I would also like to thank Dr. Matti Herranen for early collaboration in one of the research articles.
I am grateful to Prof. Mikko Laine and Prof. Aleksi Vuorinen for carefully reviewing the manuscript, and to Dr. Mathias Garny for accepting to be my opponent.

I wish to thank my previous teachers Prof. Jukka Maalampi and Prof. Kimmo Tuominen for giving me the chance to do my graduate and undergraduate theses on particle physics, and Prof. Kari J. Eskola for introducing me to this fascinating subject. Warm thanks belong also to other teachers, councellors and administrative staff at the Department of Physics for their kind help in various tasks and questions. I also thank the CERN Theory Department for hospitality during my brief visit there. 
I would also like to thank my friends from the university, classmates, colleagues at the office YFL347, and our research group for cheerful atmosphere and company.
These thanks belong especially to Pyry Rahkila, Olli Koskivaara, Tommi Alanne and Risto Paatelainen. Special thanks to Risto for helping me out with the access card at CERN.

I would also like to express my gratitude to family and friends for their support.
\begin{otherlanguage}{finnish}
Lopuksi haluan kiittää perhettä ja ystäviä tuesta. Lämpimät kiitokset sekä omille että vaimoni vanhemmille, joilta olemme saaneet lukemattomasti tukea.
Suurin kiitos kuuluu vaimolleni Johannalle ja lapsillemme. Kiitos tuesta, kärsivällisyydestä ja kannustuksesta, sekä kaikkein eniten rakkaudesta.
\end{otherlanguage}

\begin{flushleft}
\textit{To my wife Johanna and to our ``particles'': Benjamin, Albin, Adabel and Belinda.}
\end{flushleft}

\begin{flushright}
\smallskip
In Muurame, August 2022\\
Henri Jukkala
\end{flushright}

%% file: frontmatter/contribution.tex
In article~\cite{Jukkala:2019slc} the author participated in the calculation of the analytical results and the review of the numerical code. The author also participated critically in the writing and editing of the final draft of the manuscript.
In article~\cite{Jukkala:2021sku} the author produced all numerical results and plots and participated in their analysis. The author had an essential role in the theoretical work and calculated most of the analytical results in the article.
The author also had a leading role in all stages of designing and writing of the final article.

The author also wrote the numerical software package~\cite{henri_jukkala_2021_5025929} which was used in the numerical results of article~\cite{Jukkala:2021sku} and in this work.

%% file: chapters/ch1_introduction.tex
\chapter{Introduction}

Coherence is a central feature of quantum physics.
In classical physics waves are said to be coherent if they have the same wavelength and frequency but may be displaced in time or space.
Coherence is related to the ability of waves to diffract and interfere, and this description carries over to the quantum realm where also particles have wave-like properties~\cite{deBroglie}.
Indeed, according to the wave--particle duality of quantum physics, all matter and radiation have both particle-like and wave-like aspects.
This is famously shown in diffraction and interference experiments with electrons and photons, for example~\cite{LeBellac:2006qpb}.
Also, coherence is more generally related to correlation between physical quantities. Correlation functions are a central concept in quantum field theory (QFT), and mixing of quantum fields is the fundamental property which enables the formation of coherence and coherent oscillation of particles~\cite{Giunti:2007ry}.

The smaller the considered objects are, the more significant quantum coherence is, generally speaking.
It is then no surprise that it is particularly important in elementary particle physics, and by extension, in the very hot and dense state of the early universe which is governed by elementary particle interactions.
Known elementary particles and their interactions (excluding gravity) are described by the Standard Model (SM), and the evolution and properties of the observable universe are explained to a large extent by the big bang model of cosmology.
These models are immensely successful but they have left several unsolved mysteries such as the neutrino masses or the nature of dark matter and dark energy. The events of the very early universe are also still in the dark: for example, the details of cosmological inflation and the related reheating of the universe as well as the origin of the matter--antimatter asymmetry are unknown.

The universe has been observed to be practically completely devoid of antimatter, while ordinary matter is ubiquitous. This disparity is known as the matter--antimatter asymmetry of the universe.
Ordinary matter has a share of approximately 5~\% of the total energy density of the present universe~\cite{Planck:2018vyg}. It consists mostly of protons, neutrons and electrons which make up the interstellar gas and astronomical objects such as stars and planets.
A tiny fraction of cosmic rays contains antimatter and minuscule amounts of it have been produced in terrestrial experiments, but there is no evidence of any cosmologically relevant amounts of antiprotons, antineutrons, positrons, or other antimatter particles~\cite{Cohen:1997ac,Kolb:1990vq}. This is a problem because the SM lacks a mechanism for producing this asymmetry --- the theory is too symmetrical between particles and antiparticles --- so it cannot explain how the present universe could have formed from a symmetrical initial state (set by inflation).

There are many ideas and extended models that have been proposed to explain the unsolved problems mentioned above, and coherence often has a central role in them. For example, coherence between different flavor states of neutrinos is behind neutrino oscillations~\cite{Bilenky:1987ty,Giunti:2007ry}, and particle--antiparticle coherence is essential to particle production~\cite{Calzetta:2008cam} in the early universe. Coherence can also be important in the generation of particle--antiparticle asymmetries through CP-violating processes, which is relevant to the matter--antimatter problem. Prime examples are coherence between particles reflected and transmitted by phase transition walls in electroweak baryogenesis (EWBG)~\cite{Kuzmin:1985mm,Cohen:1993nk,Rubakov:1996vz} and flavor coherence of heavy Majorana neutrinos in leptogenesis~\cite{Fukugita:1986hr,Akhmedov:1998qx}.

The use of QFT is required when studying coherent systems in high temperatures and densities of particles.
Conventional quantum mechanics is not sufficient for relativistic processes where particle numbers change.
Furthermore, evolution of quantum coherence, including its formation and decoherence, is a dynamical process and its description requires a full quantum transport formalism. Non-equilibrium QFT~\cite{Calzetta:2008cam} provides a framework of transport equations for correlation functions which describe quantum processes including interactions and (de)coherence. However, the resulting equations are generally very complex and difficult to solve.
Approximations such as the Kadanoff--Baym ansatz are conventionally employed to derive simpler kinetic equations, but this reduction usually erases coherence information from the correlation functions.

In this thesis we develop approximation methods for non-equilibrium QFT which include quantum coherence. We apply our methods to the early universe and specifically baryogenesis, which are introduced in chapter~\cref{chap:cosmology}. Non-equilibrium QFT is reviewed in chapter~\cref{chap:noneq_QFT} where general quantum transport equations are presented. In this work we concentrate mainly on fermions in spatially homogeneous and isotropic systems.
Our main results, the formulation of a new local approximation method and improvements to the cQPA~\cite{Herranen:2008hi,Herranen:2008hu,Herranen:2008yg,Herranen:2008di,Herranen:2009zi,Herranen:2009xi,Herranen:2010mh,Fidler:2011yq,Herranen:2011zg} are presented in chapter~\cref{chap:cQPA_and_LA-method}. The derived equations describe both flavor coherence and particle--antiparticle coherence contained in the local two-point correlation function.
In chapter~\cref{chap:applications} we apply our methods to EWBG (in a time-dependent example) and resonant leptogenesis, and we solve and analyze the equations numerically. Finally, in chapter~\cref{chap:conclusions} we discuss the results and give the conclusions.

We employ the natural units with $c = \hbar = k_{\rm B} = 1$ throughout this work.

%% file: chapters/ch2_cosmology.tex
\chapter{The early universe}
\label{chap:cosmology}

Cosmology studies the universe as a whole.
Modern physical cosmology~\cite{Weinberg:1972kfs,Kolb:1990vq,Weinberg:2008zzc} rests on three main pillars:
\begin{enumerate*}[label=(\arabic*)]
    \item observation of the expansion of the universe (\ie~the Hubble expansion),
    \item measurement of the primordial light element abundances and explaining their formation via nuclear reactions (BBN, \ie,~big bang nucleosynthesis), and
    \item detection of the cosmic microwave background (CMB) and measurement of its properties.
\end{enumerate*}
These findings lead to the practically indisputable conclusion that in the very distant past the universe was in an extremely hot and dense state and has been expanding and cooling ever since~\cite{Kolb:1990vq}.

The hot primordial plasma of the early universe is governed by both gravity and elementary particle interactions.
The exact timeline of the earlier stages of the universe is still unknown, but there is convincing evidence that before the hot big bang there was a phase of rapid exponential expansion which is called the cosmological inflation~\cite{Guth:1980zm}. It is also conjectured that even earlier, before the inflationary period, the universe was ruled by the enigmatic quantum gravity (at the Planck scale ${\sim} \SI{1e19}{GeV}$ and above).
However, when studying cosmology below such enormous energies it is conventionally assumed that a classical treatment of gravity is sufficient.
Gravity can then be described by the Einstein field equations of general relativity, which are the basis for the big bang model. Next we introduce parts of the big bang theory, and baryogenesis, which are necessary for our applications in this work.

%
\section{Expanding spatially flat spacetime}
\label{sec:expanding_flat_universe}
%

The spatial curvature of the universe has been measured to be very close to zero at large scales~\cite{Planck:2018vyg}. Spatial flatness is thus a very good approximation when studying particle cosmology after the inflationary epoch. The curvature of the spacetime then manifests only as the expansion of the universe.
The universe may also be assumed to be homogeneous and isotropic at very large scales, as shown by the high uniformity of the CMB, for example.
The spacetime of the universe may thus be described by the spatially flat Friedmann--Lemaître--Robertson--Walker (FLRW) metric, which in conformal coordinates takes the form~\cite{Calzetta:2008cam}
\begin{equation}
    {\dd s}^2 = g_{\mu\nu} \dd x^\mu \dd x^\nu
    = a(\eta)^2 \bigl({\dd \eta}^2 - {\dd \evec{x}}^2\bigr)
    \text{.} \label{eq:flat_FLRW_metric}
\end{equation}
Here $a(\eta)$ is the dimensionless scale factor and the conformal time $\eta$ is related to the cosmic time $t$ by $\dd t = a(\eta) \dd \eta$. In the following we denote the cosmic time derivative with a dot and the conformal time derivative with a prime: for example $\dot{a} \eqdef \dd a/{\dd t}$ and $a' \eqdef \dd a/{\dd \eta}$. We use the metric signature convention $({+}, {-}, {-}, {-})$, and denote the Minkowski metric tensor by $\eta_{\mu\nu}$ so that $g_{\mu\nu} = a^2\, \eta_{\mu\nu}$ in equation~\cref{eq:flat_FLRW_metric}.

Expansion of the FLRW universe is described by the Friedmann equations. They can be expressed equivalently by the first Friedmann equation and the continuity equation~\cite{Weinberg:2008zzc}, which in the case of zero spatial curvature are given by
\begin{subequations}
\label{eq:Friedmann_and_continuity}
\begin{align}
    H^2 &= \frac{8\pi G}{3} \rho \text{,}
    \label{eq:Friedmann}
    \\*
    \dot{\rho} &= -3H (\rho + p) \text{.}
    \label{eq:continuity}
\end{align}
\end{subequations}
Here $H \eqdef \dot{a}/a$ is the Hubble parameter, $\rho$ and $p$ are respectively the energy density and pressure of the universe, and $G$ is the gravitational constant. Provided with an equation of state, $p = p(\rho)$, one can solve the time evolution of the scale factor and the energy density from equations~\cref{eq:Friedmann_and_continuity}.

After inflation the universe was reheated, resulting in the creation of the primordial plasma consisting of elementary particles in thermal equilibrium, or very close to it. The thermalization process also created an enormous amount of entropy~\cite{Guth:1980zm}. It can be shown, using basic thermodynamics, that the relativistic particle species (\ie~radiation, with equation of state $p = \rho/3$) dominate the total energy density and pressure of this plasma~\cite{Kolb:1990vq}. Hence, in the early \emph{radiation dominated} epoch it is useful to parametrize the energy density $\rho$ and the entropy density $s$, given by $s = (\rho + p)/T$, with
\begin{subequations}
\label{eq:energy_and_entropy_density_def}
\begin{align}
    \rho &\eqdef \frac{\pi^2}{30} g_*(T) T^4 \text{,}
    \label{eq:energy_density_def}
    \\*
    s &\eqdef \frac{2\pi^2}{45} g_{*s}(T) T^3 \text{.}
    \label{eq:entropy_density_def}
\end{align}
\end{subequations}
Here $T$ is the temperature and $g_*(T)$ counts the effective number of relativistic degrees of freedom of the primordial plasma. Similarly, $g_{*s}(T)$ counts the degrees of freedom for entropy. As long as there are no relativistic particle species which are decoupled from the thermal bath, $g_{*s} = g_*$~\cite{Kolb:1990vq}. These factors are normalized so that one bosonic and fermionic degree of freedom correspond to $g_* = 1$ and $g_* = 7/8$, respectively. At high temperatures before the electroweak phase transition (EWPT) all SM degrees of freedom contribute, adding up to a total of $g_* = 106.75$ plus possible beyond-the-SM degrees of freedom in extended models.

When solving transport equations for particle numbers in the early universe, it is convenient to compare them to some conserved physical quantity. Often the total entropy $S \propto s a^3$ is used for this purpose, as it is conserved in equilibrium. In this case it is said that the universe \emph{expands adiabatically}, and the entropy density scales as $s \propto a^{-3}$. Also, when there is no net annihilation or production of any particle species taking place, $g_*$ and $g_{*s}$ in equations~\cref{eq:energy_and_entropy_density_def} are constants. Using the above adiabatic entropy scaling law we then see from equation~\cref{eq:entropy_density_def} that the temperature scales as $T \propto a^{-1}$. Furthermore, when $g_*$ is constant we can use equations~\cref{eq:Friedmann,eq:energy_density_def} to express the Hubble parameter as a simple quadratic function of $T$:
\begin{equation}
    H(T) = \sqrt{\frac{4\pi^3}{45} g_*} \, \frac{T^2}{m_{\rm Pl}} \text{,}
    \label{eq:Hubble_parameter_adiabatic}
\end{equation}
where $m_{\rm Pl} \eqdef 1/\sqrt{G} \approx \SI{1.2209e19}{GeV}$~\cite{Zyla:2020zbs} is the Planck mass. Using here $T \propto a^{-1}$ and $H = \dot a/a$ we can then further solve the time dependence of the scale factor for the radiation dominated universe: $a \propto \eta \propto \sqrt{t}$.

Furthermore, after reheating (\ie~in the beginning of the hot big bang) the total entropy of the universe was so large that often it is a good approximation to treat the expansion of the universe as adiabatic even when it is strictly speaking not so~\cite{Kolb:1990vq}. Then we can continue to use the adiabatic expansion law $s \propto a^{-3}$ and temperature scaling $T \propto a^{-1}$.
This incurs only a small error to the description of a given non-equilibrium process in the early universe, granted the process does not produce too much entropy.

%
\section{Quantum fields in expanding spacetime}
\label{sec:QFT_in_expanding_spacetime}
%

The framework where QFT is studied in the presence of a classical background gravitational field is called \emph{quantum field theory in curved spacetime}~\cite{Birrell:1982cam}. As a relevant example of QFT in a general curved spacetime, we now consider a generic Lagrangian $\mathcal{L}$ with a fermion field $\psi$, a complex scalar field $\phi$ and a renormalizable interaction part $\mathcal{L}_{\rm int}$:
\begin{equation}
    \mathcal{L} = \sqrt{-g} \Bigl[ \bar\psi (\im \gamma^\mu \nabla_\mu - m_\psi) \psi
    + (\partial_\mu\phi^\dagger)(\partial^\mu\phi) - (m_\phi^2 + \xi R)\phi^\dagger\phi
    \Bigr] + \mathcal{L}_{\rm int} \text{.}
    \label{eq:generic_curved_spacetime_lagrangian}
\end{equation}
Here $g \eqdef \det(g_{\mu\nu})$ is the determinant of the metric, $\nabla_\mu$ is the covariant derivative for fermion fields given by the spin connection, $m_\psi$ and $m_\phi$ are the masses of the fermion and scalar fields, and the constant $\xi$ couples the scalar field to the scalar curvature $R$~\cite{Birrell:1982cam}. Note that we included here the volume factor $\sqrt{-g}$ which stems from the action integral. The gamma matrices $\gamma^\mu$ are defined by the standard Minkowski space relations $\anticomm{\gamma^\mu, \gamma^\nu} = 2 \eta^{\mu\nu} \idmat$ and $(\gamma^\mu)^\dagger = \gamma^0 \gamma^\mu \gamma^0$~\cite{Giunti:2007ry}.

We use the tetrad formalism~\cite{Weinberg:1972kfs,Birrell:1982cam} for the treatment of spinors in general curved spacetime. The covariant derivative for fermion fields in equation~\cref{eq:generic_curved_spacetime_lagrangian} is then given by the spin connection
\begin{equation}
    \nabla_\alpha = {V_\alpha}^\mu(\partial_\mu + \Gamma_\mu)
    \text{,} \label{eq:spin_connection}
\end{equation}
where the tetrads ${V_\alpha}^\mu$ satisfy $g_{\mu\nu} = {V^\alpha}_\mu {V^\beta}_\nu \eta_{\alpha\beta}$ and the spin connection coefficients are given by $\Gamma_\mu = \frac{1}{2} \sigma^{\alpha \beta}{V_\alpha}^\nu V_{\beta\nu;\mu}$~\cite{Weinberg:1972kfs}. Here $V_{\beta\nu;\mu} = \partial_\mu V_{\beta\nu} - {\Gamma^\lambda}_{\mu\nu} V_{\beta\lambda}$ is the covariant derivative of the tetrad with the usual Christoffel symbols ${\Gamma^\lambda}_{\mu\nu}$, and $\sigma^{\alpha\beta} = \frac{1}{4} \comm{\gamma^\alpha, \gamma^\beta}$ are the Lorentz transformation generators for Dirac spinors.

Now we restrict to the flat FLRW metric~\cref{eq:flat_FLRW_metric} with the conformal coordinates. We can then choose the tetrads as ${V^\alpha}_\mu = a(\eta) {\delta^\alpha}_\mu$, whereby ${V_\alpha}^\mu = 1/a(\eta) {\delta_\alpha}^\mu$ and $V_{\alpha\mu} = a(\eta) \eta_{\alpha\mu}$.
The results for the spin connection coefficients are
\begin{equation}
    \Gamma_0 = 0 \text{,} \qquad
    \Gamma_i = \frac{1}{2} \gamma^0 \gamma^i \, \frac{a'}{a} \text{,}
\end{equation}
where $i$ runs over the spatial indices $1,2,3$ only. Using these results we can calculate the contracted covariant derivative:
\begin{equation}
    \gamma^\mu\nabla_\mu = \frac{1}{a}\Bigl(
        \gamma^\mu \partial_\mu + \frac{3}{2} \frac{a'}{a} \gamma^0
    \Bigr) \text{.}
    \label{eq:contracted_spin_connection}
\end{equation}
We can simplify the Lagrangian further by performing a conformal scaling of the fermion and scalar fields, $\psi \to a^{-3/2} \widetilde\psi$ and $\phi \to a^{-1} \widetilde\phi$, and by using $\sqrt{-g} = a^4$~\cite{Birrell:1982cam}. We also do the usual partial integration for the scalar field kinetic term. The resulting scaled Lagrangian is (we now omit the tilde on the fields for clarity)
\begin{equation}
    \widetilde{\mathcal{L}} = \bar\psi \bigl(\im \gamma^\mu \partial_\mu - a m_\psi\bigr) \psi
    + \phi^\dagger \biggl[-\eta^{\mu\nu} \partial_\mu \partial_\nu - a^2 \biggl(m_\phi^2 + \xi R - \frac{a''}{a^3}\biggr) \biggr]\phi
    + \widetilde{\mathcal{L}}_{\rm int} \text{.}
    \label{eq:conformally_scaled_lagrangian}
\end{equation}
This takes the standard form of a Lagrangian in a Minkowski background, except that the masses $m_\psi$ and $m_\phi$ have been multiplied by the scale factor and there are additional contributions to the scalar mass term from the scalar curvature (now given by $R = 6a''/a^3$~\cite{Calzetta:2008cam}). All renormalizable interaction terms in $\widetilde{\mathcal{L}}_{\rm int}$ with dimensionless coupling constants also take the usual Minkowski form.%
\footnote{With the exception of derivative interaction terms which are not considered here.}
This is because the conformal field scalings used the same powers as the length dimensions of the fields, $[\psi] = L^{-3/2}$ and $[\phi] = L^{-1}$, so they always produce a total factor of $a^{-4}$ which cancels the overall $a^4$ from $\sqrt{-g}$.

To summarize, QFT in expanding flat spacetime is equivalent to QFT in a Minkowski spacetime with time-dependent masses $a m_i$ when using the conformal coordinates. Also, the additional contribution to the scalar field mass term vanishes in the radiation dominated universe where $a'' \equiv 0$ (or when using the conformal value $\xi = 1/6$ for the scalar curvature coupling).

%
\section{Baryogenesis}
%

In cosmology the asymmetry between matter and antimatter is called the \emph{baryon asymmetry} of the universe (BAU)~\cite{Dine:2003ax}.
The BAU is often quantified with the ratio of the baryon number density $n_B \eqdef n_{\rm b} - {\bar n}_{\rm b}$ to the photon density $n_\gamma$ (here $B$ denotes the baryon number; $n_{\rm b}, {\bar n}_{\rm b}$ respectively are the densities of baryons and antibaryons). The present value of this ratio $\eta_B$ has been measured to be~\cite{Planck:2018vyg,Fields:2019pfx}
\begin{equation}
    \eta_B \eqdef \frac{n_B}{n_\gamma}
    = (6.129 \pm 0.039) \times 10^{-10} \text{.}
    \label{eq:observed_BAU_ratio}
\end{equation}
This value can be determined from essentially two independent sources: big bang nucleosynthesis (BBN) and the CMB anisotropy power spectrum. Historically, it was first determined from BBN where $\eta_B$ is the input parameter which controls the abundances of primordial light elements~\cite{Kolb:1990vq,Fields:2019pfx}. Nowadays the CMB measurements are much more precise~\cite{Fields:2019pfx,Zyla:2020zbs}. The agreement of the results from these two different sources is a remarkable success of big bang cosmology. However, the origin of the BAU is not explained in the big bang model; it is just a parameter.

Baryogenesis~\cite{Dine:2003ax,Cline:2006ts,Cline:2018fuq,Bodeker:2020ghk} is the name for the hypothetical process that generated the BAU dynamically from a baryon symmetric initial state~\cite{Kolb:1990vq,Weinberg:2008zzc}. This is still a highly speculative field of particle cosmology as there are many proposed baryogenesis mechanisms and there is not yet strong evidence for any specific one. Strictly speaking it is also not completely ruled out that the observed BAU could have been an initial condition of the universe (or produced by quantum gravitational effects at the Planck scale ${\sim} \SI{1e19}{GeV}$). The problem with this explanation is that cosmological inflation would have diluted an initial baryon asymmetry to a completely negligible level at the end of reheating~\cite{Dine:2003ax}. Thus, a dynamical explanation is still likely to be needed for today's observed BAU.

\paragraph{Sakharov conditions}
There are three necessary generic requirements for baryogenesis. These are called the \emph{Sakharov conditions} and they are
\begin{enumerate*}[label=(\roman*)]
    \item B-violation,
    \item CP- and C-violation, and
    \item deviation from equilibrium~\cite{Sakharov:1967dj}.
\end{enumerate*}
The first condition is obvious: baryon number must not be conserved.
The second condition refers to the charge (C) and the combined charge and parity (CP) transformations.
In a CP- and C-conserving system the B-violating processes which produce baryons and the related processes which produce antibaryons would have equal rates, meaning that no net baryon number could be generated~\cite{Kolb:1990vq,Cline:2006ts}.
The third condition is needed because in thermal (and chemical) equilibrium the B-violating processes and their inverse processes would have equal rates, again implying that no net baryon number could be produced~\cite{Cline:2006ts}.
Another way to understand this is that if the system starts with $B = 0$ and stays in chemical equilibrium, the chemical potential(s) corresponding to the baryon number would be zero and hence the baryon and antibaryon distributions would be identical (ultimately due to CPT-invariance)~\cite{Kolb:1990vq,Weinberg:2008zzc}.

\medskip\noindent
It is now known that fulfilling all three Sakharov conditions for successful baryogenesis requires physics beyond the SM, even though all of the ingredients are in some capacity already there~\cite{Morrissey:2012db}. The quark sector of the SM already has CP-violation in the CKM-matrix, but it is widely accepted to be too weak for baryogenesis~\cite{Cline:2006ts}. The expansion of the universe can provide non-equilibrium conditions, but there are no suitable heavy particles in the SM that would have the required non-equilibrium decays. A strong first order EWPT could also work, but it is known that the EWPT in the SM is only a continuous cross-over with the known value of the Higgs boson mass~\cite{Kajantie:1996mn}.

One of the Sakharov conditions is still fulfilled already in the SM and that is the B-violation. This seems surprising at first because baryon number is conserved in all perturbative particle interactions in the SM. This is only an accidental global $\U(1)$ symmetry~\cite{Cline:2018fuq}, however, and the electroweak sector actually breaks $B$ via non-perturbative processes and the axial anomaly~\cite{tHooft:1976rip,tHooft:1976snw,Kuzmin:1985mm}. This \emph{electroweak B-violation} in the SM is due to the non-trivial vacuum state structure of the non-Abelian $\SU(2)$ gauge field. Under usual conditions in the present universe (zero or low temperature) these B-violating processes are extremely suppressed but at very high temperatures baryon number can be strongly violated.

Some of the most prominent baryogenesis mechanisms are GUT baryogenesis, electroweak baryogenesis and leptogenesis~\cite{Dine:2003ax,Bodeker:2020ghk}. Historically, after Sakharov's initial work the development of baryogenesis first took off in the context of grand unified theories (GUTs). Most GUTs naturally give rise to baryogenesis because they contain suitable B-violating heavy particles.
However, reconciling GUT baryogenesis with inflation is challenging because the reheating temperature is usually below the GUT scale (${\sim} \SI{1e16}{GeV}$).%
\footnote{A higher reheating temperature is problematic in GUTs because of the gravitino-overproduction problem~\cite{Cohen:1993nk,Dine:2003ax}. Most GUTs also rely on supersymmetry which still has not been observed.}
Electroweak baryogenesis~\cite{Kuzmin:1985mm} is an interesting possibility because it only involves electroweak-scale physics and should thus be testable in present and near-future experiments~\cite{Bodeker:2020ghk}. It makes use of the electroweak B-violation present already in the SM. Leptogenesis~\cite{Fukugita:1986hr} also uses the electroweak B-violation, and more specifically, the possibility that a dynamically produced \emph{lepton} number (denoted by $L$ below) may explain the observed BAU~\cite{Dine:2003ax}.
Leptogenesis is a very attractive possibility because it is linked to the see-saw mechanism which may explain the observed light neutrino masses.

%
\subsection{Electroweak B-violation}
%

The origin of the electroweak B-violation in the SM is the vacuum state structure of the non-Abelian $\SU(2)$ gauge field~\cite{Cline:2006ts,Rubakov:1996vz}. The gauge field has multiple vacuum states labelled by an integer: a topological winding number called the Chern--Simons number. The corresponding Chern--Simons current, on the other hand, is coupled to the baryon and lepton axial vector currents via the Adler--Bell--Jackiw anomaly~\cite{Adler:1969gk,Bell:1969ts,Rubakov:1996vz}. Hence, transitions of the gauge field vacuum state break $B + L$ and transform baryons into antileptons or antibaryons into leptons (and vice versa) so that the quantum numbers change as $\Delta B = \Delta L = \pm 3$~\cite{Cline:2006ts,Cline:2018fuq}.

These gauge vacuum state transitions are negligible in zero temperature as they proceed through instanton configurations (quantum tunneling through the potential barrier) and are extremely suppressed~\cite{tHooft:1976rip,Rubakov:1996vz}. However, in very high temperatures the suppression is relieved as the transitions can occur classically via macroscopic saddle-point configurations called sphalerons~\cite{Klinkhamer:1984di,Manton:1983nd,Arnold:1987mh,Arnold:1987zg} and the exponential suppression disappears completely for even higher temperatures in the symmetric electroweak phase~\cite{Cline:2006ts,Konstandin:2013caa}. These high temperature $B + L$ violating anomalous processes thus occurred very frequently in the early universe and they were in equilibrium roughly in the temperature range $\SI{100}{GeV} \lesssim T \lesssim \SIrange{1e12}{1e13}{GeV}$~\cite{Cline:2006ts,Bodeker:2020ghk}.%
\footnote{These processes are often collectively called ``sphaleron processes'' even though the sphaleron configuration only exists in the broken state of the electroweak symmetry.}

The anomalous electroweak processes still conserve $B - L$, which is non-anomalous in the SM, and they tend to reduce any excess baryon or lepton number so that both $B, L \propto B - L$ in equilibrium~\cite{Harvey:1990qw,Arnold:1987mh,Bodeker:2020ghk}. This means that these processes will wash out any baryon asymmetry if $B - L = 0$, and especially they alone cannot generate a baryon asymmetry if initially both $B$ and $L$ vanish. On the other hand, \emph{both} baryon and lepton asymmetries will be generated if $B - L \neq 0$, even if one or the other is zero initially. This is the feature utilized in leptogenesis.

%
\subsection{Electroweak baryogenesis}
%

In electroweak baryogenesis~\cite{Cohen:1993nk,Morrissey:2012db,Konstandin:2013caa,Bodeker:2020ghk} the BAU is generated by the anomalous electroweak processes during the EWPT which is required to be strongly first order. In such a transition bubbles of the electroweak broken phase nucleate in the electroweak symmetric plasma. The bubbles expand and eventually fill the universe, and during this transition there are suitable conditions for baryogenesis~\cite{Morrissey:2012db,Konstandin:2013caa}. First, particles of the plasma scatter with the expanding bubble which results in a chiral asymmetry.
A part of this source asymmetry is then converted to a baryon asymmetry by the unsuppressed electroweak processes in the symmetric phase outside the bubble. Finally, some of the generated baryon number is captured by the expanding bubble where it is protected from washout because the sphaleron rate is suppressed. Due to this last point no $B - L$ violation is needed in the process.

Electroweak baryogenesis is a theoretically very interesting process as it involves the non-per\-turbative effects and vacuum structure of the electroweak gauge sector and bubble nucleation dynamics of the EWPT. It is also attractive due to its potential for experimental testing because it only involves processes at the electroweak scale. Gravitational wave astronomy also provides an interesting new probe for the EWPT~\cite{Bodeker:2020ghk}. As was already touched on above, successful EWBG requires extending the SM in order to get a strong enough CP-violating source and a strong first order EWPT.
However, by now the results from the LHC and electron EDM experiments~\cite{ACME:2018yjb} have strongly constrained most popular models, such as the minimal supersymmetric SM, two-Higgs-doublet models and doublet-singlet models~\cite{Bodeker:2020ghk,Cline:2000nw,Cline:2000kb,Cline:2013gha,Alanne:2016wtx}.
For some recent progress in EWBG models see for example~\cite{Cline:2017qpe,Cline:2013bln,Cline:2012hg,Cline:2011mm}.

%
\subsection{Leptogenesis}
\label{sec:leptogenesis}
%

Leptogenesis~\cite{Davidson:2008bu,Pilaftsis:2009pk,Blanchet:2012bk,Bodeker:2020ghk} is a mechanism for baryogenesis in which the BAU is produced from a dynamically generated lepton asymmetry. In leptogenesis the SM is extended with (typically three) heavy singlet neutrino fields $N_i$ with masses $m_i$ (with $i = 1, \ldots$) and they have CP-violating chiral Yukawa interactions and L-violating Majorana mass terms. These Majorana neutrinos mix and their interactions with the SM leptons and Higgs boson produce the lepton asymmetry which is converted to the baryon asymmetry by the anomalous electroweak processes. Leptogenesis is essentially a consequence of the type-I see-saw mechanism (see \eg~\cite{Giunti:2007ry}) which is a potential explanation for the lightness of the SM neutrinos~\cite{Blanchet:2012bk}. The prospect of having the same origin for the light neutrino masses and the baryon asymmetry makes leptogenesis very interesting theoretically.

Leptogenesis has different variants which can be successfully realized at different temperature ranges in the early universe. The most prominent is the original scenario of thermal leptogenesis~\cite{Fukugita:1986hr} which can be roughly divided to the unflavored (\ie, one-flavor) and flavored cases~\cite{Bodeker:2020ghk}. Resonant leptogenesis (RL)~\cite{Pilaftsis:2003gt} is a further special case of thermal leptogenesis with mixing quasidegenerate Majorana neutrinos where the relevant CP-asymmetry is resonantly enhanced. Another notable scenario is the ARS mechanism of leptogenesis~\cite{Akhmedov:1998qx} where the lepton asymmetry is produced during the production and oscillations of heavy singlet neutrinos close to the sphaleron freeze-out. For other leptogenesis scenarios see for example~\cite{Davidson:2008bu,Bodeker:2020ghk}.

In thermal leptogenesis the lepton asymmetry is generated in the non-equilibrium decays of the Majorana neutrinos when they have slight over-abundances compared to equilibrium. The asymmetry then freezes out once the washout processes drop out of equilibrium. Leptogenesis has often been studied in the one-flavor approximation (\ie~the unflavored case) where it is assumed that the asymmetry is generated equally in all lepton flavors and it suffices to consider only one effective SM lepton flavor. Within this approximation and in the case of a hierarchical Majorana neutrino mass spectrum (\ie~in the ``vanilla leptogenesis'' scenario) successful leptogenesis implies a lower bound of ${\sim} \SI{1e9}{GeV}$ for the Majorana neutrino mass scale~\cite{Davidson:2002qv,Blanchet:2012bk}. Strictly speaking the one-flavor approximation is valid only when the SM lepton flavors cannot be distinguished during leptogenesis. This is the case, roughly, if the masses of the Majorana neutrinos are greater than ${\sim} \SI{1e12}{GeV}$. This corresponds to the temperature above which all SM lepton Yukawa interactions are still slow enough to be out of equilibrium~\cite{Davidson:2008bu,Blanchet:2012bk,Bodeker:2020ghk}. On the other hand, if some or all of the SM lepton Yukawa processes are in equilibrium during leptogenesis one needs to take them into account and track the evolution of individual lepton asymmetries.%
\footnote{The SM lepton Yukawa interactions are an example of ``spectator processes'' in leptogenesis, that is, processes which can affect the lepton asymmetry indirectly~\cite{Davidson:2008bu,Bodeker:2020ghk}.}
In the flavored case the lower bound for the leptogenesis scale can then be significantly lower when compared to the one-flavor case~\cite{Bodeker:2020ghk}.

The leptogenesis temperature scale can be lowered even more when the Majorana neutrino masses are almost degenerate, as is the case in (thermal) resonant leptogenesis. In RL the lepton asymmetry generation is enhanced for a quasidegenerate mass spectrum and this resonant enhancement is maximal when the mass differences of the Majorana neutrinos are comparable to their decay widths. In this case the CP-violation in the Majorana neutrino decays is dominated by the coherent flavor mixing effects (the so-called self-energy contribution in the semiclassical approach, see below)~\cite{Davidson:2008bu,Pilaftsis:2009pk}. In RL the leptogenesis temperature can then be brought down to the TeV-scale or even lower~\cite{Granelli:2020ysj,Bodeker:2020ghk}.
Low-scale leptogenesis is also possible in the related ARS scenario, which is also called ``leptogenesis via oscillations'' or ``freeze-in leptogenesis''~\cite{Drewes:2017zyw,Klaric:2020phc}. In fact, both RL and ARS scenarios require at least two quasidegenerate heavy neutrinos and there are resonances and oscillations in both~\cite{Klaric:2020phc}. However, in the ARS mechanism the BAU is frozen in during the production of the heavy neutrinos, in contrast to the standard freeze-out scenario of thermal RL. It has been shown that a unified description of both mechanisms is possible and the parameter regions where they enable successful baryogenesis are merged~\cite{Klaric:2020phc}.

\subsubsection{Semiclassical transport equations}

The non-equilibrium evolution of the lepton asymmetry must be solved from transport equations. Thermal leptogenesis is conventionally studied in a semiclassical approach with Boltzmann equations where the interaction rates are supplied by perturbative vacuum QFT~\cite{Kolb:1979qa,Luty:1992un,Basboll:2006yx}. Usually simplified momentum-integrated rate equations are also derived, which requires some additional assumptions such as kinetic equilibrium and Maxwell--Boltzmann statistics. The standard equations in the one-lepton-flavor approximation, with decay and inverse decay processes only, are~\cite{Buchmuller:2004nz,Basboll:2006yx,Blanchet:2012bk,Bodeker:2020ghk}
\begin{subequations}
\label{eq:leptogenesis_Boltzmann}
\begin{align}
    \frac{\dd n_i}{\dd t} + 3 H n_i &= -\biggl(\frac{n_i}{n_i^{\rm eq}} - 1\biggr) \gamma_i
    \text{,}
    \\*
    \frac{\dd n_L}{\dd t} + 3 H n_L &= \sum_i \biggl[
        \biggl(\frac{n_i}{n_i^{\rm eq}} - 1\biggr) \epsilon_i^{CP} \gamma_i
        - \frac{n_L}{2 n_{\ell}^{\rm eq}} \gamma_i
    \biggr] \text{.}
\end{align}
\end{subequations}
Here $n_i$ are the number densities of the Majorana neutrinos. The lepton asymmetry density is $n_L \eqdef n_{\ell} - \bar n_{\ell}$ with the lepton and antilepton densities $n_{\ell}, \bar n_{\ell}$. The corresponding equilibrium number densities are $n_i^{\rm eq}$ and $n_{\ell}^{\rm eq}$ and the reaction density for the decays and inverse decays is denoted by $\gamma_i$ (explicit formulae are given in~\cite{Jukkala:2021sku}). In the standard scenario the lepton asymmetry is generated mainly when $T \lesssim m_1$, where $m_1$ is the mass of the lightest Majorana neutrino. We have only included the decay and inverse decay processes of the Majorana neutrinos because they give the dominant contributions to Majorana neutrino production and lepton asymmetry washout in this case~\cite{Bodeker:2020ghk}. Various scattering processes are relevant in higher temperatures $T > m_1$ and also for $T \ll m_1$, and should be accounted for in more complete (or phenomenological) studies.

A crucial part of the semiclassical equations~\cref{eq:leptogenesis_Boltzmann} is the CP-asymmetry parameter $\epsilon_i^{CP}$ which quantifies the amount of CP-violation in the Majorana neutrino decays. It is defined in the one-lepton-flavor case as
\begin{equation}
    \epsilon_i^{CP} \eqdef
    \frac{\Gamma(N_i \to \ell \phi) - \Gamma(N_i \to \bar\ell \bar\phi)}
    {\Gamma(N_i \to \ell \phi) + \Gamma(N_i \to \bar\ell \bar\phi)} \text{,}
    \label{eq:leptogenesis_Boltzmann_CP-asymmetry_def}
\end{equation}
where $\Gamma(N_i \to \ell \phi)$ is the partial width for the decay of the Majorana neutrino $N_i$ to the SM lepton and Higgs doublets, and $\Gamma(N_i \to \bar\ell \bar\phi)$ is the corresponding partial decay width with antiparticles. The CP-asymmetry vanishes at tree level and its calculation involves the interference of tree level and higher order amplitudes of the decay process.
The two main contributions to $\epsilon_i^{CP}$ at one-loop order are the vertex contribution (also called $\epsilon'$-type or direct CP-violation) which arises from the one-loop correction to the Yukawa interaction vertex and the self-energy contribution (also called $\epsilon$-type or indirect CP-violation) which is related to the wavefunction renormalization of the mixing Majorana neutrinos~\cite{Pilaftsis:1997jf}. However, the self-energy contribution cannot be calculated correctly in conventional perturbation theory and a straightforward calculation with a one-loop external leg correction (\ie~a non-1PI diagram) gives a diverging result in the limit of degenerate masses. Determining the correct result for $\epsilon_i^{CP}$ in the quasidegenerate case is highly non-trivial because the Majorana neutrinos mix and are unstable~\cite{Buchmuller:1997yu,Pilaftsis:1997jf,Pilaftsis:2003gt}. Proper calculations, which usually involve resummations, have been done with different methods in the literature. However, the exact form of $\epsilon_i^{CP}$ is uncertain because different methods yield different results in the maximally resonant region~\cite{Buchmuller:1997yu,Pilaftsis:1997jf} (see also \eg~\cite{Anisimov:2005hr,Dev:2014laa}).

So far we have considered simplified transport equations for calculating the lepton asymmetry. More complete studies should also include the effects of the anomalous electroweak processes which redistribute the generated asymmetry among $B$ and $L$. A simple way to take this into account is to find the relation between the number densities $n_L$ and $n_{B-L}$. At very high temperatures when the anomalous processes are slow the relation is trivially $n_L = -n_{B-L}$ (assuming zero initial $B$). But when the anomalous processes are in equilibrium the relation is non-trivial and depends also on other spectator processes~\cite{Buchmuller:2001sr,Nardi:2005hs,Pilaftsis:2005rv}. In thermal leptogenesis these processes are conventionally neglected during the generation of the lepton asymmetry and one continues to use $n_L \simeq -n_{B-L}$ (accurate to within 10~\%~\cite{Giudice:2003jh})~\cite{Bodeker:2020ghk}. Once $n_{B-L}$ is determined one can then use the standard equilibrium relations~\cite{Harvey:1990qw,Bodeker:2020ghk}
\begin{equation}
    n_B = \frac{28}{79} n_{B-L} \text{,} \qquad
    n_L = -\frac{51}{79} n_{B-L}
\end{equation}
to estimate the final redistribution of the asymmetry by the anomalous processes.

\subsubsection{CP-asymmetry in resonant leptogenesis}

We now concentrate on RL and hence consider the self-energy contribution to $\epsilon_i^{CP}$ as it is the dominant part in this case. We further restrict to the case of two Majorana neutrinos ($i = 1,2$) which is sufficient in this work. The CP-asymmetry~\cref{eq:leptogenesis_Boltzmann_CP-asymmetry_def} in the one-lepton-flavor approximation then takes the generic form~\cite{Garny:2011hg}
\begin{equation}
    \epsilon^{CP}_{i, x} =
    \frac{\Im\bigl[(y_1^* y_2)^2\bigr]}{\abs{y_1}^2 \abs{y_2}^2}
    \frac{(m_2^2 - m_1^2) m_i \Gamma_j^{(0)}}{(m_2^2 - m_1^2)^2 + (R_{ij,x})^2}
    \text{.} \qquad \text{($j \neq i$)}
    \label{eq:Boltzmann_CP-asymmetry_self-energy-part}
\end{equation}
Here $y_i$ are the relevant Yukawa couplings and $\Gamma_i^{(0)} = \abs{y_i}^2 m_i/(8 \pi)$ is the tree-level total decay width of the Majorana neutrino. The CP-asymmetry~\cref{eq:Boltzmann_CP-asymmetry_self-energy-part} is resonantly enhanced for small mass differences $\Delta m_{21} \eqdef m_2 - m_1$, and the resonance is regulated by the factor $R_{ij,x}$ in the degeneracy limit $\Delta m_{21} \to 0$. Different approaches and approximation methods used in the literature lead to different forms of the regulator, which we have labelled by $x$. Some of the most relevant ones are
\begin{subequations}
\label{eq:CP-asymm_regulators}
\begin{align}
    R_{ij,{\rm mix}} &= m_i \Gamma_j^{(0)} \text{,}
    \label{eq:CP-asymm_mixed_regulator}
    \\*
    R_{ij,{\rm diff}} &= m_i \Gamma_i^{(0)} - m_j \Gamma_j^{(0)} \text{,}
    \label{eq:CP-asymm_difference_regulator} \displaybreak[0]
    \\
    R_{ij,{\rm sum}} &= m_i \Gamma_i^{(0)} + m_j \Gamma_j^{(0)} \text{,}
    \label{eq:CP-asymm_sum_regulator}
    \\*
    R_{ij,{\rm eff}} &= (m_i \Gamma_i^{(0)} + m_j \Gamma_j^{(0)})
    \abs{\sin\theta_{ij}} \text{,} \label{eq:CP-asymm_eff_regulator}
\end{align}
\end{subequations}
which we call the \emph{mixed} regulator, \emph{difference} regulator, \emph{sum} regulator and \emph{effective} sum regulator, respectively~\cite{Buchmuller:1997yu,Pilaftsis:1997jf,Pilaftsis:2003gt,Garny:2011hg,Dev:2017wwc}. In the effective regulator~\cref{eq:CP-asymm_eff_regulator} the angle $\theta_{ij}$ denotes the relative phase of the complex Yukawa couplings $y_i$ and $y_j$:
\begin{equation}
    \sin(\theta_{ij}) \eqdef \frac{\Im(y_i^* y_j)}{\abs{y_i}\abs{y_j}}
    \text{,} \hspace{3em}
    \cos(\theta_{ij}) \eqdef \frac{\Re(y_i^* y_j)}{\abs{y_i}\abs{y_j}}
    \text{.}
    \label{eq:leptogenesis_Yukawa_CP_phase}
\end{equation}
In the one-lepton-flavor approximation with two Majorana neutrinos the angle $\theta_{12}$ determines the overall strength of the CP-asymmetry. This can be seen by writing the prefactor in equation~\cref{eq:Boltzmann_CP-asymmetry_self-energy-part} as $\Im[(y_1^* y_2)^2]/(\abs{y_1}^2 \abs{y_2}^2) = \sin(2\theta_{12})$.

Finally, it should be noted that some of the regulators lead to spurious enhancements of the CP-asymmetry in certain cases. For example, the difference regulator~\cref{eq:CP-asymm_difference_regulator} vanishes when both $\abs{y_2} \to \abs{y_1}$ and $m_2 \to m_1$. Also, the effective regulator~\cref{eq:CP-asymm_eff_regulator} overestimates the asymmetry for small $\theta_{12}$~\cite{Dev:2017wwc}.

\subsubsection{Shortcomings of the semiclassical approach}

The application of classical Boltzmann equations to leptogenesis is known to have shortcomings~\cite{Lindner:2005kv}. It is also very non-trivial because the required CP-violation is a purely quantum effect. The source for the lepton asymmetry arises from the interference of a radiatively corrected and a tree level amplitude instead of a purely tree level process. This results in subtle issues, such as the difficulty in determining the exact form of the CP-asymmetry parameter~\cref{eq:leptogenesis_Boltzmann_CP-asymmetry_def}, as discussed above. Furthermore, the CP-asymmetry generation is fundamentally a dynamical phenomenon and not always adequately captured by a static parameter such as~\cref{eq:Boltzmann_CP-asymmetry_self-energy-part}~\cite{Dev:2017wwc}. Another fundamental problem is the double counting of real intermediate states (RIS) of the scattering processes. The on-shell part of the s-channel exchange of a Majorana neutrino is included both in the decay and inverse decay contributions and in the $\Delta L = 2$ scattering processes~\cite{Giudice:2003jh,Buchmuller:2004nz,Davidson:2008bu}. This results in a spurious CP-violating source (in the decay and inverse decay part) which does not vanish in equilibrium~\cite{Buchmuller:2004nz,Basboll:2006yx}. This is usually resolved by an ad hoc subtraction of the problematic part from the equations, but this is a delicate procedure.

These issues stem from the fact that the semiclassical approach is an ad hoc combination of vacuum QFT calculations and classical transport equations. Another shortcoming of this approach is that the radiative calculations do not include thermal corrections.
All of these problems can be avoided by using a different approach where the non-equilibrium and quantum aspects are unified in one dynamical framework, such as the CTP method. The Schwinger--Keldysh CTP method is a first principles approach to non-equilibrium QFT and it has been widely used to study leptogenesis, see for example~\cite{Buchmuller:2000nd,DeSimone:2007gkc,DeSimone:2007edo,Cirigliano:2007hb,Garny:2009qn,Garny:2009rv,Beneke:2010wd,Beneke:2010dz,Anisimov:2010dk,Garbrecht:2011aw,Garny:2011hg,Iso:2013lba,Iso:2014afa,Hohenegger:2014cpa,Garbrecht:2014aga,Dev:2014wsa,Kartavtsev:2015vto,Drewes:2016gmt,Dev:2017wwc,Dev:2017trv,Garbrecht:2018mrp,Depta:2020zmy} (a more comprehensive list is given in~\cite{Jukkala:2021sku}). As a field theoretic and fully-quantum method, it is very well suited to study the delicate features of leptogenesis such as the coherent transitions of the Majorana neutrinos and details of the CP-asymmetry generation. However, full implementation of this method is very difficult and various approximations are usually needed (see \eg~\cite{Buchmuller:2000nd,DeSimone:2007gkc,Lindner:2005kv}). Therefore, some questions about the exact form of the CP-asymmetry, especially in the case of RL, have still not been settled~\cite{Garny:2011hg,Garbrecht:2014aga,Dev:2017wwc,Dev:2017trv,Racker:2021kme}.
This shows how non-trivial the CP-violation is in leptogenesis, and it calls for clear and concise treatments of the subject.
It is our modest hope that the present work helps to clarify this important issue.

%% file: chapters/ch3_nonequilibrium.tex
\chapter{Non-equilibrium quantum field theory}
\label{chap:noneq_QFT}

Non-equilibrium quantum field theory is needed to study the time evolution of high-energy thermodynamic systems where quantum effects are essential for the dynamics. Such situations arise in cosmology in the hot and dense early universe and also during the brief existence of the extreme form of matter in high-energy heavy-ion collisions~\cite{Berges:2004vw,Berges:2001fi}. Ordinary vacuum QFT or the imaginary time formalism of thermal QFT are not sufficient for describing the time evolution of such systems because the boundary conditions are different: vacuum QFT is used to calculate vacuum-to-vacuum transition amplitudes over a large time interval and thermal QFT applies to static systems with no time-dependence.

In thermal equilibrium the quantum density operator of the system takes the form of a time evolution operator with an imaginary time variable. This enables the study of QFTs in finite temperature: thermal expectation values can be formulated as ordinary QFT transition amplitudes over the Euclidean time~\cite{LeBellac:1996cam,Kapusta:2006cam}. This approach, known as the imaginary time QFT formalism, however relies on the specific equilibrium form of the density operator.
With general density operators one needs a real time formalism instead.
There exist different formulations but common for all of them is that two real time branches are needed in the time-contour, one forward and one backward, and the imaginary part of the contour must be non-increasing~\cite{LeBellac:1996cam}.
In this work we consider the simplest realization, called the closed time path formalism, which we will turn to next.

%
\section{Schwinger--Keldysh formalism}
\label{sec:Schwinger-Keldysh_formalism}
%

\begin{figure}[t!]
    \centering
    \includegraphics{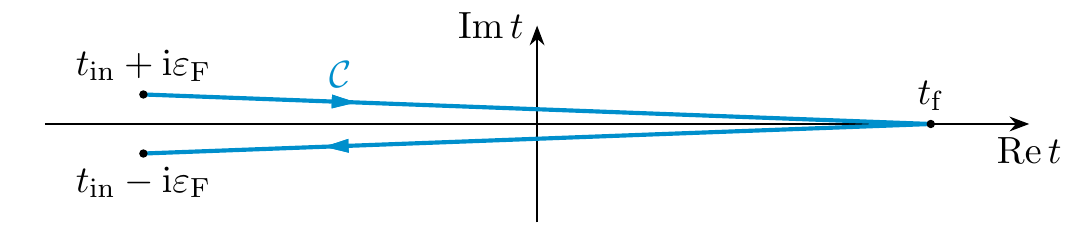}
    \caption{The Schwinger--Keldysh closed time path $\mathcal{C}$ for a considered initial time $\tin$ and a final time parameter $\tf$. The arrows on the path show the direction of increasing contour-time. We have also indicated the infinitesimal displacements $\pm \im \epsilon_{\rm F}$ required for convergence and correct boundary conditions.}
    \label{fig:Schwinger-Keldysh_path}
\end{figure}

A framework for non-equilibrium QFT is provided by the Schwinger--Keldysh closed time path (CTP) method, which was developed by Schwinger~\cite{Schwinger:1960qe}, Keldysh~\cite{Keldysh:1964ud} and many others~\cite{Bakshi:1962dv,Feynman:1963fq,Chou:1980,DeWitt:1986,Jordan:1986ug,Su:1987pi} (see also \eg~\cite{Chou:1984es} and references therein). The CTP method is a rather general idea and it has been applied to various problems in different contexts~\cite{Chou:1984es}. In the context of QFT, one way to understand the need for the closed time path (shown in figure~\cref{fig:Schwinger-Keldysh_path}) is as follows. Consider for example a two-point correlation function of a real scalar field $\phi$ in a system with a quantum density operator $\rho$:
\begin{equation}
    \expectval[\big]{\hat\phi(x_1) \hat\phi(x_2)} \eqdef
    \Tr\bigl[\rho \,\hat\phi(x_1) \hat\phi(x_2)\bigr] \text{.}
    \label{eq:two-point_function_example}
\end{equation}
The density operator $\rho$ fully describes the quantum state of the system. The operators $\mathcal{O} = \rho, \hat\phi$ in equation~\cref{eq:two-point_function_example} are written in the Heisenberg picture. In terms of Schrödinger picture operators $\mathcal{O}_{\rm S}$
\begin{equation}
    \expectval[\big]{\hat\phi(x_1) \hat\phi(x_2)} =
    \Tr\Bigl[
        \rho \, U(t_1,\tin)^\dagger \hat\phi_{\rm S}(\evec{x}_1) U(t_1,\tin) \,
        U(t_2,\tin)^\dagger \hat\phi_{\rm S}(\evec{x}_2) U(t_2,\tin)
    \Bigr] \text{,}
    \label{eq:two-point_function_with_Schroedinger}
\end{equation}
where $\tin$ is the initial time where the density operator $\rho = \rho_{\rm S}(\tin)$ is prepared at, and $U(t_1,t_2)$ is the full unitary time-evolution operator of the Schrödinger states. Now, when the density operator is generic and does not have any specific form (such as $\rho = \e^{-\beta H}$ in the canonical ensemble) the only way to evaluate~\cref{eq:two-point_function_with_Schroedinger} is to use two separate time branches or ``histories''~\cite{Chou:1984es,LeBellac:1996cam,Calzetta:2008cam}.

%
%

We can understand the need for two time branches by writing equation~\cref{eq:two-point_function_with_Schroedinger} in the functional integral form using standard QFT methods. To do this we use a basis of field eigenstates $\lvert \varphi_a \rangle$ of the Schrödinger field $\hat\phi_{\rm S}$ (\ie~$\hat\phi_{\rm S}(\evec{x}) \lvert \varphi_a \rangle = \varphi_a(\evec{x}) \lvert \varphi_a \rangle$) which is complete and orthonormal: $\int \mathcal{D}\varphi_a \, \lvert \varphi_a \rangle \langle \varphi_a \rvert = 1$ and $\langle \varphi_a | \varphi_b \rangle = \delta[\varphi_a - \varphi_b]$~\cite{Peskin:1995ev,Berges:2015kfa}. We also need the Feynman path integral formulation of the transition amplitude,%
\footnote{We use the following shorthand notation for constrained functional integrals:
\begin{equation*}
    \smash{ \int_{\varphi_a}^{\varphi_b} } \mathcal{D}\phi \cdots
    \exp\Bigl(\im S[\phi]_{t_1}^{t_2} \Bigr) \eqdef
    \smash{ \int } \mathcal{D}\phi \, \cdots \exp\Bigl(\im S[\phi]_{t_1}^{t_2} \Bigr)
    \smash{
        \bigg\rvert_{\substack{\phi(t_1,\evec{x}) = \varphi_a(\evec{x}), \\ \phi(t_2,\evec{x}) = \varphi_b(\evec{x})}}
    }
\end{equation*}
where the action integral is $S[\phi]_{t_1}^{t_2} \eqdef \smash{ \int_{t_1}^{t_2} } \dd^4 x \, \mathcal{L}[\phi(x)]$, and $\int_{t_1}^{t_2} \dd^4 x \eqdef \smash{ \int_{t_1}^{t_2} } \dd t \int \dd^3 \evec{x}$.}
\begin{equation}
    \expectval[\big]{\varphi_b \big| U(t_2,t_1) \big| \varphi_a} =
    \smash{ \int_{\varphi_a}^{\varphi_b} } \mathcal{D}\phi \,
    \exp\Bigl(\im S[\phi]_{t_1}^{t_2} \Bigr) \text{.}
\end{equation}
Using these we can express the operator trace in~\cref{eq:two-point_function_with_Schroedinger} as the functional integral
\begin{align}
    \expectval[\big]{\hat\phi(x_1) \hat\phi(x_2)} &=
    \begin{aligned}[t]
    \smash{\int} \phantom{\mathclap{\Big|}}
    \mathcal{D}\varphi_a \mathcal{D}\varphi_b \mathcal{D}\varphi_c \mathcal{D}\varphi_d \, &
    \expectval[\big]{\varphi_a \big| \rho \big| \varphi_b}
    \expectval[\big]{\varphi_b \big| U(t_1,\tin)^\dagger \big| \varphi_c} \, \varphi_c(\evec{x}_1) \notag
    \\*
    {} \times {} &
    \expectval[\big]{\varphi_c \big| U(t_1,t_2) \big| \varphi_d}
    \expectval[\big]{\varphi_d \big| U(t_2,\tin) \big| \varphi_a} \, \varphi_d(\evec{x}_2)
    \end{aligned}
    \\*
    &\overset{\mathclap{\scriptscriptstyle (t_1 > t_2)}}{=}
    \begin{aligned}[t]
    \int \mathcal{D}\varphi_a \mathcal{D}\varphi_b \mathcal{D}\varphi_c \,
    \expectval{\varphi_a | \rho | \varphi_b} &
    \int_{\varphi_b}^{\varphi_c} \mathcal{D}\phi^{-} \, \phi^{-}(x_1)
    \exp\Bigl(\im S[\phi^{-}]_\tin^{t_1} \Bigr)^* \notag
    \\*
    {} \times {} &
    \int_{\varphi_a}^{\varphi_c} \mathcal{D}\phi^{+} \, \phi^{+}(x_2)
    \exp\Bigl(\im S[\phi^{+}]_\tin^{t_1} \Bigr)
    \end{aligned}
    \\*
    & \! \begin{aligned}[b]
    {} = \smash{\int} \mathcal{D}\phi^{+} \mathcal{D}\phi^{-} \,
    \expectval{\phi^{+} | \rho | \phi^{-}} \, & \phi^{-}(x_1) \phi^{+}(x_2)
    \\*
    {} \times {} &
    \exp\Bigl(\im S[\phi^{+}]_\tin^\tf - \im S[\phi^{-}]_\tin^{\tf\,{*}} \Bigr)
    \smash{ \bigg\rvert }_{\substack{\hspace{-0.7em} \phi^{+}(\tf,\evec{x}) \\ \,{=}\, \phi^{-}(\tf,\evec{x})}}
    \text{.}
    \end{aligned}
    \label{eq:two-point_function_as_functional_integral}
\end{align}
The final time parameter $\tf$ here must be chosen so that $\tf \geq \max(t_1, t_2)$ but otherwise it is arbitrary.
Note that we have written the second equality leading to~\cref{eq:two-point_function_as_functional_integral} explicitly for the case $t_1 > t_2$ but the final result is valid more generally when $\tin < t_i < \tf$ with $i = 1,2$.
Also, the states $\lvert \phi^\pm \rangle$ in the final result correspond to the field configurations $\phi^\pm(\tin,\evec{x})$.

The result~\cref{eq:two-point_function_as_functional_integral} for the two-point function indeed contains two time branches: a forward branch for $\phi^{+}$ and a backward branch for $\phi^{-}$. Both functional integrations are independent except for the boundary condition which identifies the fields at the final time. Thus, the result can be interpreted as a single functional integral along the CTP shown in figure~\cref{fig:Schwinger-Keldysh_path} \emph{if} we also specify at which branch the fields at $x_1$ and $x_2$ are evaluated on~\cite{Calzetta:2008cam}. Finally, we note that the form of equation~\cref{eq:two-point_function_as_functional_integral} with the two time branches and the initial density matrix $\expectval{\phi^{+} | \rho | \phi^{-}}$ is generic to any $n$-point correlation function, and not only to the two-point function used in this example.

\subsection*{Generating functional}

The development of the CTP formalism then proceeds along the lines of conventional QFT where one defines the generating functional of the $n$-point functions. The generating functional for non-equilibrium systems, also called the in-in generating functional, can be defined as~\cite{Calzetta:1986cq}
\begin{equation}
    Z[J_\pm; \rho] \eqdef \Tr\biggl\{
        \rho \,
        \bar{\mathcal{T}} \exp\biggl[-\im \int_\tin^\tf \dd^4 x\, J_{-}(x) \hat\phi(x)\biggr] \,
        \mathcal{T} \exp\biggl[\im \int_\tin^\tf \dd^4 x\, J_{+}(x) \hat\phi(x)\biggr]
    \biggr\} \text{.}
    \label{eq:non-eq_generating_functional}
\end{equation}
There are contributions from both branches of the CTP which have their own external source and time-ordering: the forward branch with $J_{+}$ and usual time-ordering $\mathcal{T}$ and the backward branch with $J_{-}$ and reversed time-ordering $\bar{\mathcal{T}}$.

The generating functional of connected $n$-point functions is defined as usual as the logarithm
\begin{equation}
    W[J_\pm; \rho] \eqdef -\im \log Z[J_\pm; \rho] \text{.}
    \label{eq:non-eq_connected_generating_functional}
\end{equation}
The physical $n$-point functions can be generated from~\cref{eq:non-eq_connected_generating_functional} by functional differentiation with respect to the sources and taking the limit of vanishing sources in the end, similarly to conventional QFT. However, there are now multiple ways to generate a given $n$-point function because the amount of sources is doubled. As an example, we show how to generate the one-point function and also the two-point function~\cref{eq:two-point_function_example} used in the previous example:
\begin{subequations}
\label{eq:generating_function_examples}
\begin{align}
    -\left.\frac{\delta W[J; \rho]}{\delta J_-(x)}\right\vert_{J \to 0} =
    \left.\frac{\delta W[J; \rho]}{\delta J_+(x)}\right\vert_{J \to 0} &=
    \expectval[\big]{\hat\phi(x)} \text{,}
    \label{eq:generating_function_example-1}
    \\*
    \im \left.\frac{\delta^2 W[J; \rho]}{\delta J_+(x_2) \delta J_-(x_1)}\right\vert_{J \to 0} &=
    \expectval[\big]{\hat\phi(x_1) \hat\phi(x_2)} \text{.}
    \label{eq:generating_function_example-2}
\end{align}
\end{subequations}
In the calculation of~\cref{eq:generating_function_examples} we used the normalization of density operators, $\Tr\rho = 1$~\cite{Calzetta:2008cam}, which implies that $Z[0; \rho] = 1$. Note that equation~\cref{eq:generating_function_example-2} gives just one of four different two-point functions which can be generated from~\cref{eq:non-eq_connected_generating_functional}. Not all of them will be independent functions however, as we will see later on.

The generating functional~\cref{eq:non-eq_generating_functional} admits a path integral form which can be derived similarly to equation~\cref{eq:two-point_function_as_functional_integral}. The result is~\cite{Calzetta:1986cq,Calzetta:2008cam}
\begin{align}
    Z[J_\pm; \rho] &=
    \begin{aligned}[t]
    \smash{\int} \mathcal{D}\phi^{+} \mathcal{D}\phi^{-} \, &
    \expectval{\phi^{+} | \rho | \phi^{-}} \,  \notag
    \\*
    {} \times {} &
    \exp\Bigl[
        \im \Bigl( S[\phi^{+}]_\tin^\tf + J_{+} \phi^{+} \Bigr)
        - \im \Bigl( S[\phi^{-}]_\tin^{\tf\,{*}} + J_{-} \phi^{-} \Bigr)
    \Bigr]
    \smash{ \bigg\rvert }_{\substack{\hspace{-0.7em} \phi^{+}(\tf,\evec{x}) \\ \,{=}\, \phi^{-}(\tf,\evec{x})}}
    \end{aligned}
    \\*
    &\eqdef
    \int_{\rm c} \mathcal{D}\phi^a \, \expectval{\phi^{+} | \rho | \phi^{-}} \,
    \exp\Bigl[ \im \Bigl( S_{\rm c}[\phi^a]_\tin^\tf + J^a \phi^a \Bigr) \Bigr] \text{,}
    \label{eq:non-eq_generating_functional_PI-form}
\end{align}
where we also introduced some often used notation for the CTP components. In this notation $a = {\pm}$ is the CTP branch index, repeated indices are implicitly summed over, and one defines a ``metric'' $c^{ab} = c_{ab} \eqdef \diag(1,-1)$ using which the indices may be raised or lowered~\cite{Calzetta:1986cq}.
Furthermore, we suppress the spacetime arguments and integrations where appropriate, so that for example $J_\pm \phi^\pm \eqdef \int_\tin^\tf \dd^4 x \, J_\pm(x) \phi^\pm(x)$.
We also defined the shorthand $S_{\rm c}[\phi^a]_\tin^\tf \eqdef S[\phi^{+}]_\tin^\tf - S[\phi^{-}]_\tin^{\tf\,{*}}$ for the CTP action, and the subscript `$\mathrm{c}$' in the integral designates the closure condition $\phi^{+}(\tf,\evec{x}) = \phi^{-}(\tf,\evec{x})$.

\subsubsection{Non-local sources}

An important property of the initial density matrix $\expectval{\phi^{+} | \rho | \phi^{-}}$ is that it is a functional of the field configurations $\phi^\pm(\tin,\evec{x})$. It can then be expanded functionally using the field configurations as~\cite{Calzetta:1986cq,Berges:2004yj}
\begin{equation}
    \expectval{\phi^{+} | \rho | \phi^{-}} =
    \exp\Bigl[\im\bigl(
        K + K_a \phi^a + \sfrac{1}{2!}K_{ab} \phi^a \phi^b + \sfrac{1}{3!} K_{abc} \phi^a \phi^b \phi^c + \cdots
    \bigr)\Bigr] \text{.}
    \label{eq:initial_density_matrix_parametrization}
\end{equation}
The constant $K$ and the infinite sequence of functions $K_a$, $K_{ab}$, $K_{abc}$, $\ldots$ parametrize the density operator $\rho$.%
\footnote{More precisely: $K_a(x) = \widetilde K_a(\evec{x}) \delta(x^0 - \tin)$, $K_{ab}(x,y) = \widetilde K_{ab}(\evec{x},\evec{y}) \delta(x^0 - \tin)  \delta(y^0 - \tin)$, \etc}
They have support only at the initial time, and hence they can be viewed as initial-time sources.
Note that we used here the notation defined below equation~\cref{eq:non-eq_generating_functional_PI-form}: repeated CTP indices are implicitly summed over and $K_{ab} \phi^a \phi^b$ denotes $\int_\tin^\tf \dd^4 x \dd^4 y \, K_{ab}(x,y) \phi^a(x) \phi^b(y)$, for example.

We can now write the original non-equilibrium generating functional~\cref{eq:non-eq_generating_functional} in a remarkably simple form. By using equation~\cref{eq:initial_density_matrix_parametrization} in~\cref{eq:non-eq_generating_functional_PI-form} and absorbing the constant $K$ to the normalization of the integral measure, we get~\cite{Calzetta:1986cq,Berges:2000ur}
\begin{align}
    Z[J_\pm; \rho] &= Z[\mathfrak{J}_a, K_{ab}, K_{abc}, \ldots] \notag
    \\*
    &\eqdef \int_{\rm c} \mathcal{D}\phi^a \, \exp\Bigl[
        \im \Bigl( S_{\rm c}[\phi^a]_\tin^\tf + \mathfrak{J}_a \phi^a + \sfrac{1}{2} K_{ab} \phi^a \phi^b
        + \sfrac{1}{3!} K_{abc} \phi^a \phi^b \phi^c + \cdots \Bigr)
    \Bigr] \text{,}
    \label{eq:non-eq_generating_functional_PI-form-2}
\end{align}
where $\mathfrak{J}_a \eqdef J^a + K_a = c^{ab} J_b + K_a$. Equation~\cref{eq:non-eq_generating_functional_PI-form-2} is an exact result and it demonstrates that a general non-equilibrium field theory can be formally described in the CTP formalism by including all possible $n$-point sources in the generating functional.%
\footnote{Note that in~\cite{Berges:2004vw,Berges:2004yj,Berges:2015kfa} already the operator representation of $Z$ is given with the non-local sources, in contrast to~\cite{Calzetta:1986cq}. We have used the simpler approach in~\cref{eq:non-eq_generating_functional}, leading to~\cref{eq:non-eq_generating_functional_PI-form-2}.}
Of course, the infinite number of initial-time sources is only necessary when describing arbitrary density matrices. For specific situations like a vacuum or a thermal system one can make additional simplifications. For example, in the case of a Gaussian initial state only the sources $K_a$ and $K_{ab}$ are necessary~\cite{Berges:2015kfa}.

%
\section{Generalized effective actions}
%

Now that we have the non-equilibrium generating functional~\cref{eq:non-eq_generating_functional_PI-form} and a path integral representation for the $n$-point functions we could try to evaluate them in specific theories using the conventional methods of perturbative QFT. Here one runs into problems however. For example, perturbative expansions of the non-equilibrium $n$-point functions contain so-called secular terms~\cite{Berges:2004yj} which grow with increasing powers of time at each order, and this ruins the expansion even for small coupling constants~\cite{Berges:2003pc,Berges:2004vw}. Description of thermalization and late-time universality also generally require non-linear dynamics and conserved charges (\eg~energy conservation) which are usually absent in more conventional approaches~\cite{Berges:2004yj,Berges:2004vw,Berges:2015kfa}.

While the secularity problem can be solved by resummation, a more sophisticated and efficient solution is to use a generalized effective action from which equations for the $n$-point functions can be derived by a variational principle~\cite{Berges:2004yj,Berges:2004pu,Brown:2015zla}. The non-equilibrium generating functional~\cref{eq:non-eq_generating_functional_PI-form-2} and its logarithm $W$ serve as the starting point for this method. The generalized effective action is defined as the multiple Legendre transform of $W[J_a, K_{ab}, K_{abc}, \ldots]$ with respect to the infinite sequence of sources $J_a$, $K_{ab}$, $K_{abc}$, $\ldots$~\cite{Calzetta:1986cq}. Stationarity conditions of the effective action then lead to an infinite set of coupled equations of motion for the full $n$-point functions, which is analogous to the BBGKY hierarchy in classical statistical mechanics~\cite{Calzetta:1986cq,Calzetta:2008cam}. This hierarchy of equations then fully describes the non-equilibrium evolution of the system.

In practice the infinite hierarchy is impossible to solve exactly in realistic interacting theories and one needs to use approximations. First of all, the hierarchy of equations can be truncated by only taking into account the $n$-point sources up to some finite $n$ when doing the Legendre transformation~\cite{Calzetta:1986cq,Calzetta:2008cam,Berges:2004yj}. This leads to the $n$-particle irreducible effective action ($n$PIEA). In classical mechanics and thermodynamics different Legendre transforms lead to different but equivalent descriptions of the same physics. The $n$PIEAs for different $n \in \naturals$ are in this sense also equivalent when no other approximations are used~\cite{Berges:2015kfa}. However, further approximations \emph{are} usually needed.

For a given $n$PIEA one also usually needs some approximative method to calculate the effective action itself. This can be done, for example, by truncating a loop expansion or by using a $1/N$-expansion in theories with $N$ field components~\cite{Berges:2003pc,Berges:2015kfa}. When approximated this way the different $n$PIEAs are no longer equivalent but form equivalence hierarchies~\cite{Berges:2004yj,Berges:2004pu}. In practice different $n$PIEAs have restrictions on what kind of dynamics and initial states they can adequately describe.
Which $n$PIEA to choose then depends largely on the problem at hand and the order of the approximation. Typically the 2PI, the 3PI or the 4PI effective action already provides a complete description of the problem in feasible approximations~\cite{Berges:2004yj}.

Next we turn to the 2PIEA method which is used to derive the non-equilibrium evolution equations used in this work. The 2PIEA method~\cite{Brown:2015zla} was developed originally for non-relativistic many-body theory in~\cite{Lee:1960zza,Luttinger:1960ua,Baym:1962sx} and later in the functional formulation for relativistic QFT in~\cite{Cornwall:1974vz}. Its adaptation to non-equilibrium QFT using the CTP was given in~\cite{Chou:1984es,Calzetta:1986cq}. We present some details on how the formalism is derived, following mostly~\cite{Calzetta:2008cam}, culminating in the derivation of the general equation of motion for the two-point function.

%
\subsection{2PI effective action on the CTP}
\label{sec:CTP-2PIEA}
%

The starting point for the CTP 2PIEA method~\cite{Calzetta:2008cam} is the generating functional
\begin{equation}
    Z[J_a, K_{ab}] \eqdef \int_{\rm c} \mathcal{D}\phi^a \exp\Bigl[
        \im \Bigl( S_{\rm c}[\phi^a]_\tin^\tf + J_a \phi^a + \sfrac{1}{2} K_{ab} \phi^a \phi^b \Bigr)
    \Bigr] \text{,}
    \label{eq:2PI-generating-functional}
\end{equation}
which is a truncation of the general non-equilibrium case~\cref{eq:non-eq_generating_functional_PI-form-2}. The notation was defined below  equation~\cref{eq:non-eq_generating_functional_PI-form}. As in the previous section, we continue to use real scalar field theory as an example on the development of the formalism. The generating functional~\cref{eq:2PI-generating-functional} is equivalent to the full $Z[J_\pm; \rho]$ only for Gaussian initial density matrices or a vacuum initial state, and otherwise it must be considered an approximation. The effects of non-trivial initial correlations tend to be short-lived however~\cite{Garny:2015oza}, so this should be a sound approximation when the focus is on the dynamical evolution rather than the intricacies of the initial conditions.

The connected generating functional $W[J_a, K_{ab}] \eqdef -\im \log Z[J_a, K_{ab}]$ is defined like in equation~\cref{eq:non-eq_connected_generating_functional}, but there is now a difference when the physical $n$-point functions are calculated from $W$. When taking the limit of vanishing sources also the initial density matrix gets discarded in the process, which is also reflected in that $Z[0, 0] \neq 1$. But because the initial density matrix~\cref{eq:initial_density_matrix_parametrization} has support only at $t = \tin$ one can compensate for this loss of information by adjusting the initial conditions of the $n$-point functions suitably~\cite{Berges:2004yj}. Hence, the physical $n$-point functions can still be calculated by the standard method where the sources are removed in the end.

To formulate the 2PIEA we still need to keep the sources, of course. We first define the mean field $\bar\phi^a(x)$ and the fluctuation propagator $G^{ab}(x,y)$ by
\begin{subequations}
\label{eq:mean_field_and_G_def}
\begin{align}
    \bar\phi^a(x) &\eqdef \frac{\delta W[J, K]}{\delta J_a(x)} \text{,}
    \\*
    G^{ab}(x,y) &\eqdef 2 \frac{\delta W[J, K]}{\delta K_{ab}(x,y)} -
    \bar\phi^a(x) \bar\phi^b(y) \text{.}
    \label{eq:2PIEA_connected_propagator_def}
\end{align}
\end{subequations}
In the limit $J,K \to 0$ both of the mean fields $\bar\phi^\pm$ reduce to the physical one-point function. The function $G^{ab}$ describes the propagation of the field fluctuations around the mean value. It thus reduces to the physical two-point function when both the mean field and the sources vanish. The 2PIEA is defined as the double Legendre transform
\begin{align}
    \Gamma_{\rm 2PI}\bigl[\bar\phi^a, G^{ab}\bigr] &\eqdef
    W[J_a, K_{ab}] - J_a \frac{\delta W}{\delta J_a} - K_{ab} \frac{\delta W}{\delta K_{ab}} \notag
    \\*
    &= W[J_a, K_{ab}] - J_a \bar\phi^a - \frac{1}{2} K_{ab} \Bigl(G^{ab} + \bar\phi^a \bar\phi^b \Bigr)
    \text{.} \label{eq:2PIEA-def}
\end{align}
The meaning of the Legendre transform here is that we switch from the functional $W$ and its independent variables (functions) $J, K$ to the effective action $\Gamma_{\rm 2PI}$ and $\bar\phi, G$. The inverse relations of~\cref{eq:mean_field_and_G_def} are then given by
\begin{subequations}
\label{eq:2PI_stationarity_with_sources}
\begin{align}
    \frac{\delta \Gamma_{\rm 2PI}}{\delta \bar\phi^a} &= -J_a - K_{ab} \bar\phi^b \text{,}
    \\*
    \frac{\delta \Gamma_{\rm 2PI}}{\delta G^{ab}} &= -\frac{1}{2} K_{ab} \text{.}
\end{align}
\end{subequations}
In the first equation we also used that for a real scalar field the two-point source is symmetric: $K_{ab}(x,y) = K_{ba}(y,x)$. This follows from the corresponding property of the propagator $G^{ab}(x,y)$.

Equations~\cref{eq:2PI_stationarity_with_sources} give the stationarity conditions of the effective action in the limit of vanishing sources. That is, eventually they yield the equations of motion for the full one- and two-point functions. To this end, we need to eliminate the sources $J, K$ from $\Gamma_{\rm 2PI}$. Exponentiating equation~\cref{eq:2PIEA-def}, using~\cref{eq:2PI-generating-functional,eq:2PI_stationarity_with_sources} and shifting the functional integral variable to the fluctuation field $\varphi^a \eqdef \phi^a - \bar\phi^a$ leads to the compact equation
\begin{equation}
    \e^{\im \Gamma_{\rm 2PI}} = \int_{\rm c} \mathcal{D}\varphi^a \exp\biggl[
        \im \biggl(
            S_{\rm c}\bigl[\bar\phi^a + \varphi^a\bigr]_\tin^\tf
            - \frac{\delta \Gamma_{\rm 2PI}}{\delta \bar\phi^a} \varphi^a
            - \frac{\delta \Gamma_{\rm 2PI}}{\delta G^{ab}} \bigl(\varphi^a \varphi^b - G^{ab}\bigr)
        \biggr)
    \biggr] \text{.}
    \label{eq:2PIEA-implicit-form}
\end{equation}
Equation~\cref{eq:2PIEA-implicit-form} gives an exact self-contained formula for $\Gamma_{\rm 2PI}$ in terms of $\bar\phi, G$ only. If we could solve $\Gamma_{\rm 2PI}$ from it we could then calculate the variations on the left-hand sides of equations~\cref{eq:2PI_stationarity_with_sources} and take the limit $J,K \to 0$ to get the equations of motion. This is of course not possible to do exactly for realistic interacting theories so some approximative method is needed.

\subsubsection{Expansion of the 2PIEA}

We will now use a loop expansion to calculate $\Gamma_{\rm 2PI}$.%
\footnote{The expansion parameter, number of loops, technically corresponds to powers of $\hbar$~\cite{Cornwall:1974vz}.}
To proceed from equation~\cref{eq:2PIEA-implicit-form} we must extract the lowest order parts of $\Gamma_{\rm 2PI}$. This can be done by using the background field method~\cite{Calzetta:2008cam} where the shifted action $S_{\rm c}[\bar\phi^a + \varphi^a]_\tin^\tf$ is functionally expanded around the mean field:
\begin{equation}
    S_{\rm c}\bigl[\bar\phi^a + \varphi^a\bigr] =
    S_{\rm c}\bigl[\bar\phi^a]
    + \frac{\delta S_{\rm c}[\bar\phi^a]}{\delta \bar\phi^b} \varphi^b
    + \frac{1}{2!} \frac{\delta^2 S_{\rm c}[\bar\phi^a]}{\delta \bar\phi^b \delta \bar\phi^c} \varphi^b \varphi^c
    + S_2\bigl[\varphi^a; \bar\phi^a\bigr] \text{.}
    \label{eq:fluctuation_field_expansion_of_action}
\end{equation}
Here $S_2[\varphi^a; \bar\phi^a]$ is defined as the rest of the terms which are \emph{cubic} or higher order in the fluctuation field $\varphi$; the reason for this notation will become clear below. Next we define the function~\cite{Berges:2004yj}
\begin{equation}
    \im G_{0,ab}^{-1}(x,y) \eqdef
    \frac{\delta^2 S_{\rm c}[\bar\phi^c]_\tin^\tf}{\delta \bar\phi^a(x) \delta \bar\phi^b(y)}
    \text{,}
\end{equation}
which is a generalization of the inverse free propagator in the presence of the mean field. The quadratic term in equation~\cref{eq:fluctuation_field_expansion_of_action} can then be written as $\sfrac{\im}{2} G_{0,ab}^{-1} \varphi^a \varphi^b$.

As an effective quantum action, $\Gamma_{\rm 2PI}$ should contain the classical CTP action plus quantum corrections: $\Gamma_{\rm 2PI} = S_{\rm c}[\bar\phi^a] + \mathcal{O}(\hbar)$~\cite{Calzetta:2008cam}. We can already verify the classical part by using equation~\cref{eq:fluctuation_field_expansion_of_action} in~\cref{eq:2PIEA-implicit-form}; the term $S_{\rm c}[\bar\phi^a]$ can be factored out of the functional integral. We will now construct the rest of the terms. The first quantum correction is the ``one-loop part'', that is, the Gaussian functional integral which corresponds to the functional determinant of the propagator:
\begin{equation}
    \int_{\rm c} \mathcal{D}\varphi^a \exp\Bigl(
        -\sfrac{1}{2} G_{ab}^{-1} \varphi^a \varphi^b
    \Bigr)
    \propto (\Det G^{-1})^{-\frac{1}{2}} \text{.}
    \label{eq:functional_determinant}
\end{equation}
To get this contribution to the right-hand side of equation~\cref{eq:2PIEA-implicit-form} we need to add a term to $\Gamma_{\rm 2PI}$ which yields $-\sfrac{\im}{2} G_{ab}^{-1}$ from the variation $\delta \Gamma_{\rm 2PI}/\delta G^{ab}$. The correct term is given by the functional identity $\delta \Tr[\log G^{-1}]/\delta G^{ab} = -G_{ab}^{-1}$, which leads to $\Gamma_{\rm 2PI} = S_{\rm c}[\bar\phi^a] + \sfrac{\im}{2} \Tr[\log G^{-1}] + \mathrm{const.} + \mathcal{O}(\hbar^2)$. The addition of the one-loop term also introduces an infinite constant $-\sfrac{\im}{2} \Tr\idmat \eqdef -\sfrac{\im}{2} G_{ab}^{-1} G^{ab} = -\im \int_\tin^\tf \dd^4 x \, \delta^{(4)}(0)$ which, however, does not affect the stationarity conditions~\cite{Calzetta:2008cam}.
Lastly, we also add an extra term to $\Gamma_{\rm 2PI}$ to cancel the quadratic part of equation~\cref{eq:fluctuation_field_expansion_of_action} in~\cref{eq:2PIEA-implicit-form}. The term is $\sfrac{\im}{2} \Tr[G_{0}^{-1} G] = \sfrac{\im}{2} G_{0,ab}^{-1} G^{ab}$ and it is required for the stationarity conditions to eventually produce the correct Schwinger--Dyson equation for the propagator.

The final parametrization of the 2PIEA is then~\cite{Cornwall:1974vz,Berges:2004yj,Calzetta:2008cam}
\begin{equation}
    \Gamma_{\rm 2PI}\bigl[\bar\phi^a, G^{ab}\bigr] =
    S_{\rm c}\bigl[\bar\phi^a\bigr]_\tin^\tf + \sfrac{\im}{2} \Tr\bigl[G_{0}^{-1} G\bigr] + \sfrac{\im}{2} \Tr\bigl[\log G^{-1}\bigr]
    + \Gamma_2\bigl[\bar\phi^a, G^{ab}\bigr] + \text{const.} \quad
    \label{eq:2PIEA-parametrization-with-Gamma2}
\end{equation}
Equation~\cref{eq:2PIEA-parametrization-with-Gamma2} is essentially a definition for the non-trivial part $\Gamma_2$, that is, the higher-order quantum corrections. Due to the construction we expect it to consist of diagrams with two or more loops. To derive an equation for $\Gamma_2$ we first calculate the variations
\begin{subequations}
\label{eq:2PIEA_proto-equations-of-motion}
\begin{align}
    \frac{\delta \Gamma_{\rm 2PI}}{\delta \bar\phi^a} &=
    \frac{\delta S_{\rm c}[\bar\phi]}{\delta \bar\phi^a}
    + \frac{1}{2} \frac{\delta^3 S_{\rm c}[\bar\phi]}{\delta \bar\phi^a \delta \bar\phi^b \delta \bar\phi^c} G^{bc}
    + \frac{\delta \Gamma_2}{\delta \bar\phi^a} \text{,}
    \\*
    \frac{\delta \Gamma_{\rm 2PI}}{\delta G^{ab}} &=
    \frac{\im}{2} G_{0,ab}^{-1}
    - \frac{\im}{2} G_{ab}^{-1}
    + \frac{\delta \Gamma_2}{\delta G^{ab}} \text{.}
    \label{eq:2PIEA_proto-SD-equation}
\end{align}
\end{subequations}
By using equations~\cref{eq:2PIEA-parametrization-with-Gamma2,eq:fluctuation_field_expansion_of_action,eq:2PIEA_proto-equations-of-motion} (and the identity $\log[\Det G^{-1}] = \Tr[\log G^{-1}]$) in~\cref{eq:2PIEA-implicit-form}  we can then derive the result
\begin{subequations}
\begin{gather}
    \begin{aligned}[b]
    \e^{\im \Gamma_{\rm 2}} = (\Det G^{-1})^{\frac{1}{2}}
    \int_{\rm c} \mathcal{D}\varphi^a \exp\biggl[
        \im \biggl(&
            \sfrac{\im}{2} G_{ab}^{-1} \varphi^a \varphi^b
            + S_2\bigl[\varphi^a; \bar\phi^a\bigr]
    \\*
            & {} - \widetilde J_a \varphi^a
            - \widetilde K_{ab}\bigl(\varphi^a \varphi^b - G^{ab}\bigr)
        \smash{ \biggr) }
    \smash{ \biggr] } \text{,}
    \end{aligned} \label{eq:Gamma2_implicit_formula}
    \\*
\shortintertext{where}
    \widetilde J_a \eqdef \frac{\delta \Gamma_2}{\delta \bar\phi^a} +
    \frac{1}{2} \frac{\delta^3 S_{\rm c}[\bar\phi]}{\delta \bar\phi^a \delta \bar\phi^b \delta \bar\phi^c} G^{bc}
    \text{,} \hspace{3em}
    \widetilde K_{ab} \eqdef \frac{\delta \Gamma_2}{\delta G^{ab}} \text{.}
    \label{eq:fixed_Gamma2_sources}
\end{gather}
\end{subequations}
This is a self-contained equation for $\Gamma_2$ in terms of $\bar\phi, G$. At first glance it seems even more cryptic than equation~\cref{eq:2PIEA-implicit-form}, but it actually has a very specific meaning.

It turns out that $\im \Gamma_2$ consists of the \emph{2PI vacuum diagrams} where the propagator lines correspond to $G$, that is, the \emph{full} propagator of the original theory, and the interaction vertices are given by $S_2[\varphi^a; \bar\phi^a]$~\cite{Cornwall:1974vz,Berges:2004yj,Calzetta:2008cam}. This is a consequence of the constraining sources $ \widetilde J_a$ and $\widetilde K_{ab}$ in equation~\cref{eq:Gamma2_implicit_formula}. To understand this, consider a system with a generating functional corresponding to the functional integral in~\cref{eq:Gamma2_implicit_formula}. The form of the sources is precisely such that the exact connected propagator in this system is fixed to $G$ and the exact one-point function vanishes~\cite{Cornwall:1974vz,Jackiw:1974cv}. This can be shown from the invertibility of the Legendre transformation by taking variations of $\Gamma_2$ with respect to $\bar\phi$ and $G$~\cite{Calzetta:2008cam}. Furthermore, because the one-loop part was already extracted, the diagrams contained in $\Gamma_2$ have two or more loops, as expected.%
\footnote{The $\Det G$-factor in equation~\cref{eq:Gamma2_implicit_formula} cancels to lowest order due to~\cref{eq:functional_determinant}.}
%

\subsubsection{Physical equations of motion}

We can now finally obtain the equations of motion for the (full) physical one- and two-point functions. Combining equations~\cref{eq:2PIEA_proto-equations-of-motion} and~\cref{eq:2PI_stationarity_with_sources} and taking the limit $J,K \to 0$ results in coupled non-linear equations for $\bar\phi^a$ and $G^{ab}$ in terms of the actions $S_{\rm c}[\bar\phi^a]$ and $\Gamma_2[\bar\phi^a, G^{ab}]$. We are mainly interested in the propagator equation which can be written as
\begin{align}
    G_{ab}^{-1}(x,y) &= G_{0,ab}^{-1}(x,y) - \Pi_{ab}^{}(x,y) \text{,}
    \label{eq:2PIEA_SD-equation}
    \\*
\shortintertext{with}
    \Pi_{ab}(x,y) &\eqdef 2\im \frac{\delta \Gamma_2[\bar\phi, G]}{\delta G^{ab}(x,y)}
    \text{.} \label{eq:2PIEA_self-energy}
\end{align}
Note that equation~\cref{eq:2PIEA_SD-equation} is generally a very complicated non-linear equation for $G$ because of the dependence on $\Gamma_2$.

Now, we know that $G$ is the full connected propagator because of its definition~\cref{eq:2PIEA_connected_propagator_def}. This means that equation~\cref{eq:2PIEA_SD-equation} is nothing else but its Schwinger--Dyson equation. Therefore $\Pi_{ab}$, given by~\cref{eq:2PIEA_self-energy}, must also be equal to the proper 1PI self-energy (\ie~it does not contain propagator insertions)~\cite{Cornwall:1974vz}. There are thus two different perspectives for $\Pi_{ab}$: it is the (whole) perturbative series of proper 1PI self-energy diagrams (with internal lines corresponding to $G_0$) and, on the other hand, it can be calculated from $\Gamma_2$ (in terms of the full $G$).%
\footnote{This provides another way to see that $\Gamma_2$ can only contain 2PI diagrams (with internal lines corresponding to $G$) as otherwise $\Pi_{ab}$ would contain propagator insertions~\cite{Cornwall:1974vz,Berges:2004yj}.}
This is one of the main points of the 2PIEA method: a large class of diagrams is automatically resummed in $\Gamma_2$.
The 2PIEA method is however more than just a resummation scheme as the equations of motion are derived \emph{self-consistently} from a variational principle~\cite{Brown:2015zla}. The method is consistent with global symmetries of the theory and it provides an efficient and systematic way to derive approximations~\cite{Berges:2004pu}.

The main 2PIEA results which we use in this work are equation~\cref{eq:2PIEA_SD-equation} and the method to calculate the self-energy~\cref{eq:2PIEA_self-energy} from the 2PI loop expansion. Although equation~\cref{eq:2PIEA_SD-equation} has the form of the usual Schwinger--Dyson equation, the viewpoint is now very different compared to the standard perturbative approach:
equation~\cref{eq:2PIEA_SD-equation} is a non-linear equation of motion for the full propagator.

\subsubsection{Generalization to other types of fields}

So far we have considered only a real scalar field theory when we introduced the CTP and the 2PIEA and derived the equations of motion in this chapter. Already the derivation of the path integral form of the propagator, given by equation~\cref{eq:two-point_function_as_functional_integral}, is much more complicated for other types of fields.
The formulation of the 2PIEA method is also considerably more complicated in more realistic situations, such as in gauge theories~\cite{Reinosa:2007vi,Calzetta:2004sh}. Furthermore, in an interacting theory with different types of fields (\eg~quantum electrodynamics) one generally has to consider in the intermediate steps also all mixed propagators (like \eg~a photon-fermion propagator) and all mixed non-local sources to get the correct equations of motion~\cite{Reinosa:2007vi}.

In these more realistic cases the final \emph{physical} equations of motion are luckily not so complicated and they are mostly analogous to the real scalar field case. The main differences are that the 2PIEA parametrization~\cref{eq:2PIEA-parametrization-with-Gamma2} contains additional one-loop and inverse free propagator terms for each degree of freedom separately, the functional determinant has a different numerical coefficient depending on the type of field (like in standard QFT), and the propagators and self-energies have more complicated symmetry properties in the spacetime-arguments and CTP indices~\cite{Cornwall:1974vz,Berges:2002wr,Prokopec:2003pj,Reinosa:2007vi}.
We will not give these results explicitly but from now on just quote the relevant parts when needed. As a final remark we point out that in these more general situations the non-trivial part $\Gamma_2$ of the 2PIEA is still calculated as the sum of all 2PI vacuum diagrams but it is then a functional of all physical propagators and mean fields of the system. Also, when all mean fields of the system vanish, the action $S_2$ (used for calculating $\Gamma_2$) consists of the same interaction vertices as the classical CTP action of the theory.

%
\section{Non-equilibrium evolution equations}
%

In this section we present the non-equilibrium evolution equations based on the Schwinger--Dyson equation~\cref{eq:2PIEA_SD-equation} derived in the CTP 2PIEA formalism. From now on we are working with vanishing mean fields in which case the non-equilibrium evolution of the system is completely described by the propagators. We first give the definitions of the CTP propagators, one of which was already used as an example in equations~\cref{eq:two-point_function_example,eq:generating_function_example-2}. Then we present the general evolution equations which are written in the form of Kadanoff--Baym equations~\cite{Baym:1961zz,Kadanoff:1962book}.

%
\subsection{Closed time path propagators}
\label{sec:CTP_propagators}
%

The CTP propagators with real time arguments can be conveniently packaged into one object by using contour-ordered (complex) time arguments on the CTP~\cite{LeBellac:1996cam,Prokopec:2003pj}. Using this ``contour notation'', we define the propagators of a fermion field $\psi$ and a complex scalar boson field $\phi$ as~\cite{Prokopec:2003pj}
\begin{subequations}
\label{eq:CTP_propagator_contour_definitions}
\begin{align}
    \im S(u,v) &\eqdef \expectval[\big]{
        \mathcal{T}_\mathcal{C}\bigl[ \psi(u) \bar\psi(v) \bigr]
    } \text{,} \label{eq:CTP_fermion_propagator}
    \\*
    \im \Delta(u,v) &\eqdef \expectval[\big]{
        \mathcal{T}_\mathcal{C}\bigl[ \phi(u) \phi^\dagger(v) \bigr]
    } \text{,}
\end{align}
\end{subequations}
where $u^0,v^0$ lie on the CTP and $\mathcal{T}_\mathcal{C}$ denotes the contour-ordering (indicated in figure~\cref{fig:Schwinger-Keldysh_path}).%
\footnote{We use an explicit imaginary unit in the definitions~\cref{eq:CTP_propagator_contour_definitions}. This is different from the convention used for $G$ in section~\cref{sec:CTP-2PIEA} which is what is usually used in the 2PIEA-literature.}
Also, $\bar\psi \eqdef \psi^\dagger \gamma^0$ denotes the usual conjugated fermion field, and we now suppress the hats on the operators.
The expectation values are taken with respect to the non-equilibrium density operator of the system as in equation~\cref{eq:two-point_function_example}. In the definitions~\cref{eq:CTP_propagator_contour_definitions} the ordering of any discrete indices of the fields, such as flavor or spinor indices, follows the ordering of the spacetime arguments. For example, the fermion propagator~\cref{eq:CTP_fermion_propagator} reads $\im S_{\alpha\beta}(u,v) = \expectval{\mathcal{T}_\mathcal{C}[ \psi_\alpha(u) \bar\psi_\beta(v)]}$ when written explicitly with the Dirac spinor indices.

The component propagators with the CTP branch indices $a,b = \pm$ and real time arguments can be found by evaluating the time-ordering in equations~\cref{eq:CTP_propagator_contour_definitions} for the different cases. There are four of these real-time propagators and we denote them as
\begin{subequations}
\label{eq:fermionic_CTP_propagators}
\begin{alignat}{2}
    S^\lt &\eqdef -S^{+-} \text{,}
    &
    \im S^{+-}_{\alpha\beta}(u,v) &\eqdef
    -\expectval[\big]{\bar\psi_\beta(v) \psi_\alpha(u)} \text{,}
    \label{eq:fermionic_Wightman_function_def}
    \\*
    S^\gt &\eqdef S^{-+} \text{,} \hspace{4em}
    &
    \im S^{-+}_{\alpha\beta}(u,v) &\eqdef
    \expectval[\big]{\psi_\alpha(u) \bar\psi_\beta(v)} \text{,}
    \\*
    S^{\rm T} &\eqdef S^{++} \text{,}
    &
    \im S^{++}_{\alpha\beta}(u,v) &\eqdef
    \expectval[\big]{\mathcal{T}\bigl[\psi_\alpha(u) \bar\psi_\beta(v)\bigr]} \text{,}
    \\*
    S^{\bar{\rm T}} &\eqdef S^{--} \text{,}
    &
    \im S^{--}_{\alpha\beta}(u,v) &\eqdef
    \expectval[\big]{\bar{\mathcal{T}}\bigl[\psi_\alpha(u) \bar\psi_\beta(v)\bigr]} \text{,}
\end{alignat}
\end{subequations}
for the fermion and
\begin{subequations}
\label{eq:bosonic_CTP_propagators}
\begin{alignat}{2}
    \Delta^\lt &\eqdef \Delta^{+-} \text{,}
    &
    \im \Delta^{+-}(u,v) &\eqdef
    \expectval[\big]{\phi^\dagger(v) \phi(u)} \text{,}
    \\*
    \Delta^\gt &\eqdef \Delta^{-+} \text{,} \hspace{4em}
    &
    \im \Delta^{-+}(u,v) &\eqdef
    \expectval[\big]{\phi(u) \phi^\dagger(v)} \text{,}
    \\*
    \Delta^{\rm T} &\eqdef \Delta^{++} \text{,}
    &
    \im \Delta^{++}(u,v) &\eqdef
    \expectval[\big]{\mathcal{T}\bigl[\phi(u) \phi^\dagger(v)\bigr]} \text{,}
    \\*
    \Delta^{\bar{\rm T}} &\eqdef \Delta^{--} \text{,}
    &
    \im \Delta^{--}(u,v) &\eqdef
    \expectval[\big]{\bar{\mathcal{T}}\bigl[\phi(u) \phi^\dagger(v)\bigr]} \text{,}
\end{alignat}
\end{subequations}
for the scalar~\cite{Prokopec:2003pj,Calzetta:2008cam}. The symbols $\mathcal{T}$ and $\bar{\mathcal{T}}$ denote the usual time-ordering and reversed time-ordering metaoperators, and the minus sign in the definition of $S^{+-}$ is due to the fermionic time-ordering. The functions $S^{\lt,\gt}$ and $\Delta^{\lt,\gt}$ are called the \emph{Wightman functions}, and we use the convention where the fermionic function $S^\lt$ is defined without the sign: $\im S^\lt_{\alpha\beta}(u,v) = \expectval[\big]{\bar\psi_\beta(v) \psi_\alpha(u)}$ (\cf~\cite{Prokopec:2003pj}).

The different propagators in equations~\cref{eq:fermionic_CTP_propagators} and in~\cref{eq:bosonic_CTP_propagators} are not independent, as we briefly mentioned already below equations~\cref{eq:generating_function_examples}. Writing the time-orderings explicitly we get directly
\begin{subequations}
\label{eq:Feynman_and_Dyson_propagator_relations}
\begin{alignat}{2}
    & S^{\rm T}(u,v) &&= \hphantom{-}\theta(u^0 - v^0) S^\gt(u,v) - \theta(v^0 - u^0) S^\lt(u,v)
    \text{,}
    \\*
    & S^{\bar{\rm T}}(u,v) &&= -\theta(v^0 - u^0) S^\lt(u,v) + \theta(v^0 - u^0) S^\gt(u,v)
    \text{,}
\end{alignat}
\end{subequations}
and similarly for the scalar propagators. Here it is evident that only two of the propagators, say $S^\gt$ and $S^\lt$, are independent. It is a general feature that two independent propagators (per field) are needed to describe non-equilibrium systems~\cite{Chou:1984es,Berges:2004yj}. However, the functions we have given so far are not optimal for arranging and solving the equations of motion and often various other combinations of the propagators are introduced. Next we give these for the fermion field; the bosonic case is similar and can be obtained by replacing first $S^\lt \rightarrow -S^\lt$ and then $S \rightarrow \Delta$.

\subsubsection{Profusion of propagators}

An often used pair of independent propagators are the statistical propagator $S^{\rm F}$ and the spectral function $S^{\mathcal{A}}$~\cite{Berges:2004yj}, which we define as
\begin{equation}
    S^{\rm F} \eqdef \frac{1}{2} \bigl(S^\gt - S^\lt\bigr) \text{,}
    \hspace{3em}
    S^{\mathcal{A}} \eqdef \frac{\im}{2} \bigl(S^\gt + S^\lt\bigr) \text{.}
    \label{eq:stat_and_spectral_propagator_def}
\end{equation}
Statistical observables of the system, such as occupation numbers or current densities, can be extracted from $S^{\rm F}$. The spectral function $S^{\mathcal{A}}$ describes the spectrum of the system and it is constrained by the canonical equal-time (anti)commutation relations. It does not contain direct information on the state~\cite{Berges:2004yj,Calzetta:2008cam}.

The spectral function $S^{\mathcal{A}}$ can be used to further define the retarded and advanced pole propagators
\begin{subequations}
\label{eq:ret_and_adv_propagator_def}
\begin{alignat}{2}
    &\im S^r(u,v) &&\eqdef \hphantom{-}2\theta(u^0 - v^0) S^\mathcal{A}(u,v)
    \text{,}
    \\*
    &\im S^a(u,v) &&\eqdef            -2\theta(v^0 - u^0) S^\mathcal{A}(u,v)
    \text{.}
\end{alignat}
\end{subequations}
They are useful for deriving kinetic equations in the limit $\tin \to -\infty$ of the CTP. In this context it is also convenient to define the ``Hermitian part'' $S^{\rm H}$ of the pole propagators which satisfies $S^{r,a} \eqdef S^{\rm H} \mp \im S^\mathcal{A}$ where the minus (plus) sign corresponds to $S^r$ ($S^a$). Using the relations~\cref{eq:Feynman_and_Dyson_propagator_relations} and the definitions of the different propagators we can then derive the following general identities:
\begin{subequations}
\label{eq:CTP_propagator_relations}
\begin{alignat}{2}
    S^{\rm H} &= \frac{1}{2} \bigl(S^r + S^a\bigr) &&= \frac{1}{2} \bigl(S^{\rm T} - S^{\bar{\rm T}}\bigr) \text{,}
    \\*
    S^{\mathcal{A}} &= \frac{\im}{2} \bigl(S^r - S^a\bigr) &&= \frac{\im}{2} \bigl(S^\gt + S^\lt\bigr) \text{,}
    \\*
    S^{\rm F} &= \frac{1}{2} \bigl(S^{\rm T} + S^{\bar{\rm T}}\bigr) &&= \frac{1}{2} \bigl(S^\gt - S^\lt\bigr) \text{.}
\end{alignat}
\end{subequations}
These relations are useful for converting between different formulations of the equations and in practical calculations of the self-energies.%
\footnote{Also, the \emph{spectral relation} $S^{\rm H}(u,v) = -\im \sgn(u^0 - v^0) S^{\mathcal{A}}(u,v)$ follows directly from~\cref{eq:ret_and_adv_propagator_def}.}

We also make frequent use of the Hermiticity conditions of the various propagators, which follow directly from the definitions. For the fermion these are
\begin{subequations}
\label{eq:propagator_hermiticity_properties}
\begin{align}
    \bar S^{\lt,\gt}(u,v)^\dagger &= \bar S^{\lt,\gt}(v,u)
    \text{,}
    \\
    \bar S^{{\rm H},\mathcal{A}}(u,v)^\dagger &= \bar S^{{\rm H},\mathcal{A}}(v,u)
    \text{,}
    \\
    \bar S^{r,a}(u,v)^\dagger &= \bar S^{a,r}(v,u)
    \text{,} \label{eq:pole_propagator_hermiticity}
\end{align}
\end{subequations}
where we also defined the barred propagators $\bar S^{\lt,\gt} \eqdef \im S^{\lt,\gt} \gamma^0$, $\bar S^{r,a} \eqdef S^{r,a} \gamma^0$ and $\bar S^{{\rm H},\mathcal{A}} \eqdef S^{{\rm H},\mathcal{A}} \gamma^0$. Note that the pole propagators $S^{r,a}$ are swapped under Hermitian conjugation in equation~\cref{eq:pole_propagator_hermiticity}. The scalar propagators satisfy similar Hermiticity relations with the matrix $\gamma^0$ omitted.

%
\subsection{General CTP equations of motion}
\label{sec:general_eom}
%

The generalizations of the 2PI equation of motion~\cref{eq:2PIEA_SD-equation} for the fermion and complex scalar boson are~\cite{Berges:2002wr,Prokopec:2003pj,Reinosa:2007vi}
\begin{subequations}
\label{eq:fermionic_and_bosonic_CTP_SD-equations}
\begin{align}
    S^{-1}_{ab}(x,y) &= S^{-1}_{0,ab}(x,y) - \Sigma_{ab}(x,y) \text{,}
    \label{eq:fermionic_CTP_SD-equation}
    \\*
    \Delta^{-1}_{ab}(x,y) &= \Delta^{-1}_{0,ab}(x,y) - \Pi_{ab}(x,y) \text{,}
    \label{eq:bosonic_CTP_SD-equation}
\end{align}
\end{subequations}
with the self-energies
\begin{subequations}
\label{eq:2PIEA_fermion_and_boson_self-energies}
\begin{align}
    \Sigma_{ab}(x,y) &\eqdef -\im \frac{\delta \Gamma_2}{\delta S^{ba}(y,x)} \text{,}
    \label{eq:2PIEA_fermion_self-energy}
    \\*
    \Pi_{ab}(x,y) &\eqdef \im \frac{\delta \Gamma_2}{\delta \Delta^{ba}(y,x)} \text{.}
\end{align}
\end{subequations}
The different numerical factors in the definitions~\cref{eq:2PIEA_fermion_and_boson_self-energies} compared to the real scalar case~\cref{eq:2PIEA_self-energy} reflect the different scalings of the Gaussian path integrals in the fermionic and complex scalar cases. In the case of vanishing mean fields we can write the inverse free propagators explicitly as~\cite{Prokopec:2003pj,Berges:2015kfa}
\begin{subequations}
\begin{align}
    S^{-1}_{0,ab}(x,y) &\eqdef \bigl(\im \slashed\partial_x - m_\psi \bigr) \delta^{(4)}(x - y) \, c_{ab}
    \text{,} \label{eq:fermionic_inverse_free_propagator}
    \\*
    \Delta^{-1}_{0,ab}(x,y) &\eqdef \bigl(-\partial_x^2 - m_\phi^2\bigr) \delta^{(4)}(x - y) \, c_{ab}
    \text{,}
\end{align}
\end{subequations}
where $m_\psi$ and $m_\phi$ are the masses of the fermion and boson fields, respectively.%
\footnote{We also assumed here the standard form of the classical action in the Minkowski spacetime.}

From now on we only consider explicitly the fermionic case as we only need the fermionic equation~\cref{eq:fermionic_CTP_SD-equation} further in this work. More details on the bosonic case can be found in~\cite{Prokopec:2003pj}, for example.

\subsubsection{Self-energy functions}

The four real-time CTP self-energies~\cref{eq:2PIEA_fermion_self-energy} are similar to the propagators~\cref{eq:fermionic_CTP_propagators}. One important difference however is that the self-energies may contain a singular part~\cite{Prokopec:2003pj}. We can extract it by defining
\begin{equation}
    \Sigma_{ab}(x,y) \eqdef c_{ab} \, \delta^{(4)}(x - y) \Sigma^{\rm sg}(x)
    + \widetilde\Sigma_{ab}(x,y) \text{,}
    \label{eq:CTP_self-energy_singular_part_extraction}
\end{equation}
where $\widetilde\Sigma_{ab}$ are non-singular. Note that the singular part has the same structure as the mass term in~\cref{eq:fermionic_inverse_free_propagator}. Hence, it can be absorbed to the inverse free propagator in equation~\cref{eq:fermionic_CTP_SD-equation} as a spacetime dependent mass shift~\cite{Prokopec:2003pj,Berges:2015kfa}. We do this by \emph{formally} replacing the mass $m$ with the effective mass
\begin{equation}
    \widetilde m(x) \eqdef m + \Sigma^{\rm sg}(x) \text{.}
    \label{eq:effective_fermion_mass}
\end{equation}
Now we may further define the non-singular self-energy functions
\begin{equation}
    \Sigma^\lt \eqdef - \Sigma^{+-} \text{,} \quad
    \Sigma^\gt \eqdef \Sigma^{-+}  \text{,} \quad
    \Sigma^{\rm T} \eqdef \widetilde\Sigma^{++}  \text{,} \quad
    \Sigma^{\bar{\rm T}} \eqdef \widetilde\Sigma^{--} \text{,}
    \label{eq:fermion_CTP_self_energies}
\end{equation}
analogously to~\cref{eq:fermionic_CTP_propagators}. Note that in these definitions the CTP branch indices are raised instead of lowered which would be the ``natural'' position according to equation~\cref{eq:2PIEA_fermion_self-energy}.

Relations between the self-energy functions~\cref{eq:fermion_CTP_self_energies} depend on the approximation scheme used to calculate the 2PIEA. In reasonable approximations the following relations, which are analogous to~\cref{eq:Feynman_and_Dyson_propagator_relations}, are expected to hold:~\cite{Prokopec:2003pj}
\begin{subequations}
\begin{alignat}{2}
    & \Sigma^{\rm T}(u,v) &&= \hphantom{-}\theta(u^0 - v^0) \Sigma^\gt(u,v) - \theta(v^0 - u^0) \Sigma^\lt(u,v)
    \text{,}
    \\*
    & \Sigma^{\bar{\rm T}}(u,v) &&= -\theta(v^0 - u^0) \Sigma^\lt(u,v) + \theta(v^0 - u^0) \Sigma^\gt(u,v)
    \text{.}
\end{alignat}
\end{subequations}
Assuming this is the case we can then define the rest of the functions $\Sigma^{{\rm F},{\mathcal{A}}}$, $\Sigma^{r,a}$ and $\Sigma^{\rm H}$ like in section~\cref{sec:CTP_propagators} so that \emph{all} of the relations~\cref{eq:stat_and_spectral_propagator_def,eq:ret_and_adv_propagator_def,eq:CTP_propagator_relations} hold analogously also for the self-energies. The self-energy functions should also satisfy the same Hermiticity conditions~\cref{eq:propagator_hermiticity_properties} as the corresponding propagators.
However, we define the barred self-energies slightly differently compared to the propagators, with the matrix $\gamma^0$ on the other side: $\bar \Sigma^{\lt,\gt} \eqdef \im \gamma^0 \Sigma^{\lt,\gt}$, $\bar \Sigma^{r,a} \eqdef \gamma^0 \Sigma^{r,a}$ and $\bar \Sigma^{{\rm H},\mathcal{A}} \eqdef \gamma^0 \Sigma^{{\rm H},\mathcal{A}}$. This is convenient when manipulating the equations later on.

\subsubsection{Kadanoff--Baym equations}

In practice the Schwinger--Dyson equations in the form~\cref{eq:fermionic_and_bosonic_CTP_SD-equations} must be inverted to solve them. The equivalent integro-differential form can be obtained by contracting the equations from the right by the full propagators. Equation~\cref{eq:fermionic_CTP_SD-equation} then becomes
\begin{equation}
    \bigl(\im \slashed\partial_x - m \bigr) S^{ab}(x,y)
    = c^{ab} \delta^{(4)}(x - y) + \sum_{c = {\pm}} c \int_\tin^\tf \dd^4 z \, \Sigma^{ac}(x,z) S^{cb}(z,y) \text{.}
    \label{eq:fermionic_integro-differential_SD-equation}
\end{equation}
This is a coupled set of non-linear partial integro-differential equations for the four real-time CTP propagators~\cref{eq:fermionic_CTP_propagators}. The form~\cref{eq:fermionic_integro-differential_SD-equation} of the equations is still not very convenient for practical purposes because the statistical and spectral sectors are mixed together. This is where the various propagators defined in section~\cref{sec:CTP_propagators} (and the corresponding self-energies) come in handy. Next we give three different equivalent formulations of equations~\cref{eq:fermionic_integro-differential_SD-equation} which are all useful in their own contexts. We also introduce the notation for the generalized convolution integral
\begin{equation}
    (F \convol G)_{t_1}^{t_2}(x,y) \eqdef \int_{t_1}^{t_2} \dd^4 z \, F(x,z) G(z,y)
    \label{eq:generalized_convolution_direct_space}
\end{equation}
and the effective inverse free propagator
\begin{equation}
    \widetilde S_0^{-1}(x,y) \eqdef
    \bigl[\im \slashed\partial_x - \widetilde m(x) \bigr] \delta^{(4)}(x - y)
\end{equation}
to make for a more streamlined notation in the subsequent equations.

\paragraph{First formulation}

This is the most economic formulation with respect to the number of functions involved. It is possible to write a set of equations equivalent to~\cref{eq:fermionic_integro-differential_SD-equation} using only the statistical propagator $S^{\rm F}$ and the spectral function $S^{\mathcal{A}}$:
\begin{subequations}
\label{eq:fermionic_SD-equation_F-A-formulation}
\begin{align}
    \bigl(\widetilde S_0^{-1} \convol S^{\mathcal{A}}\bigr)_\tin^\tf(x,y)
    &= -2\im \bigl(\Sigma^{\mathcal{A}} \convol S^{\mathcal{A}}\bigr)_{y^0}^{x^0}(x,y) \text{,}
    \\*
    \bigl(\widetilde S_0^{-1} \convol S^{\rm F}\bigr)_\tin^\tf(x,y)
    &= -2\im \bigl(\Sigma^{\mathcal{A}} \convol S^{\rm F}\bigr)_\tin^{x^0}(x,y)
    + 2\im \bigl(\Sigma^{\rm F} \convol S^{\mathcal{A}}\bigr)_\tin^{y^0}(x,y) \text{,}
    \\*
    S^{\mathcal{A}}(x,y) \, \delta(x^0 - y^0) &= \frac{1}{2}\gamma^0 \, \delta^{(4)}(x - y) \text{.}
    \phantom{\bigr)_{y^0}^{x^0}}
    \label{eq:fermionic_sum_rule_in_SD-form}
\end{align}
\end{subequations}

\paragraph{Second formulation}

Here the statistical and spectral propagators are replaced by the Wightman functions $S^s$ (with $s = {<}, {>}$) and the pole propagators $S^p$ (with $p = r,a$), respectively:
\begin{subequations}
\label{eq:general_fermionic_KB_and_pole_eqs}
\begin{align}
    \bigl[\bigl(\widetilde S_0^{-1} - \Sigma^p\bigr) \convol S^p\bigr]_\tin^\tf(x,y)
    &= \delta^{(4)}(x - y) \text{,}
    \label{eq:general_fermionic_pole-equations}
    \\*
    \bigl[\bigl(\widetilde S_0^{-1} - \Sigma^r\bigr) \convol S^s\bigr]_\tin^\tf(x,y)
    &= \bigl(\Sigma^s \convol S^a\bigr)_\tin^\tf(x,y) \text{.}
    \label{eq:general_fermionic_KB-equations}
\end{align}
\end{subequations}

\paragraph{Third formulation}

The last formulation is a direct variant of equations~\cref{eq:general_fermionic_KB_and_pole_eqs} where the pole propagators have been traded for $S^{\rm H}$ and $S^{\mathcal{A}}$:
\begin{subequations}
\label{eq:general_fermionic_KB_and_pole-eqs_variant}
\begin{alignat}{3}
    \bigl[\bigl(\widetilde S_0^{-1} - \Sigma^{\rm H}\bigr) \convol & S^{\rm H} &&{}\bigr]_\tin^\tf(x,y)
    + \bigl(\Sigma^{\mathcal{A}} \convol S^{\mathcal{A}} \bigr)_\tin^\tf(x,y)
    &&= \delta^{(4)}(x - y) \text{,} \label{eq:general_fermionic_pole-eqs_var1}
    \\*
    \bigl[\bigl(\widetilde S_0^{-1} - \Sigma^{\rm H}\bigr) \convol & S^{\mathcal{A}} &&{}\bigr]_\tin^\tf(x,y)
    - \bigl(\Sigma^{\mathcal{A}} \convol S^{\rm H}\bigr)_\tin^\tf(x,y)
    &&= 0 \text{,} \label{eq:general_fermionic_pole-eqs_var2}
    \\*
    \bigl[\bigl(\widetilde S_0^{-1} - \Sigma^{\rm H}\bigr) \convol & S^s &&{}\bigr]_\tin^\tf(x,y)
    - \bigl(\Sigma^s \convol S^{\rm H}\bigr)_\tin^\tf(x,y)
    &&= \pm \mathcal{C}_{\rm coll}(x,y) \text{,} \label{eq:general_fermionic_KB-eqs_variant}
\end{alignat}
\end{subequations}
where the ${\pm}$ corresponds to $s = {\lessgtr}$, and
\begin{equation}
    \mathcal{C}_{\rm coll}(x,y) \eqdef \frac{1}{2} \bigl(
        \Sigma^\gt \convol S^\lt - \Sigma^\lt \convol S^\gt
    \bigr)_\tin^\tf(x,y) \text{.}
    \label{eq:general_KB-eq_collision_term}
\end{equation}
Equations~\cref{eq:general_fermionic_KB-eqs_variant} are known as the Kadanoff--Baym (KB) equations where~\cref{eq:general_KB-eq_collision_term} is the collision term~\cite{Danielewicz:1982kk,Prokopec:2003pj}.

\medskip\noindent
All three sets of equations~\cref{eq:fermionic_SD-equation_F-A-formulation,eq:general_fermionic_KB_and_pole_eqs,eq:general_fermionic_KB_and_pole-eqs_variant} are separately equivalent to~\cref{eq:fermionic_integro-differential_SD-equation} and hence also to each other.
As we can see the original four equations~\cref{eq:fermionic_integro-differential_SD-equation} contained redundant information: the amount of equations was reduced to three in equations~\cref{eq:fermionic_SD-equation_F-A-formulation}. Furthermore, equation~\cref{eq:fermionic_sum_rule_in_SD-form} is just the position-space representation of the \emph{spectral sum rule}~\cite{LeBellac:1996cam} which can also be derived directly from the canonical anticommutation relations.%
\footnote{Nevertheless, the result~\cref{eq:fermionic_SD-equation_F-A-formulation} shows the consistency of the formalism: the canonical (anti)commutation relations are built-in even though they were not explicitly used in the derivation.}
In practice the sum rule fixes the equal-time values of the spectral function, including the value at the initial two-time point $(\tin,\tin)$. This shows that the initial conditions of the system are not directly reflected in the spectral function.

The form~\cref{eq:fermionic_SD-equation_F-A-formulation} of the equations is useful because the structure of the convolution integrals shows that the equations are \emph{causal}: only the past values of the propagators determine their future values~\cite{Berges:2004yj}. Hence, the convolution integrals are also often called ``memory integrals'' as they integrate over the history of the time evolution.
Being manifestly causal, equations~\cref{eq:fermionic_SD-equation_F-A-formulation} are also convenient for direct numerical calculations~\cite{Berges:2004yj}. However, in realistic applications a full numerical implementation and solution can be very demanding because the complexity of the equations increases considerably with more degrees of freedom.

A commonly used approach to simplify the equations is to Fourier transform them into a mixed position-momentum representation. It is then possible to perform a gradient expansion to get more tractable equations~\cite{Prokopec:2003pj,Berges:2015kfa}. In equations~\cref{eq:fermionic_SD-equation_F-A-formulation} the time integration limits are finite and depend on the external arguments $x^0, y^0$ however, and this complicates the use of Fourier transforms.
Equations~\cref{eq:general_fermionic_KB_and_pole_eqs} and~\cref{eq:general_fermionic_KB_and_pole-eqs_variant} avoid this problem as all of the integration limits are extended to $\tin$ and $\tf$ (which can then be eventually taken to $-\infty$ and $+\infty$).
In these two formulations there are again four equations but in practice not all of them are needed due to the relations~\cref{eq:CTP_propagator_relations}.

Equations~\cref{eq:general_fermionic_KB-eqs_variant} [or~\cref{eq:general_fermionic_KB-equations}] are called the Kadanoff--Baym equations and equations~\cref{eq:general_fermionic_pole-eqs_var1,eq:general_fermionic_pole-eqs_var2} [or~\cref{eq:general_fermionic_pole-equations}] are the pole equations. The KB equations are the main statistical evolution equations while the pole equations describe the dynamics of the spectral and dispersive  properties of the system.
The variant~\cref{eq:general_fermionic_KB_and_pole_eqs} of the equations is more compact to write down but equations~\cref{eq:general_fermionic_KB_and_pole-eqs_variant} are physically more transparent because all of the interaction terms have clear interpretations. This makes the latter variant more convenient for spectral quasiparticle approximations, for example. The terms in~\cref{eq:general_fermionic_KB_and_pole-eqs_variant} which explicitly contain $\Sigma^{\rm H}$ and $\Sigma^{\mathcal{A}}$ describe the dispersive corrections and the width of the propagators, respectively. Also the $\Sigma^s \convol S^{\rm H}$-term is related to the width~\cite{Herranen:2008hu,Kainulainen:2021oqs}. The collision term $\mathcal{C}_{\rm coll}$ on the right-hand side of the KB equation consists of generalized gain and loss terms which describe the equilibration process~\cite{Prokopec:2003pj}.

The KB equations can be used as a basis for deriving simpler kinetic equations which describe the evolution of particle distribution functions.
For example, the equations can be reduced to semiclassical Boltzmann equations~\cite{Danielewicz:1982kk,Buchmuller:2000nd,Lindner:2005kv,Drewes:2012qw}.
In the context of kinetic transport theory the limit $\tin \to -\infty$ is taken for the CTP contour which enables the use of the mixed representation and the gradient expansion.%
\footnote{The upper limit $\tf$ of the CTP can always be taken to be arbitrarily large, see~equation~\cref{eq:two-point_function_as_functional_integral}.}
This limit also discards the effects of the initial conditions set at $\tin$. The propagators are then prescribed at some finite time $t_0$ instead. This is a suitable approximation for systems with efficient equilibration processes (such as the electroweak plasma)~\cite{Prokopec:2003pj} or when solving the equations for late times~\cite{Lindner:2005kv,Berges:2015kfa}.

\subsubsection{Wigner representation}

We now define the mixed momentum-position representation, called the Wigner representation. It is useful when considering kinetic equations but it is valid also more generally. The equations can be transformed to this representation by using the \emph{Wigner transform} (with $\tin \to -\infty$)~\cite{Prokopec:2003pj,Berges:2004yj}
\begin{equation}
    F(k,x) \eqdef \int_{-\infty}^{\infty} \dd^4 r \e^{\im k \cdot r} \,
    F\bigl(x + \sfrac{r}{2}, x - \sfrac{r}{2}\bigr) \text{,}
    \label{eq:general_Wigner_transform}
\end{equation}
where $F(u,v)$ is any two-point (or self-energy) function in the position representation.%
\footnote{We use the same notation for functions in both the position and the Wigner representation as these can be inferred from context.}
The inverse transform is given by
\begin{equation}
    F(u,v) \eqdef \int \frac{\dd^4 k}{(2\pi)^4} \e^{-\im k \cdot (u - v)} \,
    F\Bigl(k, \frac{u + v}{2}\Bigr) \text{.}
    \label{eq:general_Wigner_transform_inverse}
\end{equation}
Equation~\cref{eq:general_Wigner_transform} is essentially a Fourier transform in the relative coordinate $r \eqdef u - v$. In this transform the dependence on the pair of spacetime points $(u,v)$ is exchanged to the four-momentum $k$ and the average coordinate $x \eqdef (u + v)/2$. In order to transform the equations of motion we also need the Wigner transform of the generalized convolution~\cref{eq:generalized_convolution_direct_space} (where $t_{1,2} \to \mp\infty$). The result is~\cite{Calzetta:2008cam,Berges:2015kfa}
\begin{equation}
    (F \convol G)_{-\infty}^{+\infty}(k,x) = \e^{-\im\Diamond} \bigl\{F(k,x)\bigr\} \bigl\{G(k,x)\bigr\}
    \label{eq:convolution_Wigner_transform}
    \text{,}
\end{equation}
where the right-hand side is called the Moyal product~\cite{Moyal:1949sk,Groenewold:1946kp} and we defined the derivative operator $\Diamond \eqdef \sfrac{1}{2}(\partial^{(1)}_{x\vphantom{k}} \cdot \partial^{(2)}_k - \partial^{(1)}_k \cdot \partial^{(2)}_{x\vphantom{k}})$~\cite{Prokopec:2003pj}. Note that the only meaning of the curly brackets in~\cref{eq:convolution_Wigner_transform} is to denote the first and second ``slot'' on which the derivatives $\partial^{(1)}$ and $\partial^{(2)}$ operate.

The pole and KB equations~\cref{eq:general_fermionic_KB_and_pole_eqs} can now be written straightforwardly in the Wigner representation by using equations~\cref{eq:general_Wigner_transform,eq:general_Wigner_transform_inverse,eq:convolution_Wigner_transform}:
\begin{subequations}
\label{eq:pole_and_KB-equations_Wigner}
\begin{alignat}{3}
    \bigl(\slashed k + \sfrac{\im}{2} \slashed\partial_x\bigr) & S^p &&{}
    - \e^{-\im\Diamond} \{ \widetilde\Sigma^p \}\{ S^p \} &&= \idmat
    \text{,}
    \\*
    \bigl(\slashed k + \sfrac{\im}{2} \slashed\partial_x\bigr) & S^s &&{}
    - \e^{-\im\Diamond} \{ \widetilde\Sigma^r \}\{ S^s \} &&=
    \e^{-\im\Diamond} \{ \Sigma^s \}\{ S^a \} \text{.}
    \label{eq:KB-equation_Wigner}
\end{alignat}
\end{subequations}
Here we denoted $\widetilde \Sigma^p(k,x) \eqdef \widetilde m(x) + \Sigma^p(k,x)$ for both $p = r,a$ as it is now more convenient to group the mass term with the self-energy [$\widetilde m$ was defined in~\cref{eq:effective_fermion_mass}]. We also suppressed the arguments $(k,x)$ of all propagators and self-energies here for brevity. In the Wigner representation the sum rule~\cref{eq:fermionic_sum_rule_in_SD-form} also takes the more usual form
\begin{equation}
    \int_{-\infty}^{\infty} \frac{\dd k^0}{\pi} \, \bar S^{\mathcal{A}}(k,x) = \idmat
    \text{.} \label{eq:fermionic_sum_rule_Wigner}
\end{equation}
Equations~\cref{eq:pole_and_KB-equations_Wigner} are still exact and non-linear and they are equivalent to the original equations (under the simplification $\tin \to -\infty$). Instead of the non-local convolution integrals these equations contain infinite series of exponential gradient terms. The advantage of this form is that outside of periods of rapid oscillations it can often be assumed that the functions change slowly in the average coordinate $x$~\cite{Prokopec:2003pj,Berges:2015kfa}. Higher order terms in the gradient expansion can then be discarded and this reduces the complexity of the equations.

\paragraph{Translational invariance}

An important special case is when the propagators and self-energy functions are fully \emph{translationally invariant} in their spacetime arguments and the mass $m$ is constant (here we denote $\widetilde m$ by simply $m$). This implies that the Wigner representation functions have no dependence on the average coordinate $x$. The full equations~\cref{eq:pole_and_KB-equations_Wigner} are in this case reduced to the algebraic matrix equations%
\begin{subequations}
\label{eq:pole_and_KB-equations_Wigner_transl_invariant}
\begin{align}
    \bigl(\slashed k - m - \Sigma^p_{\rm ti}(k)\bigr) S^p_{\rm ti}(k) &= \idmat \text{,}
    \\*
    \bigl(\slashed k - m - \Sigma^r_{\rm ti}(k)\bigr) S^s_{\rm ti}(k) &= \Sigma^s_{\rm ti}(k) S^a_{\rm ti}(k) \text{,}
\end{align}
\end{subequations}
which have the (formal) solutions
\begin{subequations}
\label{eq:transl_invariant_solutions}
\begin{align}
    S^p_{\rm ti}(k) &= \bigl(\slashed k - m - \Sigma^p_{\rm ti}(k)\bigr)^{-1} \text{,}
    \\*
    S^s_{\rm ti}(k) &= S^r_{\rm ti}(k)\Sigma^s_{\rm ti}(k) S^a_{\rm ti}(k) \text{.}
    \label{eq:translationally_invariant_Wightmans}
\end{align}
\end{subequations}

\paragraph{Thermal equilibrium}

If the system is also in \emph{thermal equilibrium} the Wightman functions~\cref{eq:translationally_invariant_Wightmans} satisfy the Kubo--Martin--Schwinger (KMS) condition $S^\gt_{\rm eq}(k) = \e^{\beta k^0} S^\lt_{\rm eq}(k)$~\cite{LeBellac:1996cam}, where $\beta \eqdef 1/T$. The KMS condition follows from the (anti)peri\-odicity of the field configurations in the standard way when the density operator $\rho$ has the equilibrium form~\cite{LeBellac:1996cam}. It may be described by adding the imaginary branch to the CTP~\cite{Calzetta:2008cam,Berges:2015kfa}. The thermal equilibrium solutions satisfy~\cref{eq:transl_invariant_solutions} and using the KMS condition the Wightman functions can be further simplified to
\begin{subequations}
\label{eq:thermal_eq_Wightmans}
\begin{align}
    \im S^\lt_{\rm eq}(k) &= 2S^{\mathcal{A}}_{\rm eq}(k) f_{\rm FD}(k^0) \text{,}
    \\*
    \im S^\gt_{\rm eq}(k) &= 2S^{\mathcal{A}}_{\rm eq}(k) \bigl(1 - f_{\rm FD}(k^0)\bigr) \text{.}
\end{align}
\end{subequations}
Here $f_{\rm FD}(k^0) \eqdef 1/(\e^{\beta k^0} + 1)$ is the Fermi--Dirac distribution.%
\footnote{Bosons have analogous solutions with the Bose--Einstein distribution $f_{\rm BE}(k^0) \eqdef 1/(\e^{\beta k^0} - 1)$~\cite{Jukkala:2021sku}.}
Equations~\cref{eq:thermal_eq_Wightmans} are equivalent to the fluctuation--dissipation relation $\im S^{\rm F}_{\rm eq}(k) = 2S^{\mathcal{A}}_{\rm eq}(k) \bigl(\sfrac{1}{2} - f_{\rm FD}(k^0)\bigr)$ which is satisfied only in equilibrium~\cite{Berges:2004yj}.
%
%
The Kadanoff--Baym ansatz, which is often used to derive kinetic equations, is a generalization of~\cref{eq:thermal_eq_Wightmans} where the equilibrium distribution function $f_{\rm FD}$ is replaced by a more general non-equilibrium function~\cite{Berges:2004yj,Lindner:2005kv}.

%
\subsection{Spatially homogeneous and isotropic case}
%

We now specialize to time-dependent systems which are spatially homogeneous and isotropic. This means that the propagators and self-energies have translational invariance in the spatial arguments and the singular contributions such as the mass $m$ can only depend on the time variable. This special case is relevant for cosmology (\eg~leptogenesis) and other time-dependent phenomena which happen uniformly throughout space.

In this case it is useful to introduce another transformation of the two-point functions and self-energies: the \emph{two-time representation}
\begin{equation}
    F_\evec{k}(t_1,t_2) \eqdef
    \int \dd^3 \evec{r} \e^{-\im\evec{k} \cdot \evec{r}} \,
    F\bigl((t_1, \evec{r}), (t_2, \evec{0})\bigr) \text{.}
    \label{eq:two-time_representation_def}
\end{equation}
This is the spatial Fourier transform with the three-momentum $\evec{k}$. Its inverse is
\begin{equation}
    F(u,v) \eqdef
    \int \frac{\dd^3 \evec{k}}{(2\pi)^3} \e^{\im\evec{k} \cdot (\evec{u} - \evec{v})} \,
    F_\evec{k}\bigl(u^0, v^0\bigr) \text{.}
    \label{eq:two-time_representation_inverse}
\end{equation}
In the representation~\cref{eq:two-time_representation_def} the pole and KB equations~\cref{eq:general_fermionic_KB_and_pole_eqs} become
\begin{subequations}
\label{eq:pole_and_KB-equations_twotime}
\begin{alignat}{2}
    & \bigl[
        (\widetilde S_{0,\evec{k}}^{-1} - \Sigma^p_\evec{k}) \convol S^p_\evec{k}
    \bigr]_\tin^\tf(t_1, t_2) &&= \delta(t_1 - t_2) \text{,}
    \label{eq:pole-equation_twotime}
    \\*
    & \bigl[
        (\widetilde S_{0,\evec{k}}^{-1} - \Sigma^r_\evec{k}) \convol S^s_\evec{k}
    \bigr]_\tin^\tf(t_1, t_2) &&=
    \bigl(\Sigma^s_\evec{k} \convol S^a_\evec{k}\bigr)_\tin^\tf(t_1, t_2) \text{,}
    \label{eq:KB-equation_twotime}
\end{alignat}
\end{subequations}
where
\begin{equation}
    \widetilde S_{0,\evec{k}}^{-1}(t_1, t_2) = \bigl[
        \im \gamma^0 \partial_{t_1} - \evec{\gamma} \cdot \evec{k} - \widetilde m(t_1)
    \bigr] \delta(t_1 - t_2) \text{,}
\end{equation}
and the convolutions are defined with only time integrals. The sum rule~\cref{eq:fermionic_sum_rule_in_SD-form} can be written in this case as
\begin{equation}
    2 \bar S^{\mathcal{A}}_{\evec{k}}(t,t) = \idmat \quad \text{for any $t$.}
    \label{eq:fermionic_sum_rule_twotime}
\end{equation}
Equations~\cref{eq:pole_and_KB-equations_twotime} are still very complicated, despite the reduction in the dimensions of the domain when compared to the more general equations~\cref{eq:general_fermionic_KB_and_pole_eqs}.

The Wigner representation~\cref{eq:general_Wigner_transform} can also be employed in the spatially homogeneous and isotropic case. The Wigner-transformed functions then have no dependence on the spatial average coordinate $\evec{x}$ and the corresponding spatial derivatives are dropped from equations~\cref{eq:pole_and_KB-equations_Wigner}. One can move between the Wigner and two-time representation (when $\tin \to -\infty$) by doing the Wigner transform~\cref{eq:general_Wigner_transform} and its inverse~\cref{eq:general_Wigner_transform_inverse} with only the time and energy variables (see~\cite{Jukkala:2021sku} for the explicit transformations). However, the two-time representation has the advantage of being more general as the initial time $\tin$ can be kept finite.

\subsubsection{Free time evolution}

We first consider the free theory and introduce the associated free time evolution operator before analyzing the solutions in the full theory. This serves as an example of the solutions and, more importantly, it elucidates the special role of the spectral function and an important local property of the free solutions.

The free theory equations corresponding to~\cref{eq:pole_and_KB-equations_twotime} are
\begin{subequations}
\label{eq:free_pole_and_KB-equations_twotime}
\begin{alignat}{2}
    & \bigl(S_{0,\evec{k}}^{-1} \convol S^p_{0,\evec{k}}\bigr)_\tin^\tf(t_1, t_2)
    &&= \delta(t_1 - t_2) \text{,}
    \label{eq:free_pole-equations_twotime}
    \\*
    & \bigl(S_{0,\evec{k}}^{-1} \convol S^s_{0,\evec{k}}\bigr)_\tin^\tf(t_1, t_2)
    &&= 0 \text{,}
    \label{eq:free_KB-equations_twotime}
\end{alignat}
\end{subequations}
with the free pole propagators $S_0^p$ and Wightman functions $S_0^s$. The equations are decoupled so it suffices to consider only the free KB equation~\cref{eq:free_KB-equations_twotime} which can be rewritten as
\begin{equation}
    \partial_{t_1} \bar S^s_{0,\evec{k}}(t_1,t_2) =
    -\im H_{\evec{k}}(t_1) \bar S^s_{0,\evec{k}}(t_1,t_2)
    \text{.}
    \label{eq:free_KB-equation_twotime_Hamiltonian_form}
\end{equation}
Here $H_{\evec{k}}(t)$ is the Hermitian free Dirac Hamiltonian and we used the barred propagator notation defined below equations~\cref{eq:propagator_hermiticity_properties}. The solutions to the free pole equations~\cref{eq:free_pole-equations_twotime} can be obtained from the spectral function $S^{\mathcal{A}}_{0,\evec{k}}$ after solving the Wightman functions from the free KB equation.

Equation~\cref{eq:free_KB-equation_twotime_Hamiltonian_form} has a form similar to the general Schrödinger equation and thus its solutions can be found similarly by using a time-evolution operator~\cite{LeBellac:2006qpb}. We define the time-evolution operator $U_{0,\evec{k}}(t_1, t_2)$ of equation~\cref{eq:free_KB-equation_twotime_Hamiltonian_form} as the unique solution with the initial condition $U_{0,\evec{k}}(\tin, \tin) = \idmat$.
It is given explicitly by the time-ordered exponential (also called the Dyson series)~\cite{Peskin:1995ev,Weinberg:1995qft1}
\begin{equation}
    U_{0,\evec{k}}(t_1, t_2) \eqdef \mathcal{T} \exp \biggl(
        -\im \int_{t_2}^{t_1} \dd t' \, H_{\evec{k}}(t')
    \biggr) \text{.}
\end{equation}
It has the properties that $U_{0,\evec{k}}(t, t) = \idmat$, $U_{0,\evec{k}}(t_1, t_2)^\dagger = U_{0,\evec{k}}(t_2, t_1)$ and
\begin{align}
    \bar S_{\evec{k}}(t_1, t_2) =  U_{0,\evec{k}}(t_1, t) \bar S_{\evec{k}}(t, t) U_{0,\evec{k}}(t, t_2)
    \text{,} \label{eq:free_KB-eq_general_solution}
\end{align}
for any $t, t_1, t_2 \geq \tin$ and where $\bar S_{\evec{k}}$ stands for any solution of equation~\cref{eq:free_KB-equation_twotime_Hamiltonian_form}. These properties mean that $U_{0,\evec{k}}$ describes unitary time-evolution. Equation~\cref{eq:free_KB-eq_general_solution} implies that the complete solution $\bar S_{\evec{k}}$ can be calculated from its local value at \emph{any} time $t$. In particular, the solution to equation~\cref{eq:free_KB-equation_twotime_Hamiltonian_form} can be obtained from the initial value $\bar S_{\evec{k}}(\tin, \tin)$ by setting $t = \tin$ in equation~\cref{eq:free_KB-eq_general_solution}. Furthermore, because the free spectral function $\bar S^{\mathcal{A}}_{0,\evec{k}}$ is also a solution to~\cref{eq:free_KB-equation_twotime_Hamiltonian_form} and because it satisfies the sum rule~\cref{eq:fermionic_sum_rule_twotime} it actually coincides with the free time-evolution operator:
\begin{equation}
    2 \bar S^{\mathcal{A}}_{0,\evec{k}}(t_1, t_2) \eqdef U_{0,\evec{k}}(t_1, t_2)
    \text{.} \label{eq:free_spectral_function}
\end{equation}
This result then yields the solutions to the free pole equations~\cref{eq:free_pole-equations_twotime} via the general definitions~\cref{eq:ret_and_adv_propagator_def,eq:two-time_representation_def}.

\subsubsection{Formal full solutions}

We now consider some exact properties of the full solutions to equations~\cref{eq:pole_and_KB-equations_twotime}. To this end, we first rewrite the equations in the equivalent integral equation form:
\begin{subequations}
\label{eq:integral_form_of_fermionic_KB_and_pole_eqs}
\begin{align}
    S^p_\evec{k} &= S_{0,\evec{k}}^p
    + S_{0,\evec{k}}^p \nconvol \Sigma^p_\evec{k} \nconvol S^p_\evec{k} \text{,}
    \\*
    S^s_\evec{k} &= S_{0,\evec{k}}^s
    + S_{0,\evec{k}}^s \nconvol \Sigma^a_\evec{k} \nconvol S^a_\evec{k}
    + S_{0,\evec{k}}^r \nconvol \Sigma^s_\evec{k} \nconvol S^a_\evec{k}
    + S_{0,\evec{k}}^r \nconvol \Sigma^r_\evec{k} \nconvol S^s_\evec{k} \text{.}
\end{align}
\end{subequations}
Here we suppressed the external time arguments $(t_1, t_2)$ of all terms and the time integration limits $\tin, \tf$ of the convolutions. Equations~\cref{eq:integral_form_of_fermionic_KB_and_pole_eqs} can be derived straightforwardly by writing the Schwinger--Dyson equations~\cref{eq:fermionic_integro-differential_SD-equation} in the integral equation form by contracting with the free propagator from the left. Equations~\cref{eq:integral_form_of_fermionic_KB_and_pole_eqs} then result from using the definitions~\cref{eq:fermionic_CTP_propagators} and relations~\cref{eq:CTP_propagator_relations} for the different propagators (and the corresponding results for the self-energies).

We can obtain exact formal solutions to the integral equations~\cref{eq:integral_form_of_fermionic_KB_and_pole_eqs} by suitably resumming them~\cite{Jukkala:2021sku,Greiner:1998vd}. The result is
\begin{subequations}
\label{eq:formal_solutions_twotime}
\begin{align}
    S^p_\evec{k} &= \bigl(S_{0,\evec{k}}^{-1} - \Sigma^p_\evec{k}\bigr)^{-1} \text{,}
    \label{eq:formal_pole_propagator_twotime}
    \\*
    S^s_\evec{k} &= S^s_{{\rm hom},\evec{k}} + S^s_{{\rm inh},\evec{k}} \text{,}
    \label{eq:formal_Wightman_twotime}
\end{align}
\end{subequations}
with the \emph{homogeneous} and \emph{inhomogeneous} Wightman functions
\begin{subequations}
\begin{align}
    S^s_{{\rm hom},\evec{k}} &\eqdef
    \bigl(S^r_\evec{k} \nconvol S_{0,\evec{k}}^{-1}\bigr)
    \nconvol S_{0,\evec{k}}^s \nconvol
    \bigl(S_{0,\evec{k}}^{-1} \nconvol S^a_\evec{k}\bigr)
    \text{,} \label{eq:homogeneous_Wightman}
    \\*
    S^s_{{\rm inh},\evec{k}} &\eqdef
    S^r_\evec{k} \nconvol \Sigma^s_\evec{k} \nconvol S^a_\evec{k}
    \text{.} \label{eq:inhomogeneous_Wightman}
\end{align}
\end{subequations}
The homogeneous solution~\cref{eq:homogeneous_Wightman} can be further simplified to
\begin{align}
    \bar S^s_{{\rm hom},\evec{k}}(t_1, t_2) =
    2\bar S^{\mathcal{A}}_\evec{k}(t_1, \tin)
    \bar S^s_{{\rm hom}, \evec{k}}(\tin, \tin)
    2\bar S^{\mathcal{A}}_\evec{k}(\tin, t_2)
    \text{,} \label{eq:homogeneous_Wightman_2}
\end{align}
when $t_1, t_2 \geq \tin$~\cite{Jukkala:2021sku}. It is a solution to the KB equation~\cref{eq:KB-equation_twotime} with a vanishing right-hand side. The inhomogeneous solution~\cref{eq:inhomogeneous_Wightman} is a particular solution to the full KB equation.
It is notable that the homogeneous solution~\cref{eq:homogeneous_Wightman_2} has the same structure as the free solution given by equations~\cref{eq:free_KB-eq_general_solution,eq:free_spectral_function}. Even though in the full interacting case there is no such unitary time-evolution operator, the homogeneous solution is still ``evolved'' from the initial value by the full spectral function. However, a big difference to the free case is that in interacting dissipative systems the spectral function is exponentially damped in the absolute difference of its time arguments~\cite{Jukkala:2021sku}. This means that the solution~\cref{eq:homogeneous_Wightman_2} is transient in nature and vanishes quickly when $t_1,t_2 \gg \tin$.

Despite the formal solutions~\cref{eq:formal_solutions_twotime} appearing simpler than equations~\cref{eq:integral_form_of_fermionic_KB_and_pole_eqs} they are generally still implicit non-linear integral equations because the self-energies generally depend on the Wightman functions $S^s$ and the pole propagators $S^p$. However, the exact equations~\cref{eq:formal_solutions_twotime} are useful as a starting point for deriving approximative solutions. This is so especially if the self-energies $\Sigma^s$ and $\Sigma^p$ are some externally given functions or if they can be approximated so that they do not depend on the full solutions $S^s$ and $S^p$. This is the case for example when the self-energy is dominated by a thermal equilibrium part. We will use this approach and put these results to use when we present the \emph{local approximation method} in the next chapter.

%% file: chapters/ch4_coherent_QKEs.tex
%
\chapter{Coherent quantum kinetic equations}
\label{chap:cQPA_and_LA-method}
%

The KB equations~\eqref{eq:general_fermionic_KB_and_pole_eqs} describe the full non-equilibrium dynamics of fermions with all quantum effects, including quantum coherence. The price to pay for this generality is that the equations are very difficult to solve in realistic situations as they are coupled non-linear integro-differential equations for the two-point correlation functions $S^\alpha$ (with $\alpha = s, p$). The convolution integrals make the equations inherently non-local, as the evolution of the functions depends not only on their values at a local neighborhood of a given point but also on the values of the functions arbitrarily far away. In the equivalent formulation~\eqref{eq:fermionic_SD-equation_F-A-formulation} of the equations this non-local time-evolution manifests explicitly as the causal memory integrals. In the Wigner representation~\cref{eq:pole_and_KB-equations_Wigner}, on the other hand, the non-locality is encoded in the exponential gradient operators, making the equations no easier to solve without approximations.

In this chapter we introduce two approximation methods to simplify the fermionic KB equations~\cref{eq:pole_and_KB-equations_twotime}: the cQPA and the more general LA-method. The goal of these methods is to obtain quantum kinetic equations (QKEs) which are straightforward to solve while still containing all relevant quantum coherence effects. Both methods ultimately revolve around making the equations local. There is also the problem of several types of time-evolution mixed together in the propagator (\eg, there are several types of coherence). Next we describe the parametrization used to separate these different components of the propagator before getting into the approximation methods themselves.

%
\section{Projection matrix parametrization}
%

An essential part of the local quantum kinetic methods presented in this chapter is the projection matrix parametrization~\cite{Herranen:2010mh,Fidler:2011yq,Jukkala:2019slc,Jukkala:2021sku}. The fermion two-point function generally has both a Dirac and a flavor matrix structure. The components of the propagator have complicated time dependencies because the propagator describes different physical quantities and properties such as the abundance of the particle species or the flavor and particle--antiparticle oscillations. The main purpose of the projection matrix parametrization is to efficiently separate the time-scales associated with the evolution of these different features.

The projection matrix parametrization can be motivated by the generic form of the \emph{free} dynamical equation
\begin{equation}
    \im \partial_t \mathcal{S}_\evec{k}(t) =
    \comm[\big]{H_\evec{k}(t), \mathcal{S}_\evec{k}(t)} \text{.}
    \label{eq:generic_free_dynamical_equation}
\end{equation}
Here $H_{\evec{k}}$ is the free Dirac Hamiltonian and $\mathcal{S}_\evec{k}$ represents the local value of a generic fermion two-point function. As we will see later, the relevant equations of motion will take the form~\cref{eq:generic_free_dynamical_equation} in both the two-time representation (the equal-time or local equation) and the Wigner representation (the $k^0$-integrated equation). We assume that equation~\cref{eq:generic_free_dynamical_equation} is given in the mass eigenbasis, that is, $H_\evec{k}(t) = \evec{\alpha} \cdot \evec{k} + \gamma^0 m(t)$ with $\evec{\alpha} \eqdef \gamma^0 \evec{\gamma}$ and a diagonal and positive-semidefinite mass matrix $m(t)$.

The projection matrix parametrization for the flavored two-point function in a spatially homogeneous and isotropic system is then defined by~\cite{Jukkala:2021sku}
\begin{subequations}
\label{eq:projection_matrix_parametrization}
\begin{align}
    \mathcal{S}_{\evec{k}ij}(t) &=
    \sum_{\smash{\mathclap{h,s,s'=\pm}}} \, \mathcal{P}_{\evec{k}hij}^{ss'}(t) \,
    \mathcal{F}_{\evec{k}hij}^{ss'}(t) \text{,}
\shortintertext{with}
    \mathcal{P}_{\evec{k}hij}^{ss'} &\eqdef N_{\evec{k}hij}^{ss'}
    \overbrace{P_{\evec{k}h}^{} P_{\evec{k}i}^s \gamma^0 P_{\evec{k}j}^{s'}}^{
        \hspace{4ex}{} \eqdef \smash{P_{\evec{k}hij}^{ss'}}
    } \text{.}
    \label{eq:projection_matrix_basis_element}
\end{align}
\end{subequations}
Here $P_{\evec{k}h}$ and $P^s_{\evec{k}i}$ are the helicity and energy projection matrices, respectively. The basis matrices $P_{\evec{k}hij}^{ss'} \eqdef P_{\evec{k}h}^{} P_{\evec{k}i}^s \gamma^0 P_{\evec{k}j}^{s'}$ span the subalgebra of Dirac matrices consistent with homogeneity and isotropy and hence there is no loss of generality by using this parametrization~\cite{Jukkala:2021sku,Herranen:2010mh}. The quantities $\mathcal{F}_{\evec{k}hij}^{ss'}$ are generic complex-valued components of the local two-point function which can be viewed as generalized phase space distribution functions, and $N_{\evec{k}hij}^{ss'}$ are arbitrary normalization factors. The aforementioned projection matrices are given by
\begin{subequations}
\label{eq:projection_matrix_definitions}
\begin{alignat}{2}
    P_{\evec{k}h}^{} &= \frac{1}{2}\Bigl(\idmat + h \widehat{h}_{\evec{k}}\Bigr)
    \text{,} \qquad &
    P_{\evec{k}i}^s &= \frac{1}{2}\biggl(\idmat + s\frac{H_{\evec{k}i}}{\omega_{\evec{k}i}}\biggr)
    \text{,}
\shortintertext{where}
    \widehat{h}_{\evec{k}} &\eqdef \evec{\alpha} \cdot \Uevec{k} \,\gamma^5
    \text{,} \qquad &
    H_{\evec{k}i} &\eqdef \evec{\alpha} \cdot \evec{k} + \gamma^0 m_i
    \text{,}
\end{alignat}
\end{subequations}
are the helicity operator and the diagonal element of the Hamiltonian, respectively. The helicity and energy indices take the values $h,s = \pm1$, and $\omega_{\evec{k}i} \eqdef \sqrt{\abs{\evec{k}}^2 + {m_i}^2}$ is the free on-shell energy for the flavor state $i$. More details about the projection matrices and the parametrization can be found in~\cite{Jukkala:2021sku}. However, unlike in~\cite{Jukkala:2021sku}, in this chapter we keep the normalization factors $N_{\evec{k}hij}^{ss'}$ unfixed in all equations.

The main idea of the parametrization~\cref{eq:projection_matrix_parametrization} is that it separates different oscillation frequencies described by equation~\cref{eq:generic_free_dynamical_equation}. Technically this happens because of the property
\begin{equation}
    P^s_{\evec{k}i} H_{\evec{k}i}
    = H_{\evec{k}i} P^s_{\evec{k}i}
    = s \omega_{\evec{k}i} P^s_{\evec{k}i} \text{,}
\end{equation}
that is, the Hamiltonian gets projected to the frequencies $\omega_{\evec{k}i}$ when it acts on the energy projection matrices. The parametrization thus separates the fast and slowly varying parts of the two-point function: each component $\mathcal{F}_{\evec{k}hij}^{ss'}$ has a distinct well-defined oscillation frequency. Finally, we define some notation for different combinations of the frequencies which are needed later:
\begin{subequations}
\label{eq:frequency_combinations_notation}
\begin{align}
    \bar\omega_{\evec{k}ij} &\eqdef (\omega_{\evec{k}i} + \omega_{\evec{k}j})/2 \text{,} \\
    \delta\omega_{\evec{k}ij} &\eqdef (\omega_{\evec{k}i} - \omega_{\evec{k}j})/2 \text{,} \\
    \Delta\bar\omega^{ss'}_{\evec{k}ij} &\eqdef s\omega_{\evec{k}i} - s'\omega_{\evec{k}j} \text{.}
    \label{eq:frequency_diff_sum_notation}
\end{align}
\end{subequations}
%

%
\section{Coherent quasiparticle approximation}
\label{sec:cQPA}
%

The idea of cQPA, originally developed in~\cite{Herranen:2008hi,Herranen:2008hu,Herranen:2008yg,Herranen:2008di,Herranen:2009zi,Herranen:2009xi,Herranen:2010mh,Fidler:2011yq,Herranen:2011zg}, is to provide tractable QKEs where quantum coherence is included dynamically with interactions and decoherence. In practice, cQPA is a two-step method to solve the Wigner representation KB equations~\cref{eq:pole_and_KB-equations_Wigner} in a spectral approximation. First the phase space structure is determined from the free equations with all gradients neglected. The resulting \emph{spectral} (quasiparticle) shell structure contains the usual mass shells for all propagators and additional coherence shells for the Wightman functions. The second step is to use this spectral form, including coherence, as an ansatz for solving the full equations.

We consider here cQPA in the spatially homogeneous and isotropic case for fermions with a time dependent mass matrix. We present the improved way to derive both the cQPA ansatz and the resummed equations of motion given in~\cite{Jukkala:2019slc}, and review the main features of cQPA. We also generalize the treatment in~\cite{Jukkala:2019slc} slightly by including multiple fermion flavors.

%
\subsection{Spectral phase space structure}
\label{sec:cQPA_spectral_structure}
%

In the first step we derive the constraints for the phase space structure of the cQPA propagators. We start by dropping the collision terms and all gradients. The Hermitian part of equation~\cref{eq:KB-equation_Wigner}, which is also called the constraint equation, then becomes
\begin{equation}
    2 k^0 \bar S^\lt(k,t) = \anticomm[\big]{H_\evec{k}(t), \bar S^\lt(k,t)}
    \text{.} \label{eq:free_zeroth_order_Wigner_CE}
\end{equation}
We assume a flavor-diagonal free Hamiltonian as described below equation~\cref{eq:generic_free_dynamical_equation}, resulting in free theory dispersion relations. Generally in cQPA it is also possible to consider quasiparticle dispersion relations which originate from the interaction terms with the dispersive self-energy $\Sigma^{\rm H}$, but this is not necessary for the scope of this work.

Next we use the projection matrix basis~\cref{eq:projection_matrix_basis_element} to parametrize the Wightman function in terms of component functions $D^{ss'}_{hij}(k,t)$~\cite{Jukkala:2019slc}:
\begin{equation}
    \bar S^\lt_{ij}(k,t) \eqdef \sum_{h,s,s'}
    \mathcal{P}_{\evec{k}hij}^{ss'} \, D^{ss'}_{hij}(k,t) \text{.}
    \label{eq:cQPA_projection_matrix_parametrization}
\end{equation}
After using the parametrization~\cref{eq:cQPA_projection_matrix_parametrization} in equation~\cref{eq:free_zeroth_order_Wigner_CE} and taking relevant projections and traces of the resulting equation we get the constraints [with notation~\cref{eq:frequency_combinations_notation}]
\begin{subequations}
\begin{align}
   (k^0 - s \bar\omega_{\evec{k}ij}) D^{ss}_{hij}(k,t) &= 0 \text{,}
   \\*
    (k^0 - s \delta\omega_{\evec{k}ij}) D^{s,-s}_{hij}(k,t) &= 0 \text{.}
\end{align}
\end{subequations}
These equations admit the distributional solutions
\begin{subequations}
\label{eq:cQPA_distributional_solutions}
\begin{align}
    D^{ss}_{hij}(k,t) &= f^{m,s}_{\evec{k}hij} \,
    2\pi \delta(k^0 - s \bar\omega_{\evec{k}ij}) \text{,}
    \\*
    D^{s,-s}_{hij}(k,t) &= f^{c,s}_{\evec{k}hij} \,
    2\pi \delta(k^0 - s \delta\omega_{\evec{k}ij}) \text{,}
\end{align}
\end{subequations}
where the unknown quantities
\begin{equation}
    f^{m,\pm}_{\evec{k}hij} \eqdef f^{\pm\pm}_{\evec{k}hij} \text{,} \hspace{3em}
    f^{c,\pm}_{\evec{k}hij} \eqdef f^{\pm\mp}_{\evec{k}hij}
    \label{eq:cQPA_distribution-fun_def}
\end{equation}
are called the mass shell functions and the coherence shell functions, respectively~\cite{Herranen:2010mh,Fidler:2011yq}. Substituting the solutions~\cref{eq:cQPA_distributional_solutions} back to~\cref{eq:cQPA_projection_matrix_parametrization} we obtain the flavored spectral cQPA ansatz of~\cite{Fidler:2011yq} for the Wightman function:
\begin{equation}
    \bar S^\lt_{ij}(k,t) = 2\pi \sum_{h,s} \Bigl(
        \mathcal{P}^{ss}_{\evec{k}hij} f^{m,s}_{\evec{k}hij} \,
        \delta(k^0 - s \bar\omega_{\evec{k}ij})
        + \mathcal{P}^{s,-s}_{\evec{k}hij} f^{c,s}_{\evec{k}hij} \,
        \delta(k^0 - s \delta\omega_{\evec{k}ij})
    \Bigr)
    \text{.} \label{eq:the_cQPA_ansatz}
\end{equation}

The ansatz~\cref{eq:the_cQPA_ansatz} contains four different kinds of spectral shells and their corresponding phase space distribution functions $f$. These describe the usual mass shell excitations and different kinds of quantum coherence information as shown in table~\cref{tbl:phase_space_functions}.
Note that with this naming convention the mass shell functions $f^{m,\pm}_{\evec{k}hij}$ contain not only the flavor-diagonal particle and antiparticle excitations but also the off-diagonal flavor coherence which resides on the \emph{average} energy shells $k^0 = \pm\bar\omega_{\evec{k}ij}$ (with $i \neq j$) in the phase space.

%
\begin{table}[t!]
    \centering
    \begin{tabular}{ccllll}
    \toprule
    $f$ & $i,j$ & Particle type & Information & Dispersion & Osc. freq. \\
    \midrule
    $f^{m,\pm}_{\evec{k}hij}$ & $i = j$ & particle/antipart. & mass shell excit. &
        $k^0 = \pm\omega_{\evec{k}i}$ & $\nu = 0$ \\
    $f^{m,\pm}_{\evec{k}hij}$ & $i \neq j$ & particle/antipart. &  flavor coherence &
        $k^0 = \pm\bar\omega_{\evec{k}ij}$ & $\nu = \abs{2 \delta\omega_{\evec{k}ij}}$ \\
    $f^{c,\pm}_{\evec{k}hij}$ & $i = j$ & particle--antipart. & coherence &
        $k^0 = 0$ & $\nu = 2 \omega_{\evec{k}i}$ \\
    $f^{c,\pm}_{\evec{k}hij}$ & $i \neq j$ & particle--antipart. & flavor coherence &
        $k^0 = \pm\delta\omega_{\evec{k}ij}$ & $\nu = 2 \bar\omega_{\evec{k}ij}$ \\
    \bottomrule
    \end{tabular}
    \caption{Phase space distribution functions $f^{x,\pm}_{\evec{k}hij}$ ($x = m,c$) of the projection matrix parametrization~\cref{eq:projection_matrix_parametrization} and the type of information they describe (mass shell excitations or quantum coherence~\cite{Fidler:2011yq}). Given are also the dispersion relations of the corresponding (spectral) shells and the leading oscillation frequencies of the functions.}
    \label{tbl:phase_space_functions}
\end{table}
%

The same construction can naturally be done also for the other Wightman function $S^\gt(k,t)$ and the spectral function $S^{\mathcal{A}}(k,t)$, as they obey the same constraint equation~\cref{eq:free_zeroth_order_Wigner_CE}. However, the spectral function is subject to an additional constraint, the sum rule~\cref{eq:fermionic_sum_rule_Wigner}. It fixes the spectral function in cQPA to a completely non-dynamical quantity with vanishing coherence shell functions~\cite{Fidler:2011yq}. In fact, when using free dispersion relations in cQPA the full spectral function has the same form as in free theory~\cite{LeBellac:1996cam,Fidler:2011yq}.

%
\subsection{Dynamical equations}
\label{sec:cqpa_dynamical_free_equations}
%

In the second step we derive the dynamical equations for the phase space distribution functions $f^m$ and $f^c$. This is done by using the cQPA ansatz~\cref{eq:the_cQPA_ansatz} in the antihermitian part of equation~\cref{eq:KB-equation_Wigner}, including all gradients and collision terms, and integrating over $k^0$. However, in practice there is a complication which we need to handle first. To see this, let us consider the free theory where the dynamical equation is
\begin{equation}
    \im \partial_t \bar S^\lt(k,t) = \widehat H_{\evec{k}}(t) \bar S^\lt(k,t) - \text{H.c.}
    \label{eq:free_Wigner_DE}
\end{equation}
Here $\widehat H_{\evec{k}}(t) \eqdef \evec{\alpha} \cdot \evec{k} + \gamma^0 m(t) \e^{-\frac{\im}{2} \cev{\partial}_t \partial_{k^0}}$ is the free Dirac Hamiltonian where we have included the exponential gradient operator originating from equation~\cref{eq:KB-equation_Wigner}~\cite{Fidler:2011yq}.

Now we insert the spectral ansatz~\cref{eq:the_cQPA_ansatz} of the Wightman function into~\cref{eq:free_Wigner_DE} and integrate over $k^0$ (and divide by $2\pi$). Note that the resulting equation has the form~\cref{eq:generic_free_dynamical_equation}. Taking again relevant projections and traces of the equation yields the integrated free dynamical equations
\begin{subequations}
\label{eq:free_cQPA_DE}
\begin{alignat}{3}
    \partial_t f^{m,s}_{\evec{k}hij} &={}& -2 \im s \delta\omega_{\evec{k}ij} &
    f^{m,s}_{\evec{k}hij} &&- \mathcal{L}^{m,s}_{\evec{k}hij}[f] \text{,}
    \\*
    \partial_t f^{c,s}_{\evec{k}hij} &={}& -2 \im s \bar\omega_{\evec{k}ij} &
    f^{c,s}_{\evec{k}hij} &&- \mathcal{L}^{c,s}_{\evec{k}hij}[f] \text{.}
\end{alignat}
\end{subequations}
Here $\mathcal{L}^{x,s}_{\evec{k}hij}[f]$ (with $x = m, c$) contain the terms proportional to the time-derivative of the projection matrices. Using analogous notation for $\mathcal{L}$ as for $f$ in equation~\cref{eq:cQPA_distribution-fun_def} (\ie~$\mathcal{L}^{m,\pm} = \mathcal{L}^{\pm\pm}$ and $\mathcal{L}^{c,\pm} = \mathcal{L}^{\pm\mp}$), we can write
\begin{equation}
    \mathcal{L}^{ss'}_{\evec{k}hij}[f] = \sum_{r,r'}
    \frac{\tr\bigl[ P^{s's}_{\evec{k}hji} \partial_t \mathcal{P}^{rr'}_{\evec{k}hij} \bigr]}
    {\tr\bigl[ \mathcal{P}^{ss'}_{\evec{k}hij} \gamma^0 \bigr]}
    f^{rr'}_{\evec{k}hij} \text{.}
    \label{eq:cQPA_projection_matrix_gradient_terms}
\end{equation}
In the end these terms are directly proportional to $\partial_t m$ and $f$. We can now see from equations~\cref{eq:free_cQPA_DE} that the coherence functions $f^c$, and also $f^m_{ij}$ for $i \neq j$, are very rapidly oscillating and the leading oscillation frequency is given by the first term on the right-hand side (see also table~\cref{tbl:phase_space_functions}). These terms will give the leading behavior of the distribution functions also in the full interacting equations, at least in the case of weak interactions. However, the main point that we wished to make here is that the oscillation frequencies are not suppressed by time-gradients.

%
\subsection{Reorganized gradient expansion}
\label{sec:cQPA_reorganized_gradients}
%

Now that we understand the leading time-dependence of the cQPA propagator, we can see that there is a problem in the full interacting equations~\cref{eq:KB-equation_Wigner}. If we use the standard Moyal product~\cref{eq:convolution_Wigner_transform} for the interaction terms the gradient expansion will include time-derivatives which operate on the rapidly oscillating coherence functions. This results in terms that grow at each order of the expansion. It is then not guaranteed to get valid approximations for the full equations by truncating the standard gradient expansion. This problem can be solved by \emph{resumming} the leading behavior of the coherence function gradients to all orders, as was done in~\cite{Herranen:2010mh,Fidler:2011yq}. The result is that the arguments of the self-energy functions are shifted so that they are always evaluated at the ordinary flavor-diagonal mass shells $k^0 = \pm \omega_{\evec{k}i}$.

A more direct way to derive the resummed equations is to rewrite the Moyal products in equations~\cref{eq:pole_and_KB-equations_Wigner} so that the leading coherence gradient resummation is more transparent. The equations with the reorganized gradient expansion are~\cite{Jukkala:2019slc}
\begin{subequations}
\label{eq:reorganized_KB_eqs_Wigner}
\begin{alignat}{3}
    \slashed D & S^p
    &&{}- \e^{-\frac{\im}{2} \partial_x^\Sigma \cdot \partial_k^{}}
    \bigl[\widetilde\Sigma^p_{\rm out}(D,x) S^p\bigr]
    &&= \idmat \text{,}
    \\*
    \slashed D & S^s
    &&{}- \e^{-\frac{\im}{2} \partial_x^\Sigma \cdot \partial_k^{}}
    \bigl[\widetilde\Sigma^r_{\rm out}(D,x) S^s\bigr]
    &&= \e^{-\frac{\im}{2}\partial_x^\Sigma \cdot \partial_k^{}}
    \bigl[\Sigma^s_{\rm out}(D,x) S^a\bigr] \text{,}
    \label{eq:reorganized_KB-eq_Wigner}
\end{alignat}
\end{subequations}
where $D \eqdef k + \frac{\im}{2}\partial_x$ and we defined the ``out-transform''
\begin{equation}
    \Sigma_{\rm out}(k,x) \eqdef \smash{\int} \dd^4 z \,\e^{\im k \cdot (x-z)} \,\Sigma(x,z)
    = \e^{\frac{\im}{2} \partial_x \cdot \partial_k^{}} \Sigma(k,x)
    \text{.} 
\end{equation}
In equations~\cref{eq:reorganized_KB_eqs_Wigner} the Moyal products have been reorganized into total $k$-derivatives and the exponential spacetime derivatives operate only on $\Sigma$. This form is useful already for the fact that the exponential gradient operator reduces to unity when integrating over $k^0$. However, the main point is that the self-energy functions have been generalized (via their Taylor expansions) to derivative operators which operate on the propagators. This can then be used to see directly how the coherence functions in $S^s$ shift the arguments of the self-energy functions~\cite{Jukkala:2019slc}.

Next we illustrate how the gradient resummation shifts the argument of the self-energy in the multi-flavor case. Consider the $k^0$-integrated equation and a generic interaction term (in the spatially homogeneous and isotropic case)
\begin{equation}
    \int \frac{\dd k^0}{2\pi} \e^{-\frac{\im}{2} \partial_t^\Sigma \cdot \partial_{k^0}}
    \Bigl[
        \Sigma_{li}^{}\bigl(k^0 + \sfrac{\im}{2} \partial_t\bigr) S^\lt_{ij}(k,t)
    \Bigr]
    = \int \frac{\dd k^0}{2\pi} \Sigma_{li}^{}\bigl(k^0 + \sfrac{\im}{2} \partial_t\bigr) S^\lt_{ij}(k,t)
    \text{,}
\end{equation}
where $i,j,l$ are any flavor indices. When using the cQPA ansatz~\cref{eq:the_cQPA_ansatz} here and performing the $k^0$-integral using the delta functions, the terms with $f^m$ and $f^c$ can be approximated according to
\begin{subequations}
\label{eq:coherence-resummed_self-energies}
\begin{alignat}{2}
    &\Sigma_{li}^{}\bigl(k^0 + \sfrac{\im}{2} \partial_t\bigr)
    f^{m,s}_{\evec{k}hij} \,\delta(k^0 - s \bar\omega_{\evec{k}ij})
    \quad &&\longrightarrow \quad
    \Sigma_{li}^{}\bigl(\smash{
        \overbrace{s \bar\omega_{\evec{k}ij} + s \delta\omega_{\evec{k}ij}}^{\textstyle \hspace{4ex}{} = s\omega_{\evec{k}i}}
    }\bigr)
    f^{m,s}_{\evec{k}hij} \text{,}
    \\*
    &\Sigma_{li}^{}\bigl(k^0 + \sfrac{\im}{2} \partial_t\bigr)
    f^{c,s}_{\evec{k}hij} \, \delta(k^0 - s \delta\omega_{\evec{k}ij})
    \quad &&\longrightarrow \quad
    \Sigma_{li}^{}\bigl(s \delta\omega_{\evec{k}ij} + s \bar\omega_{\evec{k}ij}\bigr)
    f^{c,s}_{\evec{k}hij} \text{.}
\end{alignat}
\end{subequations}
Here we used the leading behavior of $\partial_t f^m$ and $\partial_t f^c$ from equations~\cref{eq:free_cQPA_DE}. In both cases the $k^0$-argument of the self-energy reduces to $s\omega_{\evec{k}i}$. This prescription of course does not take into account the gradients that operate on the projection matrices and delta functions, but these are suppressed by the mass gradients and are thus controlled in most situations in a standard expansion. We can now see the main result of the resummation: in all cases the self-energy functions are effectively evaluated at the flavor-diagonal mass shell $k^0 = s\omega_{\evec{k}i}$ with the middle flavor index.

%
\subsection{Resummed dynamical equations}
%

We can now write the full integrated dynamical equations where the leading coherence-gradients have been resummed. This is done like described in section~\cref{sec:cqpa_dynamical_free_equations} but now starting from the antihermitian part of the KB equation~\cref{eq:reorganized_KB-eq_Wigner}. The result is
\begin{subequations}
\label{eq:full_cQPA_DE}
\begin{alignat}{3}
    \partial_t f^{m,s}_{\evec{k}hij} &={}& -2 \im s \delta\omega_{\evec{k}ij} &
    f^{m,s}_{\evec{k}hij} &&- \mathcal{L}^{m,s}_{\evec{k}hij}[f]
    - \mathcal{C}^{m,s}_{\evec{k}hij}\bigl[\bar\Sigma^r_{\rm out}, f\bigr]
    - \mathcal{C}^{m,s}_{\evec{k}hij}\bigl[\im \bar\Sigma^\lt_{\rm out}, f^{(\mathcal{A})}\bigr]
    \text{,}
    \\*
    \partial_t f^{c,s}_{\evec{k}hij} &={}& -2 \im s \bar\omega_{\evec{k}ij} &
    f^{c,s}_{\evec{k}hij} &&- \mathcal{L}^{c,s}_{\evec{k}hij}[f]
    - \mathcal{C}^{c,s}_{\evec{k}hij}\bigl[\bar\Sigma^r_{\rm out}, f\bigr]
    - \mathcal{C}^{c,s}_{\evec{k}hij}\bigl[\im \bar\Sigma^\lt_{\rm out}, f^{(\mathcal{A})}\bigr]
    \text{,}
\end{alignat}
\end{subequations}
where the projection matrix gradient terms $\mathcal{L}^{x,s}$ (with $x = m, c$) were defined in~\cref{eq:cQPA_projection_matrix_gradient_terms}. Using similar notation ($\mathcal{C}^{m,\pm} = \mathcal{C}^{\pm\pm}$ and $\mathcal{C}^{c,\pm} = \mathcal{C}^{\pm\mp}$), the collision term projections in equations~\cref{eq:full_cQPA_DE} are given by%
\footnote{Note that the collision term coefficients $C$ and $C^\star$ used here are independent for arbitrary normalization $N$ [which enters via equation~\cref{eq:projection_matrix_basis_element}].}
\begin{subequations}
\label{eq:cQPA_collision_term_functions}
\begin{align}
    \mathcal{C}^{ss'}_{\evec{k}hij}[\Sigma, f] &= \smash{\sum_{l, r}}
    \Bigl(
        C^{srs'}_{\evec{k}hilj}[\Sigma] \,f^{rs'}_{\evec{k}hlj}
        + C^{{\star},srs'}_{\evec{k}hilj}[\Sigma] f^{sr}_{\evec{k}hil}
    \Bigr) \text{,}
    \\*
\shortintertext{with}
    C^{srs'}_{\evec{k}hilj}[\Sigma] &\eqdef
    \frac{\im \tr\bigl[
        P^{s's}_{\evec{k}hji} \Sigma_{il}(r\omega_{\evec{k}l})
        \mathcal{P}^{rs'}_{\evec{k}hlj}
    \bigr]}
    {\tr\bigl[ \mathcal{P}^{ss'}_{\evec{k}hij} \gamma^0 \bigr]} \text{,}
    \\*
    C^{{\star},srs'}_{\evec{k}hilj}[\Sigma] &\eqdef
    \frac{-\im \tr\bigl[
        \mathcal{P}^{sr}_{\evec{k}hil} (\Sigma(r\omega_{\evec{k}l})^\dagger)_{lj}
        P^{s's}_{\evec{k}hji}
    \bigr]}
    {\tr\bigl[ \mathcal{P}^{ss'}_{\evec{k}hij} \gamma^0 \bigr]} \text{.}
\end{align}
\end{subequations}
In order to be consistent with the used (quasi)particle approximation we also discarded the term $\propto \Sigma^\lt S^{\rm H}$ here~\cite{Herranen:2008hu}. This term is related to the spectral width as discussed below equation~\cref{eq:general_fermionic_KB_and_pole-eqs_variant}.
Note that equations~\cref{eq:full_cQPA_DE} also contain $\Sigma^r$ which includes the dispersive self-energy $\Sigma^{\rm H}$. It would be consistent to drop also $\Sigma^{\rm H}$ here because we have used the free dispersion relations in the cQPA ansatz. However, as we will see in the context of the more general LA-method it may still be included as a leading approximation to the dispersive corrections in these equations.

An essential feature of cQPA is that the non-local degrees of freedom are treated non-dynamically in the collision terms~\cite{Jukkala:2019slc}. This idea is further refined in the LA-method which we shall present in section~\cref{eq:the_local_approximation_method}. However, first we describe how to account for a more general mass matrix in these equations.

\subsubsection{Mixing gradient terms}

Above we assumed that the mass matrix $m(t)$ is diagonal with non-negative entries. In the more general case where it is non-diagonal and complex we first have to transform the equations to the mass eigenbasis before proceeding with the projection matrix parametrization. Here we only present the main points; details can be found in~\cite{Fidler:2011yq}. We use the singular value decomposition with unitary matrices $U$ and $V$ such that
\begin{equation}
    m_{\rm d} \eqdef U m V^\dagger
    \label{eq:mass_matrix_diagonalization}
\end{equation}
is diagonal with non-negative entries. The subscript $\mathrm{d}$ distinguishes the diagonal basis from the original where necessary. The free Hamiltonian generally has the form $H_{\evec{k}} = \evec{\alpha} \cdot \evec{k} + \gamma^0 (P_{\rm R} m + P_{\rm L} m^\dagger)$ in the original basis. Its transform is given by $H_{{\rm d},\evec{k}} \eqdef Y H_{\evec{k}} Y^\dagger = \evec{\alpha} \cdot \evec{k} + \gamma^0 m_{\rm d}$ with the transformation matrix
\begin{equation}
    Y \eqdef P_{\rm L} U + P_{\rm R} V \text{.}
    \label{eq:mass_eigenbasis_transformation}
\end{equation}
All equations can be transformed to the mass eigenbasis similarly by using the unitary matrix $Y$. Because of the time dependence of $m$ and the matrices $U$ and $V$ this induces additional \emph{mixing gradient} terms to the equations which are proportional to $\partial_t Y$. In the end the dynamical equations~\cref{eq:full_cQPA_DE} are modified to
\begin{subequations}
\label{eq:cQPA_DE_with_mixing_gradients}
\begin{alignat}{3}
    \partial_t f^{m,s}_{\evec{k}hij} &={}& -2 \im s \delta\omega_{\evec{k}ij} &
    f^{m,s}_{\evec{k}hij} &&- \mathcal{L}^{m,s}_{\evec{k}hij}[f]
    - \mathcal{X}^{m,s}_{\evec{k}hij}[f] - ({\cdots})
    \text{,}
    \\*
    \partial_t f^{c,s}_{\evec{k}hij} &={}& -2 \im s \bar\omega_{\evec{k}ij} &
    f^{c,s}_{\evec{k}hij} &&- \mathcal{L}^{c,s}_{\evec{k}hij}[f]
    - \mathcal{X}^{c,s}_{\evec{k}hij}[f] - ({\cdots})
    \text{,}
\end{alignat}
\end{subequations}
where the mixing gradient contribution is given by
\begin{subequations}
\label{eq:mixing_gradient_terms}
\begin{gather}
    \mathcal{X}^{ss'}_{\evec{k}hij}[f] = \smash{\sum_{l, r}}
    \Bigl(
        X^{srs'}_{\evec{k}hilj} \,f^{rs'}_{\evec{k}hlj}
        + X^{{\star},srs'}_{\evec{k}hilj} f^{sr}_{\evec{k}hil}
    \Bigr) \text{,}
    \\*
\shortintertext{with}
    X^{srs'}_{\evec{k}hilj} \eqdef
    \frac{-\im \tr\bigl[
        P^{s's}_{\evec{k}hji} \, \dot\Xi^{}_{il} \mathcal{P}^{rs'}_{\evec{k}hlj}
    \bigr]}
    {\tr\bigl[ \mathcal{P}^{ss'}_{\evec{k}hij} \gamma^0 \bigr]}
    \text{,}
    \qquad
    X^{{\star},srs'}_{\evec{k}hilj} \eqdef
    \frac{\im \tr\bigl[
        \mathcal{P}^{sr}_{\evec{k}hil} \, \dot\Xi^{}_{lj} P^{s's}_{\evec{k}hji}
    \bigr]}
    {\tr\bigl[ \mathcal{P}^{ss'}_{\evec{k}hij} \gamma^0 \bigr]} \text{.}
\end{gather}
\end{subequations}
Here $\dot{\Xi}(t) \eqdef \im Y(t) \partial_t Y^\dagger(t)$ is the mixing gradient matrix which is Hermitian. The ellipses in equations~\cref{eq:cQPA_DE_with_mixing_gradients} stand for the collision terms which are the same as in~\cref{eq:full_cQPA_DE} except that the self-energy functions have been transformed to the mass eigenbasis according to $\bar\Sigma \rightarrow Y \bar\Sigma Y^\dagger$.

%
\section{The local approximation method}
\label{eq:the_local_approximation_method}
%

The local approximation (LA) method which was formulated in~\cite{Jukkala:2021sku} is an improved method for obtaining straightforwardly solvable dynamical QKEs including quantum coherence. The LA-method provides a closed equation for the \emph{local} Wightman function directly in the two-time representation. This is useful when calculating physical observables from the currents or the energy-momentum tensor because then knowledge of only the equal-time, or local, correlator $S^\lt_\evec{k}(t,t)$ is needed. The LA-method also sidesteps the need to specify the phase space shells of the full propagator as is required in a Wigner representation approach. This makes it more general than the cQPA, for example. Next we present the derivation and equations of the LA-method for fermions in a spatially homogeneous and isotropic system.

%
\subsection{Dynamics of the local correlator}
%

We start from the pole and KB equations~\cref{eq:pole_and_KB-equations_twotime} which we write explicitly as
\begin{subequations}
    \label{eq:KB_with_H}
    \begin{alignat}{2}
        & \bigl[\im\partial_{t_1} - H_\evec{k}(t_1)\bigr]
        \bar S^p_\evec{k}(t_1,t_2)
        - (\bar\Sigma^p_\evec{k} \convol \bar S^p_\evec{k})(t_1, t_2)
        &&= \delta(t_1 - t_2)
        \text{,}
        \\*
        & \bigl[\im\partial_{t_1} - H_\evec{k}(t_1)\bigr]
        \bar S^s_\evec{k}(t_1,t_2)
        - (\bar\Sigma^r_\evec{k} \convol \bar S^s_\evec{k})(t_1, t_2)
        &&= (\bar\Sigma^s_\evec{k} \convol \bar S^a_\evec{k})(t_1, t_2)
        \text{.} \label{eq:KB-Wightman_with_H}
    \end{alignat}
\end{subequations}
We again assume the mass eigenbasis with the free Hamiltonian $H_\evec{k}$ given below equation~\cref{eq:generic_free_dynamical_equation}. To find the equation for the local Wightman function we first take the total derivative of $S^\lt_\evec{k}(t,t)$ and use the chain rule to get
\begin{equation}
    \partial_t S^\lt_\evec{k}(t,t) \eqdef
    \frac{\dd}{\dd t}\bigl[S^\lt_\evec{k}(t,t)\bigr]
    = \bigl(\partial_{t_1} S^\lt_\evec{k} + \partial_{t_2} S^\lt_\evec{k}\bigr)(t,t) \text{.}
    \label{eq:total_derivative_of_local_S}
\end{equation}
Equation~\cref{eq:total_derivative_of_local_S} can then be used with the KB equation~\cref{eq:KB-Wightman_with_H} and its Hermitian conjugate to obtain the equation for the local function:%
\footnote{We assume that the local limit of $S^\lt_\evec{k}(t_1, t_2)$ exists and is well-enough-behaved (\eg~continuously differentiable). However, equation~\cref{eq:exact_equation_for_local_S} may also be sensible with weaker assumptions. For example, if only one-sided limits $t_1 \to t_2$ exist, we can define $S^\lt_\evec{k}(t, t)$ as the symmetrized local limit.}
\begin{align}
    \im \partial_t \bar S^\lt_\evec{k}(t,t) =
    \comm[\big]{H_\evec{k}(t), \bar S^\lt_\evec{k}(t,t)}
    &+ (\bar\Sigma^r_\evec{k} \convol \bar S^\lt_\evec{k})(t,t)
    - (\bar S^\lt_\evec{k} \convol \bar\Sigma^a_\evec{k})(t,t) \notag
    \\*
    &+ (\bar\Sigma^\lt_\evec{k} \convol \bar S^a_\evec{k})(t,t)
    - (\bar S^r_\evec{k} \convol \bar\Sigma^\lt_\evec{k})(t,t) \text{.}
    \label{eq:exact_equation_for_local_S}
\end{align}
The total derivative was required here to eventually produce an ordinary differential equation for the local correlator as a function of the single time variable $t$. Note that equation~\cref{eq:exact_equation_for_local_S} is also consistent with the Hermiticity of $\bar S^\lt_\evec{k}(t,t)$.

Equation~\cref{eq:exact_equation_for_local_S} is the starting point of the LA-method. The equation is still exact and hence of course not closed: it couples the local function $S^\lt_\evec{k}(t,t)$ to the non-local function $S^\lt_\evec{k}(t,t')$ (with $t \neq t'$) appearing in the interaction convolutions. The pole functions $S^p$ also appear explicitly in the interaction terms. The LA-method consists of two steps to overcome these problems: (1) expansion around a suitable adiabatic background solution and (2) the local ansatz for the interaction terms.

%
\subsection{Adiabatic background solution}
\label{sec:adiabatic_background_solution}
%

The main purpose of the adiabatic background solution is to enable an approximation of equation~\cref{eq:exact_equation_for_local_S} where the pole functions can be treated non-dynamically. The basis for this approximation is the assumption that the self-energy $\Sigma$ is dominated by some known part $\Sigma_{\rm ad}$ which can be calculated non-dynamically. The adiabatic background solution $S_{\rm ad}$ for the two-point functions is then defined by a suitable approximation of the full equations where the full $\Sigma$ is replaced by $\Sigma_{\rm ad}$.

For given adiabatic two-point functions $S_{\rm ad}$ and self-energies $\Sigma_{\rm ad}$ we extract the dynamical parts (or deviations) $\delta S$ and $\delta \Sigma$ from the full functions by defining
\begin{subequations}
\label{eq:adiabatic_and_dynamical_division}
\begin{align}
    S^\alpha &\eqdef S^\alpha_{\rm ad} + \delta S^\alpha \text{,}
    \label{eq:adiabatic_and_dynamical_S_division}
    \\*
    \Sigma^\alpha &\eqdef \Sigma^\alpha_{\rm ad} + \delta \Sigma^\alpha \text{.}
\end{align}
\end{subequations}
We then substitute~\cref{eq:adiabatic_and_dynamical_S_division} into equation~\cref{eq:exact_equation_for_local_S} to derive the equation for $\delta S^\lt_\evec{k}(t,t)$:
\begin{align}
    \im \partial_t \delta \bar S^\lt_\evec{k}(t,t) ={}&
    \comm[\big]{
        H_\evec{k}(t), \delta \bar S^\lt_\evec{k}(t,t)
    } + \bigl[
        (\bar\Sigma^r_\evec{k} \convol \delta \bar S^\lt_\evec{k})(t,t)
        + \bar A^\lt_\evec{k}(t)
        - \text{H.c.}
    \bigr]
    \text{,} \label{eq:local_equation_precursor}
    \\*
\shortintertext{with}
    \bar A^\lt_\evec{k}(t) \eqdef{}&
        H_\evec{k}(t) \bar S^\lt_{{\rm ad}, \evec{k}}(t,t)
        + \bigl(\bar\Sigma^r_{{\rm ad},\evec{k}} \convol \bar S^\lt_{{\rm ad}, \evec{k}}
        + \bar\Sigma^\lt_{{\rm ad},\evec{k}} \convol \bar S^a_{{\rm ad}, \evec{k}}\bigr)(t,t)
        \notag
    \\*
    {}& +
        \bigl(\delta \bar\Sigma^r_\evec{k} \convol \bar S^\lt_{{\rm ad}, \evec{k}}
        + \delta \bar\Sigma^\lt_\evec{k} \convol \bar S^a_{{\rm ad}, \evec{k}}
        + \bar\Sigma^\lt_\evec{k} \convol \delta \bar S^a_\evec{k}\bigr)(t,t)
        \notag
    \\*
    {}& - \sfrac{\im}{2} \partial_t\bar S^\lt_{{\rm ad},\evec{k}}(t,t)
    \text{.} \label{eq:local_eq_general_source}
\end{align}
Equation~\cref{eq:local_equation_precursor} is still just a reorganization of the exact equation~\cref{eq:exact_equation_for_local_S} where $\bar A^\lt_\evec{k}(t)$ contains the terms which do not depend explicitly on $\delta S^\lt$. The idea is that the adiabatic solutions are defined so that the dynamical terms proportional to $\delta S^p$ [the last term on the second line of~\cref{eq:local_eq_general_source} minus its Hermitian conjugate] will be small corrections compared to the leading terms with the non-dynamical $S^\alpha_{\rm ad}$.

There is a lot of freedom in the definition of the adiabatic functions~\cite{Jukkala:2021sku}, but a reasonable minimum requirement is that they satisfy the same generic properties and relations as the full functions. These include for example the Hermiticity properties~\cref{eq:propagator_hermiticity_properties} and the sum rule~\cref{eq:fermionic_sum_rule_twotime}. An obvious choice with these properties is the translationally invariant solution~\cref{eq:transl_invariant_solutions}. However, because we allow for a changing background, that is, the time-dependent mass $m(t)$, it is better to define the adiabatic functions as instantaneous generalizations of the solutions~\cref{eq:transl_invariant_solutions}:
\begin{subequations}
\label{eq:adiabatic_solutions_def}
\begin{align}
    S^p_{\rm ad}(k,t) &\eqdef \bigl[\slashed k - m(t) - \Sigma^p_{\rm ad}(k,t)\bigr]^{-1}
    \text{,} \label{eq:adiabatic_solution_pole}
    \\*
    S^s_{\rm ad}(k,t) &\eqdef S^r_{\rm ad}(k,t) \Sigma^s_{\rm ad}(k,t) S^a_{\rm ad}(k,t)
    \text{.} \label{eq:adiabatic_solution_Wightman}
\end{align}
\end{subequations}
Note that the definitions~\cref{eq:adiabatic_solutions_def} are given in the Wigner representation so they must be transformed to the two-time representation when using them in equation~\cref{eq:local_equation_precursor}, for example.

\subsubsection{Expansion of the source term}

Now with the adiabatic functions defined we can find the leading approximation for the ``source term'' $\bar A^\lt_\evec{k}(t)$ in~\cref{eq:local_equation_precursor}. However, this still depends significantly on the physical situation at hand so we need to make some further assumptions. For example, the full self-energy quite generally depends on $\delta S^\lt$ itself so strictly speaking $\bar A^\lt_\evec{k}(t)$ is even not only a source term. Here we make the following assumptions which are sufficient in this work:
\begin{enumerate}
    \item $S^\alpha_{\rm ad}$ are given by equations~\cref{eq:adiabatic_solutions_def}. \label{list:LA_assum1}
    \item $\Sigma^s_{\rm ad}$ contains a part satisfying the KMS condition. \label{list:LA_assum2}
    \item Linearity: $\Sigma$ does not depend on $S$ itself at first order. \label{list:LA_assum3}
\end{enumerate}
Assumption~\cref{list:LA_assum1} implies that the adiabatic pole propagators satisfy the sum rule. A direct consequence is that the local part of the dynamical pole propagator vanishes, $\delta S^p_{\evec{k}}(t,t) = 0$. This together with dissipation has the effect that the convolution integral $\bar\Sigma^\lt_\evec{k} \convol \delta \bar S^a_\evec{k}$ in~\cref{eq:local_eq_general_source} is suppressed near both the upper and lower limits.
Furthermore, examination of the full equation of motion in the Wigner representation shows that $\delta S^p$ is also quite generally suppressed by either time gradients, coupling constants or $\delta \Sigma^p$.%
\footnote{The Wightman functions $\delta S^s_{\evec{k}}(t,t)$ are not similarly suppressed because they are not restricted by the sum rule and they may also contain large initial transients.}
When assuming weak interactions, it should then be a reasonable approximation to discard $\delta S^p$ altogether. This makes the pole propagators entirely non-dynamical. Another consequence of assumption~\cref{list:LA_assum1} is that the first line of equation~\cref{eq:local_eq_general_source} vanishes to zeroth order in gradients.
The leading non-vanishing terms are also first order in both gradients and the interactions strength.

Assumptions~\cref{list:LA_assum2,list:LA_assum3} are related to the remaining terms in equation~\cref{eq:local_eq_general_source}. Assumption~\cref{list:LA_assum2} implies that $S^s_{\rm ad}(k,t)$ contains the (instantaneous) thermal equilibrium part, making it $\mathcal{O}(1)$ instead of first order in the interaction strength as might be naively expected from~\cref{eq:adiabatic_solution_Wightman}~\cite{Jukkala:2021sku}.
Assumption~\cref{list:LA_assum3} is made so that equation~\cref{eq:local_equation_precursor} remains linear in $\delta S^\lt$ (in the leading approximation) and $\delta \Sigma$ can be more safely dropped from the equation. Together assumptions~\cref{list:LA_assum1,list:LA_assum2,list:LA_assum3} imply that we can make the leading order approximation
\begin{equation}
    \bar A^\lt_\evec{k}(t) \simeq
    - \sfrac{\im}{2} \partial_t\bar S^\lt_{{\rm ad},\evec{k}}(t,t)
    \label{eq:first_order_adiabatic_source_term}
\end{equation}
in the interaction strength. In this approximation $\bar A^\lt_\evec{k}(t)$ then acts as a source term to $\delta \bar S^\lt_\evec{k}(t,t)$ in equation~\cref{eq:local_equation_precursor}. The next-to-leading correction terms to equation~\cref{eq:first_order_adiabatic_source_term} are proportional to (schematically) $\Sigma \, \partial_t S_{\rm ad}$, $\partial_t \Sigma \, S_{\rm ad}$ or $\delta \Sigma\, S_{\rm ad}$.

It should be borne in mind that the choice of the adiabatic background solutions~\cref{eq:adiabatic_solutions_def} and the above assumptions~\cref{list:LA_assum1,list:LA_assum2,list:LA_assum3} leading to the result~\cref{eq:first_order_adiabatic_source_term} are specific to the situation of a varying background (the mass) and a weak coupling to a thermal bath. For other situations this step may not be as simple and must be done more carefully.
We also remark that the (non-local) pole propagator deviations $\delta S^p$ cannot be solved perturbatively in the same way as the Wightman deviations $\delta S^s$~\cite{Jukkala:2021sku}. Hence the only practical possibility is to set $\delta S^p \equiv 0$ in this approach. The background solution must then be calculated accurately enough for the considered application; its level of approximation essentially defines the approximation scheme~\cite{Kainulainen:2021oqs}.%
\footnote{Also, the background self-energy $\Sigma_{\rm ad}$ may even include corrections from $\delta S^\lt$ itself. For example, it can include a part of $\delta \Sigma^\lt$ which can be arranged so that it \emph{cancels} the term $\bar\Sigma^\lt_\evec{k} \convol \delta \bar S^a_\evec{k}$ in the equation (to lowest order in gradients)~\cite{Kainulainen:2021oqs}. However, the adiabatic functions then become dynamical and have to be solved in an iterative way.}
However, in the situation considered in this work the results are not very sensitive to the approximation of the background solution and the error incurred to the dispersion and width of the propagators is small even with only a tree level approximation~\cite{Jukkala:2021sku}.

%
\subsection{The local ansatz}
\label{sec:local_ansatz}
%

The core of the LA-method is the technique used to localize the dynamical interaction convolutions. The problem is that the evolution of the local perturbation $\delta S^\lt_\evec{k}(t, t)$ in equation~\cref{eq:local_equation_precursor} generally depends on the values of $\delta S^\lt_\evec{k}(t_1, t_2)$ everywhere in the $(t_1,t_2)$-plane. However, when the system has dissipation this non-local dependence is suppressed~\cite{Jukkala:2019slc} and the dominant contribution to the convolution integrals should come from the vicinity of the external local time $t$ of equation~\cref{eq:local_equation_precursor}. This is the basis for the local description of the non-equilibrium dynamics.

The idea is to use the structure of the homogeneous transient solution~\cref{eq:homogeneous_Wightman_2} to formulate the local approximation of the interaction convolutions. Suppose first that we had found the correct local value $\delta S^\lt_\evec{k}(t, t)$ for some time $t > \tin$. Because most of the inhomogeneous solution~\cref{eq:inhomogeneous_Wightman} is contained in the adiabatic solution~\cref{eq:adiabatic_solution_Wightman}, it should then be reasonable to estimate the non-local values $\delta S^\lt_\evec{k}(t_1, t_2)$ by a transient solution similar to~\cref{eq:homogeneous_Wightman_2} where the initial time $\tin$ is replaced by $t$ (especially for $t_1,t_2$ close to $t$). Of course, we do not actually know the exact local values but we use this reasoning to \emph{parametrize} the non-local values with the local ones~\cite{Jukkala:2021sku}:
\begin{equation}
    \delta\bar S^s_\evec{k}(t_1,t_2) =
    2\bar S^{\mathcal{A}}_\evec{k}(t_1,t)
    \delta\bar S^s_\evec{k}(t,t)
    2\bar S^{\mathcal{A}}_\evec{k}(t,t_2) \text{,}
    \quad \text{for any } t > \tin \text{.} \label{eq:the_local_ansatz}
\end{equation}
We emphasize that this \emph{local ansatz} is only to be used as a parametrization in the interaction convolutions. In particular, it must not be used to solve the entire time-evolution non-dynamically from the initial value. This would lead to contradictions with the local equation of motion, and it would for example exclude the use of thermal initial conditions (where $\delta S^s_\evec{k} = 0$ initially).
In the interaction convolutions the ansatz~\cref{eq:the_local_ansatz} is effectively used with $t_1,t_2$ near the external time argument $t$.

The dynamical interaction convolutions remaining in equation~\cref{eq:local_equation_precursor} can now be localized by using the ansatz~\cref{eq:the_local_ansatz}. We replace the convolutions involving $\delta \bar S^\lt_\evec{k}$ according to
\begin{subequations}
\label{eq:convolution_localization}
\begin{align}
    (\bar\Sigma^r_\evec{k} \convol \delta \bar S^\lt_\evec{k})(t,t) \, &\longrightarrow \,
    \bar\Sigma^r_{{\rm eff},\evec{k}}(t,t) \delta \bar S^\lt_\evec{k}(t,t) \text{,}
    \\*
    ( \delta \bar S^\lt_\evec{k} \convol \bar\Sigma^a_\evec{k})(t,t) \, &\longrightarrow \,
    \delta \bar S^\lt_\evec{k}(t,t) \bar\Sigma^a_{{\rm eff},\evec{k}}(t,t) \text{,}
\end{align}
\end{subequations}
where we defined the \emph{effective} self-energies
\begin{subequations}
\label{eq:effective_self_energy_def}
\begin{align}
    \bar\Sigma^r_{{\rm eff},\evec{k}}(t_1,t_2) &\eqdef
    (\bar\Sigma^r_\evec{k} \convol 2\bar S^{\mathcal{A}}_\evec{k})(t_1,t_2) \text{,}
    \\*
    \bar\Sigma^a_{{\rm eff},\evec{k}}(t_1,t_2) &\eqdef
    (2\bar S^{\mathcal{A}}_\evec{k} \convol \bar\Sigma^a_\evec{k})(t_1,t_2) \text{.}
\end{align}
\end{subequations}
Note that $\bar\Sigma^r_{{\rm eff},\evec{k}}(t_1,t_2)^\dagger = \bar\Sigma^a_{{\rm eff},\evec{k}}(t_2,t_1)$, so that the Hermiticity properties of equation~\cref{eq:local_equation_precursor} are preserved. The replacements~\cref{eq:convolution_localization} reduce the dynamical convolutions to simple matrix products of the local function $\delta S^\lt_\evec{k}(t,t)$ with the effective self-energy. The convolutions in the effective self-energy now involve the spectral function which is taken to be non-dynamical, $S^\mathcal{A} = S^\mathcal{A}_{\rm ad}$, as explained in section~\cref{sec:adiabatic_background_solution}.

%
\subsection{Closed local equation}
%

We now use the local ansatz~\cref{eq:the_local_ansatz} as described in section~\cref{sec:local_ansatz} together with equations~\cref{eq:local_equation_precursor,eq:first_order_adiabatic_source_term,eq:effective_self_energy_def} to write the first order local equation of motion for the local Wightman function $\delta \bar S^\lt_\evec{k}(t,t)$. We suppress the time arguments here for brevity as all quantities in the equation are now functions of the single time variable $t$. The equation, first presented in~\cite{Jukkala:2021sku}, takes a very simple form:
\begin{equation}
    \partial_t \delta\bar S^\lt_\evec{k} =
    -\im \comm[\big]{H_\evec{k}, \delta\bar S^\lt_\evec{k}}
    - \bigl(
        \im \bar\Sigma^r_{{\rm eff},\evec{k}} \, \delta\bar S^\lt_\evec{k} + \text{H.c.}
    \bigr)
    - \partial_t \bar S^\lt_{{\rm ad},\evec{k}} \text{.}
    \label{eq:the_closed_local_equation}
\end{equation}
Equation~\cref{eq:the_closed_local_equation} is our main quantum transport equation for the fermion propagator. The equation is an ordinary matrix differential equation for $\delta \bar S^\lt_\evec{k}(t,t)$. It is a closed local equation due to the local ansatz which manifests as the effective self-energy functions $\Sigma_{\rm eff}$. Despite its simple appearance and the used first order approximation the equation is still very general~\cite{Jukkala:2021sku}. The form of the local function $\delta \bar S^\lt_\evec{k}(t,t)$ has not been restricted in any way so the equation can describe both flavor and particle--antiparticle coherence of mixing fermions. Equation~\cref{eq:the_closed_local_equation} contains the effects of the thermal medium, including thermal width and dispersive corrections in the weak coupling expansion, via the adiabatic propagators in the effective self-energy and the source term. It also covers as special limits both the cQPA-method and the semiclassical Boltzmann equations~\cite{Jukkala:2021sku}.
We have summarized in table~\cref{tbl:LA-method_steps} the main points which were used to obtain the final closed local equation~\cref{eq:the_closed_local_equation}.

%
\begin{table}[t!]
    \centering
    \begin{tabularx}{\textwidth}{r>{\raggedright}X>{\raggedright}X}
    \toprule
    \textbf{\#} & \textbf{Step} & \textbf{Effect} \tabularnewline
    \midrule
    1. &
    Expand the full solution around a non-dynamical background solution. &
    Decoupling of the pole equations from the local dynamical equation. \tabularnewline[1.6em]
    2. &
    Model the non-local part of the propagator as a homogeneous transient. &
    Closure of the interaction terms in the local dynamical equation. \tabularnewline
    \bottomrule
    \end{tabularx}
    \caption{The steps used in the local approximation (LA) method to obtain a closed dynamical equation for the local propagator (in dissipative systems). The key point of the method is the local ansatz~\cref{eq:the_local_ansatz} used in step 2.}
    \label{tbl:LA-method_steps}
\end{table}
%

%
\subsection{Generalized density matrix equations}
%

In the LA-method the local equation~\cref{eq:the_closed_local_equation} was formulated directly in the two-time presentation at the propagator level and the projection matrix parametrization was not needed (yet). This is in contrast to the cQPA where the projection matrix parametrization is used already in finding the ansatz~\cref{eq:the_cQPA_ansatz} for the propagator. However, in the LA-method the projection matrix parametrization is equally essential when actually solving the equation as it separates the quickly and slowly varying components of the propagator. We adopt also here the terminology of mass and coherence shell functions used in cQPA, although in the more general LA-method these functions are not necessarily restricted to spectral phase space shells.

To derive the component equations we use the projection matrix parametrization~\cref{eq:projection_matrix_parametrization} for the two-time representation propagator $\delta S^\lt_{\evec{k}}$ in the form
\begin{equation}
    \delta \bar S^\lt_{\evec{k}ij}(t,t) =
    \sum_{\smash{\mathclap{h,s,s'=\pm}}} \, \mathcal{P}_{\evec{k}hij}^{ss'}(t) \,
    \delta f_{\evec{k}hij}^{ss'}(t) \text{.}
    \label{eq:delta_S_parametrization}
\end{equation}
Here $\delta f_{\evec{k}hij}^{ss'}$ are the phase space distribution functions of the non-equilibrium deviation $\delta S^\lt_{\evec{k}}(t,t)$. The mass and coherence shell functions are defined similarly to~\cref{eq:cQPA_distribution-fun_def}:
\begin{equation}
    \delta f^{m,\pm}_{\evec{k}hij} \eqdef \delta f^{\pm\pm}_{\evec{k}hij} \text{,} \hspace{3em}
    \delta f^{c,\pm}_{\evec{k}hij} \eqdef \delta f^{\pm\mp}_{\evec{k}hij} \text{.}
\end{equation}
The corresponding functions $f_{\rm ad}$ for the adiabatic function $S^\lt_{{\rm ad},\evec{k}}(t,t)$ are defined analogously~\cite{Jukkala:2021sku}. Using the parametrization~\cref{eq:delta_S_parametrization} in the local equation~\cref{eq:the_closed_local_equation} yields the equation
\begin{equation}
    \partial_t \delta f^{ss'}_{\evec{k}hij} =
    - \im \Delta\bar\omega^{ss'}_{\evec{k}ij} \delta f^{ss'}_{\evec{k}hij}
    - \mathcal{L}^{ss'}_{\evec{k}hij}[f]
    - \widetilde{\mathcal{C}}^{ss'}_{\evec{k}hij}\bigl[\bar\Sigma^r, \delta f\bigr]
    - \partial_t f^{ss'}_{{\rm ad},\evec{k}hij}\text{,}
    \label{eq:generalized_density_matrix_equation}
\end{equation}
where
\begin{subequations}
\label{eq:LA_collision_term_functions}
\begin{align}
    \widetilde{\mathcal{C}}^{ss'}_{\evec{k}hij}[\Sigma, f] &= \smash{\sum_{l, r}}
    \Bigl(
        \widetilde C^{srs'}_{\evec{k}hilj}[\Sigma] \,f^{rs'}_{\evec{k}hlj}
        + \widetilde C^{{\star},srs'}_{\evec{k}hilj}[\Sigma] f^{sr}_{\evec{k}hil}
    \Bigr) \text{,}
    \\*
\shortintertext{with}
    \widetilde C^{srs'}_{\evec{k}hilj}[\Sigma] &\eqdef
    \frac{\im \tr\bigl[
        P^{s's}_{\evec{k}hji} \Sigma_{{\rm eff},\evec{k}il}(t,t)
        \mathcal{P}^{rs'}_{\evec{k}hlj}
    \bigr]}
    {\tr\bigl[ \mathcal{P}^{ss'}_{\evec{k}hij} \gamma^0 \bigr]} \text{,}
    \label{eq:effective_self_energy_projection-1}
    \\*
    \widetilde C^{{\star},srs'}_{\evec{k}hilj}[\Sigma] &\eqdef
    \frac{-\im \tr\bigl[
        \mathcal{P}^{sr}_{\evec{k}hil} (\Sigma_{{\rm eff},\evec{k}}(t,t)^\dagger)_{lj}
        P^{s's}_{\evec{k}hji}
    \bigr]}
    {\tr\bigl[ \mathcal{P}^{ss'}_{\evec{k}hij} \gamma^0 \bigr]} \text{.}
    \label{eq:effective_self_energy_projection-2}
\end{align}
\end{subequations}
Note that in equation~\cref{eq:generalized_density_matrix_equation} $f \eqdef f_{\rm ad} + \delta f$, and the projection gradient terms $\mathcal{L}^{ss'}_{\evec{k}hij}[f]$ are the same as those already defined in~\cref{eq:cQPA_projection_matrix_gradient_terms}. Also, the collision term functions~\cref{eq:LA_collision_term_functions} have the same structure as those defined in~\cref{eq:cQPA_collision_term_functions} but here they feature the effective self-energy. We also used here the notation~\cref{eq:frequency_diff_sum_notation} for the leading term in~\cref{eq:generalized_density_matrix_equation} and wrote only a single equation which describes both $\delta f^m$ and $\delta f^c$.

Equation~\cref{eq:generalized_density_matrix_equation} is our master equation for the non-equilibrium distribution functions $\delta f^{ss'}_{\evec{k}hij}$, and it has the form of a generalized density matrix equation. This equation was presented in~\cite{Jukkala:2021sku} with the specific choice of the ``symmetric'' normalization
\begin{equation}
    \bigl(N_{\evec{k}hij}^{ss'}\bigr)_{\rm symm} \eqdef
    \tr\bigl(P_{\evec{k}hij}^{ss'} \gamma^0\bigr)^{-\frac{1}{2}}
    = \sqrt{
        \frac{2 \omega_{\evec{k}i} \omega_{\evec{k}j}}
        {\omega_{\evec{k}i} \omega_{\evec{k}j} + ss'(m_i m_j - \abs{\evec{k}}^2)}
    }
    \label{eq:symmetric_projection_normalization_flavored}
\end{equation}
which further simplifies the equation in several ways.%
\footnote{One is that when the normalization factors satisfy the condition $(N^{ss'}_{\evec{k}hij})^* = N^{s's}_{\evec{k}hji}$ then the collision term coefficients $\smash{\widetilde C}$ and $\smash{\widetilde C^{\star}}$ become related as $\smash{\widetilde C^{{\star},srs'}_{\evec{k}hilj}[\Sigma] = (\widetilde C^{s'rs}_{\evec{k}hjli}[\Sigma])^*}$.}
Here we have retained the more general form of equation~\cref{eq:generalized_density_matrix_equation} with an arbitrary normalization factor. In~\cite{Jukkala:2021sku} we have also provided a simplified version of the equation where the rapidly oscillating coherence functions $\delta f^c$ have been integrated out. This is possible when the mass shell flavor oscillations are slower than the particle--antiparticle oscillations and it greatly facilitates the practical numerical solution of these equations.

\subsubsection{Comparison and reduction to cQPA}

Equation~\cref{eq:generalized_density_matrix_equation} is very similar to the corresponding cQPA equations~\cref{eq:full_cQPA_DE} but there are some major differences. The main ones are that equation~\cref{eq:generalized_density_matrix_equation} is formulated for the deviations $\delta f$ instead of the total functions $f$ and the collision term projections involve the effective self-energy $\Sigma_{{\rm eff}}$ instead of just $\Sigma$. In fact, in the LA-method the deviations $\delta f$ are defined by the adiabatic functions $f_{\rm ad}$ which are provided externally to the equation. The evaluation of the effective self-energies~\cref{eq:effective_self_energy_def} also depends on the supplied adiabatic spectral function. This means that the specification of the adiabatic functions controls the type of approximation used in the equation. This is in contrast to cQPA where the spectral limit is assumed from the outset.

Another advantage of the LA-method compared to the cQPA is that the leading coherence gradient resummation (considered in section~\cref{sec:cQPA_reorganized_gradients}) is built in to the effective self-energy. This is because the dispersion relation used for evaluating the self-energy is imposed by the adiabatic spectral function via the convolution integral~\cref{eq:effective_self_energy_def}. Because the spectral function is non-dynamical and contains only ``mass shells'' the self-energy is then automatically evaluated with the correct dispersion.

It has been shown that the local ansatz~\cref{eq:the_local_ansatz} can be reduced to the cQPA ansatz~\cref{eq:the_cQPA_ansatz} in a spectral approximation~\cite{Jukkala:2021sku}.
Equations~\cref{eq:generalized_density_matrix_equation} can be also directly reduced to the cQPA equations~\cref{eq:full_cQPA_DE} by making the spectral approximation for the adiabatic propagators. This is done by dropping $\Sigma^{\mathcal{A}}$ in equation~\cref{eq:adiabatic_solution_pole} and using the free dispersion relations (or the quasiparticle relations if appropriate). For example, when using the free spectral function the collision term functions~\cref{eq:LA_collision_term_functions} are reduced to~\cref{eq:cQPA_collision_term_functions} at leading order in gradients~\cite{Jukkala:2021sku}. In other words, the effective self-energy is then reduced to a regular self-energy projection~\cite{Jukkala:2021sku}. There are some technical differences between the LA-equations and the cQPA-equations even after this but the difference of the equations should then be first order in both gradients and coupling expansion. Also, to get equations~\cref{eq:generalized_density_matrix_equation} explicitly into the form~\cref{eq:full_cQPA_DE} one must still eliminate $\delta f$ by using $f = f_{\rm ad} + \delta f$ and reinstate some terms proportional to the adiabatic functions which were estimated to be small [the terms corresponding to the first and second lines of~\cref{eq:local_eq_general_source}].

%% file: chapters/ch5_applications.tex
\chapter{Applications of coherent QKEs}
\label{chap:applications}

In this chapter we apply the methods we have developed in chapter~\cref{chap:cQPA_and_LA-method} to baryogenesis in the early universe setting. The cQPA is applied to a toy-model of electroweak baryogenesis as an example of features of coherent QKEs~\cite{Jukkala:2019slc}. The LA-method is applied to resonant leptogenesis where we make a detailed analysis of the generation and evolution of the CP-asymmetry in the minimal leptogenesis model~\cite{Jukkala:2021sku}.

%
\section{Electroweak baryogenesis}
\label{sec:ewbg_application}
%

As an example on how to use cQPA we now consider a simple toy-model with one fermion flavor and a time-dependent complex mass. This setup, considered in~\cite{Jukkala:2019slc}, is a time-dependent analogue of the expanding bubble wall during a first order electroweak phase transition in EWBG. The main focus here is not on phenomenology but on the interplay between CP-violation and quantum coherence, and how coherence emerges and affects the results. We also confirm the shell structure predicted by cQPA in an analysis with exact solutions in a specific free case~\cite{Jukkala:2019slc}.

In EWBG the most prominent fermionic CP-violating source, contributing to the chiral asymmetry, is the axial vector current
\begin{equation}
    j^{\mu 5}(x) \eqdef \expectval[\big]{\bar\psi(x) \gamma^\mu \gamma^5 \psi(x)}
    \text{.} \label{eq:axial_vector_current}
\end{equation}
It couples directly to the electroweak vacuum structure through the axial anomaly and can thus bias the sphaleron transitions which produce the baryon asymmetry~\cite{Prokopec:2013ax}. Here we concentrate on the axial charge density $j^{05}$ which is the zeroth component of the current~\cref{eq:axial_vector_current}. It is also more generally related to particle asymmetries. Using the definitions~\cref{eq:fermionic_CTP_propagators} we can write it in terms of the fermion CTP propagator as
\begin{equation}
    j^{05}(x) = \tr\bigl[
        \gamma^5 \bar S^\lt(x,x)
    \bigr] \text{,} \label{eq:axial_charge_density}
\end{equation}
where we used the barred propagator notation introduced below equations~\cref{eq:propagator_hermiticity_properties}.

%
\subsection{Single flavor cQPA with CP-violating mass}
%

We consider a spatially homogeneous and isotropic system and a single Dirac fermion $\psi$ with a complex time-dependent mass parameter $m_{\rm c}(t)$. The free part of the Lagrangian reads
\begin{equation}
    \mathcal{L} = \im \bar\psi \slashed\partial \psi
    - m_{\rm c} \bar{\psi}_{\rm L} \psi_{\rm R} - m_{\rm c}^* \bar{\psi}_{\rm R} \psi_{\rm L} 
    \text{.}
    \label{eq:1-flavor_complex_mass_Lagrangian}
\end{equation}
The complex mass here is an effective description which can arise from complex scalar field vacuum expectation values, see for example~\cite{Cline:2020jre}.
This system will be CP-violating if the complex phase of the mass changes in time. Generally, time-dependence of the mass can be a result of the expansion of the universe, like considered in section~\cref{sec:QFT_in_expanding_spacetime}, or it may arise from the changing Higgs vacuum expectation value in phase transitions in the early universe.

\subsubsection{Equations in the mass eigenbasis}

The cQPA equations for this system are the single flavor special case of the more general equations~\cref{eq:full_cQPA_DE}. The effect of the complex mass can be taken into account via the mixing gradient terms given in equations~\cref{eq:cQPA_DE_with_mixing_gradients,eq:mixing_gradient_terms}. We perform the transformations described in equations~\cref{eq:mass_matrix_diagonalization,eq:mass_eigenbasis_transformation,eq:cQPA_DE_with_mixing_gradients,eq:mixing_gradient_terms} using
\begin{subequations}
\label{eq:complex_mass_definitions}
\begin{align}
    m_{\rm c} &= m \e^{\im \theta} \text{,}
    \label{eq:complex_mass_notations}
    \\*
    Y &= \e^{-\im \frac{\theta}{2}} P_{\rm L} + \e^{\im \frac{\theta}{2}} P_{\rm R} = \e^{\im \frac{\theta}{2} \gamma^5}\text{,}
    \\*
    \Xi &= \sfrac{1}{2} \dot\theta \gamma^5 \text{,} 
\end{align}
\end{subequations}
where $m \eqdef \abs{m_{\rm c}}$, $\theta \eqdef \Arg(m_{\rm c})$ and $\dot\theta \eqdef \partial_t \theta$.
We also choose here the ``symmetric'' normalization~\cref{eq:symmetric_projection_normalization_flavored} for the projection matrix para\-metrization~\cref{eq:cQPA_projection_matrix_parametrization}:
\begin{equation}
    N^{ss'}_{\evec{k}h} \eqdef \tr\bigl(P_{\evec{k}h}^{ss'} \gamma^0\bigr)^{-\frac{1}{2}}
    \text{,} \qquad
    N^{ss}_{\evec{k}h} = \frac{\omega_\evec{k}}{m} \text{,} \qquad
    N^{s,-s}_{\evec{k}h} = \frac{\omega_\evec{k}}{\abs{\evec{k}}} \text{.}
    \label{eq:optimal_1-flavor_projection_normalization}
\end{equation}
The resulting single-flavor cQPA equations can then be cast into the form
\begin{subequations}
\label{eq:1-flavor_cQPA_with_complex_mass}
\begin{align}
    \partial_t f^{m,s}_{\evec{k}h} &=
    \sum_{r} \Phi^{r}_{\evec{k}h} f^{c,r}_{\evec{k}h}
    - \tr\bigl[\mathcal{C}_{\rm coll} \mathcal{P}^{m,s}_{\evec{k}h} \bigl]
    \text{,}
    \\*
    \partial_t f^{c,s}_{\evec{k}hij} &=
    -2 \im s \Bigl[
        \omega_{\evec{k}} + \sfrac{\abs{\evec{k}}}{m} \Im(\Phi^{-}_{\evec{k}h})
    \Bigr] f^{c,s}_{\evec{k}h}
    - \Phi^{-s}_{\evec{k}h} \sum_{r} f^{m,r}_{\evec{k}h}
    - \tr\bigl[\mathcal{C}_{\rm coll} \mathcal{P}^{c,-s}_{\evec{k}h} \bigl]
    \text{,}
    \\*
    \mathcal{C}_{\rm coll} &= \sum_{h, s} \biggl(
        \Bigl[
            f^{m,s}_{\evec{k}h} \bar\Sigma^{\mathcal{A}}(s\omega_\evec{k})
            - \frac{s}{2} \bar\Sigma^\lt(s\omega_\evec{k})
        \Bigr] \mathcal{P}^{m,s}_{\evec{k}h}
        + f^{c,s}_{\evec{k}h} \bar\Sigma^{\mathcal{A}}(s\omega_\evec{k}) \mathcal{P}^{c,s}_{\evec{k}h}
    \biggr) + \text{H.c.}
\end{align}
\end{subequations}
We defined here
\begin{equation}
    \Phi^s_{\evec{k}h} \eqdef \frac{1}{2} \biggl(
        \frac{\abs{\evec{k}} \dot m}{\omega_\evec{k}^2}
        + \im s h \frac{m \dot\theta}{\omega_\evec{k}}
    \biggr) \text{,}
    \label{eq:mass_basis_gradient_factor}
\end{equation}
which contains the contributions from both projection gradient and mixing gradient terms: $\mathcal{L}^{ss'}_{\evec{k}h} \propto \dot m$ and $\mathcal{X}^{ss'}_{\evec{k}h} \propto \dot\theta$. Note that we have suppressed the flavor indices in these equations because there is only one flavor. We also discarded the external gradient corrections of the self-energy by replacing $\Sigma_{\rm out}$ with $\Sigma$ and we dropped the dispersive self-energy $\Sigma^{\rm H}$ from the equations for simplicity.

Equations~\cref{eq:1-flavor_cQPA_with_complex_mass} are equivalent to the single flavor cQPA equations given in~\cite{Jukkala:2019slc}, however here we used the mass eigenbasis in the projection matrix parametrization. The exact form of the equations is then different because in~\cite{Jukkala:2019slc} the projection matrix parametrization was defined using the complex mass parameter directly. Equations~\cref{eq:1-flavor_cQPA_with_complex_mass} are also similar to the single flavor equations given in~\cite{Fidler:2011yq} but they too differ because of the different normalization~\cref{eq:optimal_1-flavor_projection_normalization}. We chose here this different approach to be consistent with the multi-flavor case presented in section~\cref{sec:cQPA}. In the multi-flavor case the diagonalization approach is mandatory if the original mass matrix is not real and diagonal. The symmetric normalization also makes the equations simpler. The resulting equations are also physically more transparent: one can see directly from the form of the equations~\cref{eq:1-flavor_cQPA_with_complex_mass} that a changing phase ($\dot\theta \neq 0$) is a source for helicity asymmetry of the distribution functions $f^{m,c}$. The helicity asymmetry in turn leads to CP-violation, as we will see below.

\paragraph{Connection to the complex mass basis}

To bridge the gap to~\cite{Jukkala:2019slc} we now give the relationship between the phase space distribution functions $f^m$ and $f^c$ used in~\cref{eq:1-flavor_cQPA_with_complex_mass} and the corresponding functions used in~\cite{Jukkala:2019slc} which we denote here by $\widehat f^m$ and $\widehat f^c$. Since in both cases the Wightman function in the original complex mass basis~\cref{eq:1-flavor_complex_mass_Lagrangian} is the same, we have $\bar{\mathcal{S}}^\lt = Y \widehat{\mathcal{S}}^\lt Y^\dagger$ where $\bar{\mathcal{S}}^\lt$ and $\widehat{\mathcal{S}}^\lt$ denote the integrated Wightman functions in the different bases.
Using the different projection matrix parametrizations on each side of the equation then yields
\begin{subequations}
\label{eq:1-flavor_distribution_transformations}
\begin{align}
    \frac{m}{\omega_\evec{k}} N^{ss}_{\evec{k}h} \; f^{m,s}_{\evec{k}h}
    &= s \widehat f^{m,s}_{\evec{k}h} \text{,}
    \\*
    \frac{\abs{\evec{k}}}{\omega_\evec{k}} N^{s,-s}_{\evec{k}h} f^{c,s}_{\evec{k}h}
    &= \frac{1}{m} \biggl(
        \frac{\abs{\evec{k}}}{\omega_\evec{k}} m_{\rm R} + \im s h m_{\rm I}
    \biggl) \widehat f^{c,s}_{\evec{k}h} \text{.}
\end{align}
\end{subequations}
The normalization $N^{ss'}_{\evec{k}h}$ of the mass eigenbasis side is still general here. Choosing~\cref{eq:optimal_1-flavor_projection_normalization} reduces the coefficients of $f^m$ and $f^c$ on the left-hand side to unity. We also defined here
\begin{equation}
    m_{\rm c} \eqdef m_{\rm R} + \im m_{\rm I} \text{,}
    \qquad
    (m_{\rm R}, m_{\rm I} \in \reals)
    \label{eq:complex_mass_Re_Im_def}
\end{equation}
and used $\e^{-\im\theta} = (m_{\rm R} - \im m_{\rm I})/m$ with the notation of equation~\cref{eq:complex_mass_notations}. We now also have the identities
\begin{align}
    \dot m = \frac{m_{\rm R} \dot m_{\rm R} + m_{\rm I} \dot m_{\rm I}}{m}
    \text{,} \qquad
    m \dot\theta = \frac{m_{\rm R} \dot m_{\rm I} - m_{\rm I} \dot m_{\rm R}}{m}
    = \frac{m_{\rm R}^2}{m} \partial_t \Bigl(\frac{m_{\rm I}}{m_{\rm R}} \Bigr)
    \text{.}
    \label{eq:mass_and_mixing_gradient_identities}
\end{align}
Using equations~\cref{eq:complex_mass_definitions,eq:optimal_1-flavor_projection_normalization,eq:1-flavor_distribution_transformations,eq:complex_mass_Re_Im_def,eq:mass_and_mixing_gradient_identities} it can be verified after a lengthy but entirely straightforward calculation that equations~\cref{eq:1-flavor_cQPA_with_complex_mass,eq:mass_basis_gradient_factor} exactly reproduce the cQPA equations given in~\cite{Jukkala:2019slc}.

%
\subsection{The mass profile}
%

So far we have considered a general complex mass $m_{\rm c}(t)$. We now choose the following $\tanh$-profile which was used in~\cite{Jukkala:2019slc}:
\begin{equation}
    m_{\rm c}(t) = m_1 + m_2 \tanh\biggl(-\frac{t}{\tau_{\rm w}}\biggr) \text{.}
    \label{eq:time-dependent_tanh_profile}
\end{equation}
This is a time-dependent analogue of the simplified tanh-profile~\cite{Cline:2017jvp,Bodeker:2004ws} for the bubble wall shape in a first order EWPT. For this reason we refer to the parameter $\tau_{\rm w}$, which describes the duration of the transition, simply as the wall width. The specific profile~\cref{eq:time-dependent_tanh_profile} was chosen because in the free theory exact analytic solutions for the fermion mode functions are then available~\cite{Prokopec:2013ax}. It also serves as a toy-model for EWBG as it features CP-violation when $m_1$ and $m_2$ are complex.

The physical CP-violation arises from the relative complex phase of the parameters $m_1$ and $m_2$ in equation~\cref{eq:time-dependent_tanh_profile}. To simplify the equations we assume that a global $\U(1)$ transformation of the chiral fermion fields has been performed which makes $m_2$ real~\cite{Prokopec:2013ax}. Hence, we can assume with no loss of generality that $m_1 = m_{1\rm{R}} + \im m_{1\rm{I}}$ and $m_2 = m_{2\rm{R}}$ with constant real parameters $m_{1\rm{R}}, m_{1\rm{I}}, m_{2\rm R}$. Using the definition~\cref{eq:complex_mass_Re_Im_def}, this implies that
\begin{subequations}
\label{eq:simplified_mass_profile}
\begin{align}
    m_{\rm R}(t) &= m_{1\rm{R}} + m_{2\rm{R}} \tanh\biggl(-\frac{t}{\tau_{\rm w}}\biggr)
    \text{,}
    \\*
    m_{\rm I}(t) &= m_{1\rm{I}} \text{.}
\end{align}
\end{subequations}
The identities~\cref{eq:mass_and_mixing_gradient_identities} are now also simplified as the terms with $\dot m_{\rm I}$ are dropped.

%
\subsection{Axial charge density with interactions}
%

Now we show an example calculation of the axial charge density evolution using cQPA with decohering interactions. The numerical results were calculated in~\cite{Jukkala:2019slc}. The spatial Fourier transform of the axial charge density~\cref{eq:axial_charge_density}, calculated with the cQPA propagator, turns out to be%
\footnote{\label{footnote:cQPA-j05}The chiral transformation $Y$ is not a symmetry of the Lagrangian so the current here is calculated in the original basis using $j^{05}_\evec{k} = \tr[\gamma^5 Y^\dagger \bar S^\lt_\evec{k}(t,t) Y]$. Equation~\cref{eq:axial_charge_density_in_cQPA} also has a corrected sign compared to~\cite{Jukkala:2019slc}, which affects the figures but has no effect on the main results.}
\begin{equation}
    j^{05}_\evec{k} = \sum_{h,s} \frac{h}{\omega_\evec{k}} \Bigl(
        \abs{\evec{k}} f^{m,s}_{\evec{k}h} - m f^{c,s}_{\evec{k}h}
    \Bigr) \text{.}
    \label{eq:axial_charge_density_in_cQPA}
\end{equation}
We can now see that this density is non-vanishing (and there is CP-violation) only if there is a helicity asymmetry in $f^{m(c)}_{\evec{k}h}$.
Furthermore, from the form of equations~\cref{eq:1-flavor_cQPA_with_complex_mass} we saw that there will generically be helicity asymmetry, even in the free theory, as long as $\dot\theta \neq 0$. These features are not as transparent in the complex mass basis as presented in~\cite{Jukkala:2019slc}. However, physical results for the current are of course the same in both conventions.

%
\begin{figure}[t!]
    \centering
    \includegraphics[width=0.9\textwidth]{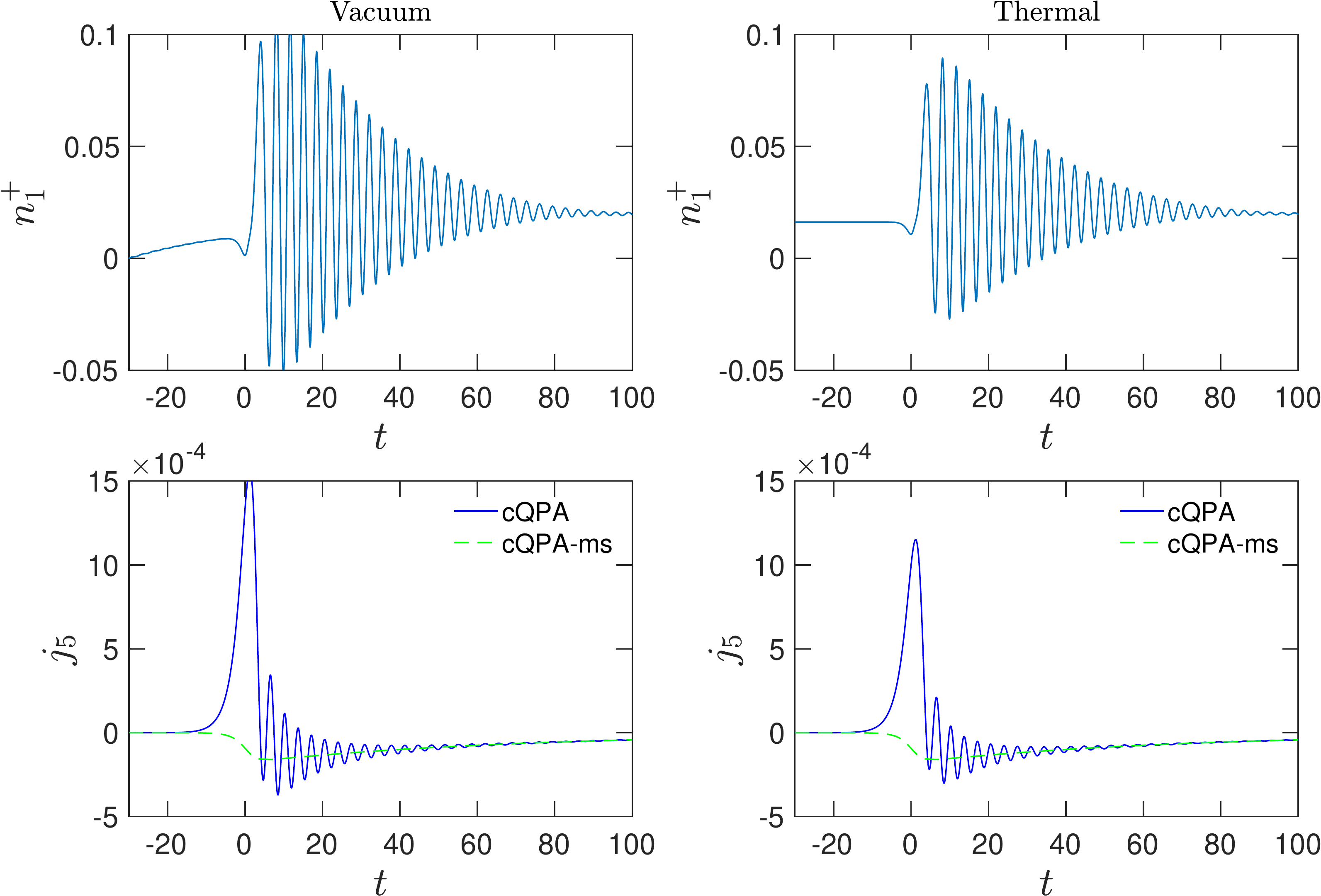}
    \caption{Time-evolution of the number density $n^{+}_{+}$ of positive helicity particles (top) and the axial charge density $j^{05}$ (bottom) from interacting single-flavor cQPA with vacuum and thermal initial conditions. Mass profile~\cref{eq:simplified_mass_profile} with parameters $m_{1\rm{R}} = 0.1$, $m_{2\rm{R}} = 1$, $m_{\rm{I}} = 0.1$ and $\tau_{\rm w} = 5$ was used. The green dashed line shows the result with only the mass-shell contribution throughout the evolution. Adapted from article~\cite{Jukkala:2019slc} (\href{https://creativecommons.org/licenses/by/4.0/}{CC BY 4.0}), with corrected sign for $j_5$.}
    \label{fig:single_flavor_cQPA_with_interactions}
\end{figure}
%

In figure~\cref{fig:single_flavor_cQPA_with_interactions} we show the results for the evolution of the total particle number density $n^{+}_{+} \eqdef \int \dd^3 \evec{k} /(2\pi)^3 n^{+}_{\evec{k}{+}}$ for particles ($s$ = +1) of positive helicity and the total axial charge density $j^{05} = \int \dd^3 \evec{k} /(2\pi)^3 j^{05}_\evec{k}$ calculated from~\cref{eq:axial_charge_density_in_cQPA}. Here we defined the phase space number densities as $n^{+}_{\evec{k}h} \eqdef f^{m,+}_{\evec{k}h}$ and $n^{-}_{\evec{k}h} \eqdef 1 + f^{m,-}_{\evec{k}h}$ and the distribution functions $f^{m,s}_{\evec{k}h}$ obey the cQPA equations~\cref{eq:1-flavor_cQPA_with_complex_mass}. As a simple example to demonstrate CP-violation and decoherence we used in the figure a chiral interaction with the following toy-model thermal equilibrium self-energy:
\begin{equation}
\begin{split}
    \bar\Sigma^{\mathcal{A}}_{\rm eq}(k) &=
    a(k^0, \abs{\evec{k}}) \, \slashed k P_{\rm L} \gamma^0 \text{,}
    \\*
    \bar\Sigma^\lt_{\rm eq}(k) &=
    2\bar\Sigma^{\mathcal{A}}_{\rm eq}(k) f_{\rm FD}(k^0) \text{,}
\end{split}
\qquad \Biggl( \text{with} \quad 
    \begin{split}
        a(\pm k^0, \abs{\evec{k}}) &= \pm a(k^0, \abs{\evec{k}}) \text{,}
        \\*
        a(\omega_\evec{k}, \abs{\evec{k}}) &= 0.03 \text{.}
    \end{split}
\Biggr)
\end{equation}
In the left panels of figure~\cref{fig:single_flavor_cQPA_with_interactions} we used the vacuum initial conditions $n^{s}_{\evec{k}h} = 0 = f^{c,s}_{\evec{k}h}$. The number density first begins to smoothly thermalize due to the interactions. In the transition region (centered around $t = 0$) it then starts oscillating rapidly due to the coherence that develops from the changing mass. The coherent oscillations also generate helicity asymmetry because of the CP-violating phase $\theta$, and this leads to the generation of the axial charge. The chiral interaction is also responsible for generating helicity asymmetry. This can be seen from the green dashed line where the $f^c$-functions have been artificially turned off. In this case the only source for helicity asymmetry is in the collision term. The decohering collisions also cause the rapid damping of the oscillations after the transition region. However, even without collisions the oscillations in these $\evec{k}$-integrated quantities would be damped (although not as strongly) because of the phase differences between different $\evec{k}$-modes~\cite{Jukkala:2019slc}. In the right panels we show the same calculations with thermal initial conditions $n^{s}_{\evec{k}h} = f_{\rm FD}(\omega_\evec{k})$, $f^{c,s}_{\evec{k}h} = 0$. The results are similar but now the particle number stays initially at the thermal value. The final thermalized particle number is slightly higher than initially because the equilibrium distribution changes across the mass transition. We used units where $T = 1$ in all plots here.

%
\subsection{Exact solution of the Dirac equation}
%

We now turn to the analysis of the free theory. One of the main results of~\cite{Jukkala:2019slc} is the independent demonstration of the shell structure, in particular the coherence shell at $k^0 = 0$, for the exact Wightman function without using cQPA. This is based on the exact mode function solutions of the free Dirac equation where the mass is a non-trivial function of time~\cite{Prokopec:2013ax}. Specifically, we use the mass profile given by equation~\cref{eq:time-dependent_tanh_profile}. We summarize here the most relevant steps and present the results. More details can be found in~\cite{Jukkala:2019slc}.

\subsubsection{Mode function equations}

We start with the canonical quantization of the fermion field operator $\hat\psi$. In a spatially homogeneous and isotropic system the field operator can be expanded using the mode functions and ladder operators as
\begin{equation}
    \hat\psi(t, \evec{x})
    = \sum_{h} \int \frac{\dd^3 \evec{k}}{(2\pi)^3 2\omega_{-}} \Bigl[
        \hat{a}_{\evec{k}h} U_{\evec{k}h}(t) \e^{\im \evec{k} \cdot \evec{x}}
        + \hat{b}^\dagger_{\evec{k}h} V_{\evec{k}h}(t) \e^{-\im \evec{k} \cdot \evec{x}}
    \Bigr] \text{.}
    \label{eq:exact_free_field_operator}
\end{equation}
Here $2\omega_{-}$ is a normalization factor and we denote $\omega_{\pm} \eqdef \omega_\evec{k}(t \to \pm\infty)$. We impose the standard canonical anticommutation relations for the field. We use the chiral representation of gamma matrices and decompose the particle and antiparticle spinors $U_{\evec{k}h}$ and $V_{\evec{k}h}$ as%
\footnote{We use the chiral representation defined \eg~in~\cite{Peskin:1995ev} where $\gamma^5 = \diag(-\idmat_2, \idmat_2)$.}
\begin{equation}
    U_{\evec{k}h}(t) = \begin{bmatrix}
        \eta_{\evec{k}h}(t) \\
        \zeta_{\evec{k}h}(t)
    \end{bmatrix} \otimes \xi_{\evec{k}h} \text{,}
    \qquad
    V_{\evec{k}h}(t) = \begin{bmatrix}
        \bar{\eta}_{\evec{k}h}(t) \\
        \bar{\zeta}_{\evec{k}h}(t)
    \end{bmatrix} \otimes \xi_{\evec{k}h} \text{.}
    \label{eq:spinor_helicity_decomposition}
\end{equation}
Here $\otimes$ is the Kronecker product and $\xi_{\evec{k}h}$ are the helicity eigenfunctions ($h = \pm1$) which satisfy
$(\evec{\sigma} \cdot \Uevec{k}) \xi_{\evec{k}h} = h \xi_{\evec{k}h}$, where further $\evec{\sigma} \eqdef (\sigma_1, \sigma_2, \sigma_3)$ and $\sigma_i$ are the standard $2 \times 2$ Pauli matrices~\cite{Peskin:1995ev}. From the Dirac equation corresponding to the Lagrangian~\cref{eq:1-flavor_complex_mass_Lagrangian} it follows that the complex \emph{mode functions} $\eta_{\evec{k}h}$ and $\zeta_{\evec{k}h}$ satisfy
\begin{subequations}
\label{eq:mode_function_equations}
\begin{align}
    \im \partial_t \eta_{\evec{k}h} + h\abs{\evec{k}} \eta_{\evec{k}h}
    &= m_{\rm c}(t) \zeta_{\evec{k}h} \text{,}
    \\*
    \im \partial_t \zeta_{\evec{k}h} - h\abs{\evec{k}} \zeta_{\evec{k}h}
    &= m_{\rm c}^*(t) \eta_{\evec{k}h} \text{.}
\end{align}
\end{subequations}
The complex mass $m_{\rm c}(t)$ is still general in these equations. The antiparticle mode functions $\bar\eta_{\evec{k}h}$ and $\bar\zeta_{\evec{k}h}$ satisfy equations analogous to~\cref{eq:mode_function_equations} with the replacements $h \rightarrow -h$ and $m_{\rm c} \rightarrow -m_{\rm c}^*$.

It turns out that for the mass profile~\cref{eq:simplified_mass_profile} the equations~\cref{eq:mode_function_equations} can be transformed to the hypergeometric differential equation~\cite{Prokopec:2013ax}. The solutions can then be written in terms of Gauss's hypergeometric functions $\prescript{}{2}{F}^{}_{1}$. There are two linearly independent solutions and for the numerical results we choose the one which corresponds to positive frequency particles (in vacuum) in the initial state~\cite{Jukkala:2019slc}. The \emph{initial} values for the particle mode functions are then fixed to
\begin{equation}
    \eta_{\evec{k}h} = \sqrt{\omega_\evec{k} - h\abs{\evec{k}}} \e^{-\im t \omega_\evec{k}}
    \text{,} \qquad
    \zeta_{\evec{k}h} = \sqrt{\omega_\evec{k} + h\abs{\evec{k}}} \e^{-\im \theta} \e^{-\im t \omega_\evec{k}}
    \text{,}
    \label{eq:mode_function_initial_conditions}
\end{equation}
where all time-dependent quantities are evaluated at $t \to -\infty$.

\subsubsection{Wightman functions from mode functions}

We can now construct the exact Wightman functions $S^{\lt,\gt}$, defined in equations~\cref{eq:fermionic_CTP_propagators}, for this system from the vacuum expectation values of field operators by using~\cref{eq:exact_free_field_operator}. The different Wightman functions contain equivalent statistical information so either one suffices for the analysis here. We choose $S^\gt$ as it is constructed from the particle mode functions $\eta_{\evec{k}h}$ and $\zeta_{\evec{k}h}$. By using the properties of the ladder operators and the decomposition~\cref{eq:spinor_helicity_decomposition} of the spinors we can express the Wightman function $S^\gt$ in the two-time representation~\cref{eq:two-time_representation_def} as
\begin{gather}
    \bar S^\gt_\evec{k}(t_1,t_2) = \sum_h M_{\evec{k}h}(t_1,t_2)
    \otimes P^{\scriptscriptstyle (2)}_{\evec{k}h} \text{,}
    \label{eq:twotime_Wightman_from_mode_functions}
\shortintertext{where}
    M_{\evec{k}h}(t_1,t_2) \eqdef
    \frac{1}{2\omega_{-}} \begin{bmatrix}
        \eta_{\evec{k}h}(t_1) \eta_{\evec{k}h}^*(t_2) &
        \eta_{\evec{k}h}(t_1) \zeta_{\evec{k}h}^*(t_2)
        \\
        \zeta_{\evec{k}h}(t_1) \eta_{\evec{k}h}^*(t_2) &
        \zeta_{\evec{k}h}(t_1) \zeta_{\evec{k}h}^*(t_2)
    \end{bmatrix} \text{.}
    \label{eq:mode_function_matrix}
\end{gather}
The two-dimensional helicity projection matrix is defined by $P^{\scriptscriptstyle (2)}_{\evec{k}h} \eqdef \frac{1}{2}(\idmat + h\evec{\sigma} \cdot \Uevec{k})$.

Now that we have equations~\cref{eq:twotime_Wightman_from_mode_functions,eq:mode_function_matrix} we can also easily calculate quantum currents from the mode functions. For example the axial charge density $j^{05}_\evec{k}$ which we considered before is given by
\begin{equation}
    j^{05}_\evec{k}(t) =  \frac{1}{2\omega_{-}} \sum_{h} \bigl(
        \abs{\eta_{\evec{k}h}(t)}^2 - \abs{\zeta_{\evec{k}h}(t)}^2
    \bigr) \text{.}
    \label{eq:axial_charge_density_from_mode_functions}
\end{equation}
To get this result we used the sum rule~\cref{eq:fermionic_sum_rule_twotime} in the form $(\bar S^\gt_\evec{k} + \bar S^\lt_\evec{k})(t,t) = \idmat$ to write the current in terms of $S^\gt$: $j^{05}_\evec{k} = \tr[\gamma^5 \bar S^\lt_\evec{k}(t,t)] = -\tr[\gamma^5 \bar S^\gt_\evec{k}(t,t)]$.

\subsubsection{Phase space structure}

%
\begin{figure}[t!]
    \centering
    \includegraphics[width=0.495\textwidth]{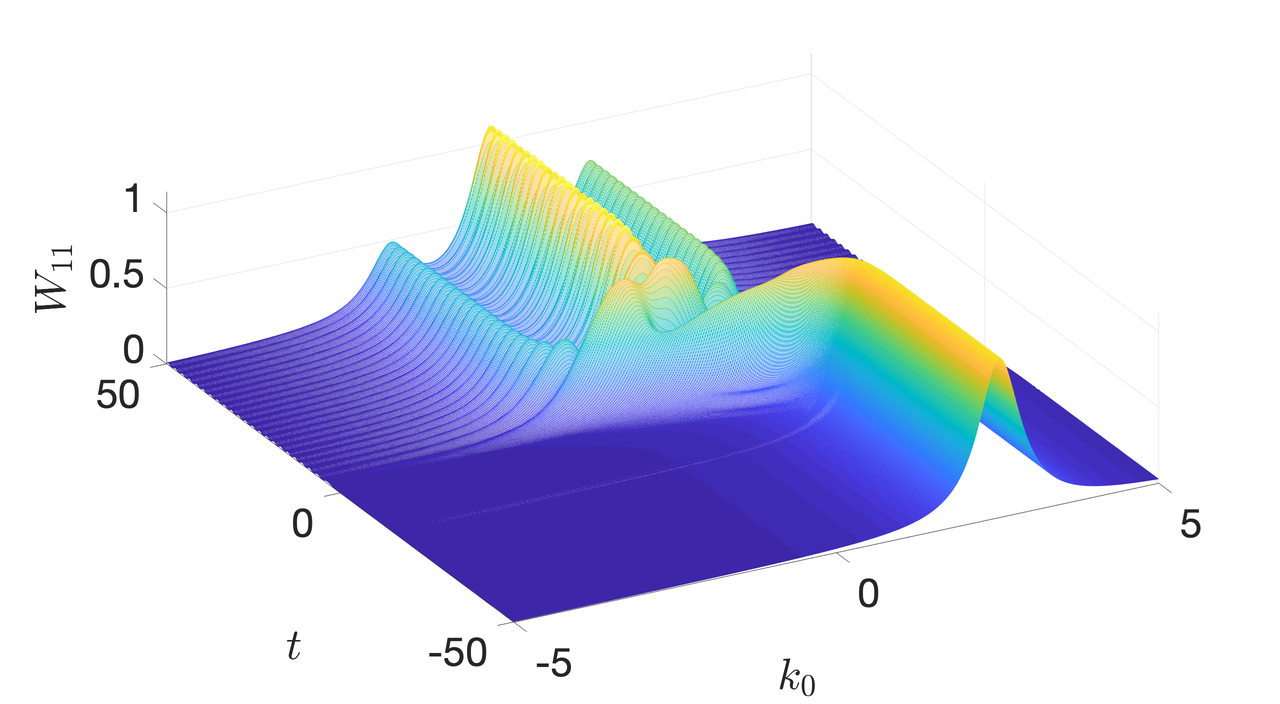}~%
    \includegraphics[width=0.495\textwidth]{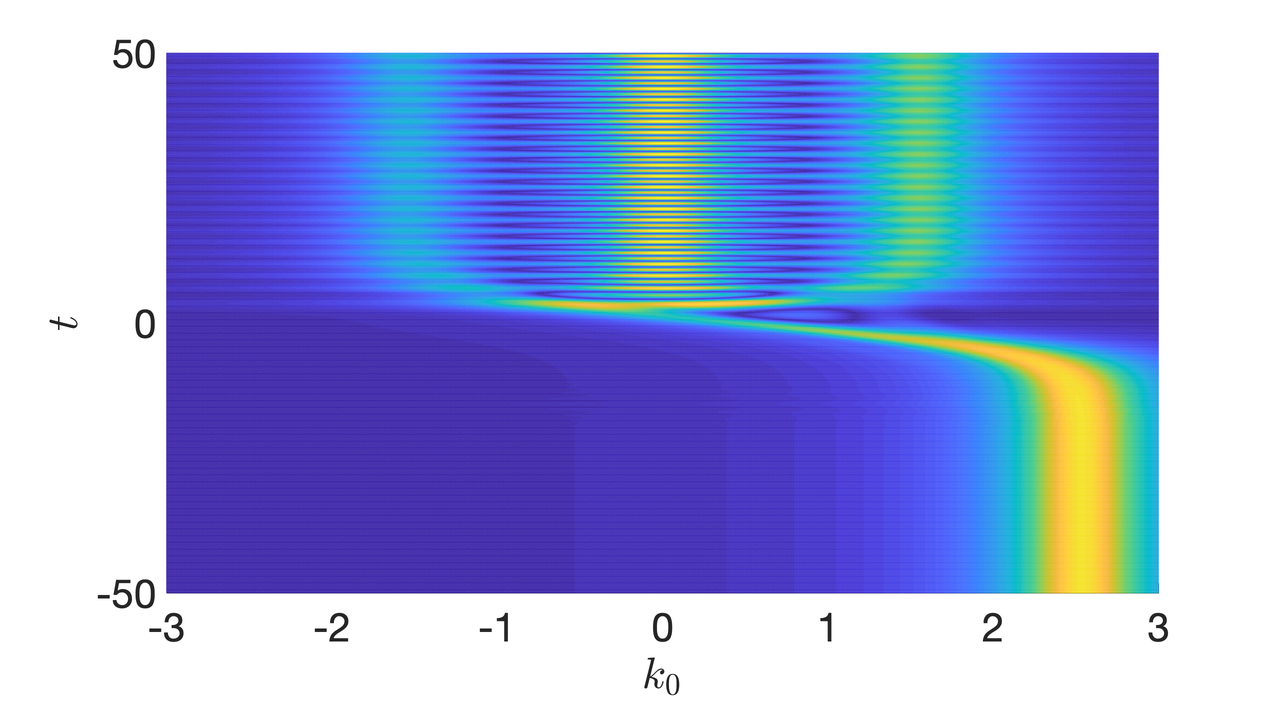}\par
    \includegraphics[width=0.495\textwidth]{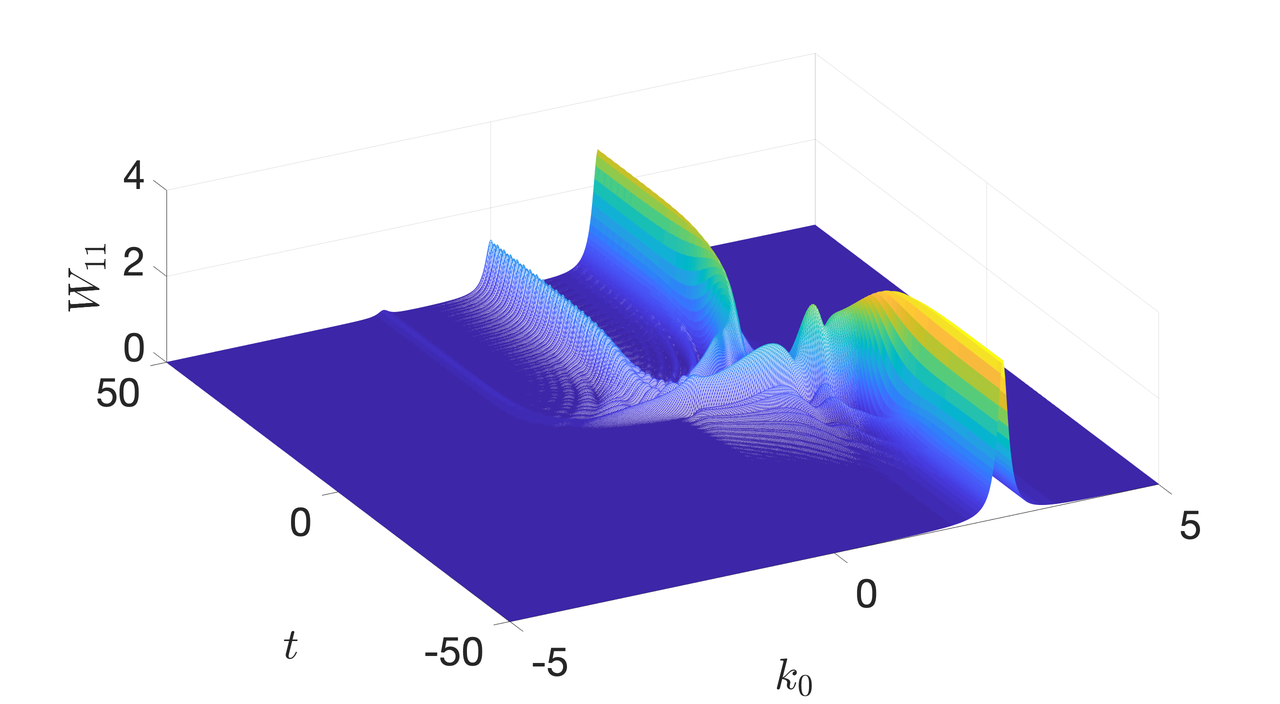}~%
    \includegraphics[width=0.495\textwidth]{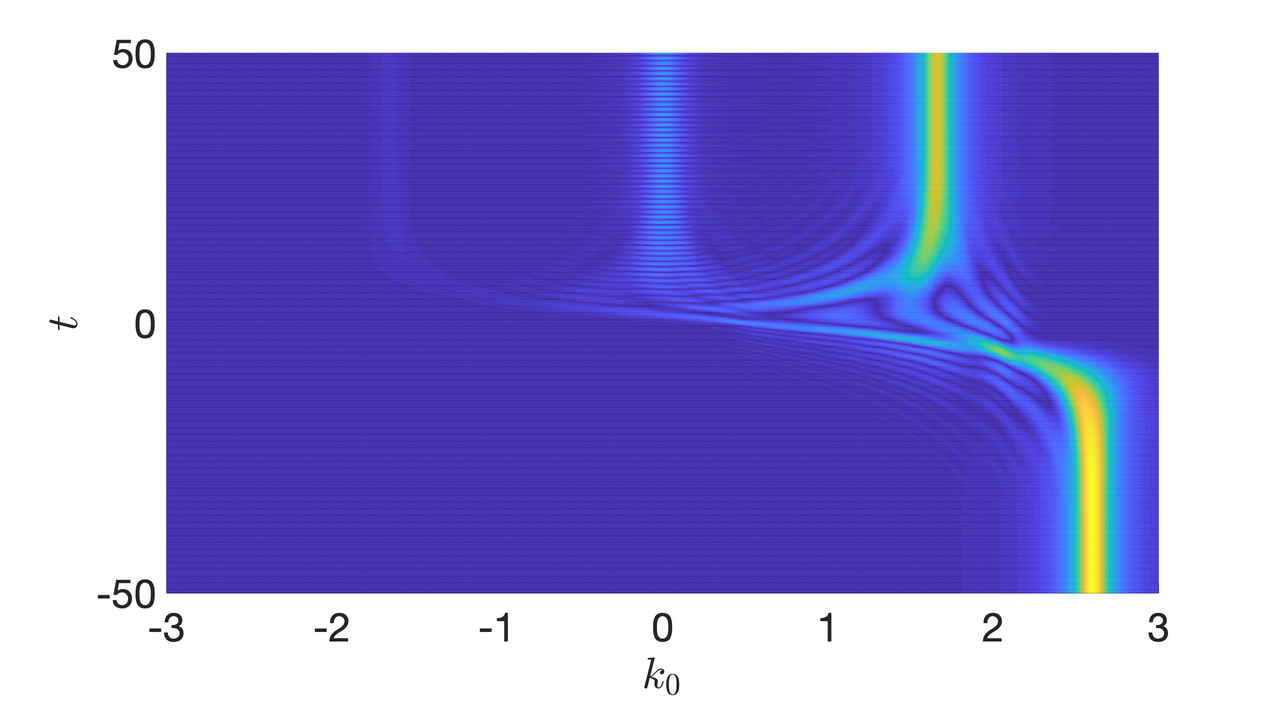}\par
    \includegraphics[width=0.495\textwidth]{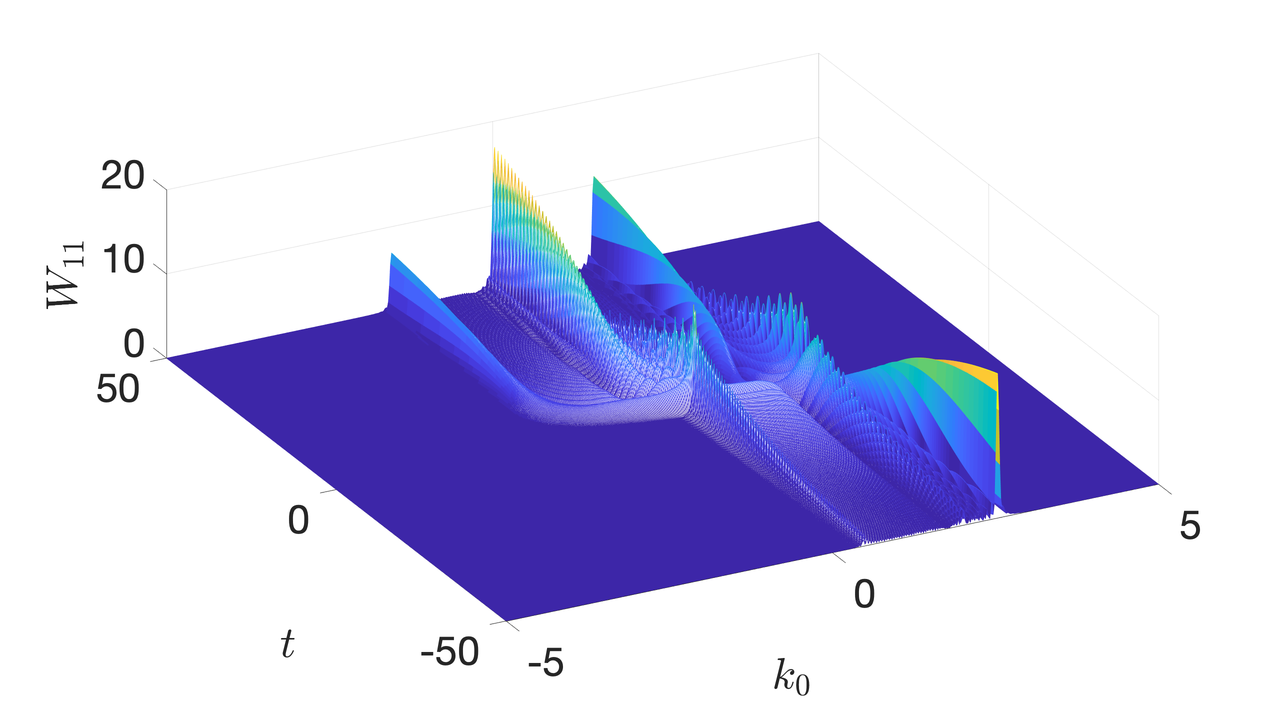}~%
    \includegraphics[width=0.495\textwidth]{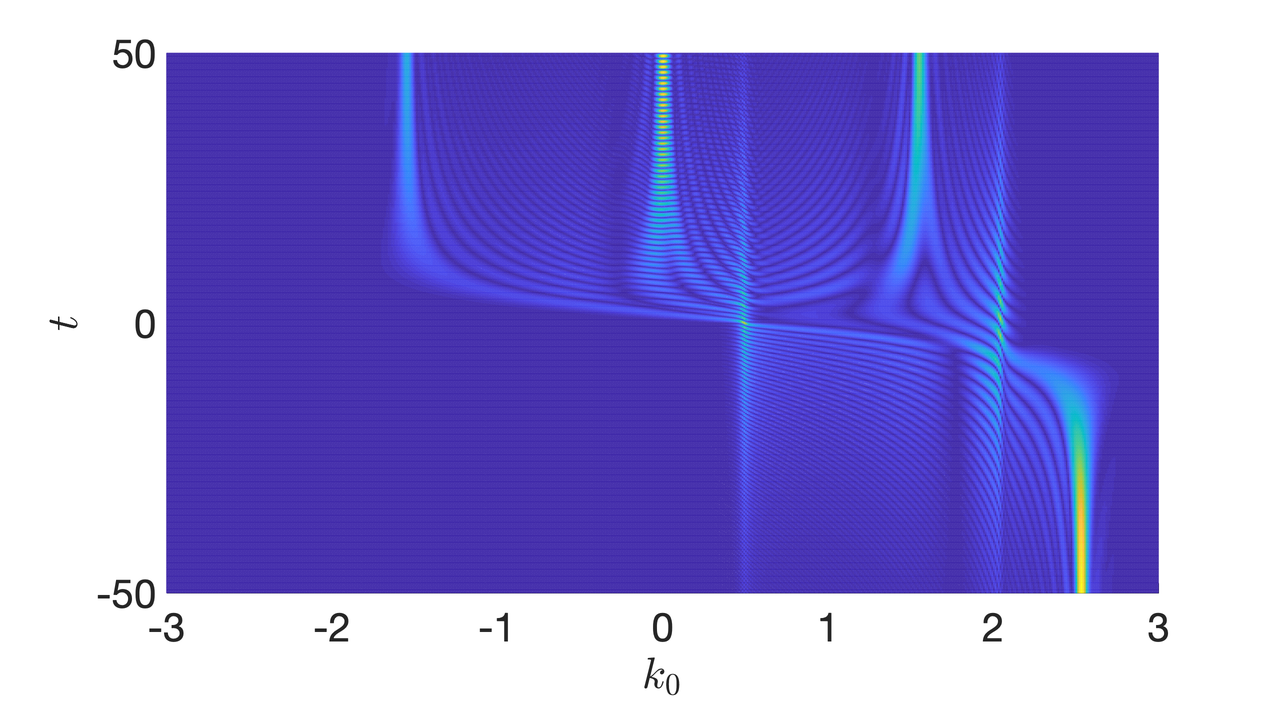}
    \caption{Phase space structure of the exact free Wightman function $S^\gt$. Shown is $\abs[\big]{W_{\evec{k}h,\Gamma}^{(11)}}$ as a function of $(k^0,t)$ for $h = +1$. The three pairs of plots have $\abs{\evec{k}} = 0.4, 0.7, 0.4$ and $\Gamma_\evec{k} = 0.4, 0.1, 0.02$ from top to bottom, respectively. We used the mass profile~\cref{eq:simplified_mass_profile} with $m_{1{\rm R}} = 0.5$, $m_{2{\rm R}} = 2$, $m_{\rm I} = -0.005$, $\tau_{\rm w} = 5$. From article~\cite{Jukkala:2019slc} (\href{https://creativecommons.org/licenses/by/4.0/}{CC BY 4.0}).}
	\label{fig:exact_Wightman_function_phase_space}
\end{figure}
%

To study the phase space structure of the Wightman function we need to calculate the Wigner transform $\bar S^\gt(k,t)$ of~\cref{eq:twotime_Wightman_from_mode_functions} by using equations~\cref{eq:general_Wigner_transform,eq:two-time_representation_inverse}. However, with free particle mode functions the Wightman function~\cref{eq:twotime_Wightman_from_mode_functions} contains correlations over arbitrarily large time intervals. This is a physical result, but it also makes the Wigner transform divergent here. In a realistic system there would be interactions which would suppress the faraway correlations. We take this into account heuristically by adding a parametric \emph{damping rate} $\Gamma_{\evec{k}h}$ and use the Wigner transform~\cite{Jukkala:2019slc}
\begin{subequations}
\begin{align}
    \bar S^\gt_\Gamma(k,t) &\eqdef \sum_h W_{\evec{k}h,\Gamma}(k^0,t)
    \otimes P^{\scriptscriptstyle (2)}_{\evec{k}h} \text{,}
    \\*
    W_{\evec{k}h,\Gamma}(k^0,t) &\eqdef
    \int_{-\infty}^{\infty} \dd r^0 \e^{\im k^0 r^0 - \Gamma_{\evec{k}h} \abs{r^0}}
    \, M_{\evec{k}h}\bigl(t + \sfrac{r^0}{2}, t - \sfrac{r^0}{2}\bigr) \text{.}
    \label{eq:Wigner_transformed_mode_matrix}
\end{align}
\end{subequations}
The form of the damping factor $\e^{-\Gamma_{\evec{k}h} \abs{r^0}}$ in equation~\cref{eq:Wigner_transformed_mode_matrix} is not arbitrary: it would result from a leading order calculation when taking the shifted poles of the interacting function into account~\cite{Jukkala:2019slc}. Setting here $\Gamma_{\evec{k}h} = 0$ recovers the usual Wigner transform of~\cref{eq:twotime_Wightman_from_mode_functions}.

In figure~\cref{fig:exact_Wightman_function_phase_space} we show the numerical results of the calculation of the absolute value of the $(1,1)$-component of $W_{\evec{k}h,\Gamma}$ [\ie~the component $\propto \eta \, \eta^*$ in~\cref{eq:mode_function_matrix}]. The other chiral components are qualitatively similar. The mode functions were calculated using the hypergeometric functions with the initial conditions~\cref{eq:mode_function_initial_conditions} and these solutions were then used in equations~\cref{eq:mode_function_matrix,eq:Wigner_transformed_mode_matrix}. The plots illustrate the phase space structure for varying momentum $\abs{\evec{k}}$ and damping $\Gamma_{\evec{k}h}$. The results verify the shell structure predicted by cQPA, albeit the shells are smeared for larger $\Gamma$, as expected. The initial state particle is described by only the mass shell and the mass transition then develops coherence and creates particles and antiparticles at their respective shells (see table~\cref{tbl:phase_space_functions}). The results also show the emergence of additional structures in the transition area and two new long-range shells in the limit of small $\Gamma$. These new structures are suppressed for larger $\Gamma$ and the shell picture of cQPA remains valid even for moderately large values of $\Gamma$, however. A more thorough analysis of the additional shells and other features of these results is presented in~\cite{Jukkala:2019slc}.

\subsubsection{Connection to cQPA}

We now present a direct connection between the exact mode functions and cQPA in the free theory, which was not explicitly given in~\cite{Jukkala:2019slc}. This can be done similarly to how we calculated the axial charge density~\cref{eq:axial_charge_density_from_mode_functions} from the mode functions. The mode functions and the cQPA distribution functions can be related via the equal-time Wightman function. The connecting relation is
\begin{equation}
    \smash{\sum_h} M_{\evec{k}h}(t,t) \otimes P^{\scriptscriptstyle (2)}_{\evec{k}h}
    = \idmat - \smash{\sum_{h,s}} Y^\dagger \bigl(
        f^{m,s}_{\evec{k}h} \mathcal{P}^{m,s}_{\evec{k}h}
        + f^{c,s}_{\evec{k}h} \mathcal{P}^{c,s}_{\evec{k}h}
    \bigr) Y \text{,} \vphantom{\sum}
\end{equation}
where we also took into account the chiral transformation $Y = \e^{\im \frac{\theta}{2} \gamma^5}$. Writing here the four-dimensional projection matrices~\cref{eq:projection_matrix_basis_element} in the chiral representation and taking projections and traces of the equation eventually yields the explicit relations
\begin{subequations}
\label{eq:fm_and_fc_relation_to_mode_functions}
\begin{align}
    f^{m,s}_{\evec{k}h} &= s - \frac{1}{2} \Bigl[{\textstyle
        \bigl(s - \frac{h \abs{\evec{k}}}{\omega_\evec{k}}\bigr) M_{\evec{k}h}^{(11)}
        + \bigl(s + \frac{h \abs{\evec{k}}}{\omega_\evec{k}}\bigr) M_{\evec{k}h}^{(22)}
        + \frac{m_{\rm c}^*}{\omega_\evec{k}} M_{\evec{k}h}^{(12)}
        + \frac{m_{\rm c}}{\omega_\evec{k}} M_{\evec{k}h}^{(21)}
    }\Bigr] \text{,}
    \\*
    f^{c,s}_{\evec{k}h} &= - \frac{h}{2} \Bigl[{\textstyle
        \frac{m}{\omega_\evec{k}} \bigl(M_{\evec{k}h}^{(11)} - M_{\evec{k}h}^{(22)}\bigr)
        - \frac{m_{\rm c}^*}{m} \bigl(s - \frac{h \abs{\evec{k}}}{\omega_\evec{k}}\bigr) M_{\evec{k}h}^{(12)}
        + \frac{m_{\rm c}}{m} \bigl(s + \frac{h \abs{\evec{k}}}{\omega_\evec{k}}\bigr) M_{\evec{k}h}^{(21)}
    }\Bigr] \text{.}
\end{align}
\end{subequations}
Here $M_{\evec{k}h}^{(ij)}$ denotes the $(i, j)$-component of the mode function matrix $M_{\evec{k}h}(t,t)$. Equations~\cref{eq:fm_and_fc_relation_to_mode_functions} can also be inverted, for a given helicity $h$, to give the four different components $M_{\evec{k}h}^{(ij)}$ in terms of $f^{m,+}_{\evec{k}h}, f^{m,-}_{\evec{k}h}$ and $f^{c,+}_{\evec{k}h}, f^{c,-}_{\evec{k}h}$.

\pagebreak 

We have thus established that the cQPA phase space distribution functions $f^{m,s}_{\evec{k}h}, f^{c,s}_{\evec{k}h}$ are in a one-to-one correspondence to the components of the \emph{equal-time} mode function matrix $M_{\evec{k}h}(t,t)$. We can go even further and actually derive the free cQPA equations of motion from the mode function equations. We just need to differentiate both sides of equations~\cref{eq:fm_and_fc_relation_to_mode_functions}, use the mode function equations~\cref{eq:mode_function_equations} with~\cref{eq:mode_function_matrix} and finally use the inverted form of equations~\cref{eq:fm_and_fc_relation_to_mode_functions} to eliminate the mode functions in the end. These steps reproduce exactly the single-flavor cQPA equations~\cref{eq:1-flavor_cQPA_with_complex_mass} without the collision term. Of course, this connection is ultimately not very surprising because the free cQPA and the mode functions are both based on the same Dirac equation. Nevertheless, it works as a consistency check and it is another verification for cQPA.

%
\subsection{Comparison to the semiclassical approximation}
%

Lastly, we compare the cQPA to the semiclassical method by calculating the axial charge density $j^{05}_\evec{k}$ numerically with both methods in the free theory. As we showed above, cQPA is exact in the free theory when calculating local quantities such as the currents. The results thus show how well the semiclassical method works outside its expected validity region $\abs{\evec{k}} \tau_{\rm w} \gg 1$.

We consider here the method of semiclassical force (see~\cite{Kainulainen:2021oqs} for a recent outline and more references) which is based on a gradient approximation of the KB equations. It was first formulated in the planar case in~\cite{Cline:1997vk,Cline:2000nw,Cline:2001rk,Kainulainen:2001cn,Kainulainen:2002th} and later adapted to the time-dependent case in~\cite{Prokopec:2013ax}. The general result for the axial charge density in this method is~\cite{Prokopec:2013ax,Jukkala:2019slc}
\begin{equation}
    j^{05}_\evec{k} = \sum_h \frac{\omega_{-} f_{3\evec{k}h}^{-}}{\omega_{3\evec{k}h}}
    \text{,} \qquad
    \omega_{3\evec{k}h} \eqdef \omega_\evec{k} + h \frac{m^2 \dot\theta}{2\abs{\evec{k}}\omega_\evec{k}}
    \text{,}
    \label{eq:semiclassical_axial_charge_density}
\end{equation}
where $\omega_{3\evec{k}h}$ is the shifted energy and $f_{3\evec{k}h}^{-} \eqdef \lim_{t \to -\infty} \tr[\gamma^5 P_{\evec{k}h} \bar S^\lt_\evec{k}(t,t)]$ denotes the initial value of the helicity-projected axial charge density. The value which corresponds to the vacuum initial conditions used in figure~\cref{fig:single_flavor_cQPA_with_interactions}, and equation~\cref{eq:mode_function_initial_conditions}, is
\begin{equation}
    f_{3\evec{k}h}^{-} = -\frac{h\abs{\evec{k}}}{\omega_{-}} \text{.}
    \label{eq:axial_charge_vacuum_initial_condition}
\end{equation}
The validity of the result~\cref{eq:semiclassical_axial_charge_density} requires that $\abs{m^2 \dot\theta/(2\abs{\evec{k}}\omega_\evec{k}^2)} \ll 1$. Using this assumption and the initial condition~\cref{eq:axial_charge_vacuum_initial_condition} we can estimate
\begin{equation}
     j^{05}_\evec{k} \simeq \frac{m^2 \dot\theta}{\omega_\evec{k}^3} \text{.}
     \label{eq:semiclassical_axial_charge_density_simplified}
\end{equation}
This is the expression that is used in the numerical comparison below. Note that the signs in equations~\cref{eq:axial_charge_vacuum_initial_condition,eq:semiclassical_axial_charge_density_simplified} are corrected here when compared to~\cite{Jukkala:2019slc} (similarly to equation~\cref{eq:axial_charge_density_in_cQPA}; see footnote~\cref{footnote:cQPA-j05}). This again has no effect on the conclusions.

%
\begin{figure}[t!]
    \centering
    \includegraphics[width=0.9\textwidth]{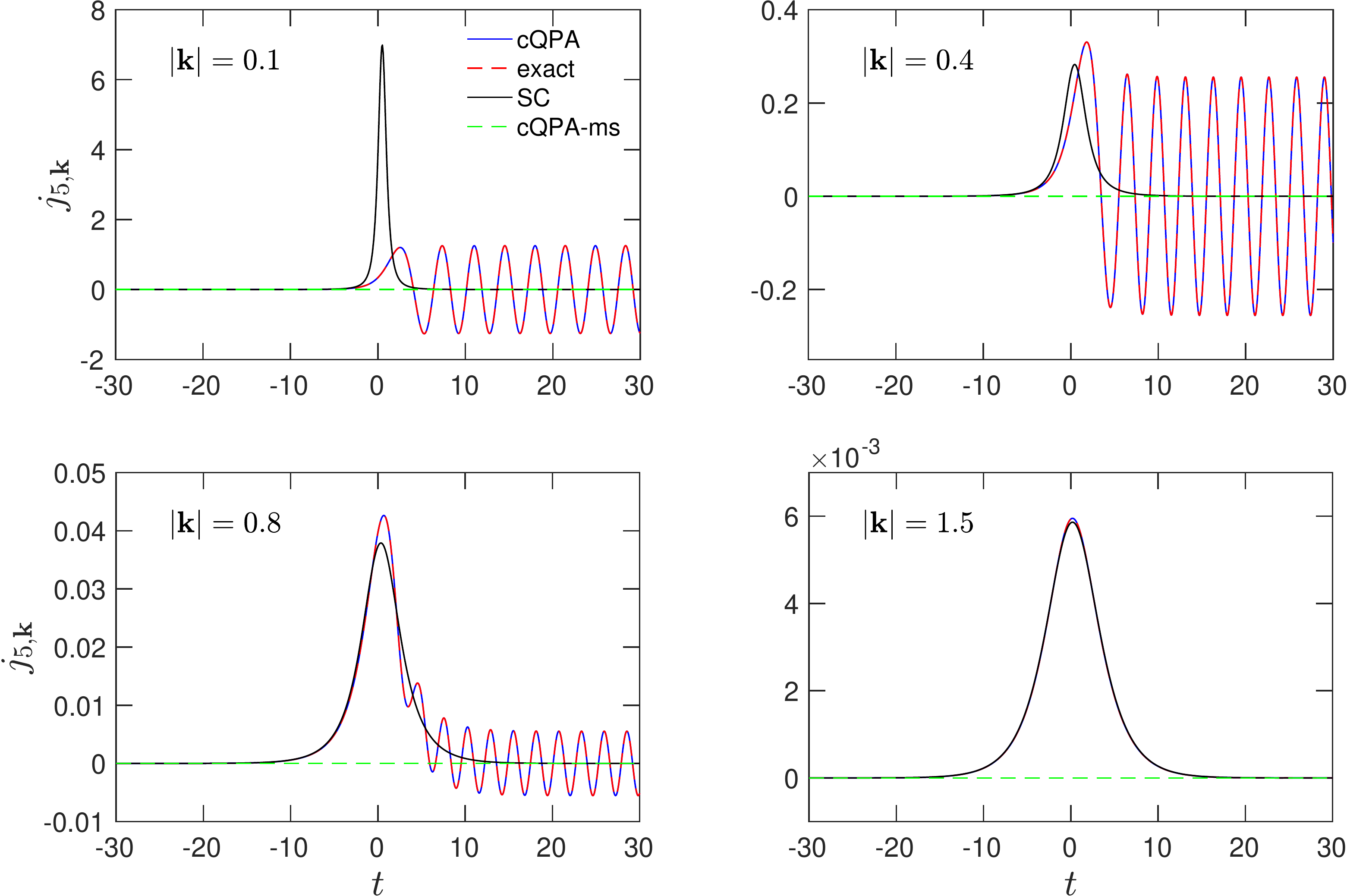}
    \caption{Time-evolution of different modes of the axial charge density $j^{05}_\evec{k}$ in free theory, calculated using single-flavor cQPA (blue line) and the semiclassical method (black line). We used vacuum initial conditions and the same mass profile parameters as in figure~\cref{fig:single_flavor_cQPA_with_interactions}. Adapted from article~\cite{Jukkala:2019slc} (\href{https://creativecommons.org/licenses/by/4.0/}{CC BY 4.0}), with corrected sign for $j_5$.}
	\label{fig:cQPA_vs_semiclassical_modes}
\end{figure}
%

In figure~\cref{fig:cQPA_vs_semiclassical_modes} we show the results for the axial charge density $j^{05}_\evec{k}$ for varying modes $\abs{\evec{k}}$ calculated from free cQPA using equation~\cref{eq:axial_charge_density_in_cQPA} and the semiclassical method using equation~\cref{eq:semiclassical_axial_charge_density_simplified}. We used again the mass profile given by equations~\cref{eq:simplified_mass_profile}. We also displayed the exact result (red dashed line) calculated from the mode functions which exactly matches the cQPA result as discussed earlier. In the free case the axial charge density is induced purely by coherence since the cQPA result vanishes when the coherence-functions are switched off (green dashed line). This is as expected from the forms of equations~\cref{eq:axial_charge_density_in_cQPA,eq:1-flavor_cQPA_with_complex_mass}.
The semiclassical result captures well the average of the oscillating current even for modes as small as $\abs{\evec{k}} \tau_{\rm w} \approx 2$ but it misses the coherent oscillations outside the transition region. For larger modes there is virtually no difference between the semiclassical and the exact result. Also, the top-left panel in figure~\cref{fig:cQPA_vs_semiclassical_modes} (with $\abs{\evec{k}} \tau_{\rm w} = 0.5$) is very far outside the validity region of the semiclassical approximation and the large spike in the result is entirely spurious and sensitive to the expression used for $j^{05}_\evec{k}$.

\newpage 

%
\section{Resonant leptogenesis}
\label{sec:RL_application}
%

In this section we apply the LA-method to resonant leptogenesis (RL), focusing on the details of the CP-asymmetry generation and evolution. We summarize the main results of the analysis done in~\cite{Jukkala:2021sku} and also provide some supplementary results. The LA-method is well-suited for studying RL because the form of the propagator is not restricted in this approach. Hence the method captures the coherent flavor oscillations during the decays of the Majorana neutrinos which dominate the generation of the lepton asymmetry in the RL case. Because of this we only consider the one-loop self-energies as they are sufficient for capturing the decay and inverse decay processes of leptogenesis. In this case we can use our first order local transport equation~\cref{eq:the_closed_local_equation}, or the equivalent equations~\cref{eq:generalized_density_matrix_equation}, to describe the Majorana neutrinos. There is also no RIS-problem because the LA-method is based on the KB equations derived from first principles in the CTP formalism.

In the SM sector we do a number of simplifications to reduce the complexity of the problem to a more manageable level. We assume that the SM lepton and Higgs fields are in kinetic equilibrium because of the electroweak gauge interactions, which are comparatively strong at the high temperature scale relevant for leptogenesis. We also neglect their decay widths and dispersive corrections and use tree level equilibrium forms for their propagators. Furthermore, we neglect the chemical potential of the Higgs field and assume a small lepton chemical potential so that its backreaction can be neglected in the Majorana neutrino equations. We also focus on the one-flavor approximation of leptogenesis where the Majorana neutrinos effectively couple to only one linear combination of the SM lepton flavors.

%
\subsection{The minimal leptogenesis model}
%

We consider the minimal leptogenesis model with two singlet Majorana neutrino fields $N_i$ ($i = 1,2$) and one lepton doublet $\ell = (\ell_1, \ell_2)$ which represents the SM leptons (one-flavor approximation). These are coupled to the SM Higgs doublet $\phi = (\phi_1, \phi_2)$ via chiral Yukawa interactions. The Lagrangian is
\begin{equation}
    \mathcal{L} =
    \frac{1}{2} \bar N_i (\im\slashed\partial - m_i) N_i
    + \bar\ell \,\im \slashed\partial \,\ell
    + (\partial_\mu\phi^\dagger)(\partial^\mu\phi)
    - \Bigl(
        y_i^* (\bar\ell \widetilde\phi) P_{\rm R} N_i + \text{H.c.}
    \Bigr) \text{,}
    \label{eq:minimal_leptogenesis_Lagrangian}
\end{equation}
with an implicit summation over the neutrino flavor $i$. The complex Yukawa couplings are denoted by $y_i$ and the $\SU(2)$-conjugate Higgs doublet is $\widetilde{\phi} \eqdef \im \sigma_2 \phi^*$. We assume the mass eigenbasis of the neutrinos where $m_i$ are the real Majorana masses. The neutrino fields satisfy the Majorana condition
\begin{equation}
    N_i = N_i^c \eqdef C \bar N_i^\transp
    \text{,} \label{eq:Majorana_condition}
\end{equation}
where $C$ is the unitary charge conjugation matrix defined by $C (\gamma^\mu)^\transp C^{-1} = - \gamma^\mu$.%
\footnote{It then follows that $C^\transp = -C$, regardless of the used Dirac matrix representation~\cite{Pal:2010ih}.}
%

\subsubsection{Propagators and self-energies}

We denote the CTP propagators of the neutrino, lepton and Higgs fields [in the contour notation defined in~\cref{eq:CTP_propagator_contour_definitions}] as
\begin{subequations}
    \begin{align}
        \im S_{ij}(x,y) &\eqdef \expectval[\big]{
            \mathcal{T}_\mathcal{C}\bigl[N_i(x) \bar N_j(y)\bigr]
        } \text{,}
        \\*
        \im S_{\ell,AB}(x,y) &\eqdef \expectval[\big]{
            \mathcal{T}_\mathcal{C}\bigl[\ell_A(x) \bar{\ell_B}(y)\bigr]
        } \text{,}
        \\*
        \im \Delta_{AB}(x,y) &\eqdef \expectval[\big]{
            \mathcal{T}_\mathcal{C}\bigl[\phi_A^{}(x) \phi_B^\dagger(y)\bigr]
        } \text{,}
    \end{align}
\end{subequations}
where $A,B = 1,2$ are the $\SU(2)$-doublet indices. We assume the $\SU(2)$-symmetric phase where~\cite{Garny:2011hg}
\begin{subequations}
\label{eq:SU2-symmetric_propagators}
\begin{align}
    S_{\ell,AB}(x,y) &\eqdef S_\ell(x,y) \delta_{AB} \text{,}
    \\*
    \Delta_{AB}(x,y) &\eqdef \Delta(x,y) \delta_{AB} \text{.}
\end{align}
\end{subequations}
In the following we write the formulae and equations directly for the $\SU(2)$-diagonal elements $S_\ell$ and $\Delta$.

We calculate the CTP self-energies for this model using the 2PI effective action method introduced in section~\cref{sec:CTP-2PIEA}. We consider the two-loop truncation of $\Gamma_2$ which, for the Lagrangian~\cref{eq:minimal_leptogenesis_Lagrangian}, can be written in the contour notation as
\begin{equation}
    \im \Gamma_2^{(2)} =
    \cw \sum_{i,j} y_i^* y_j \iint_{\mathcal{C}} \dd^4 x\dd^4 y
    \tr\bigl[
        P_{\rm R} \im S_{ij}(x,y) P_{\rm L} \im S_\ell(y,x)
    \bigr] \im \Delta(y,x)
    \text{.} \label{eq:two_loop_Gamma2}
\end{equation}
Equations~\cref{eq:SU2-symmetric_propagators} were used here to perform the sums over the $\SU(2)$-indices resulting in the multiplicity factor $\cw = 2$. The neutrino and lepton one-loop self-energies are calculated by using the functional derivative formula~\cref{eq:2PIEA_fermion_self-energy}. The results are~\cite{Jukkala:2021sku}
\begin{subequations}
\label{eq:lgen_one-loop_self-energies}
\begin{alignat}{2}
    \im \Sigma_{ij}(x,y) ={}&& \cw \Bigl[&
            y_i y_j^* P_{\rm L} \im S_\ell(x,y) P_{\rm R} \im \Delta(x,y) \notag
            \\*
            && {}+{} & y_i^* y_j P_{\rm R} C \im S_\ell(y,x)^\transp C^{-1} P_{\rm L} \im \Delta(y,x)
        \Bigr] \text{,} \label{eq:lgen_neutrino_self-energy}
    \\*
    \im \Sigma_\ell(x,y) ={}&&
    \sum_{i,j} {}& y_i^*y_j P_{\rm R} \im S_{ij}(x,y)P_{\rm L} \im \Delta(y,x)
    \text{.}
\end{alignat}
\end{subequations}
The real-time self-energies can be obtained by inserting the CTP indices which simply follow the spacetime arguments $x,y$ in equations~\cref{eq:lgen_one-loop_self-energies}.
At this point we also introduce the tree-level kinetic equilibrium ansätze for the lepton and Higgs propagators so that the self-energies can be calculated explicitly. These are defined in the Wigner representation where the Wightman functions are~\cite{Jukkala:2021sku}%
\begin{subequations}
\label{eq:lepton_and_Higgs_tree_level_propagators}
\begin{align}
    \im S_\ell^{\lessgtr}(k,t) &=
    2 \pi \sgn(k^0) P_{\rm L} \slashed k \,\delta(k^2) \,f_{\rm FD}\bigl(\pm(k^0 - \mu_\ell)\bigr)
    \text{,} \label{eq:lepton_tree_level_propagator}
    \\*
    \im \Delta^{\lessgtr}(k) &=
    \pm 2 \pi \sgn(k^0) \,\delta(k^2) \,f_{\rm BE}(\pm k^0) \eqdef \im \Delta_{\rm eq}^{\lessgtr}(k)
    \text{.}
\end{align}
\end{subequations}
The equilibrium distributions $f_{\rm FD}$ and $f_{\rm BE}$ were defined below equations~\cref{eq:thermal_eq_Wightmans}. The other real-time propagators can be recovered from~\cref{eq:lepton_and_Higgs_tree_level_propagators} by using the general relations~\cref{eq:CTP_propagator_relations}. In equations~\cref{eq:lepton_and_Higgs_tree_level_propagators} $\mu_\ell$ denotes the chemical potential of the leptons while the Higgs propagator has the thermal equilibrium form with zero chemical potential. The lepton propagator can be split into the thermal equilibrium part and the deviation,
\begin{equation}
    S^{\lt,\gt}_{\ell}(k,t) \eqdef S^{\lt,\gt}_{\ell,{\rm eq}}(k) + \delta S^{\lt,\gt}_{\ell}(k,t)
    \text{,} \label{eq:lepton_propagator_split}
\end{equation}
where the equilibrium part is defined by equation~\cref{eq:lepton_tree_level_propagator} with $\mu_\ell = 0$. The one-loop self-energies can then be calculated explicitly as functions of $k$, $T$ and $\mu_\ell$ in the Wigner representation. In particular, the Majorana neutrino self-energy~\cref{eq:lgen_neutrino_self-energy} can be divided similarly to the thermal equilibrium part $\Sigma_{\rm eq}(k)$ and the deviation $\delta \Sigma(k,t)$. The needed self-energy components can be written as~\cite{Jukkala:2021sku}
\begin{subequations}
\label{eq:Majorana_neutrino_self_energies}
\begin{align}
    \Sigma^\mathcal{A}_{{\rm eq},ij}(k) &=
    \cw \bigl(y_i y_j^* P_{\rm L} + y_i^* y_j P_{\rm R}\bigr)
    \slashed{\mathfrak{S}}^{\mathcal{A}}_{\rm eq}(k)
    \text{,}
    \\*
    \Sigma^{{\rm H}(T)}_{{\rm eq},ij}(k) &=
    \cw \bigl(y_i y_j^* P_{\rm L} + y_i^* y_j P_{\rm R}\bigr)
    \slashed{\mathfrak{S}}^{{\rm H}(T)}_{\rm eq}(k)
    \text{,}
    \\*
    \im \delta\Sigma^{\lt,\gt}_{ij}(k,t) &=
    \cw \bigl(y_i y_j^* P_{\rm L} - y_i^* y_j P_{\rm R}\bigr)
    \delta\slashed{\mathfrak{S}}^{\lt,\gt}(k) \,\beta \mu_\ell
    \text{.}
\end{align}
\end{subequations}
Explicit expressions for the different self-energy functions $\mathfrak{S}_{{\rm eq},\mu}$ and $\delta \mathfrak{S}_\mu$ are given in~\cite{Jukkala:2021sku}. Note that we assume that the lepton asymmetry and hence the chemical potential $\mu_\ell$ are small so we have expanded the deviations $\delta\Sigma$ to first order in $\beta \mu_\ell$.

%
\subsection{Lepton asymmetry}
%

The left chiral lepton asymmetry is related to the chiral current density of the SM lepton doublet, which can be defined generally as
\begin{equation}
    j^\mu_{{\rm L},\ell}(x) \eqdef \expectval[\big]{\bar \ell_A(x) \gamma^\mu P_{\rm L} \ell_A(x)}
    \text{.} \label{eq:chiral_lepton_current}
\end{equation}
Here an implicit summation over the $\SU(2)$-index $A = 1,2$ is understood.
In this work we define the the \emph{lepton asymmetry density} $n_L$ as the non-equilibrium part of (the zeroth component of) the current~\cref{eq:chiral_lepton_current}:~\cite{Jukkala:2021sku}
\begin{equation}
    n_L \eqdef \delta j^0_{{\rm L},\ell}(t)
    = \cw \int \frac{\dd^3\evec{k}}{(2\pi)^3}\tr\Bigl[
        P_{\rm L} \delta \bar S^\lt_{\ell,\evec{k}}(t,t)
    \Bigr] \text{.} \label{eq:lepton_asymmetry_def}
\end{equation}
The relation to the chemical potential $\mu_\ell$ is then given by the standard formula~\cite{Kolb:1990vq}
\begin{equation}
    n_L = \frac{\cw T^3}{6 \pi^2} \biggl[
        \pi^2 \Bigl(\frac{\mu_\ell}{T}\Bigr) + \Bigl(\frac{\mu_\ell}{T}\Bigr)^3
    \biggr]
    \simeq \frac{\cw T^3}{6} \Bigl(\frac{\mu_\ell}{T}\Bigr)
    \text{,} \label{eq:lepton_asymmetry_from_chemical_potential}
\end{equation}
once the free equilibrium form~\cref{eq:lepton_tree_level_propagator} with~\cref{eq:lepton_propagator_split} is used for the propagator.

The equation of motion for $n_L$ can be derived from the equation of the lepton propagator $S_{\ell}$ via equations~\cref{eq:lepton_asymmetry_def,eq:lepton_propagator_split}. For $S_{\ell}$ we use the KB equation~\cref{eq:general_fermionic_KB-eqs_variant} where we omit the terms with $S^{\rm H}$ and $\Sigma^{\rm H}$ on the left-hand side, because we neglect the width and dispersive corrections for the leptons. The evolution equation for the lepton asymmetry then is
\begin{equation}
    \partial_t n_L =
    -\frac{\cw}{2} \int \frac{\dd^3\evec{k}}{(2\pi)^3}
    \tr\Bigl[
        \bigl(\bar\Sigma^\gt_{\ell,\evec{k}} \convol \bar S^\lt_{\ell,\evec{k}}
        - \bar\Sigma^\lt_{\ell,\evec{k}} \convol \bar S^\gt_{\ell,\evec{k}}\bigr)(t,t)
    \Bigr]
    + \text{H.c.} \label{eq:lepton_local_equation_traced}
\end{equation}
where $\Sigma_{\ell}$ is the lepton self-energy. Next we use the relations between the self-energies~\cref{eq:lgen_one-loop_self-energies} to write the lepton collision term on the right-hand side of~\cref{eq:lepton_local_equation_traced} in terms of the Majorana neutrino collision term~\cite{Jukkala:2021sku}. This is possible because we are calculating a quantity that is local and diagonal in all indices. The result is
\begin{equation}
    \partial_t n_L =
    \frac{1}{2} \int \frac{\dd^3\evec{k}}{(2\pi)^3}
    \Tr\Bigl[
        P_{\rm R} \bigl( \bar\Sigma^\gt_\evec{k} \convol \bar S^\lt_\evec{k}
        - \bar\Sigma^\lt_\evec{k} \convol \bar S^\gt_\evec{k} \bigr)(t,t)
    \Bigr]
    + \text{H.c.} \label{eq:lepton_local_equation_v2}
\end{equation}
Here we can then use the LA-method to simplify the convolutions in the interaction terms~\cite{Jukkala:2021sku}. Note that in principle we could have used the LA-method to derive also the lepton equation but the leading contribution to the source term~\cref{eq:local_eq_general_source} would then be different from~\cref{eq:first_order_adiabatic_source_term}.
This is because the lepton is massless so $\partial_t S_{\ell,{\rm ad}} = 0$ (when choosing $S_{\ell,{\rm ad}} \eqdef S_{\ell,{\rm eq}}$).

%
\subsection{Quantum kinetic equations for leptogenesis}
\label{sec:QKEs_for_leptogenesis}
%

Next we present our leading-order QKEs for the Majorana neutrino distribution $\delta f^m$ and the lepton asymmetry $n_L$. For the Majorana neutrinos we use the mass shell equations~\cite{Jukkala:2021sku} which can be obtained from~\cref{eq:generalized_density_matrix_equation} by discarding the $f^c$-functions and choosing the symmetric normalization~\cref{eq:symmetric_projection_normalization_flavored}.
We neglect the backreaction of the lepton asymmetry to the Majorana neutrinos by using the equilibrium self-energies~\cref{eq:Majorana_neutrino_self_energies} in the neutrino equation. The lepton equation is derived from~\cref{eq:lepton_local_equation_v2} by using the division~\cref{eq:adiabatic_and_dynamical_division} (with $\Sigma_{\rm ad} = \Sigma_{\rm eq}$) and the local ansatz~\cref{eq:the_local_ansatz}, and by expanding the equation to the leading order in gradients and interaction strength.
The resulting equations are~\cite{Jukkala:2021sku}
\begin{subequations}
\label{eq:lgenQKEs}
\begin{align}
    \partial_t \delta f_{\evec{k}hij}^{m,+} ={}&
    -\im \bigl(\omega_{\evec{k}i} - \omega_{\evec{k}j}\bigr) \delta f_{\evec{k}hij}^{m,+}
    - \partial_t f_{{\rm ad},\evec{k}hij}^{m,+} \notag
    \\*
    &{} - \sum_l \Bigl[
        C^{+}_{\evec{k}hil} \delta f_{\evec{k}hlj}^{m,+}
        + \bigl(C^{+}_{\evec{k}hjl}\bigr)^* \delta f_{\evec{k}hil}^{m,+}
    \Bigr] \text{,}
    \label{eq:lgenQKE_neutrino_part}
    \\*
    \partial_t n_L ={}& S_{CP}^m + \bigl(\delta W^m + W_{\rm ad}\bigr) n_L \text{,}
    \label{eq:lgenQKE_lepton_part}
\end{align}
\end{subequations}
where in the neutrino equation~\cref{eq:lgenQKE_neutrino_part}
\begin{align}
    f_{{\rm ad},\evec{k}hij}^{m,+} ={}& f_{\rm FD}(\omega_{\evec{k}i}) \delta_{ij} \text{,}
    \\*[1ex] 
    \begin{split}
        C^{+}_{\evec{k}hil} ={}& \frac{\cw N^m_{\evec{k}il}}{2 \omega_{\evec{k}i} \omega_{\evec{k}l}}
        \Bigl[\Re(y_i^* y_l) \bigl(m_i k_{+l} + m_l k_{+i}\bigr)^\mu
            - \im h \Im(y_i^* y_l) \bigl(m_i k^\perp_{+l} + m_l k^\perp_{+i} \bigr)^\mu
        \Bigr]
        \\*[-1ex] 
        &\phantom{\frac{\cw N^m_{\evec{k}il}}{2 \omega_{\evec{k}i} \omega_{\evec{k}l}}} \times \Bigl(
            \mathfrak{S}^\mathcal{A}_{\rm eq}(k_{+l}) + \im \mathfrak{S}^{{\rm H}(T)}_{\rm eq}(k_{+l})
        \Bigr)_\mu \text{.} \raisetag{4ex} 
    \end{split}
    \label{eq:neutrino_eq_collision_term_coeff}
\end{align}
We used the free spectral approximation for the adiabatic functions and defined the four-vectors $(k_{si})^\mu \eqdef (s\omega_{\evec{k}i}, \evec{k})$ and $(k^\perp)^\mu \eqdef (\abs{\evec{k}}, k^0 \Uevec{k})$, and denoted $N^m_{\evec{k}ij} \eqdef N^{ss}_{\evec{k}hij}$ for the normalization. In the lepton equation~\cref{eq:lgenQKE_lepton_part} the CP-violating source term $S_{CP}^m$ and the washout term coefficients $\delta W^m$ and $W_{\rm ad}$ are
\begin{subequations}
\begin{alignat}{2}
    S_{CP}^m ={}&& \sum_{i,j} \int \frac{\dd^3\evec{k}}{(2\pi)^3}
    \frac{\cw N^m_{\evec{k}ij}}{2 \omega_{\evec{k}i} \omega_{\evec{k}j}}
    \Bigl[
        &\Re(y_i^* y_j) \textstyle{\sum_h} h \Re(\delta f^{m,+}_{\evec{k}hij})
        \bigl(m_i k^\perp_{+j} + m_j k^\perp_{+i}\bigr)^\mu \notag
        \\*[-2ex] 
        && {}-{} &\Im(y_i^* y_j) \textstyle{\sum_h} \Im(\delta f^{m,+}_{\evec{k}hij})
        \bigl(m_i k_{+j} + m_j k_{+i} \bigr)^{\mu}
    \Bigr] \notag
    \\*
    && {}\times{} &\Bigl(
        \mathfrak{S}^\mathcal{A}_{\rm eq}(k_{+i}) + \mathfrak{S}^\mathcal{A}_{\rm eq}(k_{+j})
    \Bigr)_{\mu} \text{,} \displaybreak[0]
    \label{eq:lepton_eq_SCP}
    \\
    \frac{\cw T^3}{6} \delta W^m ={}&& \sum_{i,j} \int \frac{\dd^3\evec{k}}{(2\pi)^3}
    \frac{\cw N^m_{\evec{k}ij}}{2 \omega_{\evec{k}i} \omega_{\evec{k}j}}
    \Bigl[
        &\Re(y_i^* y_j) \textstyle{\sum_h} \Re(\delta f^{m,+}_{\evec{k}hij})
        \bigl(m_i k_{+j} + m_j k_{+i}\bigr)^\mu \notag
        \\*[-2ex] 
        && {}-{} &\Im(y_i^* y_j) \textstyle{\sum_h} h \Im(\delta f^{m,+}_{\evec{k}hij})
        \bigl(m_i k^\perp_{+j} + m_j k^\perp_{+i} \bigr)^\mu
    \Bigr] \notag
    \\*
    && {}\times{} &\Bigl(
        \delta\mathfrak{S}^\mathcal{A}(k_{+i}) + \delta\mathfrak{S}^\mathcal{A}(k_{+j})
    \Bigr)_\mu \text{,} \displaybreak[0]
    \\
    \frac{\cw T^3}{6} W_{\rm ad} ={}&& \sum_i \int \frac{\dd^3\evec{k}}{(2\pi)^3}
    \frac{2\cw \abs{y_i}^2}{\omega_{\evec{k}i}} k&_{+i}^\mu \Bigl[
        2f_{\rm FD}(\omega_{\evec{k}i}) \delta\mathfrak{S}^{\mathcal{A}}_\mu(k_{+i})
        - \delta\mathfrak{S}^{\lt}_\mu(k_{+i})
    \Bigr] \text{.}
\end{alignat}
\end{subequations}
To obtain these results we used the symmetry properties of the self-energy functions and the constraints between different components of the Majorana distribution functions $\delta f^m$~\cite{Jukkala:2021sku}. It should also be noted that the explicit form of the collision term coefficient~\cref{eq:neutrino_eq_collision_term_coeff} is specific to the case of two Majorana neutrinos.

The Hubble expansion can be taken into account in these equations as described in sections~\cref{sec:expanding_flat_universe,sec:QFT_in_expanding_spacetime}. The equations then take the same form as given above when they are divided by appropriate powers of the scale factor, that is, when expressed in terms of physical (instead of comoving) quantities and the cosmic time $t$. The only changes to equations~\cref{eq:lgenQKEs} are that the time-derivatives are replaced by total time-derivatives, the left-hand side of the lepton equation~\cref{eq:lgenQKE_lepton_part} is changed to $[\frac{\dd}{\dd t}(n_L a^3)]/a^3$, and time-derivatives of the Majorana neutrino masses $\partial_t m_i$ are replaced by $\dot a m_i$~\cite{Jukkala:2021sku}.
In practice, when solving the lepton asymmetry we also use the entropy-normalized yield parameter $Y_L \eqdef n_L/s$.
When assuming adiabatic expansion in the radiation dominated era we then have $[\frac{\dd}{\dd t}(n_L a^3)]/a^3 = s \frac{\dd}{\dd t} Y_L$.
We also solve the equations by using the dimensionless variable $z \eqdef m_1/T$, where $m_1$ is the mass of the lightest Majorana neutrino. It is related to the cosmic time $t$ by $\dd z = H(m_1)/z \dd t$.
Note also that in leptogenesis the standard (\ie~non-coherent) gradient corrections are suppressed because $H/T \ll 1$.

%
\subsection{Helicity-symmetric approximation}
%

We now simplify the leptogenesis QKEs~\cref{eq:lgenQKE_neutrino_part,eq:lgenQKE_lepton_part} further by assuming helicity symmetry for the neutrino distribution functions $\delta f^{m,+}_{\evec{k}hij}$. This is a good approximation because the dominant contribution to the lepton asymmetry originates from the helicity-even part of the source term $S_{CP}^m$~\cref{eq:lepton_eq_SCP}. This is true especially in the non-relativistic regime, that is, when the lepton asymmetry is generated mainly when $T \lesssim m_1$. With further approximations the helicity-symmetric equations can also be reduced to the form of semiclassical Boltzmann equations.

The helicity-symmetric equations are obtained by discarding the helicity-odd part of $\delta f^{m,+}_{\evec{k}hij}$ (\ie~setting $\delta f^{m,+}_{\evec{k},+1,ij} = \delta f^{m,+}_{\evec{k},-1,ij}$). For simplicity we also drop $\Sigma^{\rm H}$ here explicitly and work with vacuum dispersion relations. The resulting neutrino equations and the source term of the lepton asymmetry can be written as~\cite{Jukkala:2021sku}
\begin{subequations}
\label{eq:lgenQKE_h-symmetric}
\begin{align}
    \partial_t \delta f_{\evec{k}11} &= -\Gamma_{\evec{k}11} \,\delta f_{\evec{k}11}
    - \partial_t f_{{\rm ad},\evec{k}11} - \Gamma_{\evec{k}12} \Re(\delta f_{\evec{k}12})
    \text{,} \label{eq:lgenQKE_diag1_h-approx}
    \\*
    \partial_t \delta f_{\evec{k}22} &= -\Gamma_{\evec{k}22} \,\delta f_{\evec{k}22}
    - \partial_t f_{{\rm ad},\evec{k}22} - \Gamma_{\evec{k}21} \Re(\delta f_{\evec{k}12})
    \text{,} \label{eq:lgenQKE_diag2_h-approx}
    \\*
    \partial_t \delta f_{\evec{k}12} &= -\bar\Gamma_{\evec{k}12} \,\delta f_{\evec{k}12}
    - \im \Delta\omega_{\evec{k}12} \,\delta f_{\evec{k}12} - \smash{\frac{1}{2}} \bigl(
        \Gamma_{\evec{k}21} \,\delta f_{\evec{k}11} + \Gamma_{\evec{k}12} \,\delta f_{\evec{k}22}
    \bigr) \text{,} \label{eq:lgenQKE_offd_h-approx}
\end{align}
\end{subequations}
and
\begin{equation}
    S_{CP}^m = -2 \,\frac{\Im(y_1^* y_2)}{\Re(y_1^* y_2)} \int \frac{\dd^3 \evec{k}}{(2\pi)^3} \,
    \bigl(\Gamma_{21} + \Gamma_{12}\bigr)_{\evec{k}} \Im\bigl(\delta f_{\evec{k}12}\bigr)
    \text{.} \label{eq:lgenQKE_SCP_h-approx}
\end{equation}
Here the rates are defined as $\Gamma_{\evec{k}il} \eqdef 2 \Re(C^{+}_{\evec{k}hil})\rvert_{\mathfrak{S}^H \to 0}$ with $C^{+}_{\evec{k}hil}$ given by equation~\cref{eq:neutrino_eq_collision_term_coeff}. The coherence damping rate is given by the average $\bar\Gamma_{\evec{k}12} \eqdef (\Gamma_{\evec{k}11} + \Gamma_{\evec{k}22})/2$. Its form agrees with the flavor coherence damping rate of mixing neutrinos found in the density matrix formalism~\cite{Enqvist:1990ad,Enqvist:1991qj,Kainulainen:1990ds}. We have suppressed the helicity indices in these equations as all quantities are now assumed to be the same for both helicities. Note that the equations could be written in a similar form in the helicity-dependent case also: there would just be additional helicity-mixing terms. It is also illustrative to integrate the off-diagonal equation~\cref{eq:lgenQKE_offd_h-approx} into the form
\begin{equation}
    \delta f_{\evec{k}12}(t) = -\frac{1}{2} \int_{t_0}^t \dd u \,
    \bigl(\Gamma_{21} \delta f_{11} + \Gamma_{12} \delta f_{22}\bigr)_{\!\evec{k}}(u) 
    \exp\biggl[
        -\int_u^t \dd v \,\Bigl(\bar\Gamma_{12} + \im \Delta\omega_{12}\Bigr)_{\!\!\evec{k}}\!(v) 
    \biggr] \text{,}
    \label{eq:lgenQKE_offd_h-approx_integrated}
\end{equation}
where it was assumed that the flavor coherence vanishes initially, $\delta f_{\evec{k}12}(t_0) = 0$. In this form the oscillations and the role of the flavor coherence damping become even more clear.

Equations~\cref{eq:lgenQKE_h-symmetric,eq:lgenQKE_SCP_h-approx} can be reduced to the semiclassical Boltzmann equations in the decoupling limit where the off-diagonal function $\delta f_{\evec{k}12}$ is discarded from the diagonal equations~\cref{eq:lgenQKE_diag1_h-approx,eq:lgenQKE_diag2_h-approx}~\cite{Jukkala:2021sku}. This is a valid approximation in the limit $\Delta m_{21} \gg \Gamma$. The equations of the diagonal functions $\delta f_{\evec{k}ii}$ then exactly match those derived in the semiclassical approach. The late-time limit of the lepton asymmetry then also matches the semiclassical result with the sum regulator~\cref{eq:CP-asymm_sum_regulator} for the CP-asymmetry. Analysis of the structure of the source term~\cref{eq:lgenQKE_SCP_h-approx} also reveals that the \emph{coherence damping rate} $\bar\Gamma_{\evec{k}12}$ is the origin of the regulator of the CP-asymmetry in the semiclassical approach~\cite{Jukkala:2021sku}.

\subsubsection{Strong washout limit}

A simplified solution to equations~\cref{eq:lgenQKE_h-symmetric,eq:lgenQKE_SCP_h-approx} can also be derived in the strong washout case $\Gamma_{ii} > H$. In this case one can derive a late-time approximation to the solution $\delta f_{ij}$ by setting the left-hand sides of equations~\cref{eq:lgenQKE_diag1_h-approx,eq:lgenQKE_diag2_h-approx,eq:lgenQKE_offd_h-approx} to zero, resulting in an algebraic equation~\cite{Garbrecht:2014aga,Dev:2017wwc,Iso:2014afa}.
The idea is that the true solution relaxes towards this late-time limit and it is a good approximation if the relaxation happens before the lepton asymmetry freeze-out~\cite{Dev:2017wwc}.
The solution to the algebraic equations leads to the approximation
\begin{align}
    S_{CP}^m \simeq \frac{\Im(y_1^* y_2)}{\Re(y_1^* y_2)} \int \frac{\dd^3 \evec{k}}{(2\pi)^3}
    & \bigl(\Gamma_{21} + \Gamma_{12}\bigr)_\evec{k}
    \bigl(\Gamma_{21} \delta f_{11} + \Gamma_{12} \delta f_{22}\bigr)_\evec{k} \notag
    \\*
    &{}\times{} \Biggl[
        \frac{-\Delta\omega_{12}}
        {(\Delta\omega_{12})^2 + (\bar\Gamma_{12})^2\Bigl(
            1 - \frac{\Gamma_{12}\Gamma_{21}}{\Gamma_{11}\Gamma_{22}}
        \Bigr)}
    \Biggr]_\evec{k} \text{.}
    \label{eq:SCP_approx_strong_washout}
\end{align}
Here the diagonal distributions $\delta f_{ii}$ are calculated without backreaction from the off-diagonal functions~\cite{Jukkala:2021sku}. The factor $(1 - \Gamma_{12}\Gamma_{21}/[\Gamma_{11}\Gamma_{22}])$ takes effectively, but in a simplified way, into account the coupling between the diagonal and off-diagonal functions.

The result~\cref{eq:SCP_approx_strong_washout} illustrates the role of the coherence damping rate $\bar\Gamma_{\evec{k}12}$ as the regulator of the CP-asymmetry in the degeneracy limit $\Delta m_{21} \to 0$. In the quasidegenerate case $m_1 \simeq m_2$ the source~\cref{eq:SCP_approx_strong_washout} is also approximated by the standard Boltzmann form of the source term with the effective CP-asymmetry parameter
\begin{equation}
    \epsilon^{CP}_{i,{\rm eff}} =
    \frac{\Im\bigl[(y_1^* y_2)^2\bigr]}{\abs{y_1}^2 \abs{y_2}^2}
    \frac{(m_2^2 - m_1^2) m_i \Gamma_j^{(0)}}
    {(m_2^2 - m_1^2)^2 + \bigl(m_1 \Gamma_1^{(0)} + m_2 \Gamma_2^{(0)}\bigr)^2 \sin^2\theta_{12}}
    \text{.} \quad (j \neq i)
    \label{eq:effective_CP-parameter}
\end{equation}
This result corresponds to the effective regulator~\cite{Dev:2017wwc} given in equation~\cref{eq:CP-asymm_eff_regulator}.

%
\subsection{Results for the lepton asymmetry}
%

%
\begin{figure}[t!]
    \centering
    \includegraphics{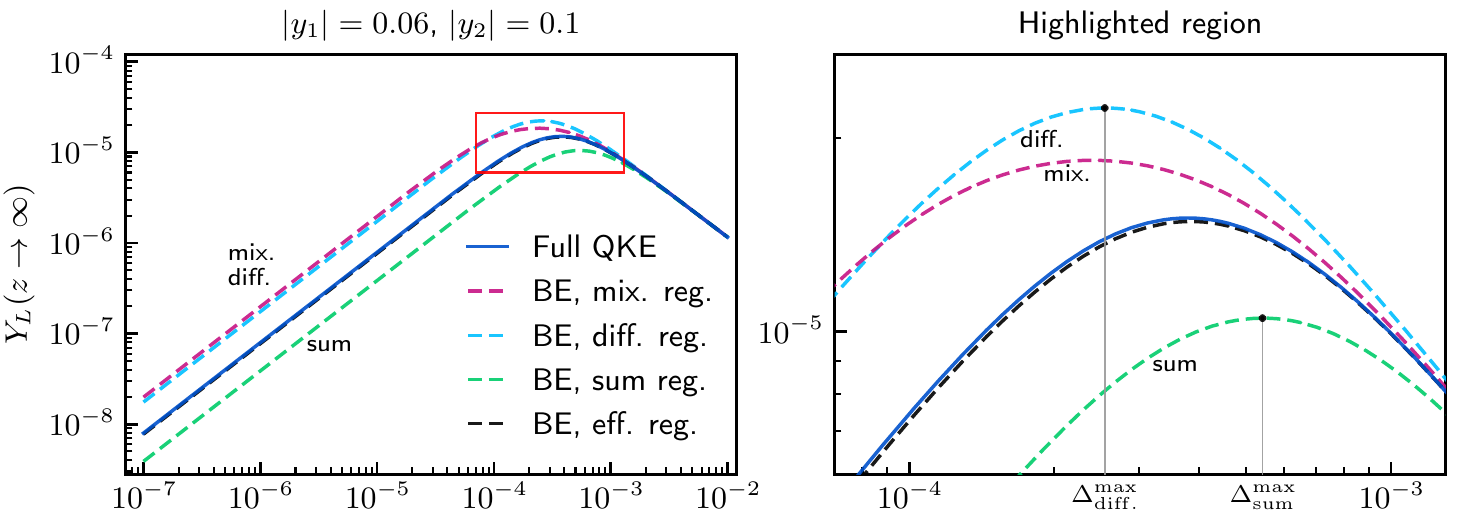}\par
    \includegraphics{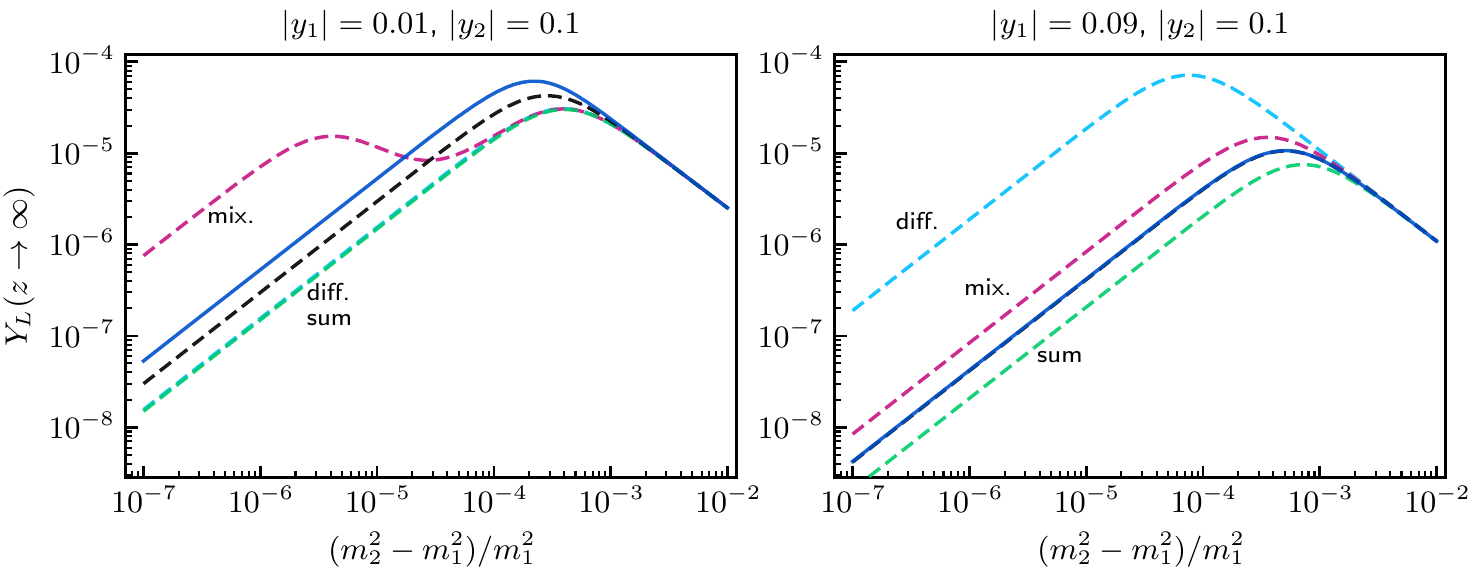}
	\caption{Final lepton asymmetry $Y_L$ as a function of the relative mass-squared difference $\Delta m^2_{21}/m_1^2$, for different Yukawa couplings $y_i$. The full QKE results are compared to the Boltzmann equation (BE) results for four different regulators of the CP-asymmetry $\epsilon^{CP}_i$. We used $m_1 = \SI{1e13}{GeV}$, $\theta_{12} = \pi/4$ and vacuum initial conditions in all plots. The top-right panel shows the region highlighted in red on the left. From article~\cite{Jukkala:2021sku} (\href{https://creativecommons.org/licenses/by/4.0/}{CC BY 4.0}).}
	\label{fig:leptogenesis_YL_vs_massdiff_comparison}
\end{figure}
%

Next we present our numerical results for the lepton asymmetry obtained from the main QKEs~\cref{eq:lgenQKE_neutrino_part,eq:lgenQKE_lepton_part}. We also compare the full results to the helicity-symmetric approximation given by equations~\cref{eq:lgenQKE_h-symmetric,eq:lgenQKE_SCP_h-approx} and to the semiclassical Boltzmann approach with the momentum-dependent version~\cite{Jukkala:2021sku} of equations~\cref{eq:leptogenesis_Boltzmann}.
We use $m_1 = \SI{1e13}{GeV}$, $g_* = 110$, and $\theta_{12} = \pi/4$ with vacuum initial conditions here; more discussion about the parameter choices is given in~\cite{Jukkala:2021sku}.
We take the Hubble expansion into account as described in the end of section~\cref{sec:QKEs_for_leptogenesis}. We assume that there is no significant entropy production and that the universe expands adiabatically as described in section~\cref{sec:expanding_flat_universe}. This is a valid approximation if the Majorana neutrinos decay when $z < g_{*s}$~\cite{Luty:1992un}, which is typically the case in thermal leptogenesis and also here. We also discard the dispersive self-energy $\Sigma^{\rm H}$ in all results for simplicity.
All of the results shown here have been calculated using the code package~\cite{henri_jukkala_2021_5025929}.

In figure~\cref{fig:leptogenesis_YL_vs_massdiff_comparison} we show the results for the asymptotic value of the lepton asymmetry yield parameter $Y_L \eqdef n_L/s$ as a function of the relative mass-squared difference $\Delta m^2_{21}/m_1^2 \simeq 2 \Delta m_{21}/m_1$. Our main QKE result is compared to the Boltzmann approach with the different CP-asymmetry regulators~\cref{eq:CP-asymm_regulators}, and the results are shown for three different choices of the Yukawa couplings of the model. The lower left and right panel correspond to hierarchical and almost degenerate Yukawa couplings, respectively, whereas the upper panels show an intermediate case. The maximal resonant saturation of the CP-asymmetry, and consequently the maximum of the lepton asymmetry, occurs in all cases roughly when $\Delta m_{21} \sim \bar \Gamma_{12}$. There are however significant qualitative and quantitative differences between the full QKE and Boltzmann approaches in the resonant and quasidegenerate regions $\Delta m_{21} \lesssim \bar \Gamma_{12}$. Both approaches converge, as expected, in the region $\Delta m_{21} > \bar \Gamma_{12}$ where the flavor oscillations are fast and the diagonal Majorana neutrino distributions effectively decouple from the off-diagonal distributions~\cite{Jukkala:2021sku}.

%
\begin{figure}[t!]
    \centering
    \includegraphics{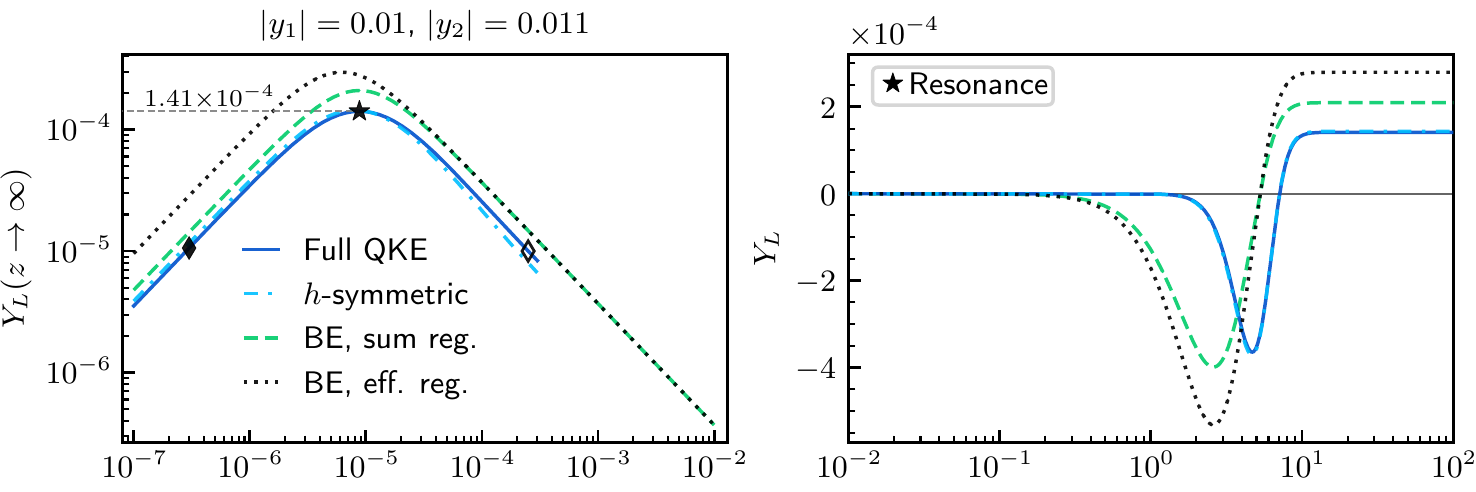}\par
    \includegraphics{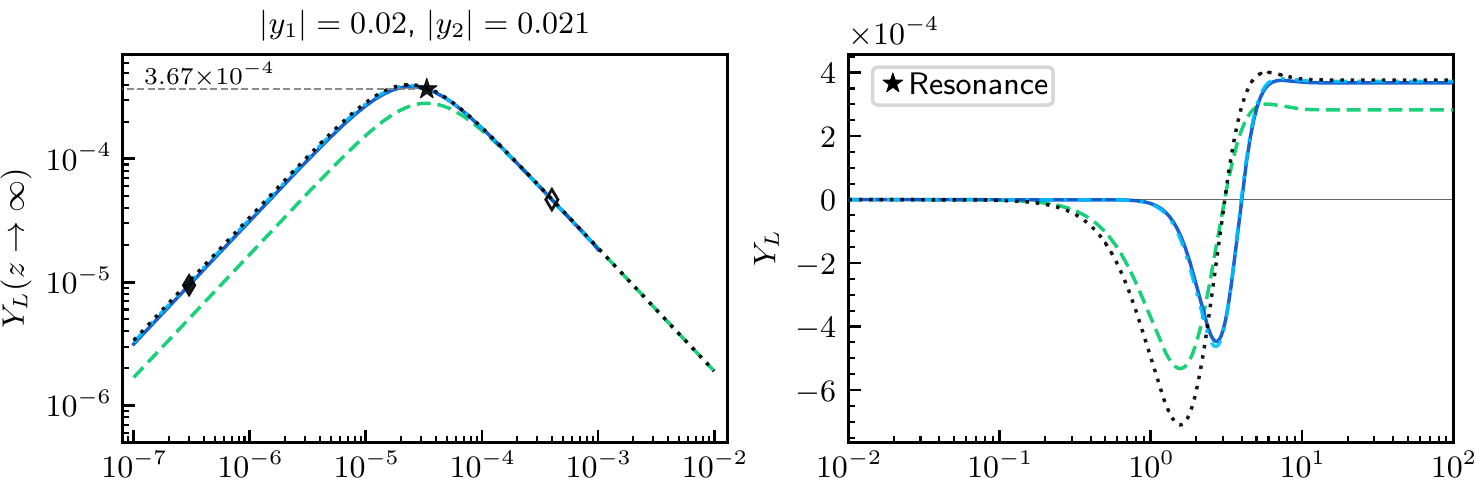}\par
    \includegraphics{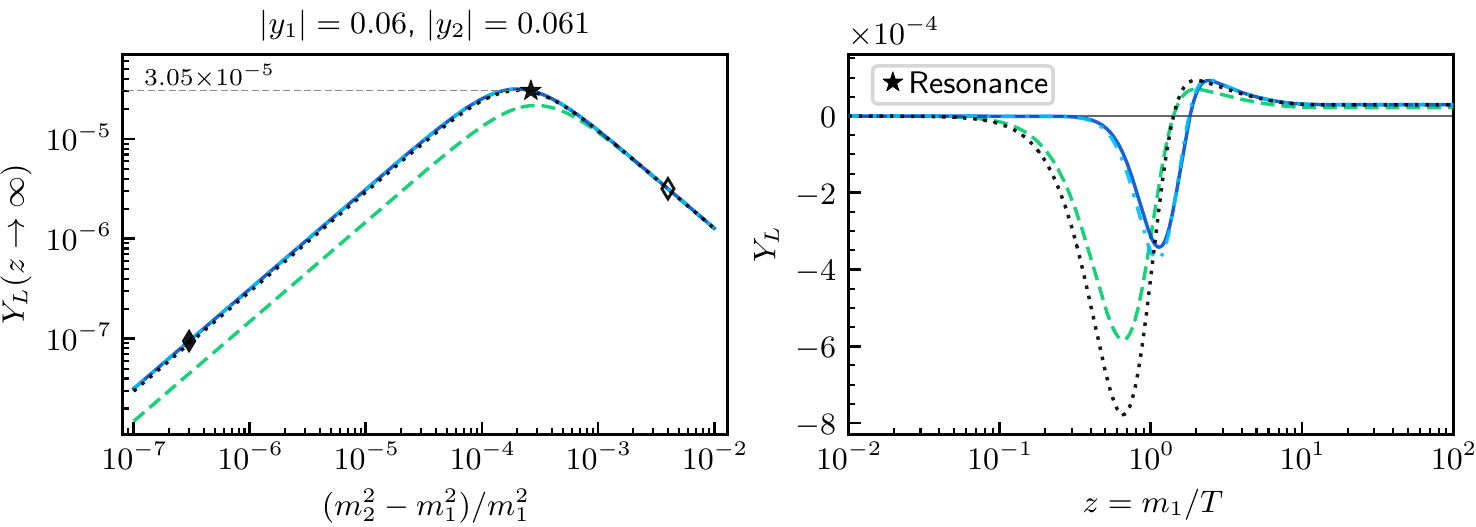}
	\caption{Final lepton asymmetry $Y_L$ as a function of $\Delta m^2_{21}/m_1^2$ (left panels) and the full evolution of $Y_L(z)$ in the maximally resonant case (right panels) indicated with a star-marker (the diamond-markers are considered in figure~\cref{fig:leptogenesis_more}). The results are shown in three different washout strength scenarios: weak (top panels), intermediate (middle panels) and strong (bottom panels).} 
	\label{fig:leptogenesis_overview}
\end{figure}
%

In figure~\cref{fig:leptogenesis_overview} we give a further comparison between three different washout strength scenarios: weak ($K_i \sim 0.1$), intermediate ($K_i \sim 1$) and strong ($K_i \sim 10$). The washout strength here is characterized by the parameter $K_i \eqdef \Gamma^{(0)}_i/H(m_1)$~\cite{Jukkala:2021sku}. Note that in figure~\cref{fig:leptogenesis_YL_vs_massdiff_comparison} (and in other results given in~\cite{Jukkala:2021sku}) the total washout was still quite strong in all cases because at least one of the Yukawa couplings was large. In figure~\cref{fig:leptogenesis_overview} we also compare the helicity-symmetric approximation given by equations~\cref{eq:lgenQKE_h-symmetric,eq:lgenQKE_SCP_h-approx} to the full QKEs and the Boltzmann equations. Furthermore, we show the full evolution of the lepton asymmetry $Y_L$ as a function of $z$ for each scenario, in addition to the plots of the asymptotic asymmetry on the left.
The mass difference parameters used in the right-hand side plots correspond to the maximal saturation points marked on the plots on the left-hand side. The left panels show that again the final lepton asymmetry is maximal at $\Delta m_{21} \sim \bar\Gamma_{12}$ and also that the largest value is obtained in the intermediate washout case (middle panels) when $\Delta m_{21} \sim \bar\Gamma_{12} \sim H$. This is as expected from general considerations~\cite{Jukkala:2021sku}.

In figure~\cref{fig:leptogenesis_more} we show the full evolution of the lepton asymmetry $Y_L(z)$ also in the overdamped ($\Delta m_{21} \ll \bar \Gamma_{12}$) and rapidly oscillating cases ($\Delta m_{21} \gg \bar \Gamma_{12}$), for the mass difference values indicated by the diamond-markers in figure~\cref{fig:leptogenesis_overview}. The full evolution plots in figures~\cref{fig:leptogenesis_overview,fig:leptogenesis_more} show how increasing the interaction strength shifts the generation of the lepton asymmetry towards the relativistic region $z < 1$ in all cases.
It can also be seen that the Boltzmann equations significantly overestimate the magnitude of the lepton asymmetry earlier during the evolution, when compared to our full QKEs. This can be explained by the static CP-asymmetry parameters $\epsilon^{\rm CP}_i$ of the Boltzmann approach which do not take into account the finite build-up time~\cite{DeSimone:2007gkc} of the CP-asymmetry. For larger mass differences $\Delta m_{21} > \bar \Gamma_{12}$ the build-up time is shorter which is shown in the right-hand side panels of figure~\cref{fig:leptogenesis_more} [see also equations~\cref{eq:lgenQKE_offd_h-approx_integrated,eq:lgenQKE_SCP_h-approx}]. There the initial lepton asymmetry spike (of the full QKE) is produced earlier when compared to the maximally resonant and overdamped cases.

The helicity-symmetric approximation tracks the full QKEs near-perfectly in most of the considered cases. It only fails somewhat in the fast oscillation regime ($\Delta m_{21} > \bar \Gamma_{12}$) as can be seen from the right-hand side panels in figure~\cref{fig:leptogenesis_more}. However, the final lepton asymmetry still converges to the full QKE result in all cases, except the case of fast oscillations with weak washout.
The effective sum regulator in the Boltzmann approach also gives a very good late-time match to our full QKEs in all considered cases except when the washout is weak. However, we do not expect the match to be this good more generally because the validity of the effective regulator is sensitive to the size of the CP-violating phase $\theta_{12}$ of the Yukawa couplings~\cite{Dev:2017wwc}. The results do not take this into account as we used only a single value for $\theta_{12}$.

%
\begin{figure}[tp!]
    \centering
    \includegraphics{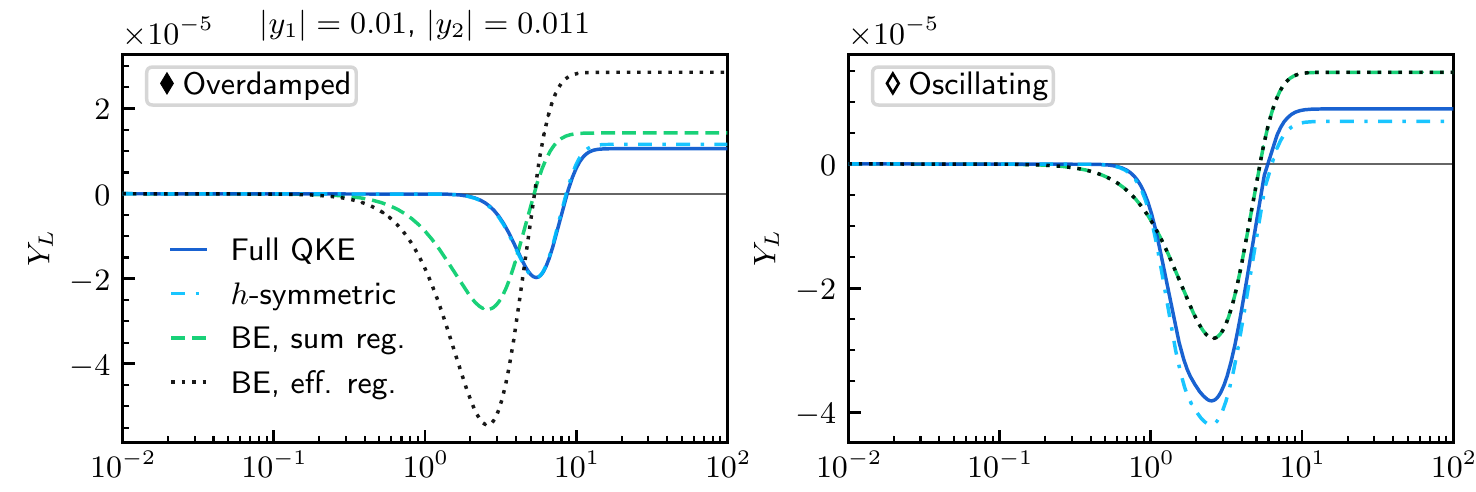}\par
    \includegraphics{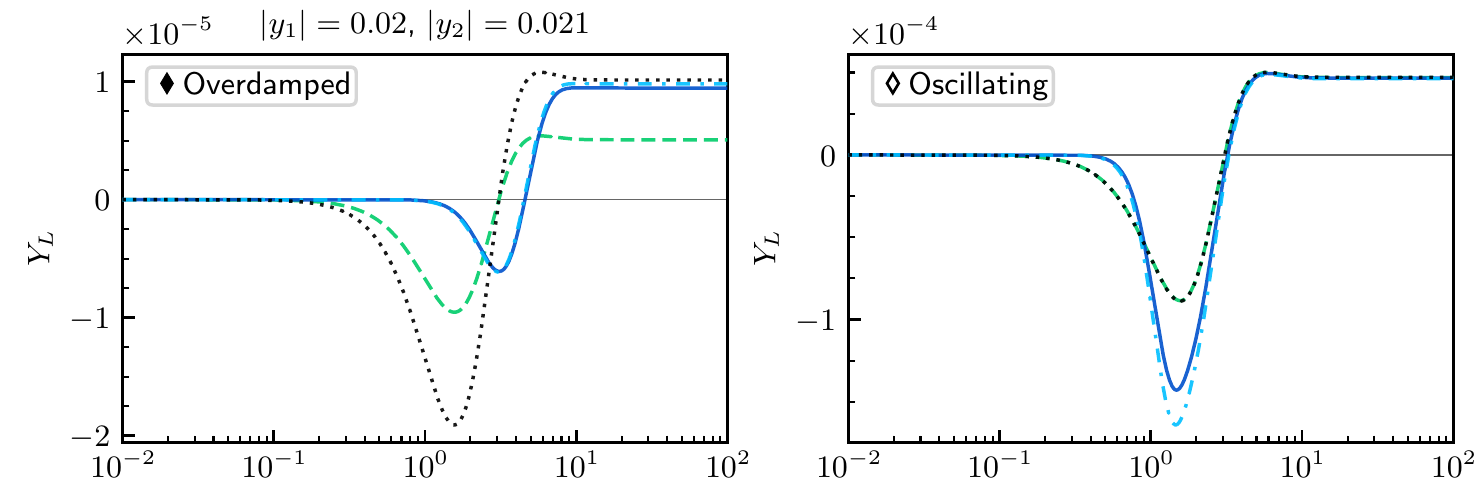}\par
    \includegraphics{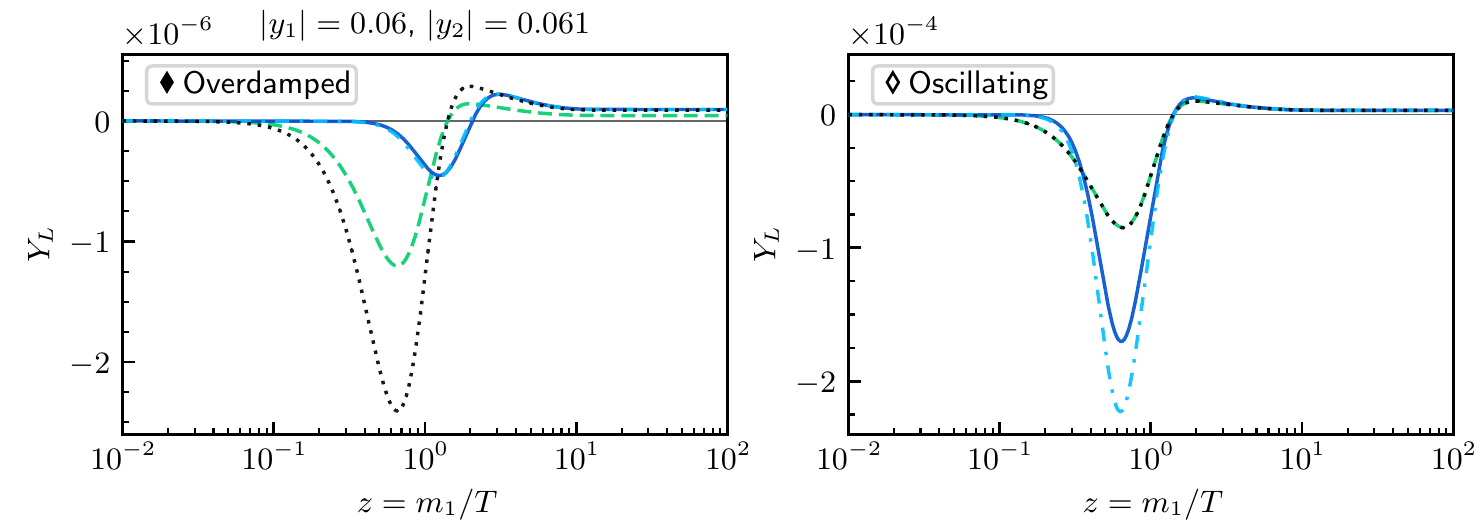}
	\caption{Continuation of figure~\cref{fig:leptogenesis_overview}, showing the full evolution of $Y_L(z)$ in the overdamped and rapidly oscillating cases for each washout scenario. The used mass difference values in each panel are indicated with the filled or empty diamond-markers (in figure~\cref{fig:leptogenesis_overview}).}
	\label{fig:leptogenesis_more}
\end{figure}
%

%% file: chapters/ch6_conclusions.tex
\chapter{Discussion and conclusions}
\label{chap:conclusions}

We have developed approximation methods in non-equilibrium QFT for deriving local quantum kinetic equations including interactions and \emph{quantum coherence}. Our main results are the formulation of the new local approximation (LA) method~\cite{Jukkala:2021sku} and also the improvement~\cite{Jukkala:2019slc} of the coherent quasiparticle approximation (cQPA) developed earlier in~\cite{Herranen:2008hi,Herranen:2008hu,Herranen:2008yg,Herranen:2008di,Herranen:2009zi,Herranen:2009xi,Herranen:2010mh,Fidler:2011yq,Herranen:2011zg}. We focused on fermions in spatially homogeneous and isotropic systems with a time-dependent mass and a weak coupling to a thermal bath. The motivation for this setting was the early universe and especially baryogenesis.
The presented flavored QKEs in these systems can describe both flavor coherence and particle--antiparticle coherence and the related oscillations (summarized in table~\cref{tbl:phase_space_functions}).

\section{Coherent quantum kinetic equations}

In section~\cref{sec:cQPA} we presented a new derivation of the cQPA equations of motion. We gave a straightforward way to derive the cQPA ansatz by using the projection matrix parametrization. We introduced a new way of organizing the gradient expansion, leading to a transparent method for performing the coherence gradient resummation~\cite{Fidler:2011yq} of the collision term. It follows that the coherence-resummed self-energy functions~\cref{eq:coherence-resummed_self-energies} are always evaluated at the conventional flavor-diagonal mass shells instead of the coherence shells. While this result is already known~\cite{Herranen:2010mh}, the more easy-to-use method presented here has been long needed (see \eg~\cite{Garbrecht:2014aga}).

In section~\cref{sec:ewbg_application} we formulated the transport equations~\cref{eq:1-flavor_cQPA_with_complex_mass} using cQPA for a generic interacting fermion with a CP-violating complex mass. These equations are a generalization of semiclassical Boltzmann equations which take into account local coherence between particles and antiparticles. The structure of the equations shows that a changing mass ($\dot m \neq 0$) induces coherent particle production while a changing complex phase of the mass ($\dot\theta \neq 0$) generates helicity asymmetry (leading to CP-asymmetry) via the coherence functions $f^c$.
We used the equations to study the evolution of axial charge density with the mass profile~\cref{eq:time-dependent_tanh_profile} which is a time-dependent analogue of a first order phase transition in EWBG. The results (figures~\cref{fig:single_flavor_cQPA_with_interactions,fig:cQPA_vs_semiclassical_modes}) indicate that in the middle of the transition (\ie~inside the transition wall), where the mass changes rapidly, almost all of the axial charge is due to coherence. Full quantum treatment of coherence can thus be important in the thin wall case~\cite{Farrar:1993hn,Joyce:1994zn,Cline:1995dg} of EWBG. In other regions the mass shell functions $f^m$ are able to track the average of the oscillating axial charge. This result also supports the exclusion of the $f^c$-functions in the leptogenesis application where the background is slowly-changing. We showed explicitly that semiclassical effects are included in cQPA.

We also verified the shell structure of cQPA by using exact solutions of the Dirac equation in a non-interacting system with the mass profile~\cref{eq:time-dependent_tanh_profile}. The results (figure~\cref{fig:exact_Wightman_function_phase_space}) additionally revealed the important twofold role of damping in the emergence of the local kinetic picture. A spectral quasiparticle description with well-distinguished shells requires that there is not too much damping as it smears the shells. On the other hand, if there is too little damping the two-point function becomes overwhelmed by non-local temporal correlations~\cite{Jukkala:2019slc}. This implies the breakdown of the simple shell picture and the description of time-evolution by local quantities.
This observation is important for the development of more general local approximation methods.

Our results also establish that cQPA is exact when calculating local quantities in non-interacting systems. We derived the one-to-one correspondence between the local mode function matrix and the cQPA phase space distribution functions, which was not done explicitly in~\cite{Jukkala:2019slc}. This is another important consistency check for the basic structure of the coherent phase space functions, even though the true utility of cQPA is of course in the interacting case. However, these developments lead to an important insight about cQPA, ultimately transcending it: the key feature is that the non-local correlations are treated non-dynamically while the local correlations related to the average time coordinate are fully included~\cite{Jukkala:2019slc}. This insight was used as the main idea for developing the more general LA-method.

\section{Improved local description of non-equilibrium dynamics}

In section~\cref{eq:the_local_approximation_method} we formulated the new LA-method and used it to derive improved coherent QKEs~\cref{eq:generalized_density_matrix_equation} where the effects of spectral width can be included. The main idea of the method is to form a closed dynamical equation directly for the local two-point function $S^\lt_{\evec{k}}(t,t)$. This can be achieved in dissipative systems by two steps (summarized in table~\cref{tbl:LA-method_steps}): (1) identification of a suitable background solution $S_{\rm ad}$ and expansion of the full solution $S$ around it, and
(2) parametrization of the non-local parts of the deviation $\delta S$ by its local values.
The second step with the local ansatz~\cref{eq:the_local_ansatz} is the core part of the method which enables the local description of the non-equilibrium dynamics.

An essential part in the derivation of the improved QKEs was also the projection matrix parametrization which separates different components of the local propagator which vary in different time-scales.
This is crucial for separating the rapid particle--antiparticle oscillations from the flavor oscillations.
Furthermore, there is no loss of generality with this parametrization in homogeneous and isotropic systems; equations~\cref{eq:generalized_density_matrix_equation} and~\cref{eq:the_closed_local_equation} are equivalent.

The LA-method can be viewed as an improved and generalized version of the cQPA. The improved QKEs~\cref{eq:generalized_density_matrix_equation} can be reduced to the cQPA equations in a spectral approximation. However, the local transport equation~\cref{eq:the_closed_local_equation} in the LA-method is formulated directly in the two-time representation at the propagator level without needing to solve the phase space structure. The method can thus be used in a spectral approximation but also more generally when accounting for the spectral width. A major practical advantage is also that the gradient resummation of the collision term is automatically imposed by the local ansatz in the LA-method.

\subsection{Resonant leptogenesis with coherent QKEs}

In section~\cref{sec:RL_application} we derived local coherent QKEs for leptogenesis by using the LA-method. The most general form of the equations contains all local quantum coherence effects of the Majorana neutrinos and also thermal medium effects in a weak coupling expansion~\cite{Jukkala:2021sku}. The simpler QKEs~\cref{eq:lgenQKEs} for the minimal leptogenesis model are still valid for arbitrary mass differences and fully include the coherent flavor oscillations of the Majorana neutrinos. Our main goal with these equations was to study the details of the CP-asymmetry generation in resonant leptogenesis.
We gave even simpler helicity-symmetric equations~\cref{eq:lgenQKE_h-symmetric,eq:lgenQKE_SCP_h-approx} which provide a good approximation to the full QKEs~\cref{eq:lgenQKEs} and from which the CP-asymmetry generation mechanism can be more easily analyzed. We also reduced the equations to the conventional Boltzmann equation form in the decoupling limit (and with further approximations)~\cite{Jukkala:2021sku}.

Our results imply that the flavor-coherence damping rate $\bar\Gamma_{\evec{k}12}$ of the Majorana neutrinos is the origin of the CP-asymmetry regulator in the Boltzmann approach. The result corresponds to the sum-regulator derived earlier in the CTP approach~\cite{Garny:2011hg,Iso:2013lba,Iso:2014afa}. The damping of flavor-coherence, rather than the oscillations averaging out (\cf~\cite{DeSimone:2007gkc}), also explains why the Boltzmann equations and the full QKEs converge to the same late-time lepton asymmetry in the decoupling limit (\ie~when $\Delta m_{21} \gg \bar\Gamma_{12}$)~\cite{Jukkala:2021sku}.
Our numerical results agree with these findings.

We performed a thorough numerical analysis of our QKEs in the resonant case and compared them to the conventional Boltzmann equations~\cite{Jukkala:2021sku}. The exact form of the CP-asymmetry regulator in the Boltzmann approach has been debated in the literature~\cite{Garny:2011hg,Garbrecht:2014aga,Dev:2017wwc,Dev:2017trv,Anisimov:2005hr,Dev:2014laa}. Therefore we included several different regulators for the Boltzmann equations in the comparison of the final lepton asymmetry calculated from the different approaches. Our results in figures~\cref{fig:leptogenesis_YL_vs_massdiff_comparison,fig:leptogenesis_overview,fig:leptogenesis_more} show that the maximal resonant saturation occurs roughly when $\Delta m_{21} \simeq \bar\Gamma_{12}$ in all approaches. However, in the quasidegenerate regime $\Delta m_{21} \lesssim \bar\Gamma_{12}$ there are significant qualitative differences between different Boltzmann regulators. We found that the sum regulator (and its effective variant) gives the best qualitative match to our full QKEs in the late-time limit, but there can still be differences by at least a factor of $2$. The late-time results of all approaches also converge in the weakly resonant regime $\bar\Gamma_{12} \ll \Delta m_{21} \ll m_1$ (except for very weak washout, $\bar\Gamma_{12} \ll H$). This provides a consistency check for our full QKEs in the validity region of the semiclassical approach.

Finally, our QKEs are in line with several other studies based on the CTP method. In particular, our RL results are similar to the results of~\cite{Garny:2011hg,Iso:2013lba,Iso:2014afa} and our equations agree with those derived in~\cite{Garbrecht:2011aw}. However, our approach is also more general in several ways: we do not need to restrict the values of the Yukawa couplings, the size of the deviation from equilibrium, or the mass difference of the Majorana neutrinos. Further comparison to other studies is given in~\cite{Jukkala:2021sku}.

\section{Outlook}

So far we have compared our methods mainly to simple semiclassical equations. More precise comparisons should be made to other methods and results found in the literature. For example, in leptogenesis with quasidegenerate Majorana neutrinos the question if the CP-asymmetry has separate contributions from mixing and oscillations remains unsettled~\cite{Dev:2017trv,Racker:2021kme}. While our results point to the interpretation that the conventional mixing source arises from the coherent oscillations in a particular limit ($\Delta m_{21} \gg \bar\Gamma_{12}$)~\cite{Jukkala:2021sku}, a more careful investigation is needed.
A thorough analysis of corrections to the adiabatic propagators and the effective self-energies is challenging but may also be required for more conclusive answers to this question.

Our leptogenesis equations could also be improved and extended in other ways. An obvious improvement would be the inclusion of scattering processes and multiple SM lepton flavors. It would then be very interesting to apply our equations to low-scale RL and ARS scenarios in a unified framework similarly to~\cite{Klaric:2020phc}. The numerical results could also be extended straightforwardly to include leading dispersive corrections for the Majorana neutrinos. Covering the hierarchical regime of leptogenesis and investigating the size of the CP-asymmetry from particle--antiparticle coherence~\cite{Jukkala:2021sku} would also be very interesting.
At a more general level, a numerical analysis to test the local ansatz~\cref{eq:the_local_ansatz} against the full memory integrals would also be desirable.

A long standing goal, which has remained elusive so far, has also been to extend the coherent quantum transport methods to the planar symmetric case~\cite{Herranen:2008hi}. This is required for a proper description of coherence in realistic EWBG models.
The transport equations developed here could also be combined with the results of~\cite{Kainulainen:2021eki} to study particle production during reheating in a realistic setup.
Another interesting application of the coherent equations is the field-theoretic description of SM neutrino oscillations.

%% file: thesis.bib
@article{Jukkala:2019slc,
    author = "Jukkala, Henri and Kainulainen, Kimmo and Koskivaara, Olli",
    title = "{Quantum transport and the phase space structure of the Wightman functions}",
    eprint = "1910.10979",
    archivePrefix = "arXiv",
    primaryClass = "hep-th",
    reportNumber = "CERN-TH-2019-173",
    doi = "10.1007/JHEP01(2020)012",
    journal = "J. High Energy Phys.",
    volume = "01",
    pages = "012",
    year = "2020",
    shorthand={\RN{1}}
}

@article{Jukkala:2021sku,
    author = "Jukkala, Henri and Kainulainen, Kimmo and Rahkila, Pyry M.",
    title = "{Flavour mixing transport theory and resonant leptogenesis}",
    eprint = "2104.03998",
    archivePrefix = "arXiv",
    primaryClass = "hep-ph",
    doi = "10.1007/JHEP09(2021)119",
    journal = "J. High Energy Phys.",
    volume = "09",
    pages = "119",
    year = "2021",
    shorthand={\RN{2}}
}

@software{henri_jukkala_2021_5025929,
    author = {Henri Jukkala},
    title = {LeptoGen},
    month = jun,
    year = 2021,
    publisher = {Zenodo},
    version = {1.1},
    %doi = {10.5281/zenodo.5025929},
    url = {https://doi.org/10.5281/zenodo.5025929},
    shorthand = {\RN{2}b}
}

@book{Kadanoff:1962book,
    author = "Kadanoff, Leo P. and Baym, Gordon",
    title = "{Quantum Statistical Mechanics}",
    doi = "10.1201/9780429493218",
    isbn = "978-0-201-41046-4",
    publisher = "W. A. Benjamin",
    year = "1962"
}

@book{Weinberg:1972kfs,
    author = "Weinberg, Steven",
    title = "{Gravitation and Cosmology}: {Principles and Applications of the General Theory of Relativity}",
    isbn = "978-0-471-92567-5",
    publisher = "John Wiley and Sons",
    year = "1972"
}

@book{Birrell:1982cam,
    author = "Birrell, N. D. and Davies, P. C. W.",
    title = "{Quantum Fields in Curved Space}",
    doi = "10.1017/CBO9780511622632",
    isbn = "978-0-521-27858-4",
    publisher = "Cambridge University Press",
    year = "1982"
}

@book{Kolb:1990vq,
    author = "Kolb, Edward W. and Turner, Michael S.",
    title = "{The Early Universe}",
    doi = "10.1201/9780429492860",
    isbn = "978-0-201-62674-2",
    publisher = "Addison-Wesley",
    year = "1990"
}

@book{Peskin:1995ev,
    author = "Peskin, Michael E. and Schroeder, Daniel V.",
    title = "{An Introduction to quantum field theory}",
    doi = "10.1201/9780429503559",
    isbn = "978-0-201-50397-5",
    publisher = "Westview Press",
    year = "1995"
}

@book{Weinberg:1995qft1,
    author = "Weinberg, Steven",
    title = "{The Quantum theory of fields. Vol. 1: Foundations}",
    doi = "10.1017/CBO9781139644167",
    isbn = "978-0-521-55001-7",
    publisher = "Cambridge University Press",
    year = "1995"
}

@book{LeBellac:1996cam,
    author = "Le Bellac, Michel",
    title = "{Thermal Field Theory}",
    doi = "10.1017/CBO9780511721700",
    isbn = "978-0-521-65477-7",
    publisher = "Cambridge University Press",
    year = "1996"
}

@book{Kapusta:2006cam,
    author = "Kapusta, Joseph I. and Gale, Charles",
    title = "{Finite-Temperature Field Theory: Principles and Applications}",
    doi = "10.1017/CBO9780511535130",
    isbn = "978-0-521-82082-0",
    publisher = "Cambridge University Press",
    year = "2006"
}

@book{LeBellac:2006qpb,
    author = "Le Bellac, Michel",
    title = "{Quantum Physics}",
    doi = "10.1017/CBO9780511616471",
    isbn = "978-0-521-85277-7",
    publisher = "Cambridge University Press",
    year = "2006"
}

@book{Giunti:2007ry,
    author = "Giunti, Carlo and Kim, Chung W.",
    title = "{Fundamentals of Neutrino Physics and Astro\-physics}",
    doi = "10.1093/acprof:oso/9780198508717.001.0001",
    isbn = "978-0-19-850871-7",
    publisher = "Oxford University Press",
    year = "2007"
}

@book{Calzetta:2008cam,
    author = "Calzetta, Esteban A. and Hu, Bei-Lok B.",
    title = "{Nonequilibrium Quantum Field Theory}",
    doi = "10.1017/CBO9780511535123",
    isbn = "978-0-521-64168-5",
    publisher = "Cambridge University Press",
    year = "2008"
}

@book{Weinberg:2008zzc,
    author = "Weinberg, Steven",
    title = "{Cosmology}",
    doi = "10.1007/s10714-008-0728-z",
    isbn = "978-0-19-852682-7",
    publisher = "Oxford University Press",
    address = "Oxford",
    year = "2008"
}

@article{Sakharov:1967dj,
    author = "Sakharov, A. D.",
    title = "{Violation of CP Invariance, C asymmetry, and baryon asymmetry of the universe}",
    doi = "10.1070/PU1991v034n05ABEH002497",
    journal = "Pisma Zh. Eksp. Teor. Fiz.",
    volume = "5",
    pages = "32--35",
    year = "1967"
}

@article{Kuzmin:1985mm,
    author = "Kuzmin, V. A. and Rubakov, V. A. and Shaposhnikov, M. E.",
    title = "{On the Anomalous Electroweak Baryon Number Nonconservation in the Early Universe}",
    reportNumber = "IC/85/8",
    doi = "10.1016/0370-2693(85)91028-7",
    journal = "Phys. Lett. B",
    volume = "155",
    pages = "36",
    year = "1985"
}

@article{Fukugita:1986hr,
    author = "Fukugita, M. and Yanagida, T.",
    title = "{Baryogenesis Without Grand Unification}",
    reportNumber = "RIFP-641",
    doi = "10.1016/0370-2693(86)91126-3",
    journal = "Phys. Lett. B",
    volume = "174",
    pages = "45--47",
    year = "1986"
}

@article{Akhmedov:1998qx,
    author = "Akhmedov, Evgeny K. and Rubakov, V. A. and Smirnov, A. Yu.",
    title = "{Baryogenesis via neutrino oscillations}",
    eprint = "hep-ph/9803255",
    archivePrefix = "arXiv",
    reportNumber = "IC-98-22, INR-98-14-T",
    doi = "10.1103/PhysRevLett.81.1359",
    journal = "Phys. Rev. Lett.",
    volume = "81",
    pages = "1359--1362",
    year = "1998"
}

@article{Dine:2003ax,
    author = "Dine, Michael and Kusenko, Alexander",
    title = "{The Origin of the matter - antimatter asymmetry}",
    eprint = "hep-ph/0303065",
    archivePrefix = "arXiv",
    reportNumber = "SCIPP-2003-08, UCLA-03-TEP-08",
    doi = "10.1103/RevModPhys.76.1",
    journal = "Rev. Mod. Phys.",
    volume = "76",
    pages = "1",
    year = "2003"
}

@inproceedings{Cline:2006ts,
    author = "Cline, James M.",
    title = "{Baryogenesis}",
    booktitle = "{Les Houches Summer School - Session 86: Particle Physics and Cosmology: The Fabric of Spacetime}",
    eprint = "hep-ph/0609145",
    archivePrefix = "arXiv",
    month = "9",
    year = "2006"
}

@article{Cline:2018fuq,
    author = "Cline, James M.",
    title = "{TASI Lectures on Early Universe Cosmology: Inflation, Baryogenesis and Dark Matter}",
    eprint = "1807.08749",
    archivePrefix = "arXiv",
    primaryClass = "hep-ph",
    journal = "PoS",
    volume = "TASI2018",
    pages = "001",
    year = "2019"
}

@article{Bodeker:2020ghk,
    author = "Bodeker, Dietrich and Buchmuller, Wilfried",
    title = "{Baryogenesis from the weak scale to the grand unification scale}",
    eprint = "2009.07294",
    archivePrefix = "arXiv",
    primaryClass = "hep-ph",
    reportNumber = "DESY 20-141, DESY-20-141",
    doi = "10.1103/RevModPhys.93.035004",
    journal = "Rev. Mod. Phys.",
    volume = "93",
    number = "3",
    pages = "035004",
    year = "2021"
}

@article{Cohen:1993nk,
    author = "Cohen, Andrew G. and Kaplan, D. B. and Nelson, A. E.",
    title = "{Progress in electroweak baryogenesis}",
    eprint = "hep-ph/9302210",
    archivePrefix = "arXiv",
    reportNumber = "UCSD-PTH-93-02, BUHEP-93-4",
    doi = "10.1146/annurev.ns.43.120193.000331",
    journal = "Ann. Rev. Nucl. Part. Sci.",
    volume = "43",
    pages = "27--70",
    year = "1993"
}

@article{Morrissey:2012db,
    author = "Morrissey, David E. and Ramsey-Musolf, Michael J.",
    title = "{Electroweak baryogenesis}",
    eprint = "1206.2942",
    archivePrefix = "arXiv",
    primaryClass = "hep-ph",
    reportNumber = "NPAC-12-08",
    doi = "10.1088/1367-2630/14/12/125003",
    journal = "New J. Phys.",
    volume = "14",
    pages = "125003",
    year = "2012"
}

@article{Konstandin:2013caa,
    author = "Konstandin, Thomas",
    title = "{Quantum Transport and Electroweak Baryogenesis}",
    eprint = "1302.6713",
    archivePrefix = "arXiv",
    primaryClass = "hep-ph",
    reportNumber = "DESY-13-036",
    doi = "10.3367/UFNe.0183.201308a.0785",
    journal = "Phys. Usp.",
    volume = "56",
    pages = "747--771",
    year = "2013"
}

@article{Davidson:2008bu,
    author = "Davidson, Sacha and Nardi, Enrico and Nir, Yosef",
    title = "{Leptogenesis}",
    eprint = "0802.2962",
    archivePrefix = "arXiv",
    primaryClass = "hep-ph",
    doi = "10.1016/j.physrep.2008.06.002",
    journal = "Phys. Rept.",
    volume = "466",
    pages = "105--177",
    year = "2008"
}

@article{Pilaftsis:2009pk,
    author = "Pilaftsis, Apostolos",
    editor = "Bernabeu, J. and Botella, F. J. and Mavromatos, N. E. and Mitsou, V. A.",
    title = "{The Little Review on Leptogenesis}",
    eprint = "0904.1182",
    archivePrefix = "arXiv",
    primaryClass = "hep-ph",
    reportNumber = "MAN-HEP-2009-14",
    doi = "10.1088/1742-6596/171/1/012017",
    journal = "J. Phys. Conf. Ser.",
    volume = "171",
    pages = "012017",
    year = "2009"
}

@article{Blanchet:2012bk,
    author = "Blanchet, Steve and Di Bari, Pasquale",
    title = "{The minimal scenario of leptogenesis}",
    eprint = "1211.0512",
    archivePrefix = "arXiv",
    primaryClass = "hep-ph",
    doi = "10.1088/1367-2630/14/12/125012",
    journal = "New J. Phys.",
    volume = "14",
    pages = "125012",
    year = "2012"
}

@article{Harvey:1990qw,
    author = "Harvey, Jeffrey A. and Turner, Michael S.",
    title = "{Cosmological baryon and lepton number in the presence of electroweak fermion number violation}",
    reportNumber = "FERMILAB-PUB-90-049-A, EFI-90-33",
    doi = "10.1103/PhysRevD.42.3344",
    journal = "Phys. Rev. D",
    volume = "42",
    pages = "3344--3349",
    year = "1990"
}

@article{Kajantie:1996mn,
    author = "Kajantie, K. and Laine, M. and Rummukainen, K. and Shaposhnikov, Mikhail E.",
    title = "{Is there a~ hot electroweak phase transition at $m_H \gtrsim m_W$?}",
    eprint = "hep-ph/9605288",
    archivePrefix = "arXiv",
    reportNumber = "CERN-TH-96-126, HD-THEP-96-15, IUHET-333",
    doi = "10.1103/PhysRevLett.77.2887",
    journal = "Phys. Rev. Lett.",
    volume = "77",
    pages = "2887--2890",
    year = "1996"
}

@article{tHooft:1976snw,
    author = "'t Hooft, Gerard",
    editor = "Shifman, Mikhail A.",
    title = "{Computation of the Quantum Effects Due to a Four-Dimensional Pseudoparticle}",
    reportNumber = "PRINT-76-0551 (HARVARD)",
    doi = "10.1103/PhysRevD.14.3432",
    journal = "Phys. Rev. D",
    volume = "14",
    pages = "3432--3450",
    year = "1976",
    note = "[Erratum: Phys.Rev.D 18, 2199 (1978)]"
}

@article{tHooft:1976rip,
    author = "'t Hooft, Gerard",
    editor = "Shifman, Mikhail A.",
    title = "{Symmetry Breaking Through Bell-Jackiw Anomalies}",
    reportNumber = "PRINT-76-0254 (HARVARD)",
    doi = "10.1103/PhysRevLett.37.8",
    journal = "Phys. Rev. Lett.",
    volume = "37",
    pages = "8--11",
    year = "1976"
}

@article{Rubakov:1996vz,
    author = "Rubakov, V. A. and Shaposhnikov, M. E.",
    title = "{Electroweak baryon number nonconservation in the early universe and in high-energy collisions}",
    eprint = "hep-ph/9603208",
    archivePrefix = "arXiv",
    reportNumber = "CERN-TH-96-13, INR-0913-96",
    doi = "10.1070/PU1996v039n05ABEH000145",
    journal = "Usp. Fiz. Nauk",
    volume = "166",
    pages = "493--537",
    year = "1996"
}

@article{Adler:1969gk,
    author = "Adler, Stephen L.",
    title = "{Axial vector vertex in spinor electrodynamics}",
    doi = "10.1103/PhysRev.177.2426",
    journal = "Phys. Rev.",
    volume = "177",
    pages = "2426--2438",
    year = "1969"
}

@article{Bell:1969ts,
    author = "Bell, J. S. and Jackiw, R.",
    title = "{A PCAC puzzle: $\pi^0 \to \gamma \gamma$ in the $\sigma$ model}",
    doi = "10.1007/BF02823296",
    journal = "Nuovo Cim. A",
    volume = "60",
    pages = "47--61",
    year = "1969"
}

@article{Klinkhamer:1984di,
    author = "Klinkhamer, Frans R. and Manton, N. S.",
    title = "{A Saddle Point Solution in the Weinberg-Salam Theory}",
    reportNumber = "NSF-ITP-84-57",
    doi = "10.1103/PhysRevD.30.2212",
    journal = "Phys. Rev. D",
    volume = "30",
    pages = "2212",
    year = "1984"
}

@article{Manton:1983nd,
    author = "Manton, N. S.",
    title = "{Topology in the Weinberg-Salam Theory}",
    reportNumber = "NSF-ITP-83-64",
    doi = "10.1103/PhysRevD.28.2019",
    journal = "Phys. Rev. D",
    volume = "28",
    pages = "2019",
    year = "1983"
}

@article{Arnold:1987mh,
    author = "Arnold, Peter Brockway and McLerran, Larry D.",
    title = "{Sphalerons, Small Fluctuations and Baryon Number Violation in Electroweak Theory}",
    reportNumber = "FERMILAB-PUB-87-034-T",
    doi = "10.1103/PhysRevD.36.581",
    journal = "Phys. Rev. D",
    volume = "36",
    pages = "581",
    year = "1987"
}

@article{Arnold:1987zg,
    author = "Arnold, Peter Brockway and McLerran, Larry D.",
    title = "{The Sphaleron Strikes Back}",
    reportNumber = "FERMILAB-PUB-87-120-T",
    doi = "10.1103/PhysRevD.37.1020",
    journal = "Phys. Rev. D",
    volume = "37",
    pages = "1020",
    year = "1988"
}

@article{Joyce:1994zn,
    author = "Joyce, Michael and Prokopec, Tomislav and Turok, Neil",
    title = "{Nonlocal electroweak baryogenesis. Part 1: Thin wall regime}",
    eprint = "hep-ph/9410281",
    archivePrefix = "arXiv",
    reportNumber = "PUPT-1495, PUP-TH-1495 (1994)",
    doi = "10.1103/PhysRevD.53.2930",
    journal = "Phys. Rev. D",
    volume = "53",
    pages = "2930--2957",
    year = "1996"
}

@article{Cline:1997vk,
    author = "Cline, James M. and Joyce, Michael and Kainulainen, Kimmo",
    title = "{Supersymmetric electroweak baryogenesis in the WKB approximation}",
    eprint = "hep-ph/9708393",
    archivePrefix = "arXiv",
    reportNumber = "MCGILL-97-26, HIP-1997-44-TH",
    doi = "10.1016/S0370-2693(97)01361-0",
    journal = "Phys. Lett. B",
    volume = "417",
    pages = "79--86",
    year = "1998",
    note = "[Erratum: Phys.Lett.B 448, 321--321 (1999)]"
}

@article{Cline:2001rk,
    author = "Cline, James M. and Joyce, Michael and Kainulainen, Kimmo",
    title = "{Erratum for 'Supersymmetric electroweak baryogenesis'}",
    eprint = "hep-ph/0110031",
    archivePrefix = "arXiv",
    month = "10",
    year = "2001"
}

@article{Kainulainen:2001cn,
    author = "Kainulainen, Kimmo and Prokopec, Tomislav and Schmidt, Michael G. and Weinstock, Steffen",
    title = "{First principle derivation of semiclassical force for electroweak baryogenesis}",
    eprint = "hep-ph/0105295",
    archivePrefix = "arXiv",
    reportNumber = "HD-THEP-01-23, NORDITA-2001-9-HE",
    doi = "10.1088/1126-6708/2001/06/031",
    journal = "J. High Energy Phys.",
    volume = "06",
    pages = "031",
    year = "2001"
}

@article{Kainulainen:2002th,
    author = "Kainulainen, Kimmo and Prokopec, Tomislav and Schmidt, Michael G. and Weinstock, Steffen",
    title = "{Semiclassical force for electroweak baryogenesis: Three-dimensional derivation}",
    eprint = "hep-ph/0202177",
    archivePrefix = "arXiv",
    reportNumber = "CERN-TH-2002-011, NORDITA-2002-5-HE, HD-THEP-02-7",
    doi = "10.1103/PhysRevD.66.043502",
    journal = "Phys. Rev. D",
    volume = "66",
    pages = "043502",
    year = "2002"
}

@article{Cline:2020jre,
    author = "Cline, James M. and Kainulainen, Kimmo",
    title = "{Electroweak baryogenesis at high bubble wall velocities}",
    eprint = "2001.00568",
    archivePrefix = "arXiv",
    primaryClass = "hep-ph",
    reportNumber = "CERN-TH-2019-227",
    doi = "10.1103/PhysRevD.101.063525",
    journal = "Phys. Rev. D",
    volume = "101",
    number = "6",
    pages = "063525",
    year = "2020"
}

@article{Cline:2017jvp,
    author = "Cline, James M.",
    editor = "Auge, Etienne and Dumarchez, Jacques and Tran Thanh Van, Jean",
    title = "{Is electroweak baryogenesis dead?}",
    eprint = "1704.08911",
    archivePrefix = "arXiv",
    primaryClass = "hep-ph",
    reportNumber = "CERN-TH-2017-096",
    doi = "10.1098/rsta.2017.0116",
    pages = "339--348",
    year = "2017"
}

@article{Bodeker:2004ws,
    author = "Bodeker, Dietrich and Fromme, Lars and Huber, Stephan J. and Seniuch, Michael",
    title = "{The Baryon asymmetry in the standard model with a low cut-off}",
    eprint = "hep-ph/0412366",
    archivePrefix = "arXiv",
    reportNumber = "BI-TP-2004-41, CERN-PH-TH-2004-258",
    doi = "10.1088/1126-6708/2005/02/026",
    journal = "J. High Energy Phys.",
    volume = "02",
    pages = "026",
    year = "2005"
}

@article{Prokopec:2013ax,
    author = "Prokopec, Tomislav and Schmidt, Michael G. and Weenink, Jan",
    title = "{Exact solution of the Dirac equation with CP violation}",
    eprint = "1301.4132",
    archivePrefix = "arXiv",
    primaryClass = "hep-th",
    doi = "10.1103/PhysRevD.87.083508",
    journal = "Phys. Rev. D",
    volume = "87",
    number = "8",
    pages = "083508",
    year = "2013"
}

@article{Cline:1995dg,
    author = "Cline, James M. and Kainulainen, Kimmo and Vischer, Axel P.",
    title = "{Dynamics of two Higgs doublet CP violation and baryogenesis at the electroweak phase transition}",
    eprint = "hep-ph/9506284",
    archivePrefix = "arXiv",
    reportNumber = "MCGILL-95-16, CERN-TH-95-136, UMN-TH-1343-94",
    doi = "10.1103/PhysRevD.54.2451",
    journal = "Phys. Rev. D",
    volume = "54",
    pages = "2451--2472",
    year = "1996"
}

@article{Farrar:1993hn,
    author = "Farrar, Glennys R. and Shaposhnikov, M. E.",
    title = "{Baryon asymmetry of the universe in the standard electroweak theory}",
    eprint = "hep-ph/9305275",
    archivePrefix = "arXiv",
    reportNumber = "CERN-TH-6732-93, CERN-TH-6734-93, RU-93-11",
    doi = "10.1103/PhysRevD.50.774",
    journal = "Phys. Rev. D",
    volume = "50",
    pages = "774",
    year = "1994"
}

@article{Kainulainen:2021oqs,
    author = "Kainulainen, Kimmo",
    title = "{CP-violating transport theory for electroweak baryogenesis with thermal corrections}",
    eprint = "2108.08336",
    archivePrefix = "arXiv",
    primaryClass = "hep-ph",
    doi = "10.1088/1475-7516/2021/11/042",
    journal = "J. Cosmol. Astropart. Phys.",
    volume = "11",
    number = "11",
    pages = "042",
    year = "2021"
}

@article{Cline:2017qpe,
    author = "Cline, James M. and Kainulainen, Kimmo and Tucker-Smith, David",
    title = "{Electroweak baryogenesis from a dark sector}",
    eprint = "1702.08909",
    archivePrefix = "arXiv",
    primaryClass = "hep-ph",
    reportNumber = "CERN-TH-2017-050",
    doi = "10.1103/PhysRevD.95.115006",
    journal = "Phys. Rev. D",
    volume = "95",
    number = "11",
    pages = "115006",
    year = "2017"
}

@article{Cline:2013bln,
    author = "Cline, James M. and Kainulainen, Kimmo",
    title = "{Improved Electroweak Phase Transition with Subdominant Inert Doublet Dark Matter}",
    eprint = "1302.2614",
    archivePrefix = "arXiv",
    primaryClass = "hep-ph",
    doi = "10.1103/PhysRevD.87.071701",
    journal = "Phys. Rev. D",
    volume = "87",
    number = "7",
    pages = "071701",
    year = "2013"
}

@article{Cline:2012hg,
    author = "Cline, James M. and Kainulainen, Kimmo",
    title = "{Electroweak baryogenesis and dark matter from a singlet Higgs}",
    eprint = "1210.4196",
    archivePrefix = "arXiv",
    primaryClass = "hep-ph",
    doi = "10.1088/1475-7516/2013/01/012",
    journal = "J. Cosmol. Astropart. Phys.",
    volume = "01",
    pages = "012",
    year = "2013"
}

@article{Cline:2011mm,
    author = "Cline, James M. and Kainulainen, Kimmo and Trott, Michael",
    title = "{Electroweak Baryogenesis in Two Higgs Doublet Models and B meson anomalies}",
    eprint = "1107.3559",
    archivePrefix = "arXiv",
    primaryClass = "hep-ph",
    doi = "10.1007/JHEP11(2011)089",
    journal = "J. High Energy Phys.",
    volume = "11",
    pages = "089",
    year = "2011"
}

@article{Cline:2000nw,
    author = "Cline, James M. and Joyce, Michael and Kainulainen, Kimmo",
    title = "{Supersymmetric electroweak baryogenesis}",
    eprint = "hep-ph/0006119",
    archivePrefix = "arXiv",
    reportNumber = "MCGILL-00-15, NORDITA-2000-38-HE, LPT-ORSAY-00-46",
    doi = "10.1088/1126-6708/2000/07/018",
    journal = "J. High Energy Phys.",
    volume = "07",
    pages = "018",
    year = "2000"
}

@article{Cline:2000kb,
    author = "Cline, James M. and Kainulainen, Kimmo",
    title = "{A New source for electroweak baryogenesis in the MSSM}",
    eprint = "hep-ph/0002272",
    archivePrefix = "arXiv",
    reportNumber = "MCGILL-00-07, NORDITA-2000-10-HE",
    doi = "10.1103/PhysRevLett.85.5519",
    journal = "Phys. Rev. Lett.",
    volume = "85",
    pages = "5519--5522",
    year = "2000"
}

@article{Cline:2013gha,
    author = "Cline, James M. and Kainulainen, Kimmo and Scott, Pat and Weniger, Christoph",
    title = "{Update on scalar singlet dark matter}",
    eprint = "1306.4710",
    archivePrefix = "arXiv",
    primaryClass = "hep-ph",
    doi = "10.1103/PhysRevD.88.055025",
    journal = "Phys. Rev. D",
    volume = "88",
    pages = "055025",
    year = "2013",
    note = "[Erratum: Phys.Rev.D 92, 039906 (2015)]"
}

@article{Alanne:2016wtx,
    author = "Alanne, Tommi and Kainulainen, Kimmo and Tuominen, Kimmo and Vaskonen, Ville",
    title = "{Baryogenesis in the two doublet and inert singlet extension of the Standard Model}",
    eprint = "1607.03303",
    archivePrefix = "arXiv",
    primaryClass = "hep-ph",
    reportNumber = "HIP-2016-22-TH, CP3-ORIGINS-2016-031",
    doi = "10.1088/1475-7516/2016/08/057",
    journal = "J. Cosmol. Astropart. Phys.",
    volume = "08",
    pages = "057",
    year = "2016"
}

@article{Davidson:2002qv,
    author = "Davidson, Sacha and Ibarra, Alejandro",
    title = "{A Lower bound on the right-handed neutrino mass from leptogenesis}",
    eprint = "hep-ph/0202239",
    archivePrefix = "arXiv",
    reportNumber = "OUTP-02-10P, IPPP-02-16, DCPT-02-32",
    doi = "10.1016/S0370-2693(02)01735-5",
    journal = "Phys. Lett. B",
    volume = "535",
    pages = "25--32",
    year = "2002"
}

@article{Anisimov:2005hr,
    author = "Anisimov, A. and Broncano, A. and Plumacher, M.",
    title = "{The CP-asymmetry in resonant leptogenesis}",
    eprint = "hep-ph/0511248",
    archivePrefix = "arXiv",
    reportNumber = "MPP-2005-147",
    doi = "10.1016/j.nuclphysb.2006.01.003",
    journal = "Nucl. Phys. B",
    volume = "737",
    pages = "176--189",
    year = "2006"
}

@article{Buchmuller:1997yu,
    author = "Buchmuller, W. and Plumacher, M.",
    title = "{CP asymmetry in Majorana neutrino decays}",
    eprint = "hep-ph/9710460",
    archivePrefix = "arXiv",
    reportNumber = "DESY-97-190",
    doi = "10.1016/S0370-2693(97)01548-7",
    journal = "Phys. Lett. B",
    volume = "431",
    pages = "354--362",
    year = "1998"
}

@article{Pilaftsis:1997jf,
    author = "Pilaftsis, Apostolos",
    title = "{CP violation and baryogenesis due to heavy Majorana neutrinos}",
    eprint = "hep-ph/9707235",
    archivePrefix = "arXiv",
    reportNumber = "MPI-PHT-97-30",
    doi = "10.1103/PhysRevD.56.5431",
    journal = "Phys. Rev. D",
    volume = "56",
    pages = "5431--5451",
    year = "1997"
}

@article{Pilaftsis:2003gt,
    author = "Pilaftsis, Apostolos and Underwood, Thomas E.J.",
    title = "{Resonant leptogenesis}",
    eprint = "hep-ph/0309342",
    archivePrefix = "arXiv",
    reportNumber = "MC-TH-2003-09",
    doi = "10.1016/j.nuclphysb.2004.05.029",
    journal = "Nucl. Phys. B",
    volume = "692",
    pages = "303--345",
    year = "2004"
}

@article{Buchmuller:2004nz,
    author = "Buchmuller, W. and Di Bari, P. and Plumacher, M.",
    title = "{Leptogenesis for pedestrians}",
    eprint = "hep-ph/0401240",
    archivePrefix = "arXiv",
    reportNumber = "DESY-03-100, UAB-FT-551, CERN-TH-2003-199",
    doi = "10.1016/j.aop.2004.02.003",
    journal = "Annals Phys.",
    volume = "315",
    pages = "305--351",
    year = "2005"
}

@article{Dev:2014laa,
    author = "Bhupal Dev, P.S. and Millington, Peter and Pilaftsis, Apostolos and Teresi, Daniele",
    title = "{Flavour Covariant Transport Equations: an Application to Resonant Leptogenesis}",
    eprint = "1404.1003",
    archivePrefix = "arXiv",
    primaryClass = "hep-ph",
    reportNumber = "MAN-HEP-2014-01, IPPP-14-20, DCPT-14-40",
    doi = "10.1016/j.nuclphysb.2014.06.020",
    journal = "Nucl. Phys. B",
    volume = "886",
    pages = "569--664",
    year = "2014"
}

@article{Drewes:2017zyw,
    author = "Drewes, M. and Garbrecht, B. and Hernandez, P. and Kekic, M. and Lopez-Pavon, J. and Racker, J. and Rius, N. and Salvado, J. and Teresi, D.",
    title = "{ARS Leptogenesis}",
    eprint = "1711.02862",
    archivePrefix = "arXiv",
    primaryClass = "hep-ph",
    doi = "10.1142/S0217751X18420022",
    journal = "Int. J. Mod. Phys. A",
    volume = "33",
    number = "05n06",
    pages = "1842002",
    year = "2018"
}

@article{Granelli:2020ysj,
    author = "Granelli, A. and Moffat, K. and Petcov, S. T.",
    title = "{Flavoured resonant leptogenesis at sub-TeV scales}",
    eprint = "2009.03166",
    archivePrefix = "arXiv",
    primaryClass = "hep-ph",
    doi = "10.1016/j.nuclphysb.2021.115597",
    journal = "Nucl. Phys. B",
    volume = "973",
    pages = "115597",
    year = "2021"
}

@article{Klaric:2020phc,
    author = "Klari\'c, Juraj and Shaposhnikov, Mikhail and Timiryasov, Inar",
    title = "{Uniting Low-Scale Leptogenesis Mechanisms}",
    eprint = "2008.13771",
    archivePrefix = "arXiv",
    primaryClass = "hep-ph",
    doi = "10.1103/PhysRevLett.127.111802",
    journal = "Phys. Rev. Lett.",
    volume = "127",
    number = "11",
    pages = "111802",
    year = "2021"
}

@article{Buchmuller:2001sr,
    author = "Buchmuller, W. and Plumacher, M.",
    title = "{Spectator processes and baryogenesis}",
    eprint = "hep-ph/0104189",
    archivePrefix = "arXiv",
    reportNumber = "DESY-01-046, OUTP-01-19-P",
    doi = "10.1016/S0370-2693(01)00614-1",
    journal = "Phys. Lett. B",
    volume = "511",
    pages = "74--76",
    year = "2001"
}

@article{Nardi:2005hs,
    author = "Nardi, Enrico and Nir, Yosef and Racker, Juan and Roulet, Esteban",
    title = "{On Higgs and sphaleron effects during the leptogenesis era}",
    eprint = "hep-ph/0512052",
    archivePrefix = "arXiv",
    doi = "10.1088/1126-6708/2006/01/068",
    journal = "J. High Energy Phys.",
    volume = "01",
    pages = "068",
    year = "2006"
}

@article{Pilaftsis:2005rv,
    author = "Pilaftsis, Apostolos and Underwood, Thomas E. J.",
    title = "{Electroweak-scale resonant leptogenesis}",
    eprint = "hep-ph/0506107",
    archivePrefix = "arXiv",
    doi = "10.1103/PhysRevD.72.113001",
    journal = "Phys. Rev. D",
    volume = "72",
    pages = "113001",
    year = "2005"
}

@article{Racker:2021kme,
    author = "Racker, J.",
    title = "{CP violation in mixing and oscillations for leptogenesis. Part II. The highly degenerate case}",
    eprint = "2109.00040",
    archivePrefix = "arXiv",
    primaryClass = "hep-ph",
    doi = "10.1007/JHEP11(2021)027",
    journal = "J. High Energy Phys.",
    volume = "11",
    pages = "027",
    year = "2021"
}

@article{Buchmuller:2000nd,
      author         = "Buchmüller, Wilfried and Fredenhagen, Stefan",
      title          = "{Quantum mechanics of baryogenesis}",
      journal        = "Phys. Lett.",
      volume         = "B483",
      year           = "2000",
      pages          = "217-224",
      doi            = "10.1016/S0370-2693(00)00573-6",
      eprint         = "hep-ph/0004145",
      archivePrefix  = "arXiv",
      primaryClass   = "hep-ph",
      reportNumber   = "DESY-00-056",
      SLACcitation   = "%%CITATION = HEP-PH/0004145;%%"
}

@article{DeSimone:2007gkc,
      author         = "De Simone, Andrea and Riotto, Antonio",
      title          = "{Quantum Boltzmann Equations and Leptogenesis}",
      journal        = "J. Cosmol. Astropart. Phys.",
      volume         = "0708",
      year           = "2007",
      pages          = "002",
      doi            = "10.1088/1475-7516/2007/08/002",
      eprint         = "hep-ph/0703175",
      archivePrefix  = "arXiv",
      primaryClass   = "hep-ph",
      reportNumber   = "DFPD-07-TH-04, MIT-CTP-3817",
      SLACcitation   = "%%CITATION = HEP-PH/0703175;%%"
}

@article{DeSimone:2007edo,
      author         = "De Simone, Andrea and Riotto, Antonio",
      title          = "{On Resonant Leptogenesis}",
      journal        = "J. Cosmol. Astropart. Phys.",
      volume         = "0708",
      year           = "2007",
      pages          = "013",
      doi            = "10.1088/1475-7516/2007/08/013",
      eprint         = "0705.2183",
      archivePrefix  = "arXiv",
      primaryClass   = "hep-ph",
      reportNumber   = "MIT-CTP-3832",
      SLACcitation   = "%%CITATION = ARXIV:0705.2183;%%"
}

@article{Cirigliano:2007hb,
      author         = "Cirigliano, Vincenzo and De Simone, Andrea and Isidori,
                        Gino and Masina, Isabella and Riotto, Antonio",
      title          = "{Quantum Resonant Leptogenesis and Minimal Lepton Flavour
                        Violation}",
      journal        = "J. Cosmol. Astropart. Phys.",
      volume         = "0801",
      year           = "2008",
      pages          = "004",
      doi            = "10.1088/1475-7516/2008/01/004",
      eprint         = "0711.0778",
      archivePrefix  = "arXiv",
      primaryClass   = "hep-ph",
      reportNumber   = "CERN-PH-TH-2007-208, LA-UR-07-7323, MIT-CTP-3889",
      SLACcitation   = "%%CITATION = ARXIV:0711.0778;%%"
}

@article{Garny:2009rv,
      author         = "Garny, M. and Hohenegger, A. and Kartavtsev, A. and
                        Lindner, M.",
      title          = "{Systematic approach to leptogenesis in nonequilibrium
                        QFT: Vertex contribution to the $CP$-violating parameter}",
      journal        = "Phys. Rev.",
      volume         = "D80",
      year           = "2009",
      pages          = "125027",
      doi            = "10.1103/PhysRevD.80.125027",
      eprint         = "0909.1559",
      archivePrefix  = "arXiv",
      primaryClass   = "hep-ph",
      reportNumber   = "TUM-HEP-735-09",
      SLACcitation   = "%%CITATION = ARXIV:0909.1559;%%"
}

@article{Garny:2009qn,
      author         = "Garny, M. and Hohenegger, A. and Kartavtsev, A. and
                        Lindner, M.",
      title          = "{Systematic approach to leptogenesis in nonequilibrium
                        QFT: Self-energy contribution to the $CP$-violating
                        parameter}",
      journal        = "Phys. Rev.",
      volume         = "D81",
      year           = "2010",
      pages          = "085027",
      doi            = "10.1103/PhysRevD.81.085027",
      eprint         = "0911.4122",
      archivePrefix  = "arXiv",
      primaryClass   = "hep-ph",
      reportNumber   = "TUM-HEP-740-09",
      SLACcitation   = "%%CITATION = ARXIV:0911.4122;%%"
}

@article{Beneke:2010wd,
      author         = "Beneke, Martin and Garbrecht, Björn and Herranen, Matti
                        and Schwaller, Pedro",
      title          = "{Finite Number Density Corrections to Leptogenesis}",
      journal        = "Nucl. Phys.",
      volume         = "B838",
      year           = "2010",
      pages          = "1-27",
      doi            = "10.1016/j.nuclphysb.2010.05.003",
      eprint         = "1002.1326",
      archivePrefix  = "arXiv",
      primaryClass   = "hep-ph",
      reportNumber   = "TTK-10-16, ZU-TH-02-10",
      SLACcitation   = "%%CITATION = ARXIV:1002.1326;%%"
}

@article{Beneke:2010dz,
      author         = "Beneke, Martin and Garbrecht, Björn and Fidler, Christian
                        and Herranen, Matti and Schwaller, Pedro",
      title          = "{Flavoured Leptogenesis in the CTP Formalism}",
      journal        = "Nucl. Phys.",
      volume         = "B843",
      year           = "2011",
      pages          = "177-212",
      doi            = "10.1016/j.nuclphysb.2010.10.001",
      eprint         = "1007.4783",
      archivePrefix  = "arXiv",
      primaryClass   = "hep-ph",
      reportNumber   = "TTK-10-44, ZU-TH-09-10",
      SLACcitation   = "%%CITATION = ARXIV:1007.4783;%%"
}

@article{Anisimov:2010dk,
      author         = "Anisimov, A. and Buchmüller, W. and Drewes, M. and
                        Mendizabal, S.",
      title          = "{Quantum Leptogenesis I}",
      journal        = "Annals Phys.",
      volume         = "326",
      year           = "2011",
      pages          = "1998-2038",
      doi            = "10.1016/j.aop.2011.02.002, 10.1016/j.aop.2013.05.00",
      note           = "[Erratum: Annals Phys.338,376(2011)]",
      eprint         = "1012.5821",
      archivePrefix  = "arXiv",
      primaryClass   = "hep-ph",
      reportNumber   = "DESY-10-218",
      SLACcitation   = "%%CITATION = ARXIV:1012.5821;%%"
}

@article{Garny:2011hg,
      author         = "Garny, Mathias and Kartavtsev, Alexander and Hohenegger,
                        Andreas",
      title          = "{Leptogenesis from first principles in the resonant
                        regime}",
      journal        = "Annals Phys.",
      volume         = "328",
      year           = "2013",
      pages          = "26-63",
      doi            = "10.1016/j.aop.2012.10.007",
      eprint         = "1112.6428",
      archivePrefix  = "arXiv",
      primaryClass   = "hep-ph",
      reportNumber   = "DESY-11-264",
      SLACcitation   = "%%CITATION = ARXIV:1112.6428;%%"
}

@article{Garbrecht:2011aw,
      author         = "Garbrecht, Björn and Herranen, Matti",
      title          = "{Effective Theory of Resonant Leptogenesis in the
                        Closed-Time-Path Approach}",
      journal        = "Nucl. Phys.",
      volume         = "B861",
      year           = "2012",
      pages          = "17-52",
      doi            = "10.1016/j.nuclphysb.2012.03.009",
      eprint         = "1112.5954",
      archivePrefix  = "arXiv",
      primaryClass   = "hep-ph",
      SLACcitation   = "%%CITATION = ARXIV:1112.5954;%%"
}

@article{Iso:2013lba,
      author         = "Iso, Satoshi and Shimada, Kengo and Yamanaka, Masato",
      title          = "{Kadanoff-Baym approach to the thermal resonant
                        leptogenesis}",
      journal        = "J. High Energy Phys.",
      volume         = "04",
      year           = "2014",
      pages          = "062",
      doi            = "10.1007/JHEP04(2014)062",
      eprint         = "1312.7680",
      archivePrefix  = "arXiv",
      primaryClass   = "hep-ph",
      reportNumber   = "KEK-TH-1697",
      SLACcitation   = "%%CITATION = ARXIV:1312.7680;%%"
}

@article{Iso:2014afa,
      author         = "Iso, Satoshi and Shimada, Kengo",
      title          = "{Coherent Flavour Oscillation and $CP$ Violating Parameter
                        in Thermal Resonant Leptogenesis}",
      journal        = "J. High Energy Phys.",
      volume         = "08",
      year           = "2014",
      pages          = "043",
      doi            = "10.1007/JHEP08(2014)043",
      eprint         = "1404.4816",
      archivePrefix  = "arXiv",
      primaryClass   = "hep-ph",
      reportNumber   = "KEK-TH-1724",
      SLACcitation   = "%%CITATION = ARXIV:1404.4816;%%"
}

@article{Hohenegger:2014cpa,
      author         = "Hohenegger, A. and Kartavtsev, A.",
      title          = "{Leptogenesis in crossing and runaway regimes}",
      journal        = "J. High Energy Phys.",
      volume         = "07",
      year           = "2014",
      pages          = "130",
      doi            = "10.1007/JHEP07(2014)130",
      eprint         = "1404.5309",
      archivePrefix  = "arXiv",
      primaryClass   = "hep-ph",
      reportNumber   = "MPP-2014-40",
      SLACcitation   = "%%CITATION = ARXIV:1404.5309;%%"
}

@article{Garbrecht:2014aga,
      author         = "Garbrecht, Björn and Gautier, Florian and Klarić, Juraj",
      title          = "{Strong Washout Approximation to Resonant Leptogenesis}",
      journal        = "J. Cosmol. Astropart. Phys.",
      volume         = "1409",
      year           = "2014",
      number         = "09",
      pages          = "033",
      doi            = "10.1088/1475-7516/2014/09/033",
      eprint         = "1406.4190",
      archivePrefix  = "arXiv",
      primaryClass   = "hep-ph",
      reportNumber   = "TUM-HEP-947-14",
      SLACcitation   = "%%CITATION = ARXIV:1406.4190;%%"
}

@article{Dev:2014wsa,
      author         = "Dev, P. S. Bhupal and Millington, Peter and Pilaftsis,
                        Apostolos and Teresi, Daniele",
      title          = "{Kadanoff–Baym approach to flavour mixing and
                        oscillations in resonant leptogenesis}",
      journal        = "Nucl. Phys.",
      volume         = "B891",
      year           = "2015",
      pages          = "128-158",
      doi            = "10.1016/j.nuclphysb.2014.12.003",
      eprint         = "1410.6434",
      archivePrefix  = "arXiv",
      primaryClass   = "hep-ph",
      reportNumber   = "MAN-HEP-2014-13, TUM-HEP-962-14",
      SLACcitation   = "%%CITATION = ARXIV:1410.6434;%%"
}

@article{Kartavtsev:2015vto,
      author         = "Kartavtsev, Alexander and Millington, Peter and Vogel,
                        Hendrik",
      title          = "{Lepton asymmetry from mixing and oscillations}",
      journal        = "J. High Energy Phys.",
      volume         = "06",
      year           = "2016",
      pages          = "066",
      doi            = "10.1007/JHEP06(2016)066",
      eprint         = "1601.03086",
      archivePrefix  = "arXiv",
      primaryClass   = "hep-ph",
      reportNumber   = "MPP-2015-235, TUM-HEP-1018-15",
      SLACcitation   = "%%CITATION = ARXIV:1601.03086;%%"
}

@article{Drewes:2016gmt,
      author         = "Drewes, Marco and Garbrecht, Björn and Gueter, Dario and
                        Klarić, Juraj",
      title          = "{Leptogenesis from Oscillations of Heavy Neutrinos with
                        Large Mixing Angles}",
      journal        = "J. High Energy Phys.",
      volume         = "12",
      year           = "2016",
      pages          = "150",
      doi            = "10.1007/JHEP12(2016)150",
      eprint         = "1606.06690",
      archivePrefix  = "arXiv",
      primaryClass   = "hep-ph",
      reportNumber   = "TUM-HEP-1050-16",
      SLACcitation   = "%%CITATION = ARXIV:1606.06690;%%"
}

@article{Dev:2017trv,
      author         = "Dev, P. S. Bhupal and Di Bari, Pasquale and Garbrecht,
                        Björn and Lavignac, Stéphane and Millington, Peter and
                        Teresi, Daniele",
      title          = "{Flavor effects in leptogenesis}",
      journal        = "Int. J. Mod. Phys.",
      volume         = "A33",
      year           = "2018",
      pages          = "1842001",
      doi            = "10.1142/S0217751X18420010",
      eprint         = "1711.02861",
      archivePrefix  = "arXiv",
      primaryClass   = "hep-ph",
      SLACcitation   = "%%CITATION = ARXIV:1711.02861;%%"
}

@article{Dev:2017wwc,
      author         = "Dev, P. S. Bhupal and Garny, Mathias and Klarić, Juraj and
                        Millington, Peter and Teresi, Daniele",
      title          = "{Resonant enhancement in leptogenesis}",
      journal        = "Int. J. Mod. Phys.",
      volume         = "A33",
      year           = "2018",
      pages          = "1842003",
      doi            = "10.1142/S0217751X18420034",
      eprint         = "1711.02863",
      archivePrefix  = "arXiv",
      primaryClass   = "hep-ph",
      reportNumber   = "TUM-HEP-1110-17",
      SLACcitation   = "%%CITATION = ARXIV:1711.02863;%%"
}

@article{Garbrecht:2018mrp,
      author         = "Garbrecht, Björn",
      title          = "{Why is there more matter than antimatter? Calculational
                        methods for leptogenesis and electroweak baryogenesis}",
      journal        = "Prog. Part. Nucl. Phys.",
      volume         = "110",
      year           = "2020",
      pages          = "103727",
      doi            = "10.1016/j.ppnp.2019.103727",
      eprint         = "1812.02651",
      archivePrefix  = "arXiv",
      primaryClass   = "hep-ph",
      reportNumber   = "TUM-HEP-1177-18",
      SLACcitation   = "%%CITATION = ARXIV:1812.02651;%%"
}

@article{Depta:2020zmy,
    author = {Depta, Paul Frederik and Halsch, Andreas and H\"utig, Janine and Mendizabal, Sebastian and Philipsen, Owe},
    title = "{Complete leading-order standard model corrections to quantum leptogenesis}",
    eprint = "2005.01728",
    archivePrefix = "arXiv",
    primaryClass = "hep-ph",
    reportNumber = "DESY-20-065, DESY 20-065",
    doi = "10.1007/JHEP09(2020)036",
    journal = "J. High Energy Phys.",
    volume = "09",
    pages = "036",
    year = "2020"
}

@article{Kolb:1979qa,
    author = "Kolb, Edward W. and Wolfram, Stephen",
    title = "{Baryon Number Generation in the Early Universe}",
    reportNumber = "Print-79-0956 (CAL TECH), OAP-579, CALT-68-754",
    doi = "10.1016/0550-3213(82)90012-8",
    journal = "Nucl. Phys. B",
    volume = "172",
    pages = "224",
    year = "1980",
    note = "[Erratum: Nucl.Phys.B 195, 542 (1982)]"
}

@article{Luty:1992un,
    author = "Luty, M.A.",
    title = "{Baryogenesis via leptogenesis}",
    doi = "10.1103/PhysRevD.45.455",
    journal = "Phys. Rev. D",
    volume = "45",
    pages = "455--465",
    year = "1992"
}

@article{Giudice:2003jh,
    author = "Giudice, G.F. and Notari, A. and Raidal, M. and Riotto, A. and Strumia, A.",
    title = "{Towards a complete theory of thermal leptogenesis in the SM and MSSM}",
    eprint = "hep-ph/0310123",
    archivePrefix = "arXiv",
    reportNumber = "IFUP-TH-2003-37, CERN-TH-2003-240",
    doi = "10.1016/j.nuclphysb.2004.02.019",
    journal = "Nucl. Phys. B",
    volume = "685",
    pages = "89--149",
    year = "2004"
}

@article{Basboll:2006yx,
    author = "Basboll, Anders and Hannestad, Steen",
    title = "{Decay of heavy Majorana neutrinos using the full Boltzmann equation including its implications for leptogenesis}",
    eprint = "hep-ph/0609025",
    archivePrefix = "arXiv",
    doi = "10.1088/1475-7516/2007/01/003",
    journal = "J. Cosmol. Astropart. Phys.",
    volume = "01",
    pages = "003",
    year = "2007"
}

@article{Lindner:2005kv,
    author = "Lindner, Manfred and Muller, Markus Michael",
    title = "{Comparison of Boltzmann equations with quantum dynamics for scalar fields}",
    eprint = "hep-ph/0512147",
    archivePrefix = "arXiv",
    reportNumber = "TUM-HEP-613-05, MPP-2005-96",
    doi = "10.1103/PhysRevD.73.125002",
    journal = "Phys. Rev. D",
    volume = "73",
    pages = "125002",
    year = "2006"
}

@article{Drewes:2012qw,
    author = "Drewes, Marco and Mendizabal, Sebastian and Weniger, Christoph",
    title = "{The Boltzmann Equation from Quantum Field Theory}",
    eprint = "1202.1301",
    archivePrefix = "arXiv",
    primaryClass = "hep-ph",
    reportNumber = "MPP-2012-3, TTK-12-03, TUM-HEP-857-12",
    doi = "10.1016/j.physletb.2012.11.046",
    journal = "Phys. Lett. B",
    volume = "718",
    pages = "1119--1124",
    year = "2013"
}

@article{Schwinger:1960qe,
      author         = "Schwinger, Julian S.",
      title          = "{Brownian motion of a quantum oscillator}",
      journal        = "J. Math. Phys.",
      volume         = "2",
      year           = "1961",
      pages          = "407-432",
      doi            = "10.1063/1.1703727",
      SLACcitation   = "%%CITATION = JMAPA,2,407;%%"
}

@article{Keldysh:1964ud,
      author         = "Keldysh, L. V.",
      title          = "{Diagram technique for nonequilibrium processes}",
      journal        = "Zh. Eksp. Teor. Fiz.",
      volume         = "47",
      year           = "1964",
      pages          = "1515-1527",
      note           = "[Sov. Phys. JETP20,1018(1965)]",
      SLACcitation   = "%%CITATION = ZETFA,47,1515;%%"
}

@article{Bakshi:1962dv,
    author = "Bakshi, Pradip M. and Mahanthappa, Kalyana T.",
    title = "{Expectation value formalism in quan\-tum field theory. 1.}",
    doi = "10.1063/1.1703883",
    journal = "J. Math. Phys.",
    volume = "4",
    pages = "1--11",
    year = "1963"
}

@article{Feynman:1963fq,
    author = "Feynman, R. P. and Vernon, Jr., F. L.",
    editor = "Brown, L. M.",
    title = "{The Theory of a general quantum system interacting with a linear dissipative system}",
    doi = "10.1016/0003-4916(63)90068-X",
    journal = "Annals Phys.",
    volume = "24",
    pages = "118--173",
    year = "1963"
}

@article{Chou:1980,
    author = "Chou, Kuang-chao and Su, Zhao-bin and Hao, Bai-lin and Yu, Lu",
    title = "{Closed time path Green's functions and critical dynamics}",
    doi = "10.1103/PhysRevB.22.3385",
    journal = "Phys. Rev. B",
    volume = "22",
    pages = "3385--3407",
    year = "1980"
}

@article{Chou:1984es,
    author = "Chou, Kuang-chao and Su, Zhao-bin and Hao, Bai-lin and Yu, Lu",
    title = "{Equilibrium and Nonequilibrium Formalisms Made Unified}",
    reportNumber = "AS-ITP-84-021",
    doi = "10.1016/0370-1573(85)90136-X",
    journal = "Phys. Rept.",
    volume = "118",
    pages = "1--131",
    year = "1985"
}

@article{Jordan:1986ug,
    author = "Jordan, R. D.",
    title = "{Effective Field Equations for Expectation Values}",
    doi = "10.1103/PhysRevD.33.444",
    journal = "Phys. Rev. D",
    volume = "33",
    pages = "444--454",
    year = "1986"
}

@inproceedings{DeWitt:1986,
    author = "DeWitt, Bryce S.",
    title = "{Effective action for expectation values}",
    booktitle = "{Quantum Concepts In Space and Time}",
    isbn = "0-19-851972-9",
    publisher = "Clarendon Press, Oxford",
    pages = "325--336",
    year = "1986"
}

@article{Su:1987pi,
    author = "Su, Zhao-bin and Chen, Liao-Yuan and Yu, Xiao-tong and Chou, Kuang-chao",
    title = "{Influence functional and closed-time-path Green's function}",
    reportNumber = "AS-ITP-87-017",
    doi = "10.1103/PhysRevB.37.9810",
    journal = "Phys. Rev. B",
    volume = "37",
    pages = "9810--9812",
    year = "1988"
}

@article{Baym:1961zz,
      author         = "Baym, Gordon and Kadanoff, Leo P.",
      title          = "{Conservation Laws and Correlation Functions}",
      journal        = "Phys. Rev.",
      volume         = "124",
      year           = "1961",
      pages          = "287-299",
      doi            = "10.1103/PhysRev.124.287",
      SLACcitation   = "%%CITATION = PHRVA,124,287;%%"
}

@article{Danielewicz:1982kk,
    author = "Danielewicz, P.",
    title = "{Quantum Theory of Nonequilibrium Processes. 1.}",
    doi = "10.1016/0003-4916(84)90092-7",
    journal = "Annals Phys.",
    volume = "152",
    pages = "239--304",
    year = "1984"
}

@article{Luttinger:1960ua,
    author = "Luttinger, J. M. and Ward, John Clive",
    title = "{Ground state energy of a many fermion system. 2.}",
    doi = "10.1103/PhysRev.118.1417",
    journal = "Phys. Rev.",
    volume = "118",
    pages = "1417--1427",
    year = "1960"
}

@article{Lee:1960zza,
    author = "Lee, T. D. and Yang, C. N.",
    title = "{Many-Body Problem in Quantum Statistical Mechanics. 4. Formulation in Terms of Average Occupation Number in Momentum Space}",
    doi = "10.1103/PhysRev.117.22",
    journal = "Phys. Rev.",
    volume = "117",
    pages = "22--36",
    year = "1960"
}

@article{Baym:1962sx,
    author = "Baym, Gordon",
    title = "{Selfconsistent approximation in many body systems}",
    doi = "10.1103/PhysRev.127.1391",
    journal = "Phys. Rev.",
    volume = "127",
    pages = "1391--1401",
    year = "1962"
}

@article{Cornwall:1974vz,
      author         = "Cornwall, John M. and Jackiw, R. and Tomboulis, E.",
      title          = "{Effective Action for Composite Operators}",
      journal        = "Phys. Rev.",
      volume         = "D10",
      year           = "1974",
      pages          = "2428-2445",
      doi            = "10.1103/PhysRevD.10.2428",
      reportNumber   = "MIT-CTP-419",
      SLACcitation   = "%%CITATION = PHRVA,D10,2428;%%"
}

@article{Jackiw:1974cv,
    author = "Jackiw, R.",
    title = "{Functional evaluation of the effective potential}",
    doi = "10.1103/PhysRevD.9.1686",
    journal = "Phys. Rev. D",
    volume = "9",
    pages = "1686",
    year = "1974"
}

@article{Calzetta:1986cq,
    author = "Calzetta, E. and Hu, B. L.",
    title = "{Nonequilibrium Quantum Fields: Closed Time Path Effective Action, Wigner Function and Boltzmann Equation}",
    reportNumber = "MDDP-PP-87-104",
    doi = "10.1103/PhysRevD.37.2878",
    journal = "Phys. Rev. D",
    volume = "37",
    pages = "2878",
    year = "1988"
}

@inproceedings{Berges:2003pc,
    author = "Berges, Jürgen and Serreau, Julien",
    title = "{Progress in nonequilibrium quantum field theory}",
    booktitle = "{5th Internationa Conference on Strong and Electroweak Matter}",
    eprint = "hep-ph/0302210",
    archivePrefix = "arXiv",
    reportNumber = "HD-THEP-03-12",
    doi = "10.1142/9789812704498_0011",
    pages = "111--126",
    year = "2003"
}

@inproceedings{Berges:2004vw,
    author = "Berges, Jürgen and Serreau, Julien",
    title = "{Progress in nonequilibrium quantum field theory II}",
    booktitle = "{6th International Conference on Strong and Electroweak Matter}",
    eprint = "hep-ph/0410330",
    archivePrefix = "arXiv",
    doi = "10.1142/9789812702159_0011",
    pages = "102--116",
    year = "2005"
}

@article{Berges:2004yj,
    author = "Berges, Jürgen",
    editor = "Bracco, Mirian and Chiapparini, Marcelo and Ferreira, Erasmo and Kodama, Takeshi",
    title = "{Introduction to nonequilibrium quantum field theory}",
    eprint = "hep-ph/0409233",
    archivePrefix = "arXiv",
    doi = "10.1063/1.1843591",
    journal = "AIP Conf. Proc.",
    volume = "739",
    number = "1",
    pages = "3--62",
    year = "2004"
}

@article{Berges:2015kfa,
    author = "Berges, Jürgen",
    title = "{Nonequilibrium Quantum Fields: From Cold Atoms to Cosmology}",
    eprint = "1503.02907",
    archivePrefix = "arXiv",
    primaryClass = "hep-ph",
    month = "3",
    year = "2015"
}

@article{Berges:2000ur,
    author = "Berges, Juergen and Cox, Jurgen",
    title = "{Thermalization of quantum fields from time reversal invariant evolution equations}",
    eprint = "hep-ph/0006160",
    archivePrefix = "arXiv",
    reportNumber = "MIT-CTP-2988",
    doi = "10.1016/S0370-2693(01)01004-8",
    journal = "Phys. Lett. B",
    volume = "517",
    pages = "369--374",
    year = "2001"
}

@article{Berges:2001fi,
    author = "Berges, Jürgen",
    title = "{Controlled nonperturbative dynamics of quantum fields out-of-equilibrium}",
    eprint = "hep-ph/0105311",
    archivePrefix = "arXiv",
    reportNumber = "HD-THEP-01-26",
    doi = "10.1016/S0375-9474(01)01295-7",
    journal = "Nucl. Phys. A",
    volume = "699",
    pages = "847--886",
    year = "2002"
}

@article{Berges:2004pu,
    author = "Berges, Jürgen",
    title = "{N-particle irreducible effective action techniques for gauge theories}",
    eprint = "hep-ph/0401172",
    archivePrefix = "arXiv",
    doi = "10.1103/PhysRevD.70.105010",
    journal = "Phys. Rev. D",
    volume = "70",
    pages = "105010",
    year = "2004"
}

@article{Moyal:1949sk,
    author = "Moyal, J. E.",
    title = "{Quantum mechanics as a statistical theory}",
    doi = "10.1017/S0305004100000487",
    journal = "Proc. Cambridge Phil. Soc.",
    volume = "45",
    pages = "99--124",
    year = "1949"
}

@article{Groenewold:1946kp,
    author = "Groenewold, H. J.",
    title = "{On the Principles of elementary quantum mechanics}",
    doi = "10.1016/S0031-8914(46)80059-4",
    journal = "Physica",
    volume = "12",
    pages = "405--460",
    year = "1946"
}

@article{Brown:2015zla,
    author = "Brown, Michael and Whittingham, Ian",
    title = "{Two-particle irreducible effective actions versus resummation: analytic properties and self-consistency}",
    eprint = "1503.08664",
    archivePrefix = "arXiv",
    primaryClass = "hep-th",
    doi = "10.1016/j.nuclphysb.2015.09.021",
    journal = "Nucl. Phys. B",
    volume = "900",
    pages = "477--500",
    year = "2015"
}

@article{Garny:2015oza,
    author = "Garny, Mathias and Reinosa, Urko",
    title = "{Renormalization out of equilibrium in a superrenormalizable theory}",
    eprint = "1504.06643",
    archivePrefix = "arXiv",
    primaryClass = "hep-ph",
    reportNumber = "CERN-PH-TH-2015-095, CPHT-RR014.0415",
    doi = "10.1103/PhysRevD.94.045012",
    journal = "Phys. Rev. D",
    volume = "94",
    number = "4",
    pages = "045012",
    year = "2016"
}

@article{Berges:2002wr,
    author = "Berges, Juergen and Borsanyi, Szabolcs and Serreau, Julien",
    title = "{Thermalization of fermionic quantum fields}",
    eprint = "hep-ph/0212404",
    archivePrefix = "arXiv",
    reportNumber = "HD-THEP-02-47",
    doi = "10.1016/S0550-3213(03)00261-X",
    journal = "Nucl. Phys. B",
    volume = "660",
    pages = "51--80",
    year = "2003"
}

@article{Reinosa:2007vi,
    author = "Reinosa, Urko and Serreau, Julien",
    title = "{Ward Identities for the 2PI effective action in QED}",
    eprint = "0708.0971",
    archivePrefix = "arXiv",
    primaryClass = "hep-th",
    reportNumber = "HD-THEP-07-20",
    doi = "10.1088/1126-6708/2007/11/097",
    journal = "J. High Energy Phys.",
    volume = "11",
    pages = "097",
    year = "2007"
}

@article{Calzetta:2004sh,
    author = "Calzetta, E. A.",
    editor = "Gunzig, E. and Mukhanov, Viatcheslav F. and Verdaguer, E.",
    title = "{The Two particle irreducible effective action in gauge theories}",
    eprint = "hep-ph/0402196",
    archivePrefix = "arXiv",
    doi = "10.1023/B:IJTP.0000048174.83795.3f",
    journal = "Int. J. Theor. Phys.",
    volume = "43",
    pages = "767--799",
    year = "2004"
}

@article{Prokopec:2003pj,
    author = "Prokopec, Tomislav and Schmidt, Michael G. and Weinstock, Steffen",
    title = "{Transport equations for chiral fermions to order h bar and electroweak baryogenesis. Part 1}",
    eprint = "hep-ph/0312110",
    archivePrefix = "arXiv",
    reportNumber = "BNL-72343-2004-JA, HD-THEP-03-62",
    doi = "10.1016/j.aop.2004.06.002",
    journal = "Annals Phys.",
    volume = "314",
    pages = "208--265",
    year = "2004"
}

@article{Bilenky:1987ty,
    author = "Bilenky, Samoil M. and Petcov, S. T.",
    title = "{Massive Neutrinos and Neutrino Oscillations}",
    doi = "10.1103/RevModPhys.59.671",
    journal = "Rev. Mod. Phys.",
    volume = "59",
    pages = "671",
    year = "1987"
    %,note = "[Erratum: Rev.Mod.Phys. 61, 169 (1989), Erratum: Rev.Mod.Phys. 60, 575--575 (1988)]"
}

@article{Enqvist:1990ad,
    author = "Enqvist, K. and Kainulainen, K. and Maalampi, J.",
    title = "{Refraction and Oscillations of Neutrinos in the Early Universe}",
    reportNumber = "HU-TFT-90-13",
    doi = "10.1016/0550-3213(91)90397-G",
    journal = "Nucl. Phys. B",
    volume = "349",
    pages = "754--790",
    year = "1991"
}

@article{Enqvist:1991qj,
    author = "Enqvist, K. and Kainulainen, K. and Thomson, Mark J.",
    title = "{Stringent cosmological bounds on inert neutrino mixing}",
    doi = "10.1016/0550-3213(92)90442-E",
    journal = "Nucl. Phys. B",
    volume = "373",
    pages = "498--528",
    year = "1992"
}

@article{Kainulainen:1990ds,
    author = "Kainulainen, Kimmo",
    title = "{Light Singlet Neutrinos and the Primordial Nucleosynthesis}",
    reportNumber = "HU-TFT-90-23",
    doi = "10.1016/0370-2693(90)90054-A",
    journal = "Phys. Lett. B",
    volume = "244",
    pages = "191--195",
    year = "1990"
}

@article{Herranen:2008hi,
    author = "Herranen, Matti and Kainulainen, Kimmo and Rahkila, Pyry Matti",
    title = "{Towards a kinetic theory for fermions with quantum coherence}",
    eprint = "0807.1415",
    archivePrefix = "arXiv",
    primaryClass = "hep-ph",
    doi = "10.1016/j.nuclphysb.2008.09.032",
    journal = "Nucl. Phys. B",
    volume = "810",
    pages = "389--426",
    year = "2009"
}

@article{Herranen:2008hu,
    author = "Herranen, Matti and Kainulainen, Kimmo and Rahkila, Pyry Matti",
    title = "{Quantum kinetic theory for fermions in temporally varying backgrounds}",
    eprint = "0807.1435",
    archivePrefix = "arXiv",
    primaryClass = "hep-ph",
    doi = "10.1088/1126-6708/2008/09/032",
    journal = "J. High Energy Phys.",
    volume = "09",
    pages = "032",
    year = "2008"
}

@article{Herranen:2008yg,
    author = "Herranen, Matti and Kainulainen, Kimmo and Rahkila, Pyry M.",
    editor = "Smit, Jan",
    title = "{Kinetic transport theory with quantum coherence}",
    eprint = "0811.0936",
    archivePrefix = "arXiv",
    primaryClass = "hep-ph",
    doi = "10.1016/j.nuclphysa.2009.01.050",
    journal = "Nucl. Phys. A",
    volume = "820",
    pages = "203C--206C",
    year = "2009"
}

@article{Herranen:2008di,
    author = "Herranen, Matti and Kainulainen, Kimmo and Rahkila, Pyry Matti",
    title = "{Kinetic theory for scalar fields with nonlocal quantum coherence}",
    eprint = "0812.4029",
    archivePrefix = "arXiv",
    primaryClass = "hep-ph",
    doi = "10.1088/1126-6708/2009/05/119",
    journal = "J. High Energy Phys.",
    volume = "05",
    pages = "119",
    year = "2009"
}

@phdthesis{Herranen:2009zi,
    author = "Herranen, Matti",
    title = "{Quantum kinetic theory with nonlocal coherence}",
    eprint = "0906.3136",
    archivePrefix = "arXiv",
    primaryClass = "hep-ph",
    reportNumber = "RESEARCH-REPORT-UNIV.-OF-JYVASKYLA",
    school = "Jyvaskyla U.",
    year = "2009"
}

@article{Herranen:2009xi,
    author = "Herranen, Matti and Kainulainen, Kimmo and Rahkila, Pyry M.",
    editor = "Bonitz, Michael and Balzer, Karsten",
    title = "{Coherent quasiparticle approximation (cQPA) and nonlocal coherence}",
    eprint = "0912.2490",
    archivePrefix = "arXiv",
    primaryClass = "hep-ph",
    doi = "10.1088/1742-6596/220/1/012007",
    journal = "J. Phys. Conf. Ser.",
    volume = "220",
    pages = "012007",
    year = "2010"
}

@article{Herranen:2010mh,
    author = "Herranen, Matti and Kainulainen, Kimmo and Rahkila, Pyry Matti",
    title = "{Coherent quan\-tum Boltz\-mann equations from cQPA}",
    eprint = "1006.1929",
    archivePrefix = "arXiv",
    primaryClass = "hep-ph",
    reportNumber = "TTK-10-34",
    doi = "10.1007/JHEP12(2010)072",
    journal = "J. High Energy Phys.",
    volume = "12",
    pages = "072",
    year = "2010"
}

@article{Fidler:2011yq,
    author = "Fidler, Christian and Herranen, Matti and Kainulainen, Kimmo and Rahkila, Pyry Matti",
    title = "{Flavoured quan\-tum Boltzmann equations from cQPA}",
    eprint = "1108.2309",
    archivePrefix = "arXiv",
    primaryClass = "hep-ph",
    reportNumber = "TTK-11-35",
    doi = "10.1007/JHEP02(2012)065",
    journal = "J. High Energy Phys.",
    volume = "02",
    pages = "065",
    year = "2012"
}

@article{Herranen:2011zg,
    author = "Herranen, Matti and Kainulainen, Kimmo and Rahkila, Pyry Matti",
    title = "{Flavour-coherent propagators and Feynman rules: Covariant cQPA formulation}",
    eprint = "1108.2371",
    archivePrefix = "arXiv",
    primaryClass = "hep-ph",
    reportNumber = "TTK-11-36",
    doi = "10.1007/JHEP02(2012)080",
    journal = "J. High Energy Phys.",
    volume = "02",
    pages = "080",
    year = "2012"
}

@article{deBroglie,
    author = "de Broglie, Louis",
    title = "{Waves and Quanta}",
    doi = "10.1038/112540a0",
    journal = "Nature",
    volume = "112",
    pages = "540",
    year = "1923"
}

@article{Planck:2018vyg,
    author = "Aghanim, N. and others",
    collaboration = "Planck Collaboration",
    title = "{Planck 2018 results. VI. Cosmological parameters}",
    eprint = "1807.06209",
    archivePrefix = "arXiv",
    primaryClass = "astro-ph.CO",
    doi = "10.1051/0004-6361/201833910",
    journal = "Astron. Astrophys.",
    volume = "641",
    pages = "A6",
    year = "2020"
    %,note = "[Erratum: Astron.Astrophys. 652, C4 (2021)]"
}

@article{Zyla:2020zbs,
    author = "Zyla, P. A. and others",
    collaboration = "Particle Data Group",
    title = "{Review of Particle Physics}",
    doi = "10.1093/ptep/ptaa104",
    journal = "Prog. Theor. Exp. Phys.",
    volume = "2020",
    number = "8",
    pages = "083C01",
    year = "2020"
}

@article{Guth:1980zm,
    author = "Guth, Alan H.",
    editor = "Fang, Li-Zhi and Ruffini, R.",
    title = "{The Inflationary Universe: A Possible Solution to the Horizon and Flatness Problems}",
    reportNumber = "SLAC-PUB-2576",
    doi = "10.1103/PhysRevD.23.347",
    journal = "Phys. Rev. D",
    volume = "23",
    pages = "347--356",
    year = "1981"
}

@article{Cohen:1997ac,
    author = "Cohen, Andrew G. and De Rujula, A. and Glashow, S. L.",
    title = "{A Matter - antimatter universe?}",
    eprint = "astro-ph/9707087",
    archivePrefix = "arXiv",
    reportNumber = "BUHEP-97-19, CERN-TH-97-154",
    doi = "10.1086/305328",
    journal = "Astrophys. J.",
    volume = "495",
    pages = "539--549",
    year = "1998"
}

@article{Fields:2019pfx,
    author = "Fields, Brian D. and Olive, Keith A. and Yeh, Tsung-Han and Young, Charles",
    title = "{Big-Bang Nucleosynthesis after Planck}",
    eprint = "1912.01132",
    archivePrefix = "arXiv",
    primaryClass = "astro-ph.CO",
    reportNumber = "UMN--TH--3902/19, FTPI--MINN--19/25",
    doi = "10.1088/1475-7516/2020/03/010",
    journal = "J. Cosmol. Astropart. Phys.",
    volume = "03",
    pages = "010",
    year = "2020",
    note = "[Erratum: JCAP 11, E02 (2020)]"
}

@article{Greiner:1998vd,
    author = "Greiner, Carsten and Leupold, Stefan",
    title = "{Stochastic interpretation of Kadanoff-Baym equations and their relation to Langevin processes}",
    eprint = "hep-ph/9802312",
    archivePrefix = "arXiv",
    reportNumber = "UGI-98-12",
    doi = "10.1006/aphy.1998.5849",
    journal = "Annals Phys.",
    volume = "270",
    pages = "328--390",
    year = "1998"
}

@article{Pal:2010ih,
    author = "Pal, Palash B.",
    title = "{Dirac, Majorana and Weyl fermions}",
    eprint = "1006.1718",
    archivePrefix = "arXiv",
    primaryClass = "hep-ph",
    doi = "10.1119/1.3549729",
    journal = "Am. J. Phys.",
    volume = "79",
    pages = "485--498",
    year = "2011"
}

@article{ACME:2018yjb,
    author = "Andreev, V. and others",
    collaboration = "ACME",
    title = "{Improved limit on the electric dipole moment of the electron}",
    doi = "10.1038/s41586-018-0599-8",
    journal = "Nature",
    volume = "562",
    number = "7727",
    pages = "355--360",
    year = "2018"
}

@article{Kainulainen:2021eki,
    author = "Kainulainen, Kimmo and Koskivaara, Olli",
    title = "{Non-equilibrium dynamics of a scalar field with quantum backreaction}",
    eprint = "2105.09598",
    archivePrefix = "arXiv",
    primaryClass = "hep-ph",
    doi = "10.1007/JHEP12(2021)190",
    journal = "J. High Energy Phys.",
    volume = "12",
    pages = "190",
    year = "2021"
}
